\documentclass[11pt,a4paper]{article}
\pdfoutput=1
\usepackage{amssymb,amsmath,amsfonts, mathtools, mathrsfs}
\usepackage[dvipsnames]{xcolor}
\usepackage{cancel}
\usepackage{graphicx}
\usepackage{caption}
\usepackage{subcaption}
\usepackage{enumerate}
\usepackage{dsfont}
\usepackage[sort,compress]{cite}
\usepackage{verbatim}
\usepackage{amsfonts,euscript,amssymb,stmaryrd,braket}
\usepackage{tikz}
\usetikzlibrary{arrows,decorations.markings,patterns}
\usepackage{slashed}
\definecolor{hyperref}{RGB}{026,028,185}
\usepackage[bookmarks=true,colorlinks=true,linkcolor=hyperref,citecolor=hyperref,urlcolor=hyperref,bookmarksnumbered]{hyperref}
\usepackage{cite}
\usepackage{gensymb}

\def\clock{{\count0=\time
           \divide\count0 60
           \ifnum\count0<10 0\fi\the\count0
           \multiply\count0 -60 \advance\count0 \time
           :\ifnum\count0<10 0\fi \the\count0
         }}
\newcommand{\timestamp}{{\small\vbox{\hbox{\tt\jobname.tex}
\hbox{\the\day/\the\month/\the\year, \clock}}}}

\newcommand{\bea}{\begin{eqnarray}}
\newcommand{\eea}{\end{eqnarray}}

\newcommand{\be}{\begin{equation}}
\newcommand{\ee}{\end{equation}}

\makeatletter
\let\old@startsection=\@startsection
\let\oldl@section=\l@section
\renewcommand{\@startsection}[6]{\old@startsection{#1}{#2}{#3}{#4}{#5}{#6\mathversion{bold}}}
\renewcommand{\l@section}[2]{\oldl@section{\mathversion{bold}#1}{#2}}
\makeatother

\numberwithin{equation}{section}
\usepackage{color}

\begin{document}
\renewcommand{\thefootnote}{\arabic{footnote}}

\overfullrule=0pt
\parskip=0pt
\parindent=12pt
\headheight=0in \headsep=0in \topmargin=0in \oddsidemargin=0in

\vspace{ -3cm} \thispagestyle{empty} \vspace{-1cm}
\begin{flushright} 
\footnotesize
\textcolor{red}{\phantom{print-report}}
\end{flushright}

\begin{center}
\vspace{1.2cm}
{\Large\bf \mathversion{bold}
Corner contributions to holographic entanglement entropy 

 \vspace{0.3cm}

in AdS$_4$/BCFT$_3$
}

 \vspace{0.8cm} {
 Domenico Seminara$^{\,a,}$\footnote[1]{seminara@fi.infn.it},
Jacopo Sisti$^{\,b,}$\footnote[2]{jsisti@sissa.it}
 and Erik Tonni$^{\,b,}$\footnote[3]{erik.tonni@sissa.it}
}
 \vskip  0.7cm

\small
{\em
$^{a}\,$Dipartimento di Fisica, Universit\'a di Firenze and INFN Sezione di Firenze, Via G. Sansone 1, 50019 Sesto Fiorentino,
Italy
  \vskip 0.05cm
$^{b}\,$SISSA and INFN, via Bonomea 265, 34136, Trieste, Italy 
}
\normalsize

\end{center}

\vspace{0.4cm}
\begin{abstract} 
We study the holographic entanglement entropy of spatial regions with corners in the AdS$_4$/BCFT$_3$ correspondence 
by considering  three dimensional boundary conformal field theories whose boundary is a timelike plane.
We compute analytically  the corner function corresponding to an infinite wedge having one edge on the boundary.
A relation between this corner function and the holographic one point function of the stress tensor is observed.
An analytic expression for the corner function of an infinite wedge having only its tip on the boundary is also provided. 
This formula requires to find the global minimum among two extrema of the area functional.
The corresponding critical configurations of corners are studied. 
The results have been checked against a numerical analysis performed by computing the area of the minimal surfaces 
anchored to some finite domains containing corners. 
\end{abstract}

\newpage

%%%%%%%%%%%%%%%%%%%%%%%%%%%%%%%%%%%%%%%%%%%%%

\tableofcontents

%%%%%%%%%%%%%%%%%%%%%%%%%%%%%%%%%%%%%%%%%%%%%

\section{Introduction}
\label{sec intro}

Entanglement entropy has been largely studied during the last two decades in
quantum field theory, quantum gravity, quantum many-body systems
and quantum information (see \cite{ent-reviews, Rangamani:2016dms} for reviews).
Nowadays it is a powerful quantity to understand some properties of quantum systems.

Given a quantum system whose Hilbert space $\mathcal{H} = \mathcal{H}_A  \otimes \mathcal{H}_B$ is bipartite,
by considering the density matrix $\rho$ characterising the state of the system, 
one can introduce the $A$'s reduced density matrix $\rho_A = \textrm{Tr}_{\mathcal{H}_B} \rho$.
We study only bipartitions associate to spatial subsystems.
The entanglement entropy between $A$ and $B$ is defined as the Von Neumann entropy of the reduced density matrix $\rho_A$, namely
\be
S_A = 
-\,\textrm{Tr} (\rho_A \log \rho_A)
\ee
In the same way, one can construct the $B$'s reduced density matrix $\rho_B = \textrm{Tr}_{\mathcal{H}_A} \rho$ and the corresponding 
entanglement entropy $S_B$.
When the state $\rho$ is pure, we have $S_A = S_B$.

In this manuscript we consider the entanglement entropy in certain conformal field theories (CFTs) at strong coupling
and we compute it through the holographic approach \cite{RT, Hubeny:2007xt, Freedman:2016zud}.

In quantum field theories, extracting information about the model from the entanglement entropy of certain kinds of domains is an important task. 
The number of spacetime dimensions plays a central role in this analysis.

In two dimensional conformal field theories on the infinite line at zero temperature, when $A$ is an interval of length $\ell$,
it is well known that the expansion of the entanglement entropy reads $S_A = (c/3) \log (\ell/\varepsilon) + O(1)$ as the ultraviolet (UV) cutoff 
$\varepsilon \to 0^+$, being $c$ the central charge of the model \cite{2dCFT-interval}.
Instead, when $A$ is made by disjoint intervals, the corresponding $S_A$ contains information also about the spectrum of the model
\cite{2dCFT-disjoint-interval}.

In a three dimensional conformal field theory (CFT$_3$), considering a two dimensional spatial region $A  \subsetneq \mathbb{R}^2$ 
whose boundary $\partial A$ is a smooth curve (which might be made by disjoint components), 
the expansion of the entanglement entropy as $\varepsilon \to 0^+$ reads $S_A = b \, P_A/\varepsilon + O(1)$, where $P_A $ is the perimeter of $A$ and $b$ is a
non universal and model dependent coefficient. 
The leading linear divergence corresponds to the area law term of the entanglement entropy in three spacetime dimensions \cite{area-law}.
When $A$ is a disk, it has been shown that the $O(1)$ term of this expansion behaves monotonically along the RG flow \cite{F-theorem}. 
For a CFT$_3$ with a holographic dual description, the holographic computation of $S_A$ at strong coupling coincides with the holographic computation
of the expectation value of a spatial Wilson loop whose contour is $\partial A$ \cite{Maldacena:1998im}.

\begin{figure}[t] 
\vspace{-.5cm}
\hspace{-.4cm}
%\begin{center}
\includegraphics[width=1.04\textwidth]{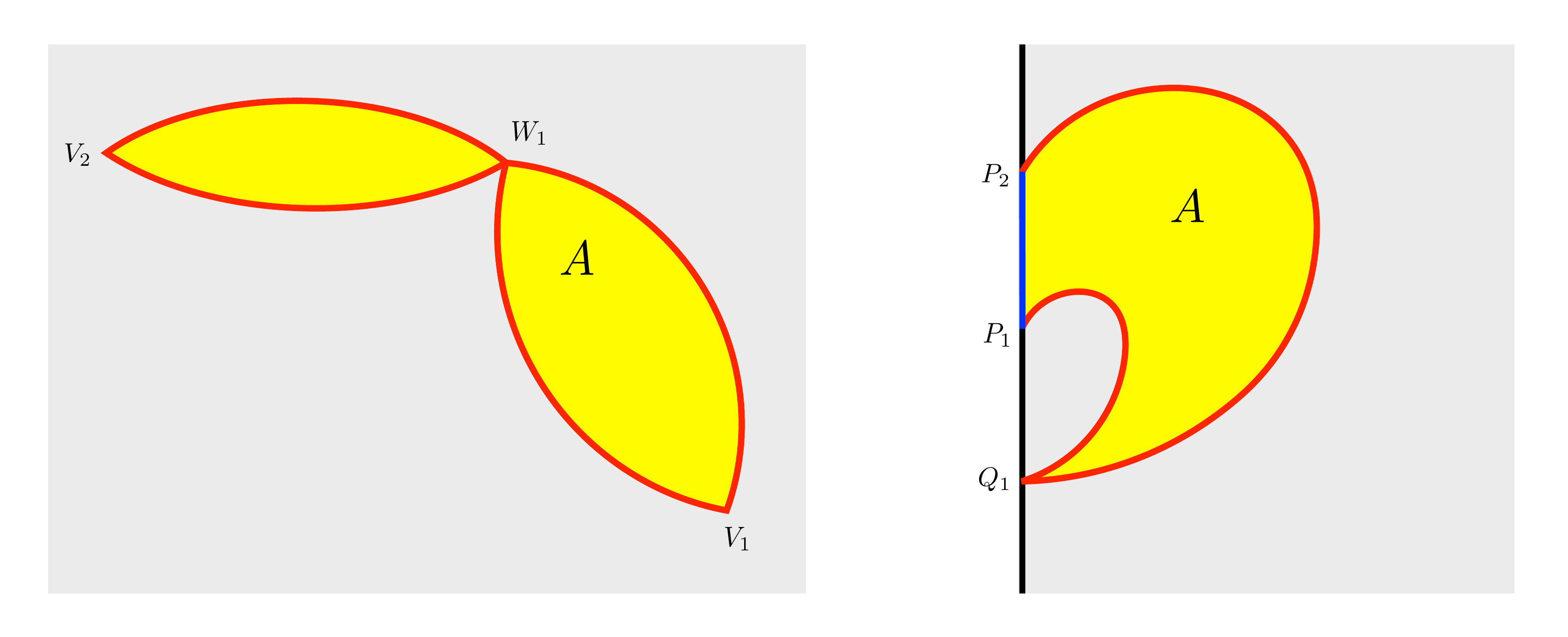}
%\end{center}
\vspace{-.3cm}
\caption{\label{fig:intro}
\small
Examples of finite two dimensional regions $A$ (yellow domains) containing the kinds of corners considered in this manuscript. 
Left: $A$ is a domain in the plane with three corners and two different kinds of vertices.
Right: $A$ is a domain in the half plane with three corners whose boundary $\partial A$ intersects the boundary of the BCFT$_3$
(solid black line). 
The three vertices in $\partial A$ are also on the boundary of the BCFT$_3$ and they belong to two different classes of vertices. 
In both panels, the red curve corresponds to the entangling curve $\partial A \cap \partial B$, whose length provides the area law term 
in (\ref{ee cft3 corner intro}) and in (\ref{ee bcft3 corner intro}).
}
\end{figure}

We are interested in two dimensional spatial regions $A  \subsetneq \mathbb{R}^2$ which contain some isolated corners, 
namely such that $\partial A$ has a finite number of isolated singular points (vertices) separated by distances much larger than the UV cutoff.
In these cases, besides the linearly diverging area law term, the expansion of the entanglement entropy includes also a subleading logarithmic divergence
\cite{Drukker:1999zq, Fradkin:2006mb, Casini:2006hu, Hirata:2006jx, Casini:2008as}
\be
\label{ee cft3 corner intro}
S_A
\,=\, 
b\,\frac{P_A}{\varepsilon} 
- \tilde{f}_{\textrm{\tiny tot}} \log(P_A / \varepsilon)
+ O(1)
\ee
where the coefficient $ \tilde{f}_{\textrm{\tiny tot}}$ is obtained by summing the contributions from all the corners occurring in $A$.
It is not difficult to construct regions whose boundaries contain vertices from which an arbitrary even number of lines depart. 
For the sake of simplicity, we consider only vertices where either two or four lines join together. 
A vertex $V$ belonging to the former class is characterised by an opening angle $\theta$,
while a vertex $W$ in the latter class is described by a vector of opening angles $\vec{\phi}_{W}$ made by three components.
In the example shown in the left panel of Fig.\,\ref{fig:intro}, the curve $\partial A$ contains two vertices of the first kind and one vertex of the second kind. 
For the domains containing these kinds of corner, $ \tilde{f}_{\textrm{\tiny tot}}$ is given by
\be
\label{f-tilde cft3 intro}
\tilde{f}_{\textrm{\tiny tot}} 
= 
\sum_{V_k} \tilde{f}(\theta_{V_k})
+
\sum_{W_r} 
\widetilde{\mathsf{F}}(\vec{\phi}_{W_r})
\ee
When $A$ has corners, we call corner functions all the functions of the opening angles in the sums occurring in 
the coefficient of the logarithmic divergence in $S_A$, whenever it is due only to the corners. 
In the case of (\ref{f-tilde cft3 intro}), the corner functions are $\tilde{f}$ and $\widetilde{\mathsf{F}}$. 
Other corner functions will be discussed in the following. 
The corner function $\tilde{f}(\theta)$ is constrained by some important properties of the entanglement entropy. 
For instance, the fact that $S_A = S_B$ for pure states implies that $\tilde{f}(\theta) = \tilde{f}(2\pi-\theta)$
and this tells us that the corner function vanishes quadratically $\tilde{f}(\theta) = \tilde{\sigma}\,(\pi - \theta)^2 + \dots$ when $\theta \to \pi$.
Other interesting features of $\tilde{f}(\theta)$ (e.g. $\tilde{f}''(\theta) \geqslant 0$) can be derived from the strong subadditivity of the entanglement entropy \cite{Lieb-Ruskai, Hirata:2006jx}.

Besides the analysis of the corner functions $\tilde{f}(\theta)$ for specific quantum field theories 
\cite{Fradkin:2006mb, Casini:2006hu, Hirata:2006jx, Casini:2008as},
many interesting results have been obtained in particular lattice models 
\cite{lattice-corner}.

The corner function $\tilde{f}(\theta)$ depends on the underlying CFT$_3$ model.
However, it has been recently found that,
by considering the coefficient $\tilde{\sigma}$ and the constant $C_T$ characterising the two point function of the stress tensor 
in the same CFT$_3$, the relation $\tilde{\sigma}/C_T = \pi^2/24$ holds for any CFT$_3$
\cite{Bueno:2015rda, Bueno:2015xda, Faulkner:2015csl}.

In this manuscript we are interested in a conformal field theory with a boundary (BCFT).
In two spacetime dimensions, the conformal field theories with boundaries have been largely studied 
\cite{bcft2}, but in higher dimensions much less is known \cite{McAvity:1993ue}. 
In this work we mainly focus on the three dimensional case (BCFT$_3$). 
The space of the boundary conditions which preserve the conformal invariance depends on the underlying model.
In BCFT$_3$, the presence of the boundary leads to a Weyl anomaly which is non vanishing only on the boundary
and this allows to introduce two boundary charges
\cite{Jensen:2015swa, Solodukhin:2015eca}.
In BCFT$_4$, the Weyl anomaly is the sum of a well known term on the bulk and a term due to the occurrence of the boundary 
\cite{Jensen:2015swa, Herzog:2015ioa, Fursaev:2015wpa, Solodukhin:2015eca}. 
Other recent related results are discussed in \cite{HuangHerzog}.

In this manuscript we mainly consider a BCFT$_3$ whose boundary is flat; 
therefore any constant time slice of the spacetime 
is the half plane which can be described by the Cartesian coordinates $(x,y) \in \mathbb{R}^2$ with $x\geqslant 0$. 
In this BCFT$_3$ setup, we study the entanglement entropy of two dimensional regions $A$ whose boundaries
$\partial A$ contain some isolated vertices which are all located on the boundary of the spatial half plane, which is the straight line at $x=0$.
A prototypical example is the yellow domain in the right panel of Fig.\,\ref{fig:intro}. 
Like for the previous regions, the distances between the vertices must be much larger than the UV cutoff.
For this kind of domains, the expansion of the entanglement entropy as $\varepsilon \to 0^+$ reads
\be
\label{ee bcft3 corner intro}
S_A
\,=\, 
b\,\frac{P_{A,B}}{\varepsilon} 
- f_{\alpha, \textrm{\tiny tot}} 
\log(P_{A,B} / \varepsilon)
+ O(1)
\ee
where $P_{A,B} \equiv \textrm{length}(\partial A \cap \partial B )$  is the length of the curve shared by $\partial A $ and the boundary $\partial B $ of its complement
(the red curve in the right panel of Fig.\,\ref{fig:intro}), which is often called entangling curve. 
Thus, $P_{A,B} \leqslant P_A$ and $P_A - P_{A,B} $ is the length of $\partial A_{\textrm{\tiny bdy}}  \equiv \partial A\, \cap \,\{(x,y)\,|\, x=0\} $ 
(in the right panel of Fig.\,\ref{fig:intro} we have $\partial A_{\textrm{\tiny bdy}} = \overline{P_1 P_2} \,\cup \,\{Q_1 \}$).
This is a consequence of the relation $S_A = S_B$, which holds whenever the whole system is in a pure state. 
The coefficient of the leading divergence in (\ref{ee bcft3 corner intro}) is the same that occurs in the leading divergence of (\ref{ee cft3 corner intro}) because
it is related to some local effects close to the entangling curve; therefore it should be independent of the occurrence of a boundary. 
Instead, the coefficient $f_{\alpha, \textrm{\tiny tot}}$ of the logarithmic divergence in (\ref{ee bcft3 corner intro}) is expected to
depend on the boundary conditions characterising the BCFT$_3$ in a highly non trivial way. 
The index $\alpha$ labels the boundary conditions allowed by the conformal invariance in the underlying model. 

Domains $A$ whose boundaries contain vertices on the $x=0$ line from which an arbitrary even number of lines belonging to $\partial A$ depart
can be easily drawn.
For the sake of simplicity, we restrict our analysis to vertices of $\partial A$ on the $x=0$ line where only two lines of $\partial A$ (the edges of the corner) 
join together. 
Given a vertex belonging to this class, there are two possibilities: either one edge or none of the two edges is on the boundary.
In the former case we denote the vertex by $P$ and the corner is characterised only by an angle $\gamma$, 
while in the latter case we label the vertex with $Q$ and the corresponding corner in $A$ is characterised by a pair $\vec{\omega}$ of opening angles.
An example of domain $A$ containing these two kinds of corners is shown in the right panel of Fig.\,\ref{fig:intro}, 
where $\partial A$ has two vertices $P_1$ and $P_2$ of the first kind and one vertex $Q_1$ of the second kind. 

For this class of regions $A$, the coefficient $f_{\alpha, \textrm{\tiny tot}}$ of the logarithmic divergence in (\ref{ee bcft3 corner intro}) is obtained 
by summing the contributions of the all corners on the boundary, namely
\be
\label{Ftot bdy corner intro cft}
f_{\alpha, \textrm{\tiny tot}} 
=
\sum_{P_i} f_\alpha(\gamma_{P_i})
+
\sum_{Q_j} \mathsf{F}_\alpha(\vec{\omega}_{Q_j})
\ee
where $f_\alpha$ and $\mathsf{F}_\alpha$ are corner functions which depend on the boundary conditions of the BCFT$_3$.
Interesting results about $f_\alpha(\gamma)$ have been obtained in \cite{Jensen:2013lxa, Fursaev:2016inw, Berthiere:2016ott}.

One can easily consider a more general class of domains $A$ whose boundaries contain also vertices of the types introduced above which have $x>0$.
In these cases the area law term of the entanglement entropy is like the one in (\ref{ee bcft3 corner intro}) and the coefficient of the logarithmic divergence is simply
the sum of (\ref{f-tilde cft3 intro}) and (\ref{Ftot bdy corner intro cft}).

In this manuscript we are interested in the corner functions occurring in (\ref{Ftot bdy corner intro cft}) for a BCFT$_3$ at strong coupling. 
The main tool employed in our analysis is the gauge/gravity correspondence and, in particular, 
the holographic prescription to compute the entanglement entropy.

Let us consider a CFT$_{d+1}$ in $d+1$ spacetime dimensions which has a gravitational dual description through an asymptotically AdS$_{d+2}$ spacetime.
In the Poincar\'e coordinates, denoting by $z$ the extra dimension of the gravitational theory with respect to the $d+1$ dimensions of the CFT$_{d+1}$,
the boundary of the gravitational spacetime, where the CFT$_{d+1}$ is defined, corresponds to $z=0$.
For the static cases, the entanglement entropy of a spatial region $A$ in a $t=\textrm{const}$ slice of the 
CFT$_{d+1}$ at strong coupling is given by  the holographic formula \cite{RT} 
(see \cite{Rangamani:2016dms} for a recent review)
\be
\label{RT formula intro}
S_A = \frac{\mathcal{A}_A}{4G_{\textrm{\tiny N}}}
\ee
where $G_{\textrm{\tiny N}}$ is the $d+2$ dimensional gravitational Newton constant and 
$\mathcal{A}_A$ is the area of the $d$ dimensional minimal area hypersurface $\hat{\gamma}_A$ 
anchored to the boundary of $A$, namely such that $\partial \hat{\gamma}_A = \partial A$.
Since the asymptotically AdS$_{d+2}$ gravitational background is a non compact space and $\hat{\gamma}_A $ reaches its boundary, the 
area of $\hat{\gamma}_A$ diverges.
In order to regulate the area $\mathcal{A}_A$, we have to introduce a cutoff $z\geqslant \varepsilon > 0$ in the holographic direction $z$ such that $\varepsilon \ll P_A$.
According to the AdS/CFT dictionary, $\varepsilon $ is the gravitational dual of the UV cutoff in the CFT$_{d+1}$.
Denoting by $\hat{\gamma}_\varepsilon \equiv \hat{\gamma}_A \cap \{  z\geqslant \varepsilon  \}$ the restriction of $\hat{\gamma}_A$ to $z\geqslant \varepsilon$, 
in (\ref{RT formula intro}) one has to consider $\mathcal{A}_A = \mathcal{A}[\hat{\gamma}_\varepsilon ]$ and then expand the resulting expression for $\varepsilon \to 0^+$.
The terms of this expansion can be compared with the ones occurring in the expansion of $S_A$ computed through CFT techniques.
Various consistency checks of (\ref{RT formula intro}) has been done (e.g. the strong subadditivity property \cite{Lieb-Ruskai, Headrick:2007km})
and nowadays the holographic formula (\ref{RT formula intro}) is largely recognised as a tool to evaluate the entanglement entropy in the strong coupling 
regime of CFTs with a gravitational dual description. 
In this manuscript we are mainly interested in the $d=2$ case, but some results are obtained for a generic $d$.

The holographic prescription (\ref{RT formula intro}) has been applied also for domains $A$ whose boundaries contain singular points 
\cite{Hirata:2006jx, Myers:2012vs, Fonda:2014cca, Fonda:2015nma, Mozaffar:2015xue}.
We are interested in the $d=2$ case, where the expansion of $\mathcal{A}[\hat{\gamma}_\varepsilon]$ as $\varepsilon \to 0^+$ 
for domains $A$ with corners reads
\be
\label{hee ads4 corner intro}
\mathcal{A}[\hat{\gamma}_\varepsilon]
= 
L^2_{\textrm{\tiny AdS}}
\bigg(
\frac{P_A}{\varepsilon} 
- \widetilde{F}_{\textrm{\tiny tot}} \log(P_A / \varepsilon)
+ O(1)
\bigg) 
\ee
being $L_{\textrm{\tiny AdS}}$ the AdS radius of the gravitational background.
Considering the two classes of vertices discussed below (\ref{ee cft3 corner intro}), 
the coefficient of the logarithmic divergence in (\ref{hee ads4 corner intro}) reads
\be
\label{Ftilde cft intro}
\widetilde{F}_{\textrm{\tiny tot}} 
= 
\sum_{V_k} \widetilde{F}(\theta_{V_k})
+
\sum_{W_r} \widetilde{\mathcal{F}}(\vec{\phi}_{W_r})
\ee
Comparing (\ref{ee cft3 corner intro}) and (\ref{f-tilde cft3 intro}) with (\ref{RT formula intro}), (\ref{hee ads4 corner intro}) and (\ref{Ftilde cft intro}), 
we have that the holographic corner functions are proportional to $\widetilde{F}(\theta)$ and $\widetilde{\mathcal{F}} (\vec{\phi}\,)$
through the positive constant $\tfrac{L^2_{\textrm{\tiny AdS}}}{4G_{\textrm{\tiny N}}}$. 
Nonetheless, in the following we mainly refer to the latter ones as the holographic corner functions,
unless stated otherwise.
The analytic expression of the corner function $\widetilde{F}(\theta)$ has been found in \cite{Drukker:1999zq}.
The corner function $\widetilde{\mathcal{F}} (\vec{\phi}\,)$ can be easily written once $\widetilde{F}(\theta)$ is known \cite{Mozaffar:2015xue}.

In this manuscript we study the corner functions $f_\alpha(\gamma)$ and $\mathsf{F}_\alpha(\vec{\omega})$ for a BCFT$_3$ at strong coupling,
assuming that a holographic dual exists.
We consider the AdS$_{d+2}$/BCFT$_{d+1}$ setup introduced in \cite{Takayanagi:2011zk} and further developed in \cite{Fujita:2011fp, Nozaki:2012qd},
where the $d+2$ dimensional bulk is limited by the occurrence of a $d+1$ dimensional hypersurface $\widetilde{\mathcal{Q}}$ in the bulk 
which has the same boundary of the BCFT$_{d+1}$.
Also previous works have considered the occurrence of a hypersurface in the bulk \cite{pre-ads/bcft}.
An interesting application of the AdS/BCFT setup has been discussed in \cite{holog-kondo}.
The boundary conditions determining $\widetilde{\mathcal{Q}}$ are crucial in the construction of \cite{Takayanagi:2011zk,Fujita:2011fp, Nozaki:2012qd}.
A setup of the correspondence based on a different boundary condition has been recently suggested 
\cite{Miao:2017gyt, Chu:2017aab, Astaneh:2017ghi, FarajiAstaneh:2017hqv}.

In this manuscript we mainly focus on the simplest situation of a BCFT$_3$ with a flat boundary.
In this case $\widetilde{\mathcal{Q}} = \mathcal{Q}$ is the same half hyperplane for both the prescriptions mentioned above.
Its slope $\alpha$ is related to a real parameter occurring in the boundary term of the gravitational action in the bulk. 

In a BCFT$_3$ with a holographic dual description, 
we consider two dimensional domains $A$ described above, namely 
the ones  with isolated corners whose vertices belong to the boundary of the BCFT$_3$.
For these domains the entanglement entropy is given by (\ref{ee bcft3 corner intro}) and (\ref{Ftot bdy corner intro cft}).
By employing the holographic formula (\ref{RT formula intro}) properly adapted to the AdS/BCFT setup, we find 
\be
\label{hee bdy corner intro}
\mathcal{A}[\hat{\gamma}_\varepsilon]
= 
L^2_{\textrm{\tiny AdS}}
\bigg(
\frac{P_{A,B}}{\varepsilon} 
- F_{\alpha, \textrm{\tiny tot}}  \log(P_{A,B} / \varepsilon)
+ O(1)
\bigg) 
\ee
where $P_{A,B} \leqslant P_A$ is the length of the entangling curve in the boundary at $z=0$.
We are mainly interested in the coefficient of the logarithmic divergence, which given by the sum of the contributions from all the vertices of $\partial A$, namely
\be
\label{Ftot bdy corner intro}
F_{\alpha, \textrm{\tiny tot}} 
=
\sum_{P_i} F_\alpha(\gamma_{P_i})
+
\sum_{Q_j} \mathcal{F}_\alpha(\omega_{Q_j} ,\gamma_{Q_j})
\ee
where the functions occurring in the sums depend on the slope $\alpha$ of the half plane $\mathcal{Q}$.
We mainly refer to $F_\alpha(\gamma)$ and $\mathcal{F}_\alpha(\omega ,\gamma)$ as the holographic corner functions in the presence of a boundary,
although the proportionality constant $\tfrac{L^2_{\textrm{\tiny AdS}}}{4G_{\textrm{\tiny N}}}$ should be taken into account. 

In this manuscript we find analytic expressions for the corner functions $F_\alpha(\gamma)$ and $\mathcal{F}_\alpha(\omega ,\gamma)$.
Numerical checks of these results are performed by constructing the minimal area surfaces corresponding to some finite domains containing corners.
In the numerical analysis we have employed {\it Surface Evolver} \cite{evolverpaper, evolverlink} to construct the minimal area surfaces,
a software which has been already used in  \cite{Fonda:2014cca, Fonda:2015nma} to study the shape dependence of the holographic 
entanglement entropy in AdS$_4$/CFT$_3$.

We also compute the holographic entanglement entropy of an infinite strip, which can be either  adjacent or parallel to the boundary.
This is done because the result for the infinite strip adjacent to the boundary can be related to the limit $\gamma \to 0$ of the corner function $F_\alpha(\gamma)$.
The holographic entanglement entropy of an infinite strip adjacent or parallel to the boundary is computed also for a generic number of spacetime dimensions. 

We find it worth mentioning here that, within the AdS$_4$/BCFT$_3$ setup of \cite{Takayanagi:2011zk}, 
we find a proportionality relation between the coefficient  $f''_\alpha(\pi/2) = \tfrac{L^2_{\textrm{\tiny AdS}}}{4G_{\textrm{\tiny N}}} F''_\alpha(\pi/2)$ 
occurring in the expansion of the holographic corner function when $\gamma \to \pi/2$
and the holographic result for a coefficient which characterises the behaviour 
of the one point function of the stress tensor in the proximity of the curved boundary of a BCFT$_3$.

The paper is organised as follows. 
In Sec.\,\ref{sec constraints} the strong subadditivity is employed to find constraints for the corner functions $f_\alpha(\gamma)$ and $\mathsf{F}_{\alpha}(\omega  ,\gamma )$ in a BCFT$_3$.
In Sec.\,\ref{sec drop} we test our numerical approach based on Surface Evolver on the corner functions in AdS$_4$/CFT$_3$.
This is a good benchmark for our method because well known analytic expressions are available for the corner functions occurring in (\ref{Ftilde cft intro}).
The AdS/BCFT setup is briefly reviewed in Sec.\,\ref{sec HEE bdy}, where we also discuss the prescription to compute the holographic entanglement entropy 
of a domain with generic shape. 
This prescription for the holographic entanglement entropy is applied for two simple domains in Sec.\,\ref{sec two simple cases}:
 the half disk centered on the boundary and the infinite strip, which can be either adjacent or parallel to the boundary at finite distance from it. 
 In Sec.\,\ref{sec:wedge} we describe the main result of this manuscript, namely
 the holographic entanglement entropy of the infinite wedge adjacent to the boundary,
 which provides the analytic expression of the corner function $F_\alpha(\gamma)$.
In Sec.\,\ref{sec wedge tip bdy} this corner function is employed to find an analytic formula for the corner function $\mathcal{F}_\alpha(\omega ,\gamma)$.
Some conclusions and open problems are discussed in Sec.\,\ref{sec:conclusions}.

The  main text of this manuscript contains only the description of the main results. 
All the computational details underlying their derivations and also some generalisations to an arbitrary number of spacetime dimensions
have been collected and discussed in the appendices 
\ref{app numerics}, 
\ref{sec app half disk}, 
\ref{sec app strip}, 
\ref{sec app strip disjoint}, 
\ref{sec app ssa}, 
\ref{sec app wedge}, \ref{app A_T} 
and \ref{app constraints}.

\newpage

%%%%%%%%%%%%%%%%%%%%%%%%%%%%%%%%%%%%%%%%%%%%%%%%%%%%%
\section{Constraining the corner functions}
\label{sec constraints}

In this section we employ the strong subadditivity of the entanglement entropy \cite{Lieb-Ruskai} to constrain the corner functions introduced in (\ref{Ftot bdy corner intro cft}).
Our analysis is similar to the one performed in \cite{Hirata:2006jx} for the corner function $\tilde{f}(\theta)$ in (\ref{f-tilde cft3 intro}).

Let us consider a BCFT$_3$ in its ground state and the domain $A$ given by the infinite wedge adjacent to the boundary whose opening angle is $\gamma$.
The complementary domain $B$ is the infinite wedge adjacent to the boundary sharing with $A$ its edge which is not on
the boundary and its opening angle is $\pi- \gamma$.
Since the ground state is a pure state, we have $S_A = S_B$.
Combining this property with (\ref{ee cft3 corner intro}) specialised to these complementary domains, we have
\be
\label{purity F_alpha}
f_\alpha(\pi - \gamma) = f_\alpha(\gamma) 
\ee
namely the corner function $f_\alpha(\gamma)$ is symmetric with respect to $\gamma=\pi/2$;
therefore we are allowed to study this corner function for $0< \gamma \leqslant \pi/2$.
Hereafter we mainly consider $\gamma \in (0,\pi/2]$ for the argument of this corner function.
Nonetheless, whenever $\gamma \in (0,\pi)$ in the following, we always mean $f_\alpha(\gamma)=f_\alpha(\textrm{min}[\gamma, \pi-\gamma]) $.

By assuming that $f_\alpha(\gamma)$ is smooth for $\gamma \in (0,\pi)$, the symmetry (\ref{purity F_alpha}) implies that its expansion around $\gamma = \pi/2$ 
includes only even powers of $\gamma-\pi/2$, namely
\be
\label{Falpha expansion}
f_\alpha(\gamma) = f_\alpha(\pi/2) + \frac{f''_\alpha(\pi/2)}{2}\, \big(\gamma-\pi/2\big)^2 + \dots 
\qquad
\gamma \to \frac{\pi}{2}
\ee

\begin{figure}[t] 
\vspace{-.5cm}
%\hspace{-.25cm}
\begin{center}
\includegraphics[width=1.\textwidth]{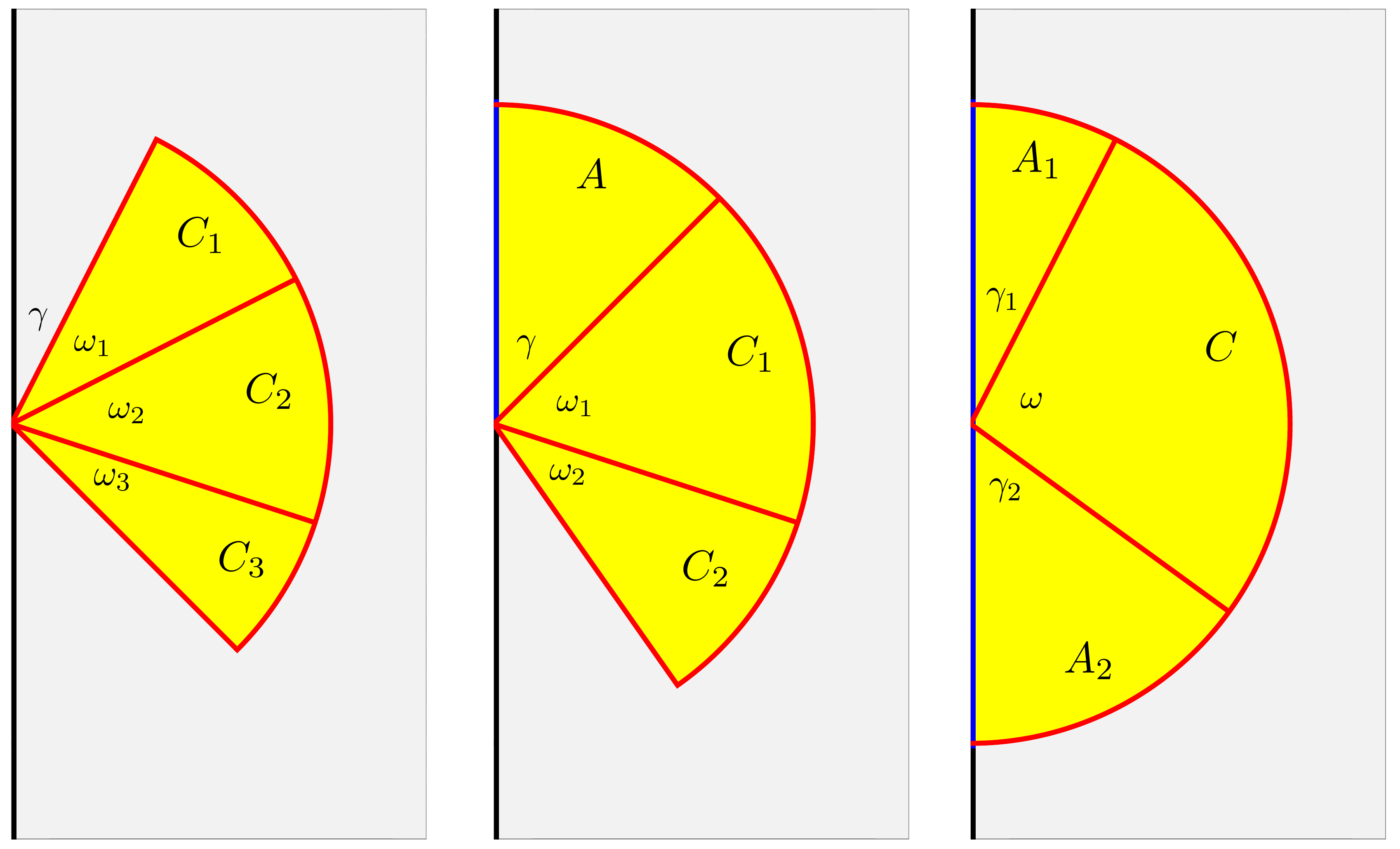}
\end{center}
\vspace{-.3cm}
\caption{\label{fig:ssa}
\small
Configurations of adjacent domains containing corners (yellow regions) in the half plane $x\geqslant 0$ (grey region)
which have been used in Sec.\,\ref{sec constraints}
to constrain the corner functions through the strong subadditivity.
}
\end{figure}

In the remaining part of this section we discuss some constraints for the corner functions in (\ref{Ftot bdy corner intro cft}) obtained by imposing that the
strong subadditivity of the entanglement entropy is valid for particular configurations of adjacent domains. 

Consider the configuration of adjacent regions shown in the left panel of Fig.\,\ref{fig:ssa}.
The strong subadditivity inequality specialised to this case states that
\be
\label{ssa config 1}
S_{C_1 \cup C_2} + S_{C_2 \cup C_3}  \geqslant  S_{C_1 \cup C_2 \cup C_3} + S_{C_2}
\ee
By employing the expressions (\ref{ee bcft3 corner intro}) and (\ref{Ftot bdy corner intro cft}), which provide the entanglement entropy
of the domains occurring in this inequality,
one observes that the area law terms and the logarithmic divergencies corresponding to vertices which are not on the boundary simplify.
The remaining terms at leading order provide the following inequality
\be
\label{ssa ineq config 1}
    \mathsf{F}_{\alpha}(\omega_1 + \omega_2 +\omega_3 ,\gamma) 
    -  \mathsf{F}_{\alpha}(\omega_1 + \omega_2 ,\gamma) 
    \geqslant
  \mathsf{F}_{\alpha}(\omega_2 + \omega_3 ,\gamma + \omega_1)
 -   \mathsf{F}_{\alpha}(\omega_2 ,\gamma + \omega_1)   
\ee
Multiplying both sides of this inequality by $1/\omega_3 > 0$ first and then taking the limit $\omega_3 \to 0^+$, one finds
\be
\label{ssa ineq2 config 1}
\partial_\omega\, \mathsf{F}_{\alpha}(\omega_2 + \omega_1 ,\gamma) 
    \,\geqslant\,
   \partial_\omega  \, \mathsf{F}_{\alpha}(\omega_2  ,\gamma + \omega_1)
\ee
Next we add  $ - \,\partial_\omega  \mathsf{F}_{\alpha}(\omega_2,\gamma) $ to both sides of (\ref{ssa ineq2 config 1}), then we multiply them by $1/\omega_1 > 0$ 
and finally take the limit $\omega_1 \to 0^+$.
The resulting inequality reads
\be
\label{ssa config 1}
\partial^2_\omega\, \mathsf{F}_{\alpha}(\omega ,\gamma) 
    \,\geqslant\,
   \partial_\omega \partial_\gamma  \, \mathsf{F}_{\alpha}(\omega  ,\gamma )
\ee
This property resembles to $\tilde{f}''(\omega) \geqslant 0$ for the corner function $\tilde{f}(\omega)$ in CFT$_3$ \cite{Hirata:2006jx}.

The second configuration of adjacent domains that we consider is the one depicted in the middle panel of Fig.\,\ref{fig:ssa}.
In this case, the constraint given by the strong subadditivity reads $S_{A \cup C_1} + S_{C_1 \cup C_2}  \geqslant  S_{A \cup C_1 \cup C_2} + S_{C_1}$ 
and simplifications similar to  the ones discussed in the previous case occur.
In particular, the leading non vanishing terms correspond to the vertex shared
by the three domains. 
The resulting inequality reads
\be
\label{ssa_2_step0}
f_\alpha(\gamma + \omega_1 + \omega_2)  - f_\alpha(\gamma + \omega_1 )
     \,\geqslant\,
\mathsf{F}_{\alpha}(\omega_1 + \omega_2  ,\gamma ) -  \mathsf{F}_{\alpha}(\omega_1   ,\gamma )    
\ee
Multiplying both sides of this relation by $1/\omega_1>0$ and taking the limit $\omega_1 \to 0^+$, one obtains
\be
\label{ssa config 2}
\partial_\omega\, \mathsf{F}_{\alpha}(\omega ,\gamma) 
    \,\leqslant\,
   \partial_\gamma  f_{\alpha}(\gamma + \omega)
\ee

Let us study also the configuration shown in the right panel of Fig.\,\ref{fig:ssa}, where $\gamma_1 +\omega + \gamma_2 = \pi$
and the strong subadditivity property provides the constraint $S_{A_1 \cup C} + S_{A_2 \cup C}  \geqslant  S_{A_1 \cup A_2 \cup C} + S_{C}$.
By using (\ref{ee bcft3 corner intro}) and (\ref{Ftot bdy corner intro cft}) as done in the previous cases, we get
another inequality among the corner functions corresponding to the vertex shared by the three adjacent domains
\be
\label{ssa config 3 v0}
f_{\alpha}(\gamma_1+\omega) + f_{\alpha}(\gamma_2+\omega) 
\,\leqslant \,
\mathsf{F}_{\alpha}(\omega,\gamma_1)
\qquad
\gamma_1 \leqslant \gamma_2
\ee
Since $\gamma_2+\omega = \pi -\gamma_1$, we can employ (\ref{purity F_alpha}), finding that (\ref{ssa config 3 v0}) can be written as
\be
\label{ssa config 3}
f_{\alpha}(\gamma+\omega) + f_{\alpha}(\gamma) \leqslant \mathsf{F}_{\alpha}(\omega,\gamma)
\qquad
\gamma \leqslant \frac{\pi -\omega}{2}
\ee

We remark that the constraints (\ref{ssa config 1}), (\ref{ssa config 2}) and (\ref{ssa config 3}) hold whenever the entanglement entropy is given by
(\ref{ee bcft3 corner intro}) and (\ref{Ftot bdy corner intro cft}), with corner functions which are regular enough to define the derivatives occurring in these
inequalities.

%%%%%%%%%%%%%%%%%%%%%%%%%%%%%%%%%%%%%%%%%%%%%%%%%%%%%
\section{Warming up: corner functions in AdS$_4$/CFT$_3$}
\label{sec drop}

In this section we consider the holographic entanglement entropy of domains with corners
in AdS$_4$/CFT$_3$.
The aim of this analysis is to test our numerical approach 
to the corner functions for the holographic entanglement entropy on the well known case of AdS$_4$/CFT$_3$, 
in order to apply it to the corner functions in AdS$_4$/BCFT$_3$ in the following sections.
Our numerical data are obtained by employing Surface Evolver \cite{evolverpaper, evolverlink},
a software developed by Ken Brakke, and the method is briefly discussed in the appendix\;\ref{app numerics}.

Given a CFT$_3$ in the three dimensional Minkowski space parameterised by $(t,x,y)$ which has a dual holographic description, 
the bulk metric dual to its ground state is AdS$_4$.
In Poincar\'e coordinates $(t,z,x,y)$, the metric induced on a constant time slice of AdS$_4$ is 
\be
\label{AdS4 metric}
ds^2 = \frac{L^2_{\textrm{\tiny AdS}}}{z^2} \,\big( dz^2 + dx^2 + dy^2 \big)
\qquad
z>0
\ee
where $L_{\textrm{\tiny AdS}}$ is the AdS radius.
The metric (\ref{AdS4 metric}) characterises the three dimensional hyperbolic space $\mathbb{H}_3=\,$AdS$_4\big|_{t \,=\, \textrm{const}}$.

The holographic entanglement entropy (\ref{RT formula intro}) of a two dimensional domain $A$ in the plane $(x,y) \in \mathbb{R}^2$
is obtained by finding the minimal area surface $\hat{\gamma}_A$ anchored to $\partial A$ first and then computing the area of 
$\hat{\gamma}_\varepsilon = \hat{\gamma}_A \cap \{ z \geqslant \varepsilon \}$ for $\varepsilon \to 0^+$.
When $\partial A$ contains isolated vertices and each of them has an arbitrary even number of edges, the expansion of the area of $\hat{\gamma}_\varepsilon $ is given by (\ref{hee ads4 corner intro}).

The analytic expression for the coefficient $\widetilde{F}_{\textrm{\tiny tot}} $ of the logarithmic divergence can be found by using the corner function $\widetilde{F}(\theta) $ found in \cite{Drukker:1999zq}.
This function reads
\be
\label{cusp exact result}
\widetilde{F}(\theta) 
\,\equiv\,
2\, F(q_0)
\ee
where
\be
\label{F(q0) def main}
F(q_0) \equiv \frac{\mathbb{E}(\tilde{q}_0^2)-\big(1-\tilde{q}_0^2\big)\mathbb{K}(\tilde{q}_0^2)}{\sqrt{1-2\tilde{q}_0^2}} 
\ee
and the opening angle $\theta$ of the wedge is given by
\be
\label{theta_V expression}
\frac{\theta}{2}
\,=\,
\tilde{q}_0 \, \sqrt{\frac{1-2\tilde{q}^2_0}{1-\tilde{q}^2_0}}\; \Big[ \,\Pi\big(1-\tilde{q}^2_0,\tilde{q}^2_0\big) - \mathbb{K}\big(\tilde{q}^2_0\big) \,\Big]
\,\equiv\,
P_0(q_0)
\ee
where the positive parameter $\tilde{q}_0 \in (0,1/2)$ is related to a positive parameter $q_0$ as 
\be
\label{q0tilde def}
\tilde{q}^2_0 \equiv \frac{q_0^2}{1+2q_0^2} 
\qquad
q_0 > 0
\ee
The geometric meaning of $q_0$ will be discussed in Sec.\,\ref{sec:wedge}.
The functions $\mathbb{K}(m)$, $\mathbb{E}(m)$ and $\Pi(n,m)$ are the complete elliptic integrals of the first, second and third kind respectively.
From (\ref{cusp exact result}) and (\ref{theta_V expression}), one can plot the curve $\widetilde{F}(\theta)$ parametrically in terms of $q_0 > 0$,
finding the blue curve shown in Fig.\,\ref{fig:drukker_data}.

Since $S_A = S_B$ for pure states, for the argument of the corner function $\widetilde{F}(\theta)$ we have $\theta \in (0,\pi]$.
Hereafter, whenever $\theta \in (0,2\pi)$  we mean $\widetilde{F}(\theta) = \widetilde{F}(\textrm{min}[\theta, 2\pi -\theta])$.

In the remaining part of this section we study two simple domains whose holographic entanglement entropy is given by (\ref{hee ads4 corner intro}) and (\ref{Ftilde cft intro}).
In the first example $\partial A$ has a single vertex with two edges and the second case $\partial A$ has a single vertex with four edges. 
Thus, only one term occurs in (\ref{Ftilde cft intro}) specialised to these domains.

\subsection{Single drop} 
\label{sec single drop}

\begin{figure}[t] 
\vspace{-1cm}
%\hspace{-.25cm}
\begin{center}
\includegraphics[width=.8\textwidth]{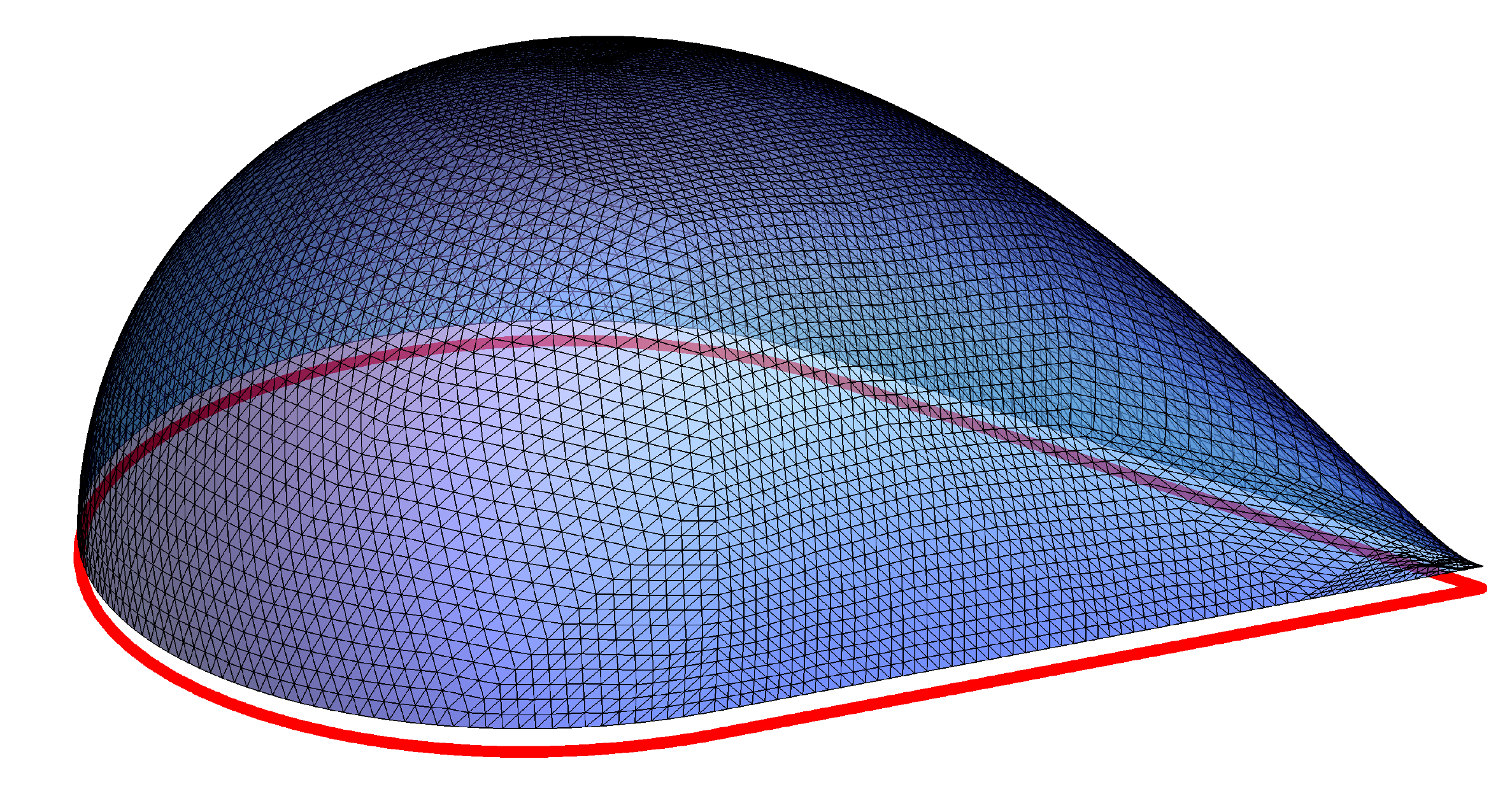}
\end{center}
\vspace{-.1cm}
\caption{\label{fig:single_drop}
\small
Triangulated surface in $\mathbb{H}_3$ which approximates the minimal area surface $\hat{\gamma}_A$ corresponding to a single drop region $A$ 
in the $z=0$ plane, as discussed in Sec.\,\ref{sec single drop}.
The boundary $\partial A$ (red curve) lies in the $z=0$ plane and it is characterised by $L=1$ and $\theta = \pi/3$.
The UV cutoff is $\varepsilon = 0.03$.
The triangulation has been obtained with Surface Evolver by setting $\partial A$ at $z=\varepsilon $.
}
\end{figure}

The first simply connected finite domain $A$ that we consider is similar to a two dimensional drop.
It is constructed by taking the infinite wedge with opening angle $\theta < \pi$ (whose tip is denoted by $P$) 
and the disk of radius $R$ which is tangent to both the edges of the wedge.
The distance between the two intersection points and $P$ is $L=R \cot(\theta/2)$.
Considering the circular sector given by the intersection of the infinite wedge with the disk centered in $P$ with radius $L$,
our drop region $A$ is obtained as the union between this circular sector and the disk of radius $R$  tangent to the edges of the infinite wedge introduced above. 
This domain can be characterised by the parameters $L$ and $\theta$.
Its boundary $\partial A$ is a smooth curve except for the vertex $P$, where two edges join, whose length is $P_A = 2L + R(\pi+\theta)$.
An example of drop domain is the region in the plane enclosed by red curve in Fig.\,\ref{fig:single_drop}.

The holographic entanglement entropy of a drop region $A$ in the $z=0$ plane is obtained by computing the 
area $\mathcal{A}[\hat{\gamma}_\varepsilon]$ from the minimal surface $\hat{\gamma}_A$ embedded in $\mathbb{H}_3 $ which is anchored to $\partial A$,
as prescribed by (\ref{RT formula intro}).
The result is (\ref{hee ads4 corner intro}) with $\widetilde{F}_{\textrm{\tiny tot}}  = \widetilde{F}(\theta)$,
being $\widetilde{F}(\theta)$ the corner function given by (\ref{cusp exact result}) and (\ref{theta_V expression}).
The main advantage of our choice for $A$ is that we can vary the opening angle $\theta$ in a straightforward way. 
The minimal area surfaces $\hat{\gamma}_A$ corresponding to regular polygons and other finite domains with three or more vertices have been studied in \cite{Fonda:2014cca}.

Finding analytic expressions for $\hat{\gamma}_A$ and for the area of $\hat{\gamma}_\varepsilon $ when $A$ is a finite region without particular symmetries is very difficult. 
We perform a numerical analysis by employing Surface Evolver, which provides approximate solutions for $\hat{\gamma}_A$ and 
the corresponding value for $\mathcal{A}[\hat{\gamma}_\varepsilon]$.
In Fig.\,\ref{fig:single_drop} we show a refined triangulation  which approximates the minimal surface $\hat{\gamma}_A$ anchored to a single drop domain.
Some technical details about the construction of this kind of triangulations are discussed in the appendix\;\ref{app numerics} (see also \cite{Fonda:2014cca}).

\begin{figure}[t] 
\vspace{-1cm}
%\hspace{-.25cm}
\begin{center}
\includegraphics[width=.9\textwidth]{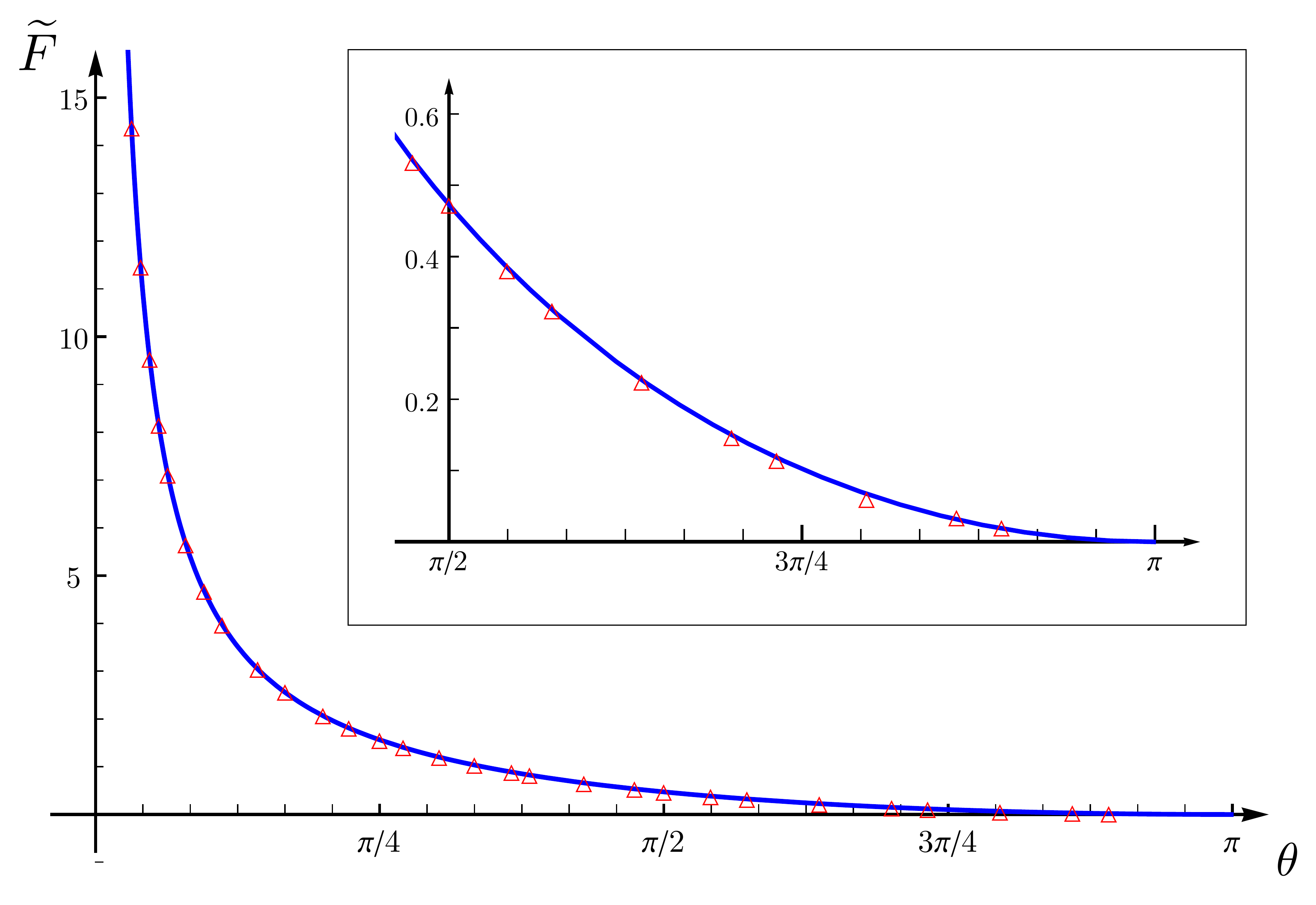}
\end{center}
\vspace{-.3cm}
\caption{\label{fig:drukker_data}
\small
Corner function for a vertex with two edges in AdS$_4$/CFT$_3$.
The blue curve corresponds to the analytic expression given by  (\ref{cusp exact result}) and (\ref{theta_V expression})
found in \cite{Drukker:1999zq}.
The points labeled by the red triangles have been found through the numerical analysis based on Surface Evolver (see Sec.\,\ref{sec single drop} and the appendix\;\ref{app numerics}).
The inset highlights the domain corresponding to opening angles close to $\pi$.
}
\end{figure}

As briefly explained in the appendix\;\ref{app numerics}, by fitting the numerical data for $\mathcal{A}[\hat{\gamma}_\varepsilon]$ obtained for various $\varepsilon$ at fixed values of $\theta$ and $L$, 
we find a numerical value for the corner function which can be compared to the corresponding value coming from the analytic expression of $\widetilde{F}(\theta)$ given by  (\ref{cusp exact result}) and (\ref{theta_V expression}).
Repeating this analysis for different values of $\theta$, we have obtained the results shown in Fig.\,\ref{fig:drukker_data}, 
where the blue solid curve is the analytic curve $\widetilde{F}(\theta)$ found in \cite{Drukker:1999zq},
while the points marked by the red triangles have been found through our numerical analysis.
The agreement is exceptionally good in the range of $\theta$ which has been explored.

\subsection{Two drops with the same tip} 
\label{sec double drop ads4}

The second region that we consider can be obtained as the union $A= A_1 \cup A_2$ of two single drop regions $A_1$ and $A_2$,
where $A_1$ and $A_2$ have the same tip $W$, which is also the only element of their intersection, i.e. $A_1 \cap A_2= \{W\}$.
The boundary $\partial A$  is smooth except at the vertex $W$, where four lines join together. 
Considering the four adjacent corners with the common vertex $W$, let us denote by $\phi_1$ and $\phi_2$ the opening angles of the corners in $A_1$ and $A_2$ respectively
and by $\varphi_1$ and $\varphi_2$ the opening angles of the other two corners which do not belong to $A$.
We can assume $0<\phi_{1} \leqslant  \phi_{2}$ and $0<\varphi_{1} \leqslant  \varphi_{2}$ without loss of generality.
The configuration of the corners around $W$ can be characterised by the three angles $\vec{\phi} =(\phi_{1}, \varphi_{1} , \phi_{2})$
because the remaining one can be determined by the consistency condition $\phi_{1} + \phi_{2} + \varphi_{1} + \varphi_{2} = 2\pi$, 
where $\phi_{1} < \pi$ and $\varphi_{1} < \pi$.

\begin{figure}[t] 
\vspace{-1cm}
%\hspace{-.25cm}
\begin{center}
\includegraphics[width=.9\textwidth]{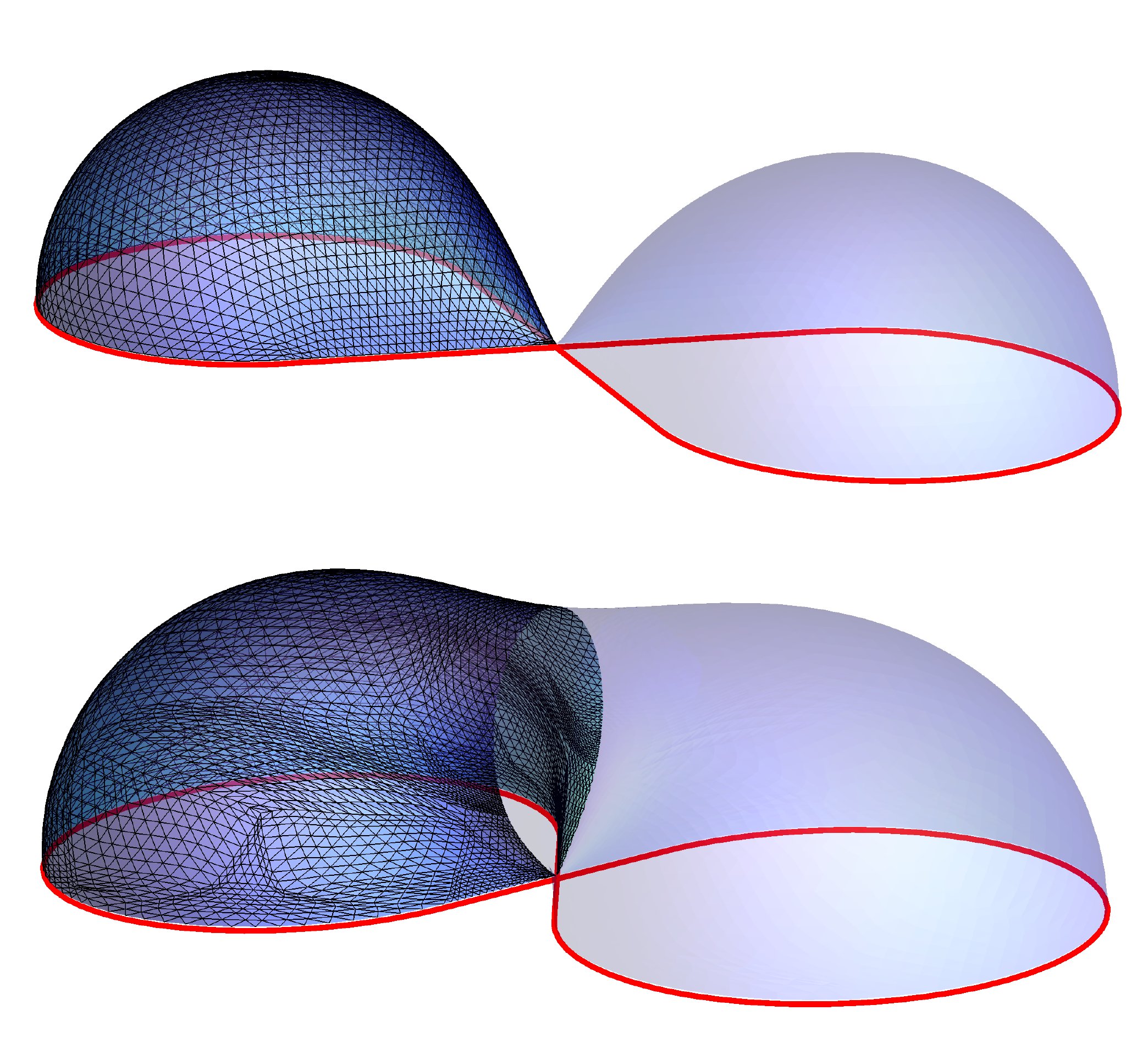}
\end{center}
\vspace{-.3cm}
\caption{\label{fig:double_drop}
\small
Triangulated surfaces in $\mathbb{H}_3$ approximating the minimal area surfaces $\hat{\gamma}_A$ which correspond to two different double drop regions $A$ described in Sec.\,\ref{sec double drop ads4}.
For these domains $\phi_1 = \phi_2 \equiv \phi$ and $\varphi_1 = \varphi_2  = \pi - \phi$.
The boundary $\partial A$ (red curve) belongs to the $z=0$ plane and the UV cutoff is $\varepsilon = 0.03$.
Top: $L=2$ and $\phi=1.4$ (below $\phi_c=\pi/2$).
Bottom: $L=1$ and $\phi=2.2$ (above $\phi_c = \pi/2$).
}
\end{figure}

The holographic entanglement entropy for this ``double drop'' region $A$ can be found from the general expressions  (\ref{hee ads4 corner intro}) and (\ref{Ftilde cft intro}).
The coefficient of the logarithmic divergence of $\mathcal{A}[\hat{\gamma}_\varepsilon]$ 
comes from the contribution of the vertex $W$ and it is given by $\widetilde{F}_{\textrm{\tiny tot}}  = \widetilde{\mathcal{F}}\big(\vec{\phi}\, \big)$.

Symmetric configurations can be considered by imposing constraints among the components of $\vec{\phi} $.
For instance, we can study domains such that $A_1$ and $A_2$ coincide after a proper rotation of one of them. 
In this cases the configuration of the corners at the common tip $W$ is determined by two parameters: the opening angle $\phi_1 = \phi_2 \equiv \phi$ and the relative orientation given by $\varphi_1$.
Let us stress that the coefficient of the logarithmic term is determined by the local configuration of corners around the vertex $W$ and it is not influenced by the shape of the entire domain $A$.

We consider first the configuration where the two drop regions $A_1$ and $A_2$ are symmetric with respect to their common tip $W$.
This means that $\phi_1 = \phi_2 \equiv \phi$ and also $\varphi_1 = \varphi_2 \equiv \varphi$.
The resulting domain $A$ is symmetric w.r.t. two orthogonal straight lines whose intersection point is $W$.
Since $\varphi + \phi= \pi$, the coefficient of the logarithmic divergence in (\ref{hee ads4 corner intro}) is determined only by the angle $\phi$ for these cases, namely
$\widetilde{F}_{\textrm{\tiny tot}}  = \widetilde{\mathcal{F}}(\phi) $.
In particular, it is not difficult to realise that for these configurations the corner function is given by \cite{Mozaffar:2015xue}
\be
\label{holog corner function W vertex symm}
\widetilde{\mathcal{F}}(\phi) 
\,=\, 
2\, \textrm{max} \Big\{ \widetilde{F}(\phi)  \,,\, \widetilde{F}(\pi- \phi)  \Big\}
\ee
being $\widetilde{F}(\phi) $ the corner function given by (\ref{cusp exact result}) and (\ref{theta_V expression}).
The factor $2$ in (\ref{holog corner function W vertex symm}) is due to the fact that the two opposite wedges provide the same contribution. 

A critical value $\phi_c$ for the common opening angle occurs when the two functions compared in (\ref{holog corner function W vertex symm}) takes the same value.
From  the arguments of the $\widetilde{F}$'s in (\ref{holog corner function W vertex symm}), it is straightforward to find that $\phi_c =\pi/2$.

In Fig.\,\ref{fig:double_drop} we show two triangulations obtained with Surface Evolver which approximate the corresponding minimal surface $\hat{\gamma}_A$ 
in the two cases of $\phi < \phi_c$ (top panel) and $\phi > \phi_c$ (bottom panel).
The crucial difference between them can be appreciated by focussing around the common tip $W$. 
Indeed, when $\phi < \phi_c$ the points of $\hat{\gamma}_A$ close to the tip have coordinates $(x,y) \in A$
and $\hat{\gamma}_A$ is made by the union of two minimal surfaces like the one in Fig.\,\ref{fig:single_drop}
which have the same tip.
Instead, when $\phi > \phi_c$ the points of $\hat{\gamma}_A$ close to the tip have coordinates $(x,y) \notin A$.
This leads to the expression (\ref{holog corner function W vertex symm}) for the coefficient of the logarithmic divergence in the expansion of $\mathcal{A}[\hat{\gamma}_\varepsilon]$.
The minimal surface $\hat{\gamma}_A$ is symmetric w.r.t. two half planes orthogonal to the $z=0$ plane whose boundaries are the two straight lines
which characterise the symmetry of $A$.
In Fig.\,\ref{fig:double_drop} the symmetry w.r.t. one of these two half planes is highlighted by the fact that the triangulation is shown only for half of the surface,
while the remaining half surface is shaded. This choice makes evident the curve given by the intersection between this half plane and $\hat{\gamma}_A$ when $\phi > \phi_c$.

\begin{figure}[t] 
\vspace{-1cm}
%\hspace{-.25cm}
\begin{center}
\includegraphics[width=.85\textwidth]{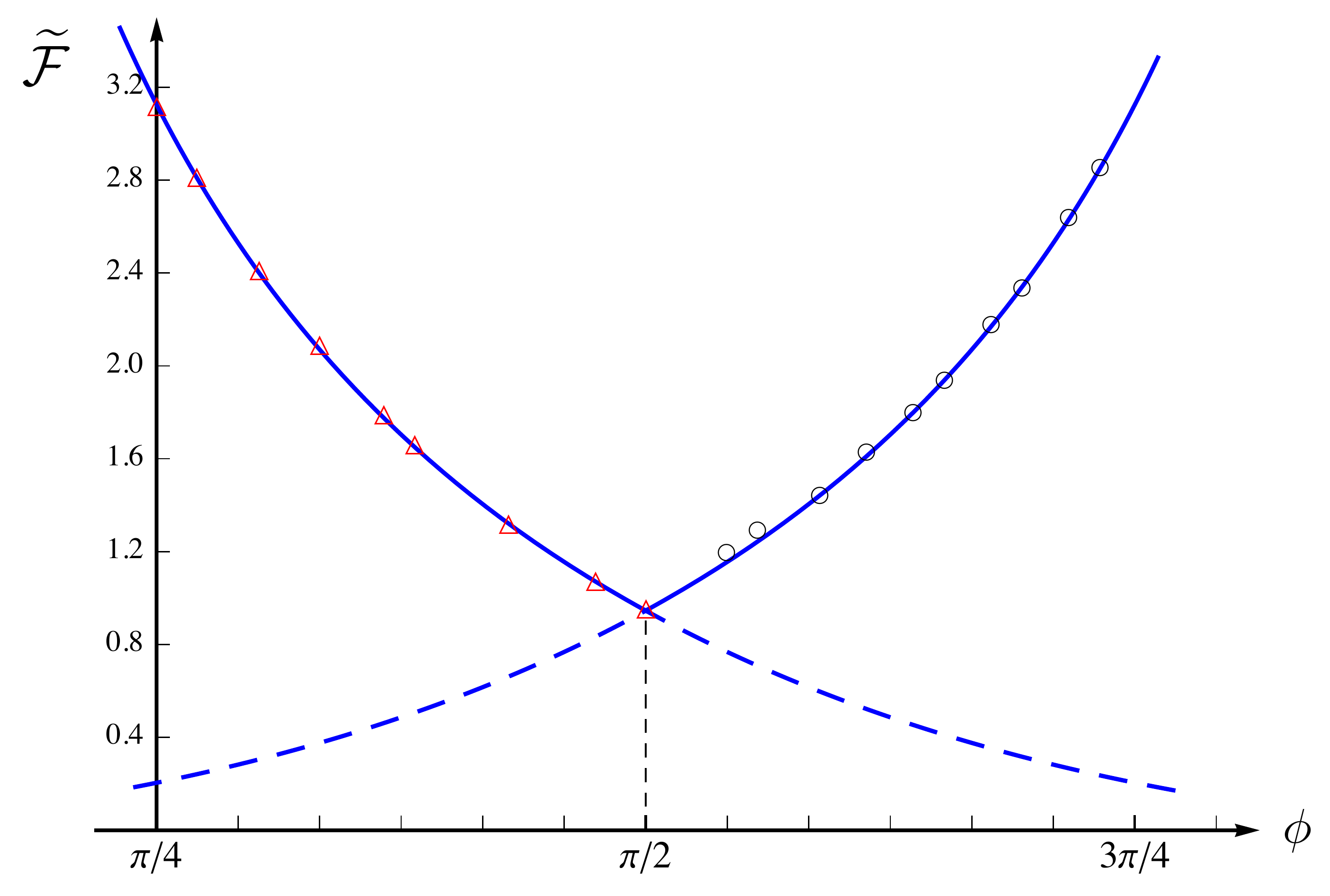}
\end{center}
\vspace{-.5cm}
\caption{\label{fig:drukker_double}
\small
Corner function for a vertex with four edges in AdS$_4$/CFT$_3$ in the symmetric case where
$\phi_1 = \phi_2 \equiv \phi$ and $\varphi_1 = \varphi_2 = \pi - \phi$ (see Sec.\,\ref{sec double drop ads4}).
The points labeled by the red triangles come from surfaces like the one in the top panel of Fig.\,\ref{fig:double_drop},
while the points labeled by the black empty circles are obtained from surfaces like the one in the bottom panel of Fig.\,\ref{fig:double_drop}.
The solid curve corresponds to the analytic expression (\ref{holog corner function W vertex symm}).
}
\end{figure}

In Fig.\,\ref{fig:drukker_double} we show the results of our numerical analysis for this kind of symmetric regions.
The points labeled by red triangles are obtained from triangulated surfaces like the one in the top panel of Fig.\,\ref{fig:double_drop},
while the points labeled by black circles correspond to triangulated surfaces like the one in the bottom panel of the same figure.
The solid blue curve in Fig.\,\ref{fig:drukker_double} is obtained from the analytic expression (\ref{holog corner function W vertex symm}).
The agreement of our numerical results with the expected analytic curve is very good. 
This strongly encourages us to apply this numerical method to study more complicated configurations.

\begin{figure}[t] 
\vspace{-1cm}
%\hspace{-.25cm}
\begin{center}
\includegraphics[width=.9\textwidth]{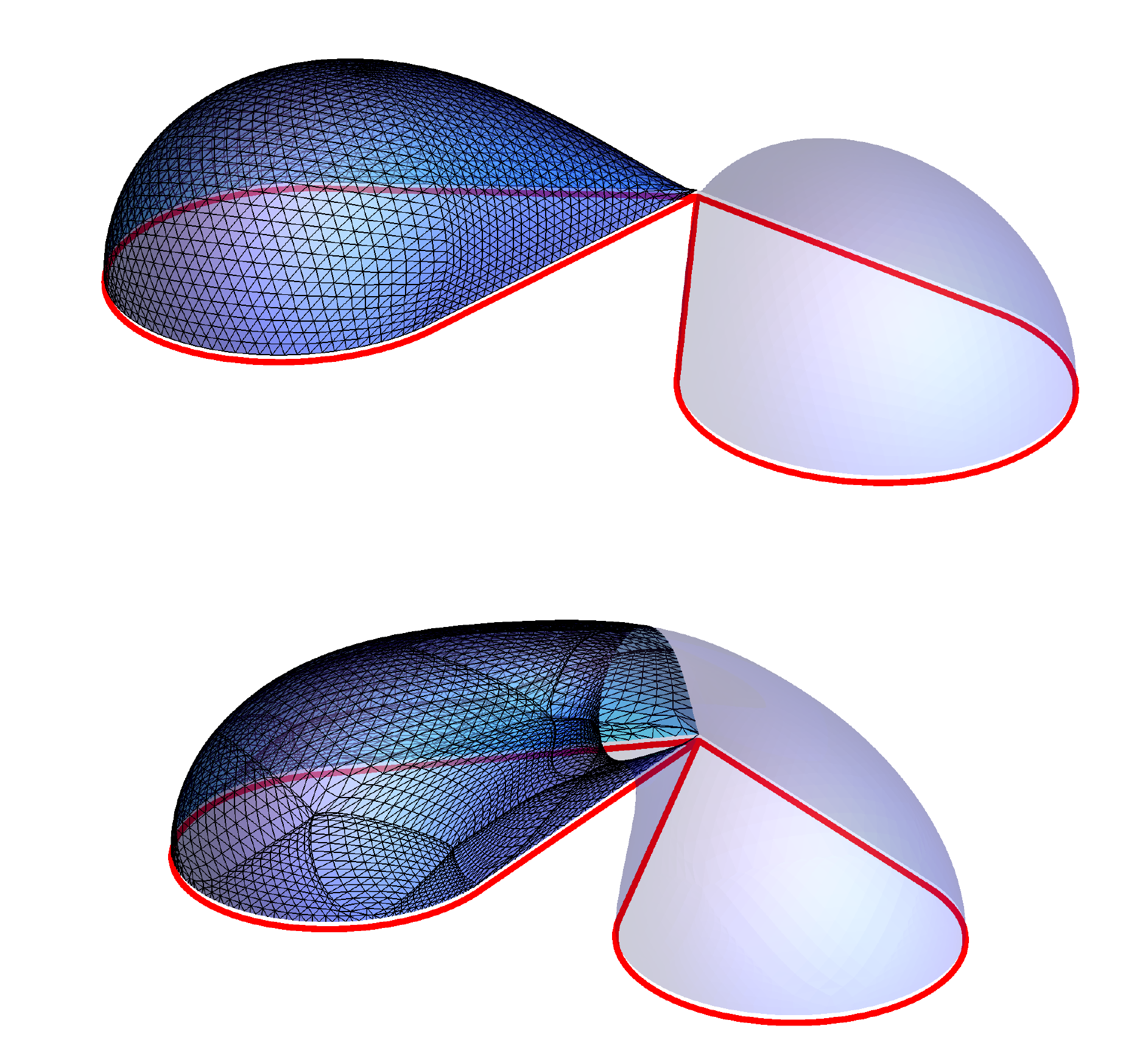}
\end{center}
\vspace{-.3cm}
\caption{\label{fig:double_drop_asym}
\small
Triangulated surfaces in $\mathbb{H}_3$ which approximate the minimal area surfaces $\hat{\gamma}_A$ 
corresponding to two different double drop regions $A$ 
which are symmetric w.r.t. a straight line passing through the vertex. 
For these domains $\phi_1 = \phi_2 \equiv \phi$ (see Sec.\,\ref{sec double drop ads4}).
The boundary $\partial A$ (red curve) belongs to the $z=0$ plane and the UV cutoff is $\varepsilon = 0.03$.
Top: $L=1.5$ with $\phi=0.9$ and $\varphi_1=0.671$.
Bottom: $L=1.5$ with $\phi=0.8$ and $\varphi_1=0.378$.
}
\end{figure}

Another class of symmetric configurations is made by double drop regions $A$ which are symmetric with respect to a straight line passing through the vertex $W$.
There are two possibilities: 
either the intersection between this straight line and $A$ is only the common tip (in this case $\phi_1 = \phi_2 \equiv \phi$)
or such intersection is given by a finite segment belonging to $A$ (in this case $\varphi_1 = \varphi_2 \equiv \varphi$).
In both these cases a constraint reduces the number of independent opening angles to two.
Focussing on the coefficient of the logarithmic divergence, one can consider the limit of infinite wedges and employ the property $S_A = S_B$ of the pure states
in this regime. This leads to conclude that these two options are equivalent and that the corresponding corner functions become the same because 
the property  $S_A = S_B$  allows to exchange $\varphi_j \leftrightarrow \phi_j $. 
Nonetheless, we find it instructive to discuss both of them separately because they look very different when $A$ is a finite domain.

As for the former class of configurations, by choosing the angles $\vec{\phi} = (\phi, \varphi_1)$ as independent variables, 
the remaining angle $\varphi_{2}$ is determined by the consistency condition $2\phi+ \varphi_{1} + \varphi_{2} =2\pi$.
The area of the minimal surface anchored to this kind of regions is given by (\ref{hee ads4 corner intro}) where 
$\widetilde{F}_{\textrm{\tiny tot}}  = \widetilde{\mathcal{F}}\big(\vec{\phi}\, \big)$ 
and the corner function reads
\be
\label{holog corner function W vertex symm3}
\widetilde{\mathcal{F}}\big(\vec{\phi}\,\big) 
\,=\, 
\textrm{max} \,\Big\{ 2\,\widetilde{F}(\phi) \,,\, \widetilde{F}(\varphi_{1} ) + \widetilde{F}(\varphi_{2} )  \Big\}
\ee
where we remind that $\widetilde{F}(\varphi_{2} )  = \widetilde{F}(\textrm{min}[\varphi_2 \,, 2\pi- \varphi_2] ) $.
Also this case has been considered in \cite{Mozaffar:2015xue}.
When the two expressions occurring in the r.h.s. of (\ref{holog corner function W vertex symm3}) are equal, a transition occurs. 
This condition determines a critical value $\varphi_{1,c}=\varphi_{1,c}(\phi)$ in terms of $\phi < \pi$.
In Fig.\,\ref{fig:double_drop_asym} we show two examples of minimal surfaces anchored to double drop regions which have this kind of symmetry.
In particular $\varphi_{1} > \varphi_{1,c} $ in the top panel and $\varphi_{1} < \varphi_{1,c} $ in the bottom panel.

Considering the second class of configurations introduced above, where $\varphi_1 = \varphi_2 \equiv \varphi$,
we have that $\phi_1 + \phi_2 + 2\varphi = 2\pi$ and therefore two angles fix the configurations of the
corners in the neighbourhood of the common tip.
One can choose e.g. $\vec{\phi} = (\phi_1, \phi_2)$.
For this kind of double drop domains the coefficient of the logarithmic divergence in the area (\ref{hee ads4 corner intro}) is 
$\widetilde{F}_{\textrm{\tiny tot}}  = \widetilde{\mathcal{F}}\big(\vec{\phi}\, \big)$  with
\be
\label{holog corner function W vertex symm2}
\widetilde{\mathcal{F}}\big(\vec{\phi}\,\big) 
\,=\, 
\textrm{max} \, \Big\{ \widetilde{F}(\phi_{1} ) + \widetilde{F}(\phi_{2} ) \,,\, 2 \,\widetilde{F}(\varphi)  \Big\}
\ee
As expected, also in this case two local solutions for the minimal surface exist and the global minimum provides the holographic entanglement entropy. 
The transition between the two kinds of solutions occurs when the two expressions in the r.h.s. of (\ref{holog corner function W vertex symm2}) are equal 
and this corresponds to a critical value for $\phi_{1,c} =\phi_{1,c}  (\varphi)$.
Notice that (\ref{holog corner function W vertex symm3}) and (\ref{holog corner function W vertex symm2}) exchange if $\phi_j \leftrightarrow \varphi_j$, as observed above.

For a generic double drop region $A$ we cannot employ symmetry arguments.
Only the constraint $\phi_{1} + \phi_{2} + \varphi_{1} + \varphi_{2} = 2\pi$ holds; 
therefore the configuration of corners at $W$ is determined by three independent angles, which are e.g. $\vec{\phi} = (\phi_1, \varphi_1, \phi_2)$.
The expansion of the area of the corresponding $\hat{\gamma}_\varepsilon$ is  (\ref{hee ads4 corner intro}) with
$\widetilde{F}_{\textrm{\tiny tot}}  = \widetilde{\mathcal{F}}\big(\vec{\phi}\, \big)$, with the corner function given by
\be
\label{holog corner function W vertex generic}
\widetilde{\mathcal{F}}\big(\vec{\phi}\,\big) 
\,=\, 
\textrm{max} \,\Big\{ 
\widetilde{F}(\phi_{1} ) + \widetilde{F}(\phi_{2} )  \,, 
\, \widetilde{F}(\varphi_{1} ) + \widetilde{F}(\varphi_{2} ) 
\Big\}
\ee
The transition occurs when the two expressions in the r.h.s. of (\ref{holog corner function W vertex generic}) are equal.
This condition provides a critical surface in the parameter space described by $(\phi_1, \varphi_1, \phi_2)$ with $\phi_1 \leqslant \phi_2$.

%%%%%%%%%%%%%%%%%%%%%%%%%%%%%%%%%%%%%%%%%%%%%%%%%%%%%
\section{Holographic entanglement entropy in AdS$_4$/BCFT$_3$}
\label{sec HEE bdy}

In this section we briefly review the AdS/BCFT 
construction introduced in \cite{Takayanagi:2011zk} and further expanded in \cite{Fujita:2011fp, Nozaki:2012qd}.
We mainly focus on the computation of the holographic entanglement entropy in the simplest setup 
where the boundary of the BCFT is a flat hyperplane.

Given a BCFT$_{d+1}$ in $d+1$ spacetime dimensions, the dual gravitational background proposed in \cite{Takayanagi:2011zk}  
is an asymptotically AdS$_{d+2}$ spacetime $\mathcal{M}$  restricted by the occurrence of a $d+1$ dimensional hypersurface $\widetilde{\mathcal{Q}}$ 
whose boundary coincides with the boundary of the BCFT$_{d+1}$.
We consider the simplified setup where the gravitational action reads \cite{Takayanagi:2011zk, Fujita:2011fp}
\be
\label{gravi-action}
\mathcal{I} = 
\frac{1}{16 \pi  G_{\textrm{\tiny N}}} 
\int_{\mathcal{M}} \sqrt{-g}\, \big( R- 2\Lambda \big) 
+
\frac{1}{8 \pi  G_{\textrm{\tiny N}}} 
\int_{\widetilde{\mathcal{Q}}} \sqrt{-h} \,\big( K- T\big)
\ee
being $\Lambda = - \tfrac{d(d+1)}{2 L^2_{\textrm{\tiny AdS}}}$ the negative cosmological constant,
$h_{ab}$ the induced metric on $\widetilde{\mathcal{Q}}$
and $K = h^{ab} K_{ab}$ the trace of the extrinsic curvature $K_{ab}$ of $\widetilde{\mathcal{Q}}$.
In our analysis the constant $T$ is a real parameter characterising the hypersurface $\widetilde{\mathcal{Q}}$.
We have omitted the boundary term due to the fact that 
the boundary of the gravitational spacetime is non smooth \cite{Hayward:1993my, Hawking:1996ww} along the boundary of the BCFT$_{d+1}$
and also the boundary terms introduced by the holographic renormalisation procedure \cite{holog-ren, Balasubramanian:1999re, Myers:1999psa, deHaro:2000vlm} 
because they are not relevant in our analysis.

In the AdS/BCFT setup of \cite{Takayanagi:2011zk}, the Neumann boundary conditions $K_{ab} = (K - T) h_{ab}$
have been proposed to determine the boundary $\widetilde{\mathcal{Q}}$.
Instead, in \cite{Miao:2017gyt, Chu:2017aab, Astaneh:2017ghi} it has been suggested that a consistent AdS/BCFT setup can be defined also by considering
the less restrictive boundary condition $ K = \frac{d+1}{d}\, T$ to find $\widetilde{\mathcal{Q}}$, obtained by taking the trace of the above Neumann boundary conditions.
When the boundary of the BCFT$_{d+1}$ is a flat $d$ dimensional hyperplane, these two prescriptions provide the same $\widetilde{\mathcal{Q}} \equiv \mathcal{Q}$.

In this manuscript we focus on the simplest case of a BCFT$_{d+1}$ in its ground state whose boundary is a flat $d$ dimensional hyperplane.
Hence, we find it convenient to introduce Cartesian coordinates $(t, x , \vec{y}\,)$ in the $d+1$ dimensional Minkowski spacetime such that
the BCFT$_{d+1}$ is defined in $x\geqslant 0$ and its boundary corresponds to $x=0$.
In \cite{Takayanagi:2011zk, Fujita:2011fp} it has been discussed that the gravitational spacetime $\mathcal{M}$ in the bulk 
dual to the ground state is AdS$_{d+2}$, whose metric in Poincar\'e coordinates reads
\be
\label{AdS metric ddim}
ds^2 \,=\, 
\frac{L^2_{\textrm{\tiny AdS}}}{z^2} \,\Big( -dt^2 + dz^2 + dx^2 + d\vec{y}^{\,2} \, \Big)
\qquad
z>0
\ee
where $d\vec{y}^{\,2}$ is the metric of $\mathbb{R}^{d-1}$, restricted by the half hyperplane $\mathcal{Q}$ given by\footnote{
Comparing our notation with the one adopted in \cite{Takayanagi:2011zk, Fujita:2011fp}, 
we have $\tan \alpha = 1/\sinh(\rho_\ast/L_{\textrm{\tiny AdS}})$, being $\rho_\ast \in \mathbb{R}$.
%\textcolor{red}{$\bullet$ Comparison with \cite{FarajiAstaneh:2017hqv}: find the relation between their $m$ and our $\alpha$ (see Jacopo's notes)}
}
\be
\label{brane-profile}
\mathcal{Q} \, :  \;\;\; 
x  =  - \,(\cot \alpha ) \,z
\qquad
\alpha \in (0,\pi)
\ee
whose boundary at $z=0$ coincides with the boundary of the BCFT$_{d+1}$, which is $x=0$.
The slope $\alpha \in (0,\pi)$ of the half hyperplane $\mathcal{Q} $ is related to the parameter $T$ in the gravitational action (\ref{gravi-action}) as $T= (d/ L_{\textrm{\tiny AdS}} ) \cos\alpha$.

In our analysis we mainly focus on a BCFT$_3$ defined for $x\geqslant 0$.
Hence, a $t=\textrm{const}$ slice of the gravitational bulk is (\ref{AdS4 metric}) constrained by the following condition 
\be
\label{AdS4 regions}
x\geqslant - \,(\cot \alpha ) \,z
\ee
which guarantees that the half plane defined by $z=0$ and $x\geqslant 0$ belongs to the boundary of the bulk spacetime.

Given this simple AdS$_4$/BCFT$_3$ setup, 
in this manuscript we are interested in the holographic entanglement entropy 
of some spatial simply connected regions $A$ defined in the spatial half plane $\{ (x,y) \,|\, x \geqslant 0 \}$.
We compute the holographic entanglement entropy by adapting the prescription (\ref{RT formula intro}) in the most natural way. 
Given a region $A$ in the $t=\textrm{const}$ section of the BCFT$_3$, let us split its boundary $\partial A$ as the union of 
$\partial A_{\textrm{\tiny bdy}}$ (see below (\ref{ee bcft3 corner intro})) and its complementary curve $\partial A \setminus \partial A_{\textrm{\tiny bdy}}$, 
which corresponds to the entangling curve.
In the bulk (\ref{AdS4 metric}) restricted by (\ref{AdS4 regions}), the holographic entanglement entropy is determined by 
the two dimensional minimal area surface $\hat{\gamma}_A$ anchored to the entangling curve. 

The latter condition becomes relevant whenever $ \partial A_{\textrm{\tiny bdy}}$ contains one dimensional curves.
When $\partial A_{\textrm{\tiny bdy}}$ is either empty or made by isolated points, 
the minimal surface $\hat{\gamma}_A$ anchored to the entire $\partial A$ must be considered. 
It is straightforward to find domains $A$ such that $\partial A_{\textrm{\tiny bdy}}$ contains both one dimensional lines and isolated points. 
For instance, for the domain $A$ in the right panel of Fig.\,\ref{fig:intro} the set $\partial A_{\textrm{\tiny bdy}}$ is made by the segment
$\overline{P_1 P_2}$ and the isolated point $Q_1$.

We find it worth remarking that for some domains $\hat{\gamma}_A \cap \mathcal{Q}$ is a non trivial curve. 
in these cases, since we do not impose any restriction on the intersection between $\hat{\gamma}_A$ and $\mathcal{Q}$, 
it is not difficult to show that $\hat{\gamma}_A$ intersects $\mathcal{Q}$ orthogonally along the curve $ \hat{\gamma}_A \cap \mathcal{Q}$.

Since the minimal surface $\hat{\gamma}_A$ constructed in this way reaches the half plane $z=0$, its area is infinite.
Thus, the holographic UV cutoff $\varepsilon$ must be introduced and the part of $\hat{\gamma}_A$ given by  $\hat{\gamma}_\varepsilon = \hat{\gamma}_A \cap \{ z \geqslant \varepsilon \}$
must be considered.
Indeed, the holographic entanglement entropy is obtained from the area of $\hat{\gamma}_\varepsilon $ as follows
\be
\label{RT formula bdy}
S_A = \frac{ \mathcal{A}[\hat{\gamma}_\varepsilon ] }{4 G_{\textrm{\tiny N}}}
\ee
where $G_{\textrm{\tiny N}}$ is the gravitational Newton constant corresponding to four dimensional spacetimes.

The generalisation of (\ref{RT formula bdy}) to a generic boundary and to a generic number of spacetime dimensions is straightforward
and the holographic entanglement entropy computed in this way gives $S_A = S_B$ for pure states.
Furthermore, the argument of \cite{Headrick:2007km} can be adapted to show that this prescription for the 
holographic entanglement entropy satisfies the strong subadditivity.

Focussing on the case of AdS$_4$/BCFT$_3$, we find it worth anticipating that for the domains $A$ in the $z=0$ half plane 
considered in the following, we find that the corresponding minimal surfaces  $\hat{\gamma}_A$ are part of auxiliary minimal surfaces 
$\hat{\gamma}_{A,\textrm{\tiny aux}} \subset \mathbb{H}_3$ anchored to the boundary of suitable auxiliary domains $A_{\textrm{\tiny \,aux}} \subset \mathbb{R}^2 = \partial  \mathbb{H}_3$.
In particular $\hat{\gamma}_A$ is the part of $\hat{\gamma}_{A,\textrm{\tiny aux}} $ identified by the constraint (\ref{AdS4 regions}).
We remark that $A \subsetneq A_{\textrm{\tiny \,aux}} $ for the spatial domains,
 $\hat{\gamma}_A \subsetneq \hat{\gamma}_{A,\textrm{\tiny aux}} $ for the minimal surfaces 
 and $( \partial A \setminus \partial A_{\textrm{\tiny bdy}} ) \subsetneq \partial A_{\textrm{\tiny \,aux}}$ for the entangling curves.

As anticipated in Sec.\,\ref{sec intro}, within the AdS$_4$/BCFT$_3$ setup described above, 
we are mainly interested in regions $A$ such that the expansion of the corresponding $\mathcal{A}[\hat{\gamma}_\varepsilon ] $ as $\varepsilon \to 0^+$ 
is given by (\ref{hee bdy corner intro}) (an example is depicted in the right panel of Fig.\,\ref{fig:intro}) because interesting pieces of information
could be encoded in the corner functions $F_\alpha(\gamma)$ and $\mathcal{F}_\alpha(\omega ,\gamma)$.
In this holographic context $\alpha$ has a geometrical meaning  because it provides the slope of $\mathcal{Q}$ in (\ref{brane-profile}).
Let us remark that understanding the possible holographic relation between the angle $\alpha$ and the conformally invariant boundary conditions of the BCFT$_3$ 
in the boundary is still an interesting question that deserves further analysis.

In the following we provide analytic expressions for the corner functions $F_\alpha(\gamma)$ and $\mathcal{F}_\alpha(\omega ,\gamma)$.
These functions are checked against numerical results obtained through an independent numerical analysis based on Surface Evolver.
Our confidence in this tool relies on the very good results obtained in Sec.\,\ref{sec drop} for the corner functions in AdS$_4$/CFT$_3$.
In the appendix\;\ref{app numerics} we briefly discuss some peculiar features which distinguish our numerical analysis 
from the one performed in \cite{Fonda:2014cca, Fonda:2015nma}, 
where Surface Evolver has been employed to study the shape dependence of the holographic entanglement entropy in AdS$_4$/CFT$_3$.

In \cite{FarajiAstaneh:2017hqv} the interesting possibility of including also the length of the curve $\hat{\gamma}_A \cap \mathcal{Q}$
in the definition of the holographic entanglement entropy in AdS$_4$/BCFT$_3$ has been explored.
This proposal is discussed in the Appendix\;\ref{sec app ssa}.

%%%%%%%%%%%%%%%%%%%%%%%%%%%%%%%%%%%%%%%%%%%%%%%%%%%%%
\section{The half disk and the infinite strip}
\label{sec two simple cases}

In this section we compute analytically the holographic entanglement entropy of two regions which are highly symmetric:
the half disk $A$ centered on the boundary (i.e. such that its diameter belonging to $\partial A$ lies on the boundary at $x=0$)
and an infinite strip parallel to the boundary, either adjacent to it or at a finite distance from it.

\subsection{Half disk centered on the boundary}
\label{sec:half disk}

Let us consider the half disk $A$ of radius $R$ whose center is located on the boundary of the BCFT$_3$.
In the Cartesian coordinates introduced above, where the boundary of the $z=0$ half plane is $x=0$, 
the translation invariance along the $y$ direction allows to choose the center of the half disk as the origin of the 
coordinates system. Thus, $A=\{ (x,y) \in \mathbb{R}^2 \, | \,x^2+y^2\leqslant R^2 ,  \,x\geqslant 0\}$.
In BCFT$_3$ the entanglement entropy of this domain has been studied in \cite{Jensen:2013lxa}, 
by using the method of \cite{Casini:2011kv}.

In our AdS$_4$/BCFT$_3$ setup the constraint (\ref{AdS4 regions}) due to the occurrence of the half plane $\mathcal{Q}$ must be taken into account. 
The key observation is that the hemisphere $x^2+y^2+z^2 = R^2$ in $\mathbb{H}_3$ intersects orthogonally the half plane $\mathcal{Q}$
along an arc of circumference of radius $R$ centered in the origin with opening angle equal to $\pi$. 
It is well known that this hemisphere is the minimal area surface anchored to the circular curve $x^2+y^2 = R^2$ in the $z=0$ plane
\cite{Maldacena:1998im, RT, Erickson:2000af}.
Thus, the minimal surface  $\hat{\gamma}_A$ corresponding to the half disk $A$ in presence of the brane $\mathcal{Q}$
is part of the minimal area surface $\hat{\gamma}_{A,\textrm{\tiny aux}}=\{ (x,y,z) \in \mathbb{H}_3 \,|\,x^2+y^2+z^2 = R^2 \}$ 
anchored to the boundary of the auxiliary domain $A_{\textrm{\tiny \,aux}} \subset \mathbb{R}^2 = \partial  \mathbb{H}_3$ given by a disk  of radius $R$ 
which includes $A$ as a proper subset. 
In particular $\hat{\gamma}_A$ is the part of $\hat{\gamma}_{A,\textrm{\tiny aux}} $ identified by the constraint (\ref{AdS4 regions}).

\begin{figure}[t] 
\vspace{-.8cm}
\hspace{-1cm}
%\begin{center}
\includegraphics[width=1.1\textwidth]{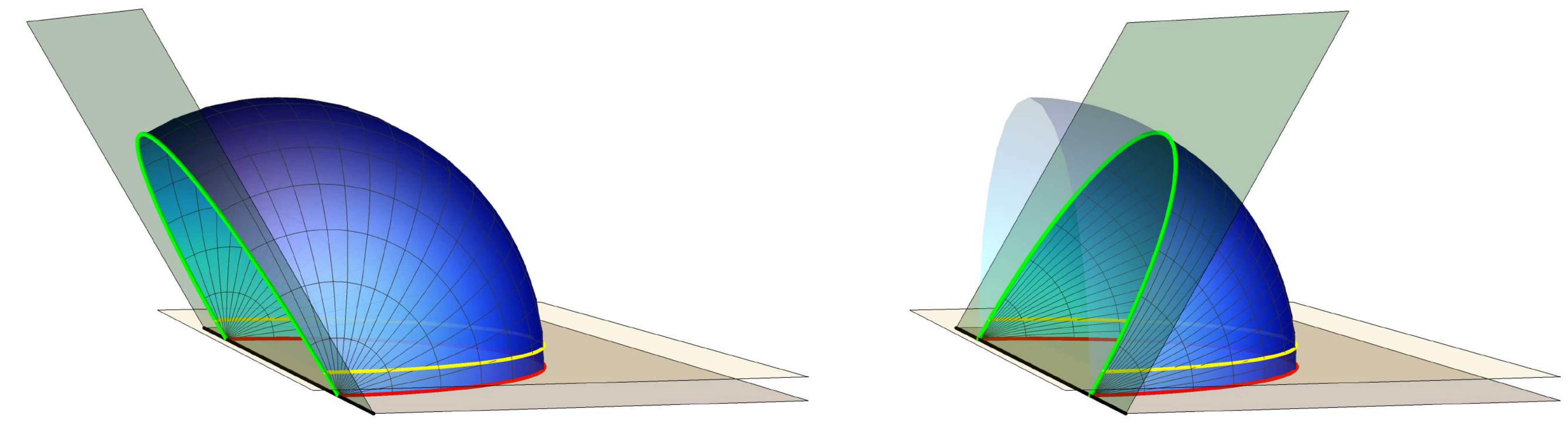}
%\end{center}
\vspace{-.1cm}
\caption{\label{fig:3Dhalf-disk}
\small
Minimal surfaces $\hat{\gamma}_A$ corresponding to the half disk centered on the boundary. 
The green half plane is $\mathcal{Q}$ in (\ref{brane-profile}), while the grey half plane is $z=0$.
In the left panel $\alpha < \pi/2$, while in the right panel $\alpha > \pi/2$.
The green curve is $\hat{\gamma}_A \cap \mathcal{Q}$ and the red curve is the entangling curve $\partial A \cap \partial B$, whose length
enters in the area law term of (\ref{area half-disk final main}).
The yellow half plane is defined by $z=\varepsilon$ and the yellow curve corresponds to its intersection with $\hat{\gamma}_A$.
}
\end{figure}

In Fig.\,\ref{fig:3Dhalf-disk} we show $\hat{\gamma}_A$ for a case having $\alpha < \pi/2$ in the left panel and for a case with $\alpha > \pi/2$ in the right panel.
Notice that the boundary of $\hat{\gamma}_A$ is a continuous curve made by two arcs whose opening angles are equal to $\pi$: 
the arc in the $z=0$ half plane defined by $\{ (x,y) \, |\, x^2+y^2 = R^2 , \, x\geqslant 0\}$ 
and the arc given by $\partial \hat{\gamma}_{\mathcal{Q}} \equiv \hat{\gamma}_A \cap \mathcal{Q}$.

Since $\hat{\gamma}_A$ reaches the boundary at $z=0$, its area is infinite; therefore we have to introduce the cutoff $\varepsilon >0$ and consider the area of the restricted surface
$\hat{\gamma}_\varepsilon = \hat{\gamma}_A \cap \{ z\geqslant \varepsilon \}$ as $\varepsilon \to 0^+$.
The details of this computation have been reported in the appendix\;\ref{sec app half disk}.
For a given $\alpha \in (0,\pi)$ we find
\be
\label{area half-disk final main}
\mathcal{A} [\hat{\gamma}_\varepsilon ] 
\,=\, 
 L^2_{\textrm{\tiny AdS}} \bigg(\frac{\pi R}{\varepsilon} + 2 (\cot \alpha) \log(R/\varepsilon) + O(1) \bigg) 
\ee
This expression is a special case of (\ref{hee bdy corner intro}) corresponding to $P_{A,B} = \pi R$ and $F_{\alpha, \textrm{\tiny tot}}  = 2F_\alpha(\pi/2)  $.
Thus, we have
\be
\label{Falpha pi/2 main}
F_\alpha(\pi/2)   = -  \cot\alpha
\ee

As consistency check, we observe that $F_{\pi/2}(\pi/2)  = 0$.
This is expected because (\ref{area half-disk final main}) for $\alpha = \pi/2$ gives half of the area of the hemisphere $x^2 + y^2 + z^2 = R^2$ restricted to $z\geqslant\varepsilon $ 
in $\mathbb{H}_3$.
Furthermore, by increasing the slope $\alpha$ of $\mathcal{Q}$ while $A$ is kept fixed, the area
$\mathcal{A} [\hat{\gamma}_\varepsilon ] $ in (\ref{area half-disk final main}) decreases because of the coefficient of the logarithmic divergence, as expected.

The result (\ref{Falpha pi/2 main}) can be obtained also by considering a bipartition whose entangling curve is a half straight line 
orthogonal to the boundary  \cite{FarajiAstaneh:2017hqv}.

\subsection{Infinite strip adjacent to the boundary}
\label{sec:strip}

A simple domain which plays an important role in our analysis is the infinite strip of finite width $\ell$ adjacent to the boundary,
namely such that one of its two edges coincides with the boundary $x=0$.
This region has been considered also in \cite{Chu:2017aab}.
In the following we present only the main results about the holographic entanglement entropy of this region in AdS$_4$/BCFT$_3$.
Their detailed derivation in AdS$_{d+2}$/BCFT$_{d+1}$ is reported in the appendix\;\ref{sec app strip}.

Considering the rectangular domain $A=\{(x,y) \in \mathbb{R}^2\, |\,0\leqslant x \leqslant \ell \,, \, 0\leqslant y \leqslant L_\parallel  \}$,
the infinite strip adjacent to the boundary is obtained by taking $L_\parallel  \gg \ell \gg \varepsilon$.
These assumptions allow to assume the invariance under translations in the $y$ direction and this symmetry
drastically simplifies  the problem of finding the minimal surface $\hat{\gamma}_A$ and its area because
$\hat{\gamma}_A$ is completely characterised by its profile $z=z(x)$ obtained through a section at $y=\textrm{const}$.

The minimal area surface $\hat{\gamma}_A$ intersects the $z=0$ half plane orthogonally along the line $x=\ell$ and this leads to the linear divergence $L_\parallel / \varepsilon$ (area law term) in its area.
Let us stress that the logarithmic divergence does not occur in this case.

When $\alpha \leqslant \pi/2$, two surfaces $\hat{\gamma}_{A}^{\textrm{\tiny \,dis}}$ and $\hat{\gamma}_{A}^{\textrm{\tiny \,con}}$ are local extrema of the area functional and 
the minimal surface $\hat{\gamma}_A$ is given by the global minimum. 
In particular, $\hat{\gamma}_{A}^{\textrm{\tiny \,dis}}$ is the half plane given by $x=\ell$, therefore it remains orthogonal to the $z=0$ plane and it does not intersect $\mathcal{Q}$ at a finite value of $z$,
while $\hat{\gamma}_{A}^{\textrm{\tiny \,con}}$ bends in the bulk towards the half plane $\mathcal{Q}$ until it intersects it orthogonally
at a finite value $z_\ast$ of the coordinate $z$.
It is straightforward to observe that the solution $\hat{\gamma}_{A}^{\textrm{\tiny \,dis}}$ does not exist for $\alpha > \pi/2$.

The surface $\hat{\gamma}_{A}^{\textrm{\tiny \,con}}$ can be also viewed as the part identified by the constraint (\ref{AdS4 regions})
of the auxiliary minimal surface $\hat{\gamma}_{A,\textrm{\tiny aux}} \subset \mathbb{H}_3$ anchored to the auxiliary infinite strip $A_{\textrm{\tiny \,aux}} \subset \mathbb{R}^2$ 
which includes $A$ and has one of its edges at $x=\ell$.
In the appendix\;\ref{sec app strip} the width of $A_{\textrm{\tiny aux}}$ has been computed (see (\ref{ell_aux strip ddim}) specialised to $d=2$).

\begin{figure}[t] 
\vspace{-.8cm}
%\hspace{-.25cm}
\begin{center}
\includegraphics[width=1.\textwidth]{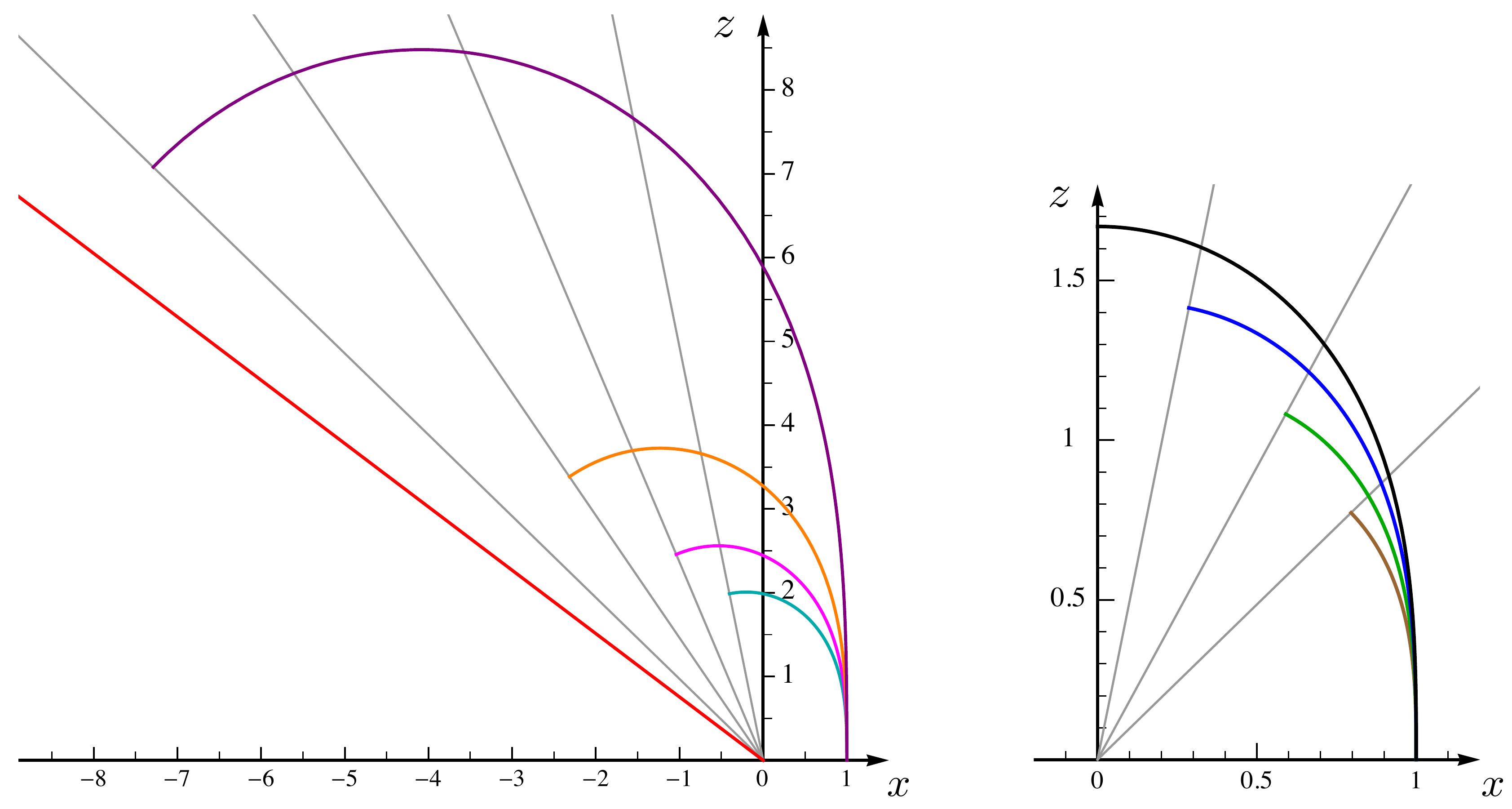}
\end{center}
\vspace{-.3cm}
\caption{\label{fig:strip_profile}
\small
Sections of minimal surfaces $\hat{\gamma}_A$ corresponding to an infinite strip adjacent to the boundary whose width is $\ell =1$
for different values of $\alpha > \alpha_c$, where $\alpha_c$ is given by (\ref{alpha_crit}).
This curves are obtained from (\ref{strip profile final}).
The grey half lines correspond to the sections of $\mathcal{Q}$ at $y=\textrm{const}$ obtained from (\ref{brane-profile}) 
and the red one is associated to $\alpha = \alpha_c$.
Each curve intersects orthogonally the corresponding section of $\mathcal{Q}$ at the point $P_\ast$, whose coordinate $z_\ast$ along the $z$ axis is (\ref{z_ast strip main}).
}
\end{figure}

Focussing on a section at $y=\textrm{const}$ of $\hat{\gamma}_{A}^{\textrm{\tiny \,con}}$, 
which is characterised by the profile $z(x)$, let us denote by $P_\ast = (x_\ast , z_\ast)$ 
 the intersection between this curve and the half line (\ref{brane-profile}) corresponding to $\mathcal{Q}$.
In the half plane described by the pair $(z,x)$, we find it convenient to write the curve $z(x)$ of $\hat{\gamma}_A$ in a parametric form $P_\theta = (x(\theta) , z(\theta))$ 
in terms of the angular variable $\theta \in [0,\pi -\alpha]$.
The angular variable $\theta$ corresponds to the angle between the outgoing vector normal to the curve given by $P_\theta$ and the $x$ semi-axis with $x\geqslant 0 $.
The parametric expressions $P_\theta$ must satisfy the boundary conditions $P_{0} = (\ell, 0)$ and $P_{\pi -\alpha} = P_\ast$.
Since $P_\ast$ lies on $\mathcal{Q}$, we have $x_\ast = - \,z_\ast \cot \alpha $; therefore we can write its position as $P_\ast = z_\ast ( -\cot\alpha \,, 1 )$.
In Fig.\,\ref{fig:strip_profile} we show the profile $z(x)$ corresponding to a given strip adjacent to the boundary for  different values of the slope $\alpha$ of $\mathcal{Q}$.
Notice that $z_\ast$ is a decreasing function of $\alpha$.

\begin{figure}[t] 
\vspace{-.8cm}
%\hspace{-.25cm}
\begin{center}
\includegraphics[width=.95\textwidth]{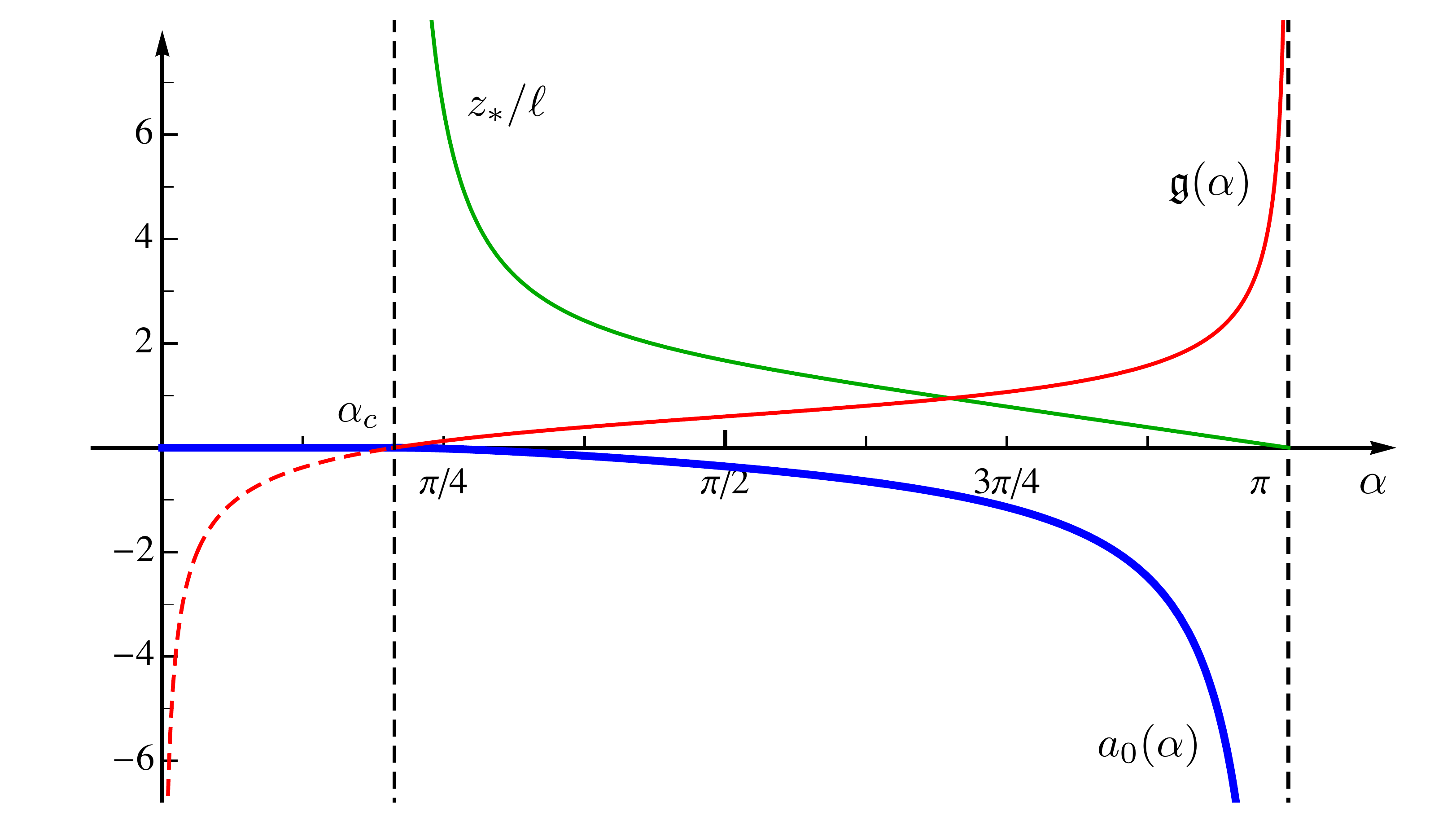}
\end{center}
\vspace{-.3cm}
\caption{\label{fig:strip_area}
\small
Infinite strip adjacent to the boundary:
The red curve is $\mathfrak{g}(\alpha) $ in (\ref{g function def main}), which is positive for $\alpha \geqslant \alpha_c$
and negative for $\alpha \leqslant \alpha_c$, being $\alpha_c$ given by (\ref{alpha_crit}).
The solid green curve corresponds to $z_\ast / \ell$ obtained from (\ref{z_ast strip main}) and it diverges as $\alpha \to \alpha_c^+$.
The solid blue line is the $O(1)$ term in the expansion (\ref{area strip final main all}) of the area $\mathcal{A} [\hat{\gamma}_\varepsilon ] $.
}
\end{figure}

In the appendix\;\ref{sec app strip} we find that, for any given slope $\alpha \in (0,\pi)$, the coordinate $z_\ast$ of $P_\ast$ 
is related to the width $\ell$ of the strip as follows
\be
\label{z_ast strip main}
z_\ast = \frac{\sqrt{\sin \alpha}}{\mathfrak{g}(\alpha)}\; \ell
\ee
where we have introduced
\be
\label{g function def main}
\mathfrak{g}(\alpha) 
\,\equiv\,
\mathbb{E}\big( \pi/4-\alpha/2 \, |\, 2 \big) - \frac{\cos\alpha}{\sqrt{\sin \alpha}} 
+ 
\frac{\Gamma\big(\tfrac{3}{4}\big)^2}{\sqrt{2\pi}}
\ee
being $\mathbb{E}( \phi | m ) $ the elliptic integral of the second kind. 
The expressions (\ref{z_ast strip main}) and (\ref{g function def main}) correspond respectively to (\ref{z_ast ddim}) and (\ref{g def ddim}) specialised to $d=2$.
In order to enlighten the notation, in the main text we slightly change the notation with respect to the appendix\;\ref{sec app strip} 
by setting $\mathfrak{g}(\alpha) \equiv \mathfrak{g}_2(\alpha) $ (see (\ref{g def ddim})).
In Fig.\,\ref{fig:strip_area} the function $\mathfrak{g}(\alpha) $ and the ratio $z_\ast / \ell$ are shown in terms of $\alpha \in (0,\pi)$.

As for the function $\mathfrak{g}(\alpha) $ in (\ref{g function def main}), we find $\mathfrak{g}(\alpha)  = -1/\sqrt{\alpha} +O(1)$ 
when $\alpha \to 0^+$ and $\mathfrak{g}(\alpha)  = 1/\sqrt{\pi- \alpha} +O(1)$ as $\alpha \to \pi^-$.
Moreover $\mathfrak{g}'(\alpha) = (\sin\alpha)^{-3/2}/2$ is positive in the whole domain $\alpha \in (0,\pi)$.
These observations imply that $\mathfrak{g}(\alpha) $ has a unique zero, namely
\be
\label{def alpha_c g}
\mathfrak{g}(\alpha_c)  = 0
\ee
where we have introduced $\alpha_c$ to label the unique solution of this transcendental equation.
Solving (\ref{def alpha_c g}) numerically, we find
\be
\label{alpha_crit}
\alpha_c \simeq \frac{\pi}{4.8525821} \simeq 0.647406
\ee
Since $z_\ast > 0$ in (\ref{z_ast strip main}), the condition (\ref{def alpha_c g}) defines the critical value for the slope $\alpha$
characterising the range of validity of (\ref{z_ast strip main}), which is well defined only for $\alpha \in (\alpha_c, \pi)$.
Thus, for $\alpha \leqslant \alpha_c$ the solution $\hat{\gamma}_{A}^{\textrm{\tiny \,con}}$ does not exist 
and therefore $\hat{\gamma}_A = \hat{\gamma}_{A}^{\textrm{\tiny \,dis}}$.
This is confirmed also by the fact that, by taking  $\alpha \to \alpha_c^+$ in (\ref{z_ast strip main}) we have $z_\ast \to + \infty$.
The occurrence of the critical value (\ref{alpha_crit}) has been observed also in \cite{Chu:2017aab}.

When $  \alpha > \alpha_c$, the extremal surface $\hat{\gamma}_{A}^{\textrm{\tiny \,con}}$ is parametrically described by the following curve 
\be
\label{strip profile final}
P_\theta 
=
\big( x(\theta)\,, z(\theta)   \big)  
\,=\,
\frac{\ell}{\mathfrak{g}(\alpha)} 
\left(
   \mathbb{E}\big( \pi/4-\alpha/2 \, |\, 2 \big) - \frac{\cos\alpha}{\sqrt{\sin \alpha}} 
   +
  \mathbb{E}\big( \pi/4-\theta/2 \, |\, 2 \big) 
  \, , \, \sqrt{\sin \theta} \;
  \right)
\ee
where the independent angular parameter is $0\leqslant \theta \leqslant \pi -\alpha$.
The profile (\ref{strip profile final}) corresponds to (\ref{profile strip adj ddim}) specialised to $d=2$.
It is straightforward to check that (\ref{strip profile final}) fulfils the required boundary conditions $P_{0} = (\ell, 0)$ and $P_{\pi -\alpha} = P_\ast = z_\ast ( -\cot\alpha \,, 1 )$,
with $z_\ast$ given by (\ref{z_ast strip main}). 
In Fig.\,\ref{fig:strip_profile} we show the profiles $z(x)$ for $\hat{\gamma}_A$ obtained from (\ref{strip profile final}) which correspond to the same strip adjacent to the boundary in the
$z=0$ half plane ($\ell =1$ in the figure) and different values of $\alpha$.
As for the maximum value $z_{\textrm{\tiny max}}$ reached by the coordinate $z$ along the curve (\ref{strip profile final}), 
we observe that $z_{\textrm{\tiny max}} = z_\ast$ when $\alpha \in [\pi/2, \pi)$, while $z_{\textrm{\tiny max}} > z_\ast$ for $\alpha \in (\alpha_c , \pi/2)$.

The expansion for $\varepsilon \to 0^+$ of the area of the extremal surface corresponding to the infinite strip adjacent to the boundary and 
characterised by the curve (\ref{strip profile final}) restricted to $z\geqslant \varepsilon$  reads
\be
\label{area strip final main}
\mathcal{A} [\hat{\gamma}_\varepsilon ] 
\,=\,
L^2_{\textrm{\tiny AdS}} \, L_\parallel \,
\bigg(\,
\frac{1}{\varepsilon}
-
\frac{\mathfrak{g}(\alpha)^2}{ \ell } 
+ O(\varepsilon^3) 
\bigg)
  \qquad
  \alpha > \alpha_c
\ee
This expression is the special case $d=2$ of (\ref{area strip d-dim app final}).
Comparing (\ref{area strip final main}) with (\ref{hee bdy corner intro}), we have that
in this case $P_{A,B} = L_\parallel$, the logarithmic divergence does not occur and the $O(1)$ term is negative.
The result (\ref{area strip final main}) restricted to $\alpha \in (\pi/2, \pi)$ has been first found in \cite{Nagasaki:2011ue}\footnote{
Comparing with the notation of \cite{Nagasaki:2011ue}, we find that $ \kappa_{\textrm{\tiny there}} = -\cot\alpha\,$.
}.

An important role in our analysis is played by the extremal surface $\hat{\gamma}_{A}^{\textrm{\tiny \,dis}}$ given by the vertical half plane at $x=\ell$.
By computing its area restricted to $\varepsilon \leqslant z \leqslant z_{\textrm{\tiny IR}}$, being $z_{\textrm{\tiny IR}} \gg \ell$ an infrared cutoff, one easily finds that
$\mathcal{A} [\hat{\gamma}_\varepsilon ]  = L^2_{\textrm{\tiny AdS}} L_\parallel (1 / \varepsilon - 1/z_{\textrm{\tiny IR}} )$.
Notice that the $O(1)$ term of this expression vanishes in the limit $z_{\textrm{\tiny IR}}  \to + \infty$.
This extremal surface exists only for $\alpha \leqslant \pi/2$ because when $\alpha > \pi/2$ the half plane $\mathcal{Q}$ and the vertical infinite strip
$x=\ell$ do not intersect orthogonally.

Summarising, for the minimal area surface $\hat{\gamma}_A $ we have that 
$\hat{\gamma}_A = \hat{\gamma}_{A}^{\textrm{\tiny \,dis}}$ when $\alpha \leqslant \alpha_c$ because (\ref{z_ast strip main}) is not well defined.
When $\alpha \in (\alpha_c, \pi/2]$, two extremal surfaces $\hat{\gamma}_{A}^{\textrm{\tiny \,dis}}$ and $\hat{\gamma}_{A}^{\textrm{\tiny \,con}}$ compete 
(the vertical half plane at $x=\ell$ and the surface characterised by (\ref{strip profile final}) respectively),
while for $\alpha > \pi/2$ we have $\hat{\gamma}_A = \hat{\gamma}_{A}^{\textrm{\tiny \,con}}$ because $\hat{\gamma}_{A}^{\textrm{\tiny \,dis}}$ does not exist. 
As for the regime $\alpha \in (\alpha_c, \pi/2]$, since the $O(1)$ term in (\ref{area strip final main}) is negative 
while it vanishes for  $\hat{\gamma}_{A}^{\textrm{\tiny \,dis}}$, 
we conclude that $\hat{\gamma}_A = \hat{\gamma}_{A}^{\textrm{\tiny \,con}}$, given by (\ref{strip profile final}).

Combining the above observations, we find that the expansion as $\varepsilon \to 0^+$ of the area of the minimal surface $\hat{\gamma}_A \cap \{z\geqslant \varepsilon\}$ 
corresponding to an infinite strip of width $\ell$ adjacent to the boundary for $\alpha \in (0,\pi)$ is
\be
\label{area strip final main all}
\mathcal{A} [\hat{\gamma}_\varepsilon ] 
\,=\,
L^2_{\textrm{\tiny AdS}} \, L_\parallel \,
\bigg(\,
\frac{1}{\varepsilon}
+
\frac{a_0(\alpha)}{ \ell } 
+ o(1)
\bigg)
  \hspace{.2cm}  \qquad \hspace{.2cm} 
a_0(\alpha) =
\left\{\begin{array}{ll}
-\,\mathfrak{g}(\alpha)^2 \hspace{.7cm} &   \alpha \geqslant \alpha_c
\\
\rule{0pt}{.5cm}
\;\;\;\; 0 &   \alpha  \leqslant \alpha_c
\end{array}\right.
\ee
where $\mathfrak{g}(\alpha)$ has been defined in (\ref{g function def main}) and $\alpha_c$ is its unique zero (\ref{alpha_crit}).
The result (\ref{area strip final main all}) is the special case $d=2$ of the expressions (\ref{area strip d-dim app final}) and (\ref{a0 ddim def}).
Since $\alpha_c$ is defined by (\ref{def alpha_c g}), the function $a_0(\alpha) $ in (\ref{area strip final main all}) is continuous
and it corresponds to the blue solid curve in Fig.\,\ref{fig:strip_area}.
Let us also observe  that $\mathfrak{g}'(\alpha)$ is continuous but $\mathfrak{g}''(\alpha)$ is not continuous at $\alpha = \alpha_c$.

\subsection{Infinite strip parallel to the boundary}
\label{sec:strip distant}

The results for the infinite strip adjacent to the boundary discussed  in Sec.\,\ref{sec:strip} 
allow to address also the holographic entanglement entropy 
of an infinite strip $A$ parallel to the boundary and at finite distance from it. 
In the appendix\;\ref{sec app strip disjoint} we discuss the analogue case in a BCFT$_{d+1}$.
In the following we report only the results of that analysis for $d=2$.

\begin{figure}[t] 
\vspace{-1cm}
%\hspace{-.25cm}
\begin{center}
\includegraphics[width=.8\textwidth]{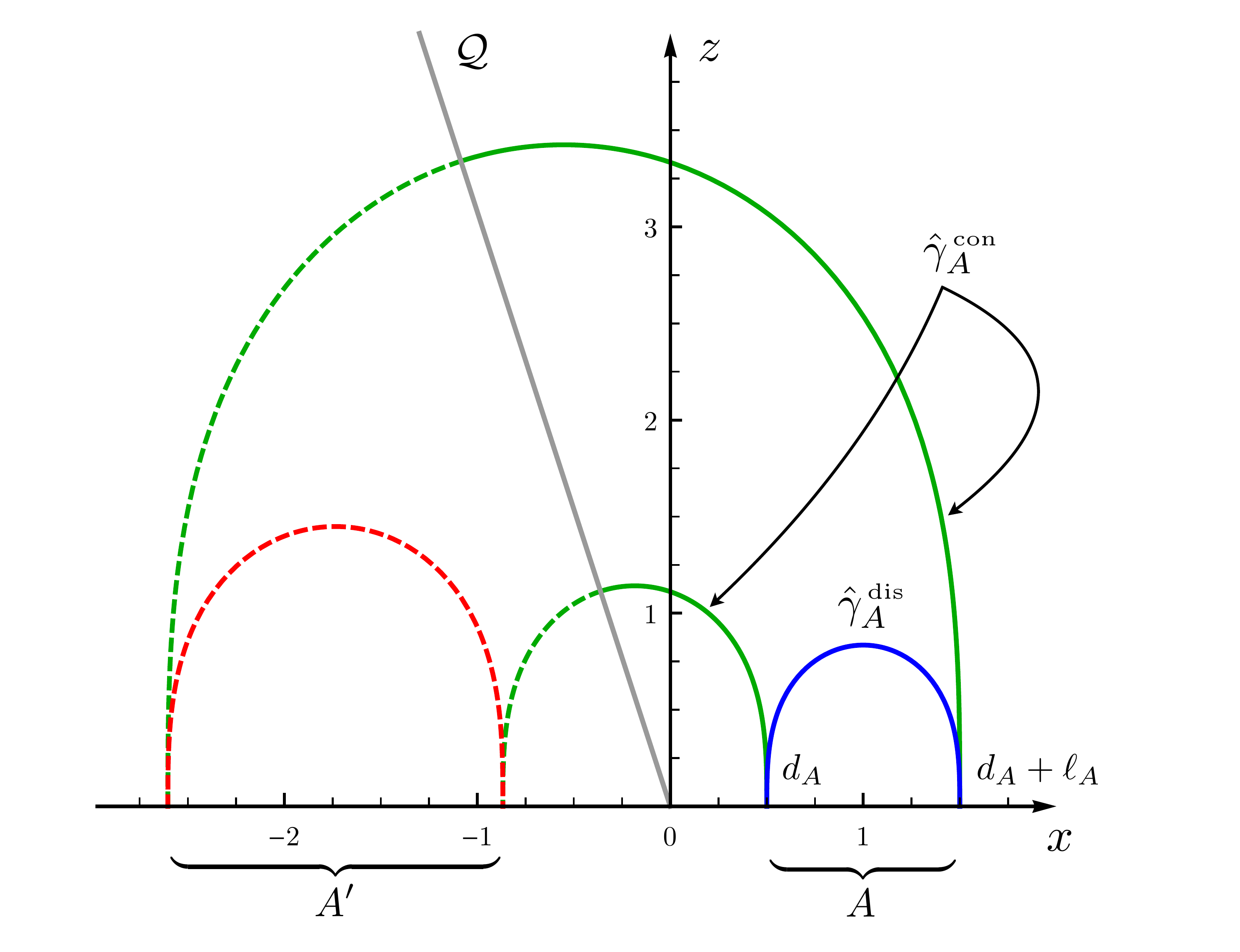}
\end{center}
\vspace{-.6cm}
\caption{\label{fig:2strips}
\small
Infinite strip of width $\ell_A$ parallel to the boundary at distance $d_A$:
Section of the surfaces  $\hat{\gamma}_{A}^{\textrm{\tiny \,dis}}$ and $\hat{\gamma}_{A}^{\textrm{\tiny \,con}}$ (blue and green solid curve respectively) 
which are local extrema of the area functional.
In this plot $\alpha > \alpha_c$.
The auxiliary domain $A_{\textrm{\tiny \,aux}} = A\cup A'$ in $\mathbb{R}^2$ is made by two parallel infinite strips $A$ and $A'$.
The green dashed curves together with $\hat{\gamma}_{A}^{\textrm{\tiny \,con}}$ provide $\hat{\gamma}_{A,\textrm{\tiny aux}}$  
when $\hat{\gamma}_{A} = \hat{\gamma}_{A}^{\textrm{\tiny \,con}}$,
while the red dashed curve together with $\hat{\gamma}_{A}^{\textrm{\tiny \,dis}}$ 
gives $\hat{\gamma}_{A,\textrm{\tiny aux}}$  when $\hat{\gamma}_{A} = \hat{\gamma}_{A}^{\textrm{\tiny \,dis}}$.
}
\end{figure}

The configuration of an infinite strip parallel to the boundary is characterised by the width $\ell_A$ of the strip and by its distance $d_A$ from the boundary.
By employing the translation invariance and  the results of Sec.\,\ref{sec:strip},
one realises that $\hat{\gamma}_A$ is the global minimum obtained by comparing the area of two possible configurations $\hat{\gamma}_{A}^{\textrm{\tiny \,dis}}$ and $\hat{\gamma}_{A}^{\textrm{\tiny \,con}}$.
The surface $\hat{\gamma}_{A}^{\textrm{\tiny \,dis}}$ is disconnected from $\mathcal{Q}$ and it connects the two parallel lines of $\partial A$ through the bulk,
while $\hat{\gamma}_{A}^{\textrm{\tiny \,con}}$ is made by two disjoint surfaces such that each of them connects an edge of $\partial A$ to $\mathcal{Q}$.
The two disjoint surfaces occurring in $\hat{\gamma}_{A}^{\textrm{\tiny \,con}}$ are like the ones described in Sec.\,\ref{sec:strip}; therefore $\hat{\gamma}_{A}^{\textrm{\tiny \,con}} \cap \mathcal{Q}$ 
is made by two parallel lines.
The two configurations $\hat{\gamma}_{A}^{\textrm{\tiny \,dis}}$ and $\hat{\gamma}_{A}^{\textrm{\tiny \,con}}$ are depicted in Fig.\,\ref{fig:2strips} for a given value of $\alpha$

For an infinite strip $A$ at a finite distance from the boundary, 
$\hat{\gamma}_{A,\textrm{\tiny aux}}$ is the minimal surface in $\mathbb{H}_3$ anchored to $A_{\textrm{\tiny \,aux}} = A\cup A' \subset \mathbb{R}^2$, 
which is the union of two parallel and disjoint infinite strips in $\mathbb{R}^2$  \cite{Tonni:2010pv}.
The minimal surface $\hat{\gamma}_{A}$ is the part of $\hat{\gamma}_{A,\textrm{\tiny aux}}$ identified by the constraint (\ref{AdS4 regions}).
The width of $A'$ and the separation between $A$ and $A'$ are given by (\ref{ell_aux strip ddim disjoint}) specialised to the case $d=2$.

As for the area of $\hat{\gamma}_\varepsilon $, we find
\be
\label{area strip disjoint}
\mathcal{A} [\hat{\gamma}_\varepsilon ]
\,=\,
L^2_{\textrm{\tiny AdS}} \, 
L_\parallel  \left(\,
 \frac{2}{\varepsilon}
 + \frac{1}{\ell_A}\,
 \textrm{min} \bigg[\,h_2 \, , \, a_{0}(\alpha) \bigg( \frac{1}{\delta_A} + \frac{1}{\delta_A + 1} \bigg)
\bigg]
+ o(1)
\right)
\qquad
\delta_A \equiv \frac{d_A}{\ell_A}
\ee
where $a_{0}(\alpha)$ has been introduced in (\ref{area strip final main all}) and 
$h_2 \equiv -\, 4\pi \big[\Gamma(\tfrac{3}{4}) / \Gamma(\tfrac{1}{4}) \big]^2$ comes from  the $O(1)$ term of the holographic entanglement entropy of an infinite strip in CFT$_3$ \cite{RT}.
The expression (\ref{area strip disjoint}) corresponds to the special case $d=2$ of (\ref{area strip disjoint d-dim}).
When $\alpha \leqslant \alpha_c$, we have that $\hat{\gamma}_A = \hat{\gamma}_{A}^{\textrm{\tiny \,dis}}$ because $a_{0}(\alpha)  =0$ and $h_2 < 0$.

The critical configuration corresponds to the value $\delta_A =\delta_{A,c}$ such that the two terms occurring in the minimisation procedure in (\ref{area strip disjoint}) provide the same result.
By imposing this condition, one finds an algebraic equation of second order with only one positive root given by\footnote{
 The formula (\ref{delta_c d=2}) (first presented in \cite{talk}) restricted to $\alpha \geqslant \pi/2$
 corresponds to a special case of a result concerning 
 the expectation value of antiparallel Wilson lines in $\mathcal{N} = 4$ SYM at strong coupling in the presence of a defect, 
 which has appeared later in \cite{Preti:2017fhw}.
 In particular, (\ref{delta_c d=2}) can be obtained from Eq.\,(4.15) of \cite{Preti:2017fhw} by setting $\phi\big|_{\textrm{\tiny there}} = \pi/2$ and 
 $\chi_+\big|_{\textrm{\tiny there}} = \chi_-\big|_{\textrm{\tiny there}} = \pi/2$.
}
\be
\label{delta_c d=2}
\delta_{A,c}
= \frac{1}{2} \bigg(  \sqrt{4 \,\big[ a_{0}(\alpha)/h_2 \big]^2 + 1 } + 2 \,a_{0}(\alpha)/h_2  -1  \bigg)
\ee
When $\delta_A \leqslant \delta_{A,c}$ the minimal surface is $\hat{\gamma}_A = \hat{\gamma}_{A}^{\textrm{\tiny \,con}}$, 
while for $\delta_A \geqslant \delta_{A,c}$ it is given by $\hat{\gamma}_A = \hat{\gamma}_{A}^{\textrm{\tiny \,dis}}$.
The function (\ref{delta_c d=2}) corresponds to the red curve in Fig.\,\ref{fig:delta_critical} and it is meaningful for $\alpha \geqslant \alpha_c$.

%%%%%%%%%%%%%%%%%%%%%%%%%%%%%%%%%%%%%%%%%%%%%%%%%%%%%
\section{Infinite wedge adjacent to the boundary}
\label{sec:wedge}

In this section we discuss the main result of this manuscript.
In the AdS$_4$/BCFT$_3$ setup that we are considering, 
we compute the minimal surface $\hat{\gamma}_A$ corresponding to an infinite wedge with opening angle $\gamma \in (0, \pi/2]$
having one of its edges on the boundary of the BCFT$_3$.
By evaluating the area of $\hat{\gamma}_\varepsilon$, 
an analytic expression for the corner function $F_\alpha(\gamma)$ occurring in (\ref{Ftot bdy corner intro}) is obtained. 
In the following we report only the main results of our analysis, while the technical details of their derivations are collected in the appendix\;\ref{sec app wedge}.

Let us adopt the polar coordinates $(\rho,\phi)$  given by  $x= \rho \, \sin\phi $ and $ y= \rho \, \cos\phi $, 
being $(x,y)$ the Cartesian coordinates employed in Sec.\,\ref{sec two simple cases} for the $t=\textrm{const}$ slice of the BCFT$_3$.
In terms of these polar coordinates, we consider the domain  $A = \big\{ (\rho, \phi) \,|\, 0\leqslant \phi \leqslant \gamma\,, \rho \leqslant L \big\}$ 
with $L \gg \varepsilon$, which is  an infinite wedge with one of its two edges on the boundary.
Since the wedge is infinite, we can look for the corresponding minimal surface $\hat{\gamma}_A$ among the surfaces described by the following ansatz
\be
\label{cusp profile main}
z=\frac{\rho}{q(\phi)}
\ee
where $q(\phi)>0$, as already done in \cite{Drukker:1999zq} to get the minimal surface in $\mathbb{H}_3$ anchored to an infinite wedge in $\mathbb{R}^2$.

The minimal surface $\hat{\gamma}_A$ can be found as part of an auxiliary minimal surface $\hat{\gamma}_{A,\textrm{\tiny aux}}$ 
embedded in $\mathbb{H}_3$ and anchored to an auxiliary infinite wedge $\hat{\gamma}_{A,\textrm{\tiny aux}}$ containing $A$ and having the same edge $\{(\rho,\phi)\,|\, \phi=\gamma \}$.
The minimal surface $\hat{\gamma}_A$ intersects orthogonally  
the half plane at $z=0$ along the edge $\{(\rho,\phi)\,|\, \phi=\gamma \}$ of $A$
and the half plane $\mathcal{Q}$ along the half line given by $\phi = \phi_\ast$.
As remarked for the previous cases, $\hat{\gamma}_A$  is the part of $\hat{\gamma}_{A,\textrm{\tiny aux}}$ identified by the constraint (\ref{AdS4 regions}).
For the infinite wedge $A$ that we are considering, $A_{\textrm{\tiny \,aux}}$ is a suitable infinite wedge in $\mathbb{R}^2$ and $\hat{\gamma}_{A,\textrm{\tiny aux}}$ is the 
corresponding minimal surface found in \cite{Drukker:1999zq}.
In the Fig.\,\ref{fig_app:wedge} described in the appendix\;\ref{sec app wedge} the auxiliary wedge $A_{\textrm{\tiny \,aux}}$ is shown.

Given the half plane $\mathcal{Q}$ described by (\ref{brane-profile}), whose slope is $\alpha \in (0, \pi)$, 
the angle $\phi_\ast$ which identifies the half line $\hat{\gamma}_A \cap \mathcal{Q}$ can be defined by introducing the following positive function
\be
\label{s_ast def main}
s_\ast(\alpha,q_0) 
 \equiv\,
 -\,\eta_\alpha  \frac{\cot \alpha}{\sqrt{2}}   
 \left\{ \frac{\sqrt{1+4 (\sin\alpha)^2 (q_0^4+q_0^2) } - \cos (2 \alpha) }{(\cos\alpha)^2 + q_0^4+q_0^2}
 \right\}^{\frac{1}{2}}
 \qquad
\eta_\alpha \equiv - \, \textrm{sign}(\cot\alpha)
\ee
where $q(\phi_0) \equiv q_0 >0 $ is the value of the function $q(\phi)$ at the angle $\phi=\phi_0$ corresponding to the bisector of the auxiliary wedge $A_{\textrm{\tiny \,aux}}$.
We find it convenient to adopt $q_0$ as parameter to define various quantities in the following. 
From (\ref{s_ast def main}), we find $\phi_\ast$ as 
\be
\label{phi_ast q0 main}
\phi_\ast(\alpha,q_0) =     \, \eta_\alpha \arcsin [  s_\ast(\alpha, q_0) ]
\ee
This result encodes the condition that $\hat{\gamma}_A$ intersects $\mathcal{Q}$ orthogonally, 
as explained in the appendix\;\ref{sec app wedge intersection}.

\begin{figure}[t] 
\vspace{-.8cm}
\hspace{-1.05cm}
%\begin{center}
\includegraphics[width=1.1\textwidth]{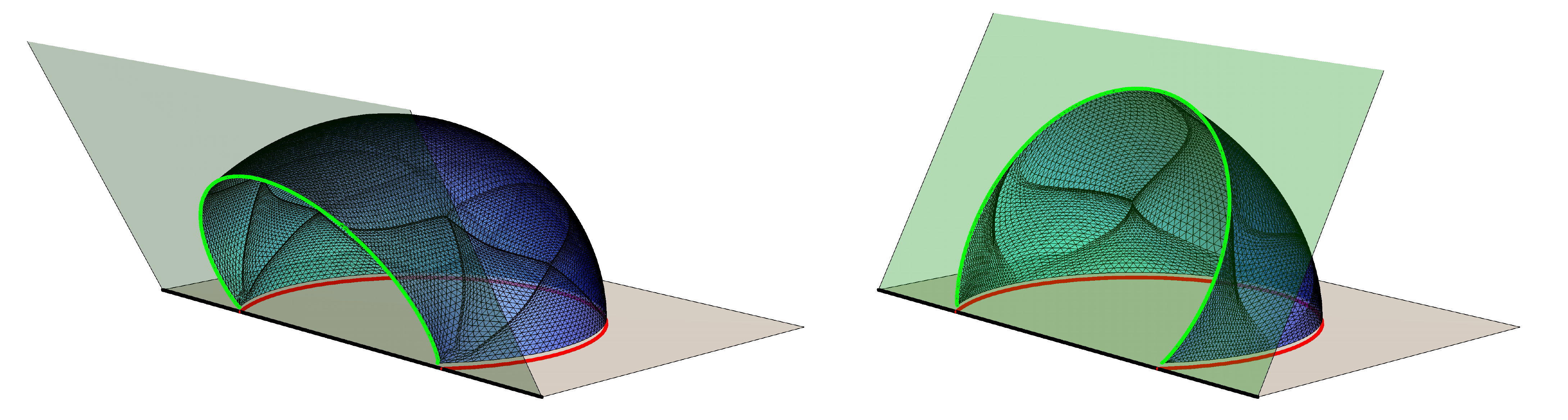}
%\end{center}
\vspace{-.1cm}
\caption{\label{fig:3dBdyDisk}
\small
Minimal surfaces $\hat{\gamma}_A$ obtained with Surface Evolver corresponding to a region $A$ 
given by the intersection between the grey half plane at $z=0$ and a disk of radius $R$ whose center has coordinate $x>0$.
The entangling curve $\partial A \cap \partial B$ (red line) is an arc of circumference. 
The green half plane is $\mathcal{Q}$ defined  by (\ref{brane-profile}) and the green curve corresponds to $\hat{\gamma}_A \cap \mathcal{Q}$.
In the figure $\varepsilon = 0.03$, $R=1$ and the center of the disk has coordinate $x=0.6$.
In the left panel $\alpha =\pi/3$, while in the right panel $\alpha =2\pi/3$.
The numerical data of the corner function $F_\alpha(\gamma)$ corresponding to this kind of domains are
labeled by empty circles in Fig.\,\ref{fig:main_plot}.
}
\end{figure}

In order to write the analytic expression for the opening angle $\gamma$ of the infinite wedge in terms of the positive parameter $q_0$,
let us introduce
\be
\label{qast(q0,alpha) main}
q_\ast(\alpha,q_0) = 
\frac{| \cot\alpha \,|}{s_\ast(\alpha,q_0)} 
\ee
where $s_\ast(\alpha,q_0)>0$ is given by (\ref{s_ast def main}).
For the opening angle $\gamma$ of $A$ we find
\be
\label{gamma q0 main}
\gamma 
\,=\,
P_0(q_0) + 
\eta_\alpha \Big( \arcsin [  s_\ast(\alpha, q_0) ] - P\big(q_\ast(\alpha,q_0) ,q_0\big) \Big)
\ee
where the function $P(q,q_0) $ is defined as 
\be
\label{Pfunction def main}
P(q,q_0) 
\equiv 
\frac{1}{q_0(1+q_0^2)}\,
\bigg\{
(1+2q_0^2) \; \Pi\big( -1/Q_0^2\, , \, \sigma(q,q_0)  \, \big|  - Q_0^2\big) 
- q_0^2\;
\mathbb{F}\big( \sigma(q,q_0)\, \big| - Q_0^2 \big)
\bigg\}
\ee
(in (\ref{Pintegration def}) we give the integral representation)
being $\mathbb{F}(\phi | m)$ and $\Pi(n,\phi | m)$ the incomplete elliptic integrals of the first and third kind respectively, with
\be
\label{Q0^2 def main}
\sigma(q,q_0) \equiv \,\arctan \sqrt{\frac{q^2-q_0^2}{1+2q_0^2}}
\hspace{.4cm} \qquad \hspace{.4cm} 
Q_0^2 \equiv \frac{q_0^2}{1+q_0^2} \in (0,1)
\ee
The function $P_0(q_0)$ in (\ref{gamma q0 main}) is the limit $P(q,q_0) \to P_0(q_0) $ as $q\to +\infty$.
The explicit expression of $P_0(q_0)$ in terms of the complete elliptic integrals has been written in (\ref{theta_V expression}), 
but  we find it convenient to provide here also an equivalent form coming directly from (\ref{Pfunction def main}), namely
\be
\label{P0 expression v2}
P_0(q_0) 
\, = \,
\frac{1}{q_0(1+q_0^2)}\;
\bigg\{
(1+2q_0^2) \; \Pi\left( -1/Q_0^2 \, ,  - Q_0^2 \right) 
- q_0^2\;
\mathbb{K}\left( -Q_0^2 \right)
\bigg\}
\ee
being $\mathbb{K}(m)$ and $\Pi(n | m)$ the complete elliptic integrals of the first and third kind respectively.

\begin{figure}[t] 
\vspace{-.8cm}
\hspace{-1cm}
%\begin{center}
\includegraphics[width=1.1\textwidth]{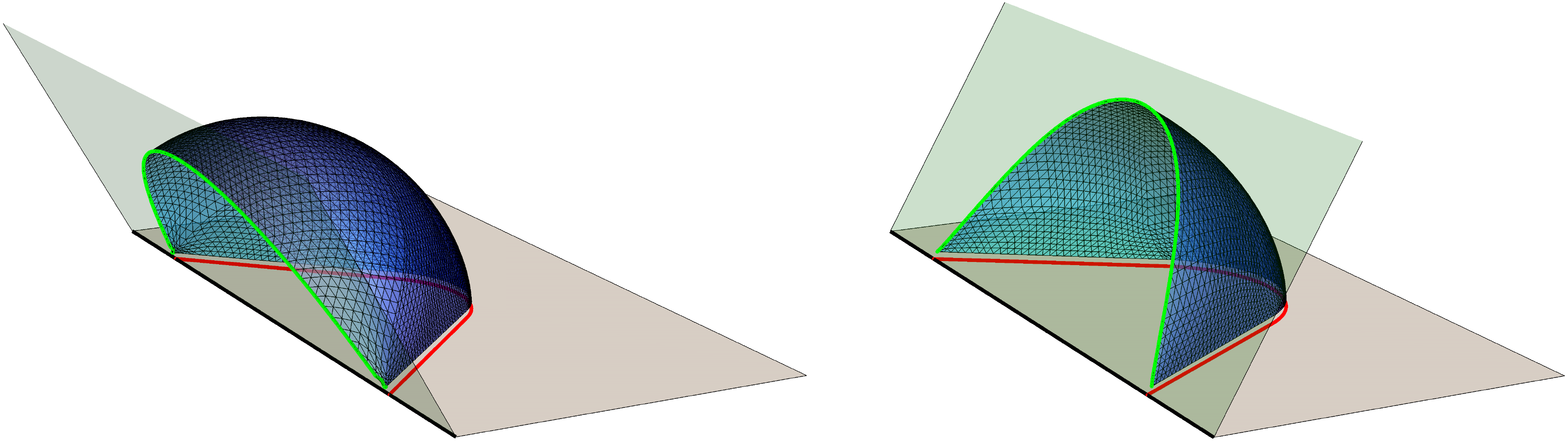}
%\end{center}
\vspace{.1cm}
\caption{\label{fig:3dBdyDrop}
\small
Minimal surfaces $\hat{\gamma}_A$ obtained with Surface Evolver corresponding to a region $A$ 
delimited by the red curve (entangling curve $\partial A \cap \partial B$) in the grey half plane at $z=0$,
which has been obtained by smoothly joining two segments of equal length $L$ forming two equal corners with the boundary, whose opening angle is $\gamma$. 
The green half plane is $\mathcal{Q}$ defined  by (\ref{brane-profile}) and the green curve corresponds to $\hat{\gamma}_A \cap \mathcal{Q}$.
In the left panel $\alpha = \pi/3$, $L=1$ and $\gamma = 0.8$,
while in the right panel $\alpha = 2\pi/3$, $L=1$ and $\gamma = 1$.
The numerical data of the corner function $F_\alpha(\gamma)$ corresponding to this kind of domains are
labeled by empty triangles in Fig.\,\ref{fig:main_plot}.
}
\end{figure}

As for the holographic entanglement (\ref{RT formula bdy}) of the infinite wedge $A$ adjacent to the boundary, since $\hat{\gamma}_A $ reaches the boundary $z=0$, its area is infinite;
therefore we have to consider its restriction $\hat{\gamma}_\varepsilon = \hat{\gamma}_A \cap \{ z\geqslant \varepsilon \}$ and take the limit $\varepsilon \to 0^+$, 
as required by the prescription (\ref{RT formula bdy}).

We find that the expansion of the area $\mathcal{A}[\hat{\gamma}_\varepsilon]$ of $\hat{\gamma}_\varepsilon$
as $\varepsilon \to 0$ reads
\be
\label{area wedge v0 main}
\mathcal{A}[\hat{\gamma}_\varepsilon]
= 
L^2_{\textrm{\tiny AdS}}
\bigg(\,
\frac{L}{\varepsilon} 
- F_{\alpha}(\gamma) \, \log(L / \varepsilon)
+ O(1)\bigg)
\ee
which is a special case of (\ref{hee bdy corner intro}) and (\ref{Ftot bdy corner intro}) with $P_{A,B} = L$ and $F_{\alpha, \textrm{\tiny tot}}  = F_{\alpha}(\gamma)$.
The leading linear divergence in (\ref{area wedge v0 main}) is the expected area law term and it comes from the part of $\hat{\gamma}_\varepsilon$ close the edge of $A$ at $\phi=\gamma$.
The occurrence of the wedge leads to the important logarithmic divergence, whose coefficient provides the corner function $F_{\alpha}(\gamma) $ we are interested in.

\begin{figure}[t] 
\vspace{-.8cm}
\hspace{-.5cm}
%\begin{center}
\includegraphics[width=1.1\textwidth]{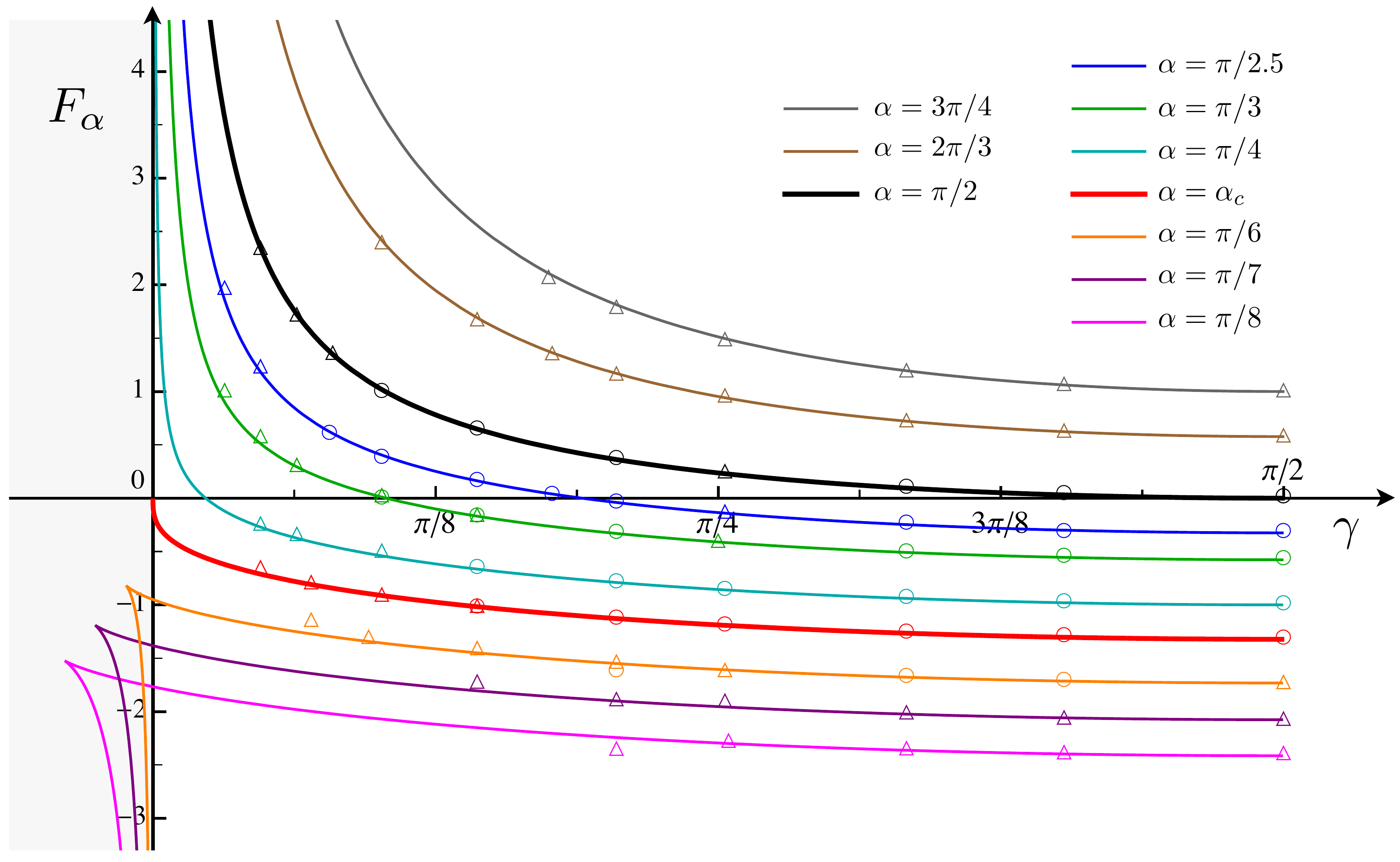}
%\end{center}
\vspace{-.2cm}
\caption{\label{fig:main_plot}
\small
The corner function $F_{\alpha}(\gamma)$ for some values of the slope $\alpha$ of the half plane $\mathcal{Q}$.
The solid curves are obtained from the analytic expressions (\ref{gamma q0 main}) and (\ref{total corner func bdy main}),
which provide the corner function parametrically in terms of $q_0>0$ (see also Fig.\,\ref{fig:main_plot_3D}).
The marked points have been  found through our numerical analysis based on Surface Evolver.
The empty circles label the data points obtained from the domain $A$ in Fig.\,\ref{fig:3dBdyDisk},
while empty triangles label the data points found by employing the domain $A$ in Fig.\,\ref{fig:3dBdyDrop}. 
The same color has been adopted for the analytic curves and the data points corresponding to the same $\alpha$.
}
\end{figure}

\begin{figure}[t] 
\vspace{-1.1cm}
\hspace{-.2cm}
%\begin{center}
\includegraphics[width=1.1\textwidth]{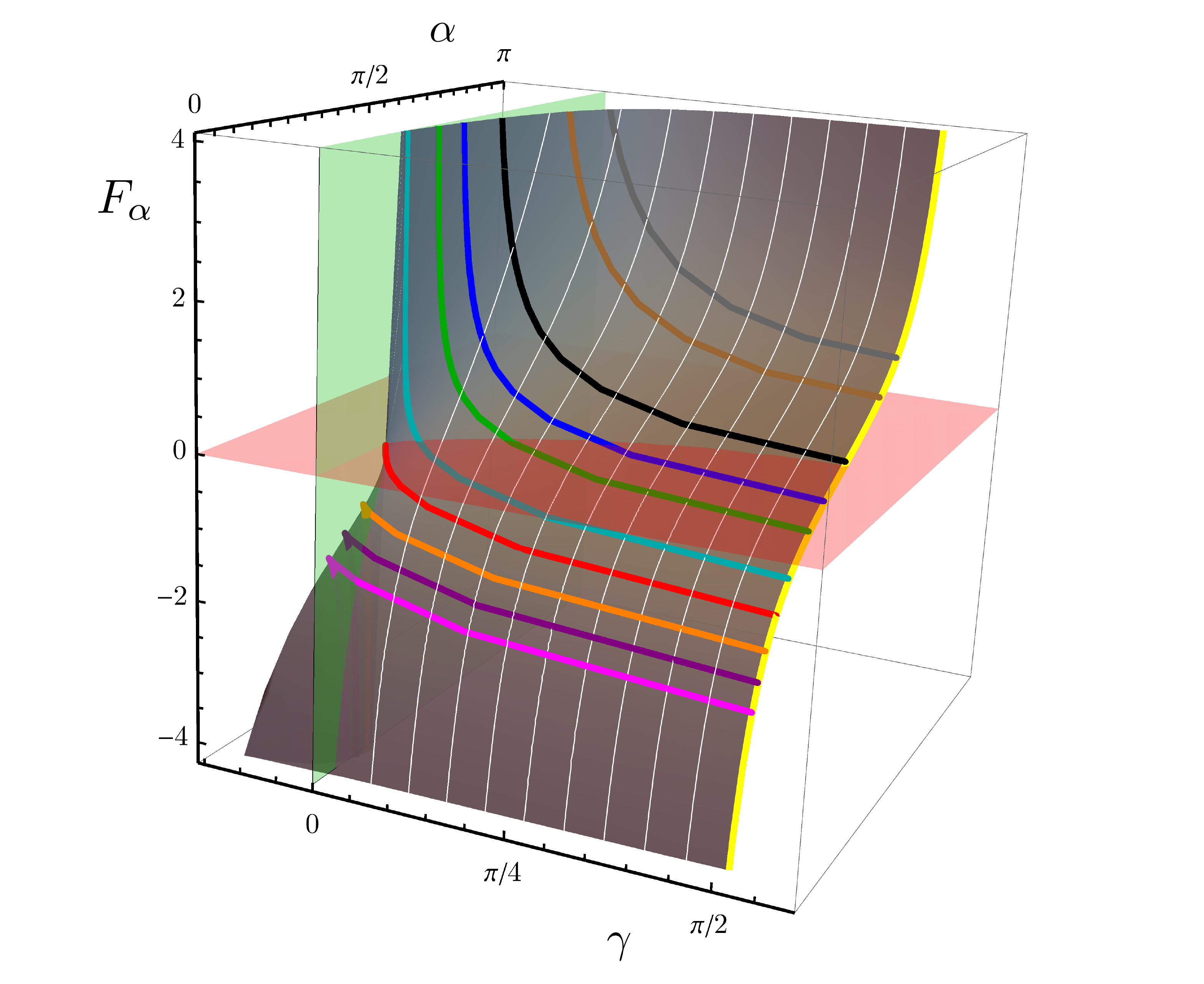}
%\end{center}
\vspace{-.6cm}
\caption{\label{fig:main_plot_3D}
\small
The corner function $F_{\alpha}(\gamma)$ given by (\ref{gamma q0 main}) and (\ref{total corner func bdy main}) in terms of $\gamma \in (0,\pi/2]$
and $\alpha \in (0,\pi)$ (grey surface).
The solid curves corresponding to the $\alpha=\textrm{const}$ sections are the same ones shown in Fig.\,\ref{fig:main_plot}, with the same colour code. 
The section $\gamma=\pi/2$ (yellow solid curve) is (\ref{Falpha pi/2 main}).
In the left panel of Fig.\,\ref{fig_app:zeros} we depict the intersection between the grey surface and the red plane 
and in the right panel of Fig.\,\ref{fig_app:zeros} the intersection of the grey surface with the green plane is shown. 
}
\end{figure}

The corner function $F_{\alpha}(\gamma) $ has been computed in the appendix\;\ref{sec app wedge area} and the result is 
\be
\label{total corner func bdy main}
F_\alpha
=
F(q_0) + \eta_\alpha\, \mathcal{G}\big( q_\ast(\alpha,q_0) ,q_0\big) 
\ee
where $F(q_0)$ has been introduced in (\ref{F(q0) def main}) and the function $\mathcal{G}(q,q_0) $ is 
\be
\label{mathcalG def}
\mathcal{G}(q,q_0) 
\, \equiv \,
\sqrt{1+q_0^2} \;
\left\{
\mathbb{F}\big( \sigma(q,q_0)\, \big| - Q_0^2 \big)
-
\mathbb{E}\big( \sigma(q,q_0)\, \big| - Q_0^2 \big)
+
\sqrt{\frac{(q^2+1)(q^2-q_0^2)}{(q_0^2+1)(q^2+q_0^2+1)}}
\;\right\}
\ee
The expression for $q_\ast(\alpha,q_0)$ to use in (\ref{total corner func bdy main}) is (\ref{qast(q0,alpha) main}).

The main result of this manuscript are (\ref{gamma q0 main}) and (\ref{total corner func bdy main}),
which provide the analytic expression of the corner function $F_\alpha(\gamma)$ in a parametric form in terms of $q_0 >0$.

In Fig.\,\ref{fig:main_plot} the solid curves corresponds to the corner function $F_\alpha(\gamma)$ for some values of $\alpha$.
As for the argument of the corner function $F_\alpha(\gamma)$, we remind that $\gamma \in (0,\pi/2]$.
Nonetheless, whenever $\gamma \in (0,\pi)$ we mean $F_\alpha(\textrm{min}[\gamma, \pi-\gamma])$.

In Fig.\,\ref{fig:main_plot_3D} we show the surface given by the corner function $F_\alpha(\gamma) $ in terms of the opening angle $\gamma$ and the slope $\alpha \in (0,\pi)$.
In this figure we have highlighted the sections corresponding to the curves reported in Fig.\,\ref{fig:main_plot} and also the curve $F_\alpha(\pi/2) $ (yellow curve).

We have employed Surface Evolver to find an important numerical evidence of our analytic result.
In this numerical analysis we have chosen domains $A$ whose entangling curves $\partial A \cap \partial B$ correspond to the red solid curves in the 
$z=0$ half plane shown in Fig.\,\ref{fig:3dBdyDisk} and in Fig.\,\ref{fig:3dBdyDrop}.
In particular, in Fig.\,\ref{fig:3dBdyDisk} we have that $A$ is part of a disk which is not centered on the boundary and
in Fig.\,\ref{fig:3dBdyDrop} the region $A$ is made by two finite wedges with an edge on the boundary and the same opening angle whose remaining edges are joined smoothly. 
These domains  are simple finite regions with the smallest number of corners providing the corner function $F_\alpha(\gamma)$ we are interested in. 
In Fig.\,\ref{fig:3dBdyDisk} and in Fig.\,\ref{fig:3dBdyDrop} we show also the corresponding minimal surface $\hat{\gamma}_A$ constructed with Surface Evolver
for a value $\alpha < \pi/2$ (left panels) and for a value $\alpha > \pi/2$ (right panels).

The marked points in Fig.\,\ref{fig:main_plot} are the numerical values of the corner function $F_\alpha(\gamma)$ obtained 
through the numerical analysis based on the data obtained from Surface Evolver, as briefly explained in the appendix\;\ref{app numerics}.
In particular, the empty circles and the empty triangles correspond to the domains $A$ shown in Fig.\,\ref{fig:3dBdyDisk} and in Fig.\,\ref{fig:3dBdyDrop} respectively.
It turns out that the domain $A$ in Fig.\,\ref{fig:3dBdyDrop} is more suitable to deal with small values of $\gamma$ in our numerical approach.
Excellent agreement is obtained with the analytic result for the values of $\alpha$ and $\gamma$ considered in Fig.\,\ref{fig:main_plot}.

From Fig.\,\ref{fig:main_plot} and Fig.\,\ref{fig:main_plot_3D}  we observe  that for the holographic corner function given by 
(\ref{gamma q0 main}) and (\ref{total corner func bdy main}) we have that $F_\alpha'(\gamma) \leqslant 0$ and also  $F_\alpha''(\gamma) \geqslant 0$
for any fixed value of the slope $\alpha \in (0,\pi)$.
Furthermore, from Fig.\,\ref{fig:main_plot_3D} we also notice that $\partial_\alpha F_\alpha(\gamma)  > 0$ for any fixed value of $\gamma \in (0,\pi/2]$.
It would be interesting to understand whether these properties come from some more fundamental principles. 

In Fig.\,\ref{fig:main_plot} the curves corresponding to the critical value $\alpha = \alpha_c$ (red curve) given by (\ref{alpha_crit}) 
and to $\alpha = \pi/2$ (black curve) have been highlighted by employing thicker lines 
because these values separate the range of $\alpha \in (0,\pi)$ into three intervals for $\alpha$ where the corner function $F_\alpha(\gamma) $
has different features. 
In particular, when $\alpha \geqslant \pi/2$ we have that $F_\alpha(\gamma) \geqslant 0$, 
while when $\alpha \leqslant \alpha_c$ we have that $F_\alpha(\gamma) \leqslant 0$.
In the intermediate range $\alpha \in (\alpha_c, \pi/2)$ the corner function does not have a definite sign in the whole range $\gamma\in (0,\pi/2]$ 
and, being $F_\alpha'(\gamma) < 0$, it has a unique zero $\gamma = \gamma_0$.
The value $\gamma_0$ in terms of $\alpha \in [\alpha_c, \pi/2]$ found numerically is shown in the left panel of Fig.\,\ref{fig_app:zeros}.

From Fig.\,\ref{fig:main_plot} we observe that the corner function $F_\alpha(\gamma) $ displays two qualitative different behaviours as $\gamma \to 0^+$.
Indeed, $F_\alpha(\gamma) \to + \infty$ when $\alpha > \alpha_c$, while it reaches a finite (non positive) value when $\alpha \leqslant \alpha_c$.
In Sec.\,\ref{sec:limiting regimes} more quantitative results about the regimes $\gamma \to 0^+$ and $\gamma \to \pi/2$ of  $F_\alpha(\gamma) $ are obtained.

It would be interesting to get a direct numerical confirmation of the occurrence of $\alpha_c$ through Surface Evolver or other methods.
Unfortunately, we have not been able to push our numerical analysis to values of $\gamma$ small enough to appreciate the qualitatively different 
behaviour of the corner function for $ \alpha \leqslant \alpha_c$ and $\alpha > \alpha_c$.
Hopefully, this gap will be fixed in future studies.

\subsection{Limiting regimes of the corner function}
\label{sec:limiting regimes}

It is worth studying the corner function $F_\alpha(\gamma)$ in some particular regimes. 
In the following we report only the main results of our analysis, referring the reader to the appendices\;\ref{sec app wedge area}
and \ref{app wedge limits}  for a detailed discussion of their derivations. 

An important special value to consider  is $\alpha =\pi/2$. 
In this case it is straightforward to realise that the minimal surface $\hat{\gamma}_A$ is half of the auxiliary minimal surface $\hat{\gamma}_{A,\textrm{\tiny aux}}$ in $\mathbb{H}_3$, 
which is anchored to the auxiliary infinite wedge $A_{\textrm{\tiny \,aux}}$ with opening angle $2\gamma$.
Indeed, for every $\alpha $ we have that $\hat{\gamma}_{A,\textrm{\tiny aux}}$ in $\mathbb{H}_3$ is smooth and symmetric 
with respect to the half plane orthogonal to $z=0$ passing through the bisector of $A_{\textrm{\tiny \,aux}}$; 
therefore $\hat{\gamma}_{A,\textrm{\tiny aux}}$ intersects this half plane orthogonally. 
When $\alpha=\pi/2$ the half plane characterising this reflection symmetry coincides with $\mathcal{Q}$.
Thus, $\hat{\gamma}_{A,\textrm{\tiny aux}}$ is the union of  $\hat{\gamma}_A$  and its reflected image with respect to $\mathcal{Q}$ obtained by sending $x \to -\,x$.

As for the corner function at $\alpha =\pi/2$, from the analytic expression (\ref{gamma q0 main}) and (\ref{total corner func bdy main}) we find  respectively that
\be
\label{limit alpha = pi/2}
\lim_{\alpha \,\to\, \pi/2} 
\gamma  \,=\, P_0(q_0)
\hspace{.4cm} \qquad \hspace{.4cm} 
\lim_{\alpha \,\to\, \pi/2} 
F_\alpha \,=\, F(q_0)
\ee
Further comments can be found in the closing remarks of the appendix\;\ref{sec app wedge area}.
Comparing (\ref{limit alpha = pi/2}) with (\ref{cusp exact result}) and (\ref{theta_V expression}) respectively, we obtain 
\be
\label{corner func at pi/2}
\widetilde{F}(2\gamma) 
\,=\,
2\,F_{\pi/2}(\gamma)
\ee
Thus, the corner function found in \cite{Drukker:1999zq} and discussed in Sec.\,\ref{sec drop} is recovered as the special case
$\alpha=\pi/2$ of the corner function $F_\alpha(\gamma)$ given by (\ref{gamma q0 main}) and (\ref{total corner func bdy main}).

We find it worth considering the corner function $F_\alpha(\gamma)$ in the limiting regimes of $\gamma \to 0$ and $\gamma \to \pi/2$,
which correspond to $q_0 \to + \infty$ and $q_0 \to 0^+$ respectively, as discussed in the appendix\;\ref{app wedge limits}.

Taking the limit $q_0 \to + \infty$ of (\ref{gamma q0 main}) and (\ref{total corner func bdy main}), we obtain 
\be
\label{gamma expansion q0=inf main}
\gamma = \frac{\mathfrak{g}(\alpha)}{q_0} + O(1/q_0^3)
\qquad
 F_\alpha \, = \, \mathfrak{g}(\alpha) \, q_0 + O(1/q_0)
  \hspace{.4cm} \qquad \hspace{.4cm}
  q_0 \to + \infty
\ee
where $\mathfrak{g}(\alpha) $ is the function (\ref{g function def main}), given by the red curve in Fig.\,\ref{fig:strip_area}.
In particular, $\gamma \to 0$ for large $q_0$.
We remark that we have different behaviours of the corner function 
$F_\alpha(\gamma)$ as $\gamma \to 0^+$, depending on whether $\alpha \in (0, \alpha_c]$ or  $\alpha \in (\alpha_c, \pi)$.
Indeed, $\mathfrak{g}(\alpha) $ changes its sign at the critical value $\alpha = \alpha_c$ defined by (\ref{def alpha_c g}), whose numerical value is (\ref{alpha_crit}).
Since $\gamma $ and $q_0$ must be strictly positive, while $\mathfrak{g}(\alpha) \leqslant $ for $\alpha \in (0, \alpha_c]$, 
the expansion of $\gamma$ in (\ref{gamma expansion q0=inf main}) is meaningful in our setup only when $\alpha \in (\alpha_c, \pi)$.
In this range, from the first expansion in (\ref{gamma expansion q0=inf main}) we find that $q_0  = \mathfrak{g}(\alpha)/\gamma + O(\gamma)$ as $\gamma \to 0$.
Then, plugging this result into the second expansion of (\ref{gamma expansion q0=inf main}), we obtain
\be
\label{limit gamma=0 main}
F_\alpha \, = \, \frac{\mathfrak{g}(\alpha)^2}{\gamma} + O(\gamma)
\hspace{.3cm} \qquad \hspace{.3cm} 
\gamma \to 0^+
\hspace{.3cm} \qquad \hspace{.3cm} 
\alpha \in (\alpha_c, \pi)
\ee

When $\alpha = \alpha_c$ the second expansion in (\ref{limit gamma=0 main}) tells us that
\be
\label{alpha_c def zeroF}
F_{\alpha_c}(0) = 0
\ee
We can interpret this observation as a possible definition of $\alpha_c$ in terms of the corner function.

Notice that (\ref{alpha_c def zeroF}) suggests the following way to find $\alpha_c$ by employing finite domains. 
Consider for simplicity a domain $A$ with only two equal corners  adjacent to the boundary, like e.g. in Fig.\,\ref{fig:3dBdyDrop}, 
and send their opening angles to zero simultaneously. 
In this limit the coefficient of the logarithmic divergence diverges when $\alpha > \alpha_c$  and tends to a finite value $F_\alpha(0) <0$
for  $\alpha \leqslant \alpha_c$.
The value $F_\alpha(0)$ corresponds to a finite value $\hat{q}_0$ of the parameter $q_0$ for a given $\alpha$.
The function $F_\alpha(0)$ can be found numerically in terms of $\alpha \in (0, \alpha_c)$ and the result of this analysis is shown in the right panel 
of Fig.\,\ref{fig_app:zeros} in the appendix\;\ref{sec app wedge}.
In particular, when $\alpha = \alpha_c$ we have (\ref{alpha_c def zeroF}).

From Fig.\,\ref{fig:main_plot} we observe that in the range $\alpha \in [\alpha_c, \pi/2)$ the function $F_\alpha(\gamma)$ vanishes at a 
positive value $\gamma_0$ of the opening angle. When $\alpha = \alpha_c$ we have $\gamma_0 = 0$, as written in (\ref{alpha_c def zeroF}).
By solving numerically the equation $F_\alpha(\gamma_0) = 0$ for $\alpha \in [\alpha_c, \pi/2)$, we find the function $\gamma_0(\alpha)$ shown in the left panel of Fig.\,\ref{fig_app:zeros}.

When $\alpha \in (0, \alpha_c)$ we have $\mathfrak{g}(\alpha)<0$; therefore the expansions in (\ref{gamma expansion q0=inf main})
imply that $\gamma \to 0^-$ and $F_\alpha \to - \infty$ as $q_0 \to + \infty$.
Negative values of $\gamma$ are meaningless in our context. 
Nonetheless, from a mathematical perspective, we find it worth extending the domain of $\gamma$ to negative values. 
When $\gamma<0$ the parametric curve given by (\ref{gamma q0 main}) and (\ref{total corner func bdy main}) does not provide a function of $\gamma$, but it is still a well defined curve. 
Indeed, from Fig.\,\ref{fig:main_plot} it is straightforward to observe that, in the regime $\alpha < \alpha_c$, we have that $\gamma \to 0^-$ when either $q_0 \to \hat{q}_0$ or $q_0 \to + \infty$.
In the latter case we have that $F_\alpha = \mathfrak{g}(\alpha)^2/\gamma + O(\gamma)$.

In the appendix\;\ref{app map strip-cusp} we explain the relation between the regime $\gamma \to 0^+$ of the corner function $F_\alpha(\gamma)$
and the holographic entanglement entropy (\ref{area strip final main all}) of the infinite strip adjacent to the boundary. 
This is due to the existence of a conformal map which relates the infinite wedge with  $\gamma \to 0^+$  to an infinite strip.
The discussion reported in the appendix\;\ref{app map strip-cusp} is a  modification of the analogue one in AdS$_4$/CFT$_3$ 
\cite{stripcusp, Myers:2012vs, Bueno:2015xda}, 
obtained by taking into account the presence of the boundary in a straightforward way.

As for the regime $q_0 \to 0^+$, in the appendix\;\ref{app small q0} we have computed the expansions 
of the opening angle $\gamma$ and of the corner function $F_\alpha$,
which are given by (\ref{gamma q0 main}) and (\ref{total corner func bdy main}) respectively,
finding  (\ref{gamma expansion q0 final}) and (\ref{F expansion q0=0 final}) respectively.
From these results we can conclude that $\gamma \to \pi / 2$ and also that
\be
\label{limit gamma=pi/2 main}
F_\alpha(\gamma) 
\,=\, 
- \cot\alpha 
+ \frac{(\pi/2-\gamma)^2}{2(\pi-\alpha)} 
% + \frac{5(\pi-\alpha + \cos \alpha \, \sin \alpha) }{16(\pi-\alpha)^4} \, (\pi/2-\gamma)^4
+ O\big((\pi/2-\gamma)^4\big) 
\ee
which agrees with the general expansion (\ref{Falpha expansion}) for this kind of corner function.
In particular, we have that with $F_\alpha(\pi/2) = - \cot\alpha$ and $F''_\alpha(\pi/2) =1/(\pi-\alpha)$.
The expression for $F_\alpha(\pi/2) $ confirms the expected result (\ref{Falpha pi/2 main}) obtained in Sec.\;\ref{sec:half disk} by considering the half disk centered on the
boundary. 
Let us remark that the method discussed in the appendix\;\ref{app wedge limits} allows to computed also higher orders in (\ref{limit gamma=pi/2 main}).
For instance, in (\ref{limit gamma=pi/2 app}) also the $O((\pi/2-\gamma)^4) $ term has been reported.

\subsection{Relations with the stress tensor}
\label{sec stress}

We find it worth exploring possible universal relations among the corner functions and other quantities of the underlying BCFT$_3$ model. 

In CFT$_3$, an important example of universal relation involves the corner function $\tilde{f}(\theta) $ and the two point function of the stress tensor $T_{\mu\nu}$,
which is given by $\langle T_{\mu\nu}(x)\, T_{\rho\sigma}(0)\rangle = (C_T / |x|^6) \,\mathcal{I}_{\mu\nu , \rho\sigma}(x)$, 
being $\mathcal{I}_{\mu\nu , \rho\sigma}(x)$ a dimensionless tensor structure fixed by symmetry.
In particular, by considering the coefficient $\tilde{\sigma} = \tilde{f}''(\pi) / 2$ of the leading term in the expansion $\tilde{f}(\theta) = \tilde{\sigma} (\pi -\theta)^2 + \dots$ as $\theta \to \pi^-$,
it has been found that \cite{Bueno:2015rda, Bueno:2015xda}
\be
\label{bueno-myers}
\frac{\tilde{\sigma}}{C_T} = \frac{\pi^2}{24}
\ee

In AdS$_4$/CFT$_3$ the holographic corner function is $\tilde{f}(\theta) =\tfrac{L^2_{\textrm{\tiny AdS}}}{4 G_{\textrm{\tiny N}}}\, \widetilde{F}(\theta)  $,
as discussed in Sec.\,\ref{sec intro}.
Denoting by $\tilde{\sigma}_{\textrm{\tiny E}}$ the coefficient $\tilde{\sigma}$ for this holographic corner function in a bulk theory described by Einstein gravity,
we have that $\tilde{\sigma}_{\textrm{\tiny E}} = \tfrac{L^2_{\textrm{\tiny AdS}}}{8 G_{\textrm{\tiny N}}} \widetilde{F}''(\pi)$.
Considering the corner function $F_\alpha(\gamma)$ in AdS$_4$/BCFT$_3$ given by (\ref{gamma q0 main}) and (\ref{total corner func bdy main}), 
in Sec.\,\ref{sec:limiting regimes} the relation (\ref{limit alpha = pi/2})  has been observed when $\alpha = \pi/2$.
Taking the limit $\gamma \to \pi/2$ of (\ref{limit alpha = pi/2}) by employing (\ref{Falpha expansion}) and $F_{\pi/2}(\tfrac{\pi}{2}) = 0$, 
one finds that $2 \widetilde{F}''(\pi)= F''_{\pi/2}(\tfrac{\pi}{2})$. 
The latter relation and $F''_\alpha(\tfrac{\pi}{2})   = 1/(\pi-\alpha)$ (see (\ref{limit gamma=pi/2 main})) evaluated for $\alpha = \pi/2$ provide
$ \sigma_{\textrm{\tiny E}}  = \tfrac{L^2_{\textrm{\tiny AdS}}}{16 G_{\textrm{\tiny N}}} \; F''_{\pi/2}(\tfrac{\pi}{2})
= \tfrac{L^2_{\textrm{\tiny AdS}}}{16 \pi G_{\textrm{\tiny N}}} \; 2$.
Then, by employing the holographic result $ C_T  = 3 L^2_{\textrm{\tiny AdS}} / (\pi^3 G_{\textrm{\tiny N}} ) = \tfrac{L^2_{\textrm{\tiny AdS}}}{16\pi G_{\textrm{\tiny N}}} (48 / \pi^2 ) $ 
found in \cite{CTholog}, one obtains $\tilde{\sigma}_{\textrm{\tiny E}} / C_T = \pi^2 / 24$, 
which corresponds to (\ref{bueno-myers}) in the holographic setup determined by the Einstein gravity in the bulk.
Thus, consistency has been found between (\ref{limit alpha = pi/2}) and the ratio (\ref{bueno-myers}).

We find it interesting to explore the possibility that universal relations exist also for BCFT$_3$.

Considering the  two dimensional manifold $\partial \mathcal{B}$ given by the boundary of a BCFT$_3$ defined in the spacetime $\mathcal{B}$,
let us denote by $b_{ij}$ and $k_{ij}$ the  metric induced on $\partial \mathcal{B}$ and its extrinsic curvature respectively.
By introducing the trace $k$ of the extrinsic curvature, the combination $\kappa_{ij} \equiv k_{ij} - (k/2) b_{ij}$ gives the traceless part of $k_{ij}$.

In a BCFT$_3$, the presence of the boundary leads to a non trivial Weyl anomaly localised on the boundary.
It is given by \cite{Jensen:2015swa, Solodukhin:2015eca}
\be
\label{trace T}
\langle \, T^{\,i}_{\;\,\,i} \, \rangle
\,=\,
\frac{1}{4\pi}\big( - \mathfrak{a} \,\mathcal{R} + \mathfrak{q}  \, \textrm{Tr} \kappa^2  \, \big) \, \delta(\partial \mathcal{B})
\ee
where $\delta(\partial \mathcal{B})$ is the Dirac delta whose support is $\partial \mathcal{B}$.
In (\ref{trace T}) we have that $\mathcal{R}$ is the Ricci scalar corresponding to the metric $b_{ij}$ induced on $\partial \mathcal{B}$ and $ \textrm{Tr} \kappa^2  \equiv  \kappa^{ij} \kappa_{ij}$.
The constants $\mathfrak{a}$ and $ \mathfrak{q}$ are the boundary central charges, 
which depends on the underlying model and also on the conformally invariant boundary conditions characterising the BCFT$_3$.
They have been computed for some free models in \cite{Nozaki:2012qd, Jensen:2015swa, Fursaev:2016inw}. 
The quantity $\langle \, T^{\,i}_{\;\,\,i} \, \rangle$ has been studied also in BCFT$_4$  \cite{Herzog:2015ioa, Fursaev:2015wpa}.

We also need to consider the behaviour of the one point function of the stress tensor in the BCFT$_3$ near $\partial \mathcal{B}$.
In terms of the proper distance $X$  from $\partial \mathcal{B}$, it is given by \cite{Deutsch:1978sc}
\be
\label{A_T def}
\langle \, T_{ij} \, \rangle
\,=\,
\frac{ A_T}{X^2} \, \kappa_{ij} + \dots
\qquad
X\to 0^+
\ee
where $ \kappa_{ij} $ is the traceless part of the extrinsic curvature of the boundary 
and the coefficient $A_T$ depends on the conformally invariant boundary conditions of the underlying BCFT$_3$.

Notice that in the BCFT$_4$ given by a scalar field, the coefficient $A_T$ has been computed in \cite{Deutsch:1978sc} 
and the same negative value has been obtained for both Dirichlet and Neumann boundary conditions.
We have not found in the literature an explicit computation of $A_T$ for a BCFT$_3$.

Let us focus on the holographic corner function $f_\alpha(\gamma)  =\tfrac{L^2_{\textrm{\tiny AdS}}}{4 G_{\textrm{\tiny N}}}\, F_\alpha(\gamma) $,
where $F_\alpha(\gamma) $ is given by (\ref{gamma q0 main}) and (\ref{total corner func bdy main}).

Let us recall that in the AdS/BCFT construction discussed in \cite{Takayanagi:2011zk}, the Neumann boundary conditions given by
$K_{ab} = (K - T) h_{ab}$ have been imposed to define the hypersurface $\widetilde{\mathcal{Q}}$ in the bulk delimiting the gravitational spacetime.
Instead, in  \cite{Miao:2017gyt, Chu:2017aab, Astaneh:2017ghi} it has been proposed to employ the less restrictive boundary condition
$K = \frac{d+1}{d}\, T$ to find $\widetilde{\mathcal{Q}}$.
When the boundary of the BCFT$_3$ is flat, both these prescriptions provides the half plane $\widetilde{\mathcal{Q}} = \mathcal{Q}$ given by (\ref{brane-profile}).

In the AdS/BCFT setup of \cite{Takayanagi:2011zk}, by considering a BCFT$_3$ defined on the three dimensional sphere (in the Euclidean signature), 
whose boundary is a two dimensional sphere for which $\kappa_{ij}$ vanishes, it has been found that \cite{Fujita:2011fp}\footnote{
Comparing with the notation of \cite{Fujita:2011fp}, we find that $ (- \,c_{bdy}/6)\big|_{\textrm{\tiny there}} = \mathfrak{a}\,$.
}
\be
\label{a-charge-holog}
\mathfrak{a} =  \frac{L^2_{\textrm{\tiny AdS}}}{4 G_{\textrm{\tiny N}}}\, ( - \cot \alpha )
\ee
which means that $\mathfrak{a} = f_\alpha(\pi/2)$.
By using instead the boundary conditions $K = \frac{d+1}{d}\, T$, the relations
$ \mathfrak{q}  =  \mathfrak{a} = f_\alpha(\pi/2)$ have been obtained \cite{Miao:2017gyt, Chu:2017aab, Astaneh:2017ghi}.
Notice that the relation $\mathfrak{q}= \mathfrak{a}$ is not true for a free scalar \cite{Nozaki:2012qd, Jensen:2015swa, Fursaev:2016inw}.

We remark that, since the holographic corner function given by (\ref{gamma q0 main}) and (\ref{total corner func bdy main})
has been found for a flat boundary, it should be same for both the above AdS$_4$/BCFT$_3$ constructions, 
once the prescription (\ref{RT formula bdy}) for the holographic entanglement entropy is accepted.

Given the holographic result (\ref{a-charge-holog}), one could wonder whether  $  f_\alpha(\pi/2) = \mathfrak{a} $ holds for any BCFT$_3$.
 In \cite{Fursaev:2016inw} it has been shown that this relation fails for the scalar field because 
 of the occurrence of a non minimal coupling to the curvature. 
Checking the validity of $  f_\alpha(\pi/2) = \mathfrak{a} $ for other models is an interesting issue for future studies.

In the remaining part of this section we explore a relation involving the coefficient $f''_\alpha(\pi/2) $ of the expansion (\ref{Falpha expansion}) 
of the holographic corner function as $\gamma \to \pi/2$ and the coefficient $A_T$ introduced in (\ref{A_T def}) by considering the one point
function of the stress tensor close to the boundary $\partial \mathcal{B}$.

In AdS$_4$/BCFT$_3$ we found that $F''_\alpha(\pi/2)   = 1/(\pi-\alpha)$ for $\alpha \in (0,\pi)$ (see  (\ref{limit gamma=pi/2 main}));
therefore  we have 
\be
\label{f'' pi/2 - holog}
f''_\alpha(\pi/2) 
= 
\frac{L^2_{\textrm{\tiny AdS}}}{16\pi G_{\textrm{\tiny N}}}\; \frac{4\pi}{\pi - \alpha}  
\ee

By employing the AdS/BCFT construction of \cite{Takayanagi:2011zk} and the standard approach to
 the holographic stress tensor discussed in \cite{Balasubramanian:1999re, Myers:1999psa, deHaro:2000vlm},
in the appendix\;\ref{app A_T} we have revisited the analysis of \cite{Miao:2017aba}\footnote{
In the appendix\;\ref{app A_T} the differences between our results and the ones obtained in \cite{Miao:2017aba} are discussed. 
} 
finding the expression of $A_T$ in AdS$_{d+2}$/BCFT$_{d+1}$ with the boundary conditions of \cite{Takayanagi:2011zk} (see (\ref{A_T ddim})).
In the special case of $d=2$, for $\alpha \in (0,\pi)$ we obtain
\be
\label{A_T holog}
A_T = -\,
\frac{L^2_{\textrm{\tiny AdS}}}{16\pi G_{\textrm{\tiny N}}} 
\; \frac{2}{\pi - \alpha} 
\ee

From (\ref{f'' pi/2 - holog}) and (\ref{A_T holog}), we find it interesting to observe that in the AdS$_4$/BCFT$_3$ setup of \cite{Takayanagi:2011zk} 
the ratio $f''_\alpha(\pi/2) / A_T$ is independent of the slope $\alpha$, which could be related to the conformally invariant boundary conditions allowed for the dual BCFT$_3$.
In particular this ratio reads
\be
\label{rel}
\frac{f''_\alpha(\pi/2)}{A_T} = -\,2\pi
\ee

We find it very interesting to compute the ratio (\ref{rel}) also for explicit models of 
three dimensional conformal field theories with boundary and for different boundary conditions. 
Free quantum field theories are the simplest models to address in this direction. 
\\

%%%%%%%%%%%%%%%%%%%%%%%%%%%%%%%%%%%%%%%%%%%%%%%%%%%%%
\section{Infinite wedge with only the tip on the boundary}
\label{sec wedge tip bdy}

In this section we consider the domain given by an infinite wedge having its tip on the boundary whose edges do not belong to it. 
As discussed in Sec.\,\ref{sec intro}, in a generic BCFT$_3$ the entanglement entropy of this region contains a logarithmic divergence 
whose coefficient provides a corner function $\mathsf{F}_\alpha(\vec{\omega})$ which cannot be determined 
from the corner function $f_\alpha(\gamma)$ corresponding to the infinite wedge adjacent to the boundary.
In the following we explain that for the holographic entanglement entropy in AdS$_4$/BCFT$_3$ 
this analysis significantly simplifies and the corner function $\mathcal{F}_\alpha(\omega,\gamma)$ corresponding to this kind of wedge 
(see (\ref{Ftot bdy corner intro})) can be written in a form which involves the corner function $F_\alpha(\gamma)$ presented in Sec.\,\ref{sec:wedge} 
and the corner function $\widetilde{F}(\theta)$ reviewed in Sec.\,\ref{sec drop}.

\begin{figure}[t!] 
\vspace{-.8cm}
\hspace{-1.1cm}
%\begin{center}
\includegraphics[width=1.1\textwidth]{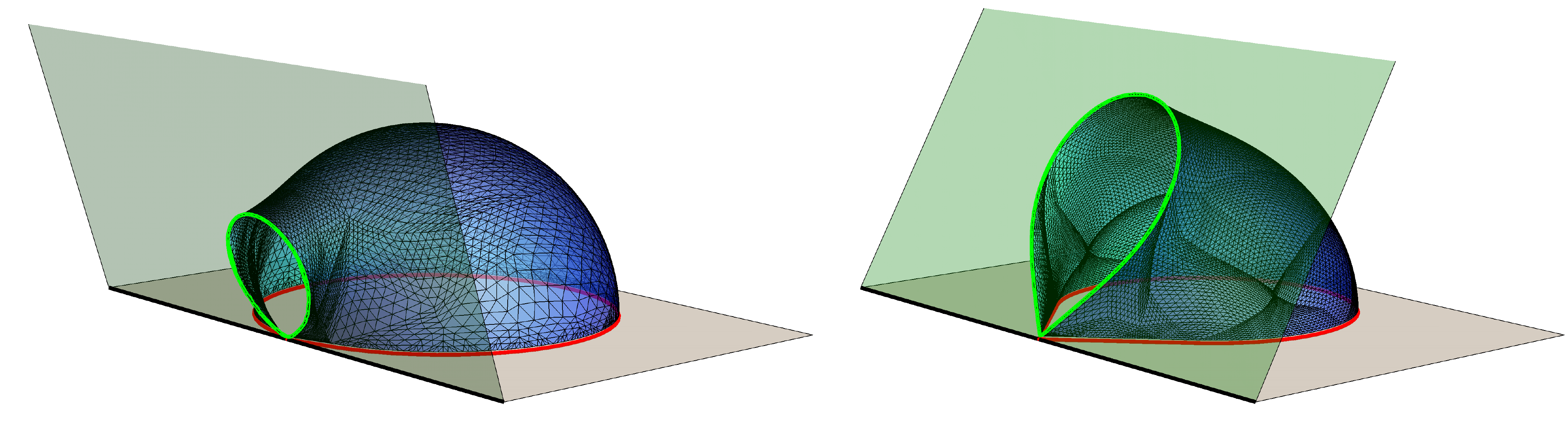}
%\end{center}
\vspace{-.1cm}
\caption{\label{fig:3dBdyTipSym}
\small
Minimal surfaces $\hat{\gamma}_A$ obtained with Surface Evolver and anchored to a single drop $A$
(whose boundary is the red solid curve in the $z=0$ grey half plane)
which has only the tip on the boundary. 
Here $A$ has been chosen in a symmetric way (i.e. $\tilde{\gamma} = \gamma$).
In the left panel $\alpha = \pi/2.5$, $\omega = 2.6$ and $L = 0.5$,
while in the right panel $\alpha = 2\pi/3$, $\omega = \pi/2$ and $L = 1.5$.
In both panels $\varepsilon = 0.03$.
This kind of minimal surfaces have been constructed to find the data corresponding 
to $\omega > \omega_c$ in Fig.\,\ref{fig:bdyDropTrans}, which have been labeled by empty black circles.
}
\end{figure}

Let us consider the infinite wedge $A$ with opening angle $\omega < \pi$ which has only the tip on the boundary $x=0$. 
Domains containing this kind of corner occur in Fig.\,\ref{fig:ssa}, where they are labeled by $C$ and $C_j$.
Setting the origin of the Cartesian coordinates in the tip of the wedge $A$, 
we have that the boundary $x=0$ is splitted into two half lines corresponding to $y<0$ and $y>0$.
Denoting by $\gamma<\pi$ and $\tilde{\gamma}<\pi$ the opening angles of the corners in $B$,
the supplementarity condition $\omega+ \gamma + \tilde{\gamma}=\pi$ holds. 
We can assume that $\gamma \leqslant \tilde{\gamma}  $ without loss of generality.
Combining this inequality with the supplementarity condition, it is straightforward to observe that $\gamma \leqslant (\pi-\omega)/2$.
Instead, since $\tilde{\gamma}$ is not restricted, we have  that $\tilde{\gamma} \in (0, \pi)$.
In the following we denote by $L \gg \varepsilon$ the length of the edges of $A$, as done in Sec.\,\ref{sec:wedge} for the wedge adjacent to the boundary.

Since the edges of $A$ do not belong to the boundary $x=0$, the minimal surface $\hat{\gamma}_A$ is anchored to both of them.
Moreover, the expansion of the area of $\hat{\gamma}_\varepsilon$ is (\ref{hee bdy corner intro}) with $P_{A,B} = 2L$ 
and  the coefficient of the logarithmic divergence (\ref{Ftot bdy corner intro}) given by $F_{\alpha, \textrm{\tiny tot}} = \mathcal{F}_\alpha(\omega , \gamma) $.

It is not difficult to realise that there are two candidates for $\hat{\gamma}_A$ which are local solutions of the minimal area condition
in presence of $\mathcal{Q}$.
The first one is a surface $\hat{\gamma}_{A}^{\textrm{\tiny \,dis}}$ which connects the two edges of $A$ through the bulk and is disconnected from the half plane $\mathcal{Q}$.
Since $\hat{\gamma}_{A}^{\textrm{\tiny \,dis}} \cap \mathcal{Q} = \emptyset$, we have that $\hat{\gamma}_{A}^{\textrm{\tiny \,dis}}$ is the minimal area surface found in \cite{Drukker:1999zq}, 
which has been discussed in Sec.\,\ref{sec drop}.
The second solution is a surface $\hat{\gamma}_{A}^{\textrm{\tiny \,con}}$ which connects the two edges of $A$ to $\mathcal{Q}$ through the bulk.
It is given by the union of two disjoint surfaces where each of them is like the one found in Sec.\,\ref{sec:wedge}; therefore 
$\hat{\gamma}_{A}^{\textrm{\tiny \,con}} \cap \mathcal{Q}$ is made by two half lines departing from the tip of the wedge.

The area $\mathcal{A}[\hat{\gamma}_\varepsilon]$, which provides the holographic entanglement entropy for this infinite wedge $A$, is the minimum between 
the area of $\hat{\gamma}_{A}^{\textrm{\tiny \,dis}} \cap \{ z\geqslant \varepsilon \}$ and  the area of $\hat{\gamma}_{A}^{\textrm{\tiny \,con}}\cap \{ z\geqslant \varepsilon \} $.
Being $P_{A,B} = 2L$ for both $\hat{\gamma}_{A}^{\textrm{\tiny \,dis}}$ and $\hat{\gamma}_{A}^{\textrm{\tiny \,con}}$, 
the minimal area surface $\hat{\gamma}_A$ must be found by comparing the coefficients of the subleading logarithmic divergence.
This comparison leads to the following corner function 
\be
\label{bdy Fcal generic}
\mathcal{F}_\alpha(\omega , \gamma) 
\,=\,
 \textrm{max}\, \Big\{  \widetilde{F}(\omega)\,, \, F_\alpha(\gamma) + F_\alpha(\tilde{\gamma})  \Big\}
\qquad
\tilde{\gamma} = \pi - (\omega + \gamma)
\ee
where the first function within the parenthesis corresponds to $\hat{\gamma}_{A}^{\textrm{\tiny \,dis}}$
and the second one to $\hat{\gamma}_{A}^{\textrm{\tiny \,con}}$.
The corner function $\widetilde{F}(\omega)$ is the one found in 
\cite{Drukker:1999zq} and reviewed in Sec.\,\ref{sec drop},
while $F_\alpha(\gamma) $ is the corner function discussed in Sec.\,\ref{sec:wedge}.
Let us remind that, since $\tilde{\gamma} \in (0,\pi)$ in (\ref{bdy Fcal generic}) we mean $ F_\alpha(\tilde{\gamma}) = F_\alpha(\textrm{min}[\,\tilde{\gamma} \,, \, \pi - \tilde{\gamma} \,]) $,
as stated in Sec.\,\ref{sec constraints}.

It could be useful to compare (\ref{bdy Fcal generic}) with (\ref{holog corner function W vertex symm3}).
Indeed, by extending the half plane $x\geqslant 0$ to the whole $\mathbb{R}^2$ 
and including the reflected image of $A$ obtained by sending $x \to -x$, one obtains the symmetric configuration of corners underlying 
(\ref{holog corner function W vertex symm3}).
Nonetheless, let us stress that (\ref{bdy Fcal generic}) with (\ref{holog corner function W vertex symm3}) are not equivalent because
in (\ref{bdy Fcal generic}) the boundary conditions (which correspond to $\alpha$ in this holographic setup) play a central role.

The corner function (\ref{bdy Fcal generic}) occurs in the constraints from the strong subadditivity found
in Sec.\,\ref{sec constraints}. 
In the appendix\;\ref{app constraints} we show that the holographic corner functions $F_\alpha(\gamma)$ and $\mathcal{F}_\alpha(\omega , \gamma) $ 
fulfils these constraints.

\begin{figure}[t!] 
\vspace{-.8cm}
\hspace{-1.3cm}
%\begin{center}
\includegraphics[width=1.15\textwidth]{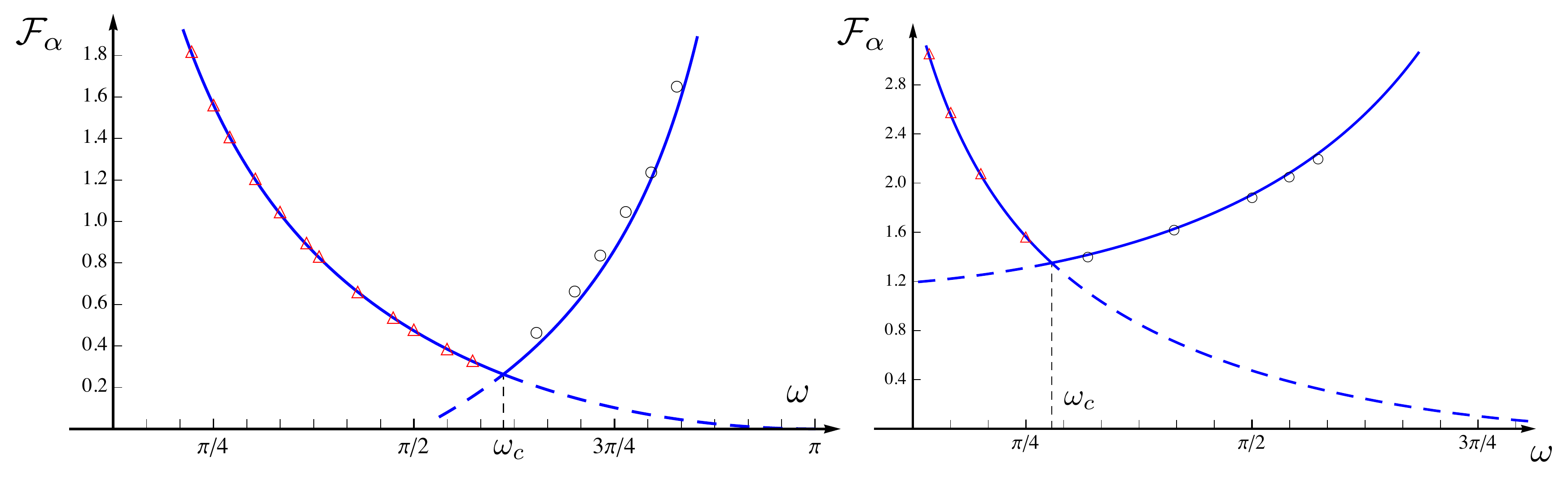}
%\end{center}
\vspace{-.1cm}
\caption{\label{fig:bdyDropTrans}
\small 
The corner function (\ref{bdy Fcal symm}) for symmetric configurations  of the infinite wedge (i.e. $\tilde{\gamma} = \gamma$).
The slope $\alpha$ of $\mathcal{Q}$ is different in the two panels: $\alpha = \pi/2.3$ (left) and $\alpha = 2\pi/3$ (right).
The solid blue line is obtained from the analytic expression (\ref{bdy Fcal symm}).
The data points have been found by constructing minimal surfaces with Surface Evolver anchored to single drop domains
whose opening angle of the corner is $\omega$.
The minimal surfaces corresponding to the empty black circles are connected to $\mathcal{Q}$ (see e.g. Fig.\,\ref{fig:3dBdyTipSym}),
while the ones corresponding to the empty red triangles are disconnected from $\mathcal{Q}$.
The critical value $\omega_c$ is defined by (\ref{omega_c def symm}).
Notice that $\omega_c > \pi/2$ when $\alpha < \pi/2$ and $\omega_c < \pi/2$ when $\alpha > \pi/2$.
}
\end{figure}

For the sake of simplicity, let us consider first the subclass of infinite wedges which are symmetric 
with respect to the half line departing from the tip and orthogonal to the boundary.
For these wedges $\tilde{\gamma} = \gamma$; therefore the supplementarity condition implies that $\gamma = (\pi - \omega)/2$.
Thus, these configurations are fully determined by $\omega$ (equivalently, one can adopt $\gamma$ as independent variable).
By substituting $\omega = \pi -2\gamma$ into (\ref{bdy Fcal generic}), 
we find that for these symmetric wedges the corner function simplifies to
\be
\label{bdy Fcal symm}
\mathcal{F}_\alpha(\omega , \gamma) 
\,=\,
 \textrm{max}\Big\{  \widetilde{F}(\omega)\,, 2 F_\alpha(\gamma)  \Big\}
 \qquad
 \gamma = \frac{\pi - \omega}{2}
\ee

The maximisation procedure occurring in (\ref{bdy Fcal generic}) and (\ref{bdy Fcal symm}) chooses the first function for some configurations
and the second function for other ones. 
In particular, there exist critical configurations such that the two functions in the r.h.s.'s of (\ref{bdy Fcal generic}) and (\ref{bdy Fcal symm}) provide the same result,
namely both $\hat{\gamma}_{A}^{\textrm{\tiny \,dis}}$ and $\hat{\gamma}_{A}^{\textrm{\tiny \,con}}$ have the same coefficient of the logarithmic divergence.

In Fig.\,\ref{fig:3dBdyTipSym} we show two examples of minimal area surfaces obtained with Surface Evolver
which correspond to single drop domains $A$ (see Sec.\,\ref{sec single drop}) whose corners have the tip on the boundary and belong to this 
class of symmetric wedges having $\tilde{\gamma} = \gamma$.
In a neighbourhood of the tips of these two domains the minimal area surface $\hat{\gamma}_A$ is given by $\hat{\gamma}_{A}^{\textrm{\tiny \,con}}$.

In Fig.\,\ref{fig:bdyDropTrans} the corner function (\ref{bdy Fcal symm}) is plotted as function of $\omega$ for two particular values of $\alpha$.
The critical value $\omega_c$, where the two functions in the r.h.s. of (\ref{bdy Fcal symm}) are equal, 
is highlighted by the vertical dashed segments and it depends on the slope $\alpha$.
For $\omega < \omega_c$ the minimal surface $\hat{\gamma}_A$ is disconnected from $\mathcal{Q}$ and it is like the one shown in Fig.\,\ref{fig:single_drop},
while for $\omega > \omega_c$ it is connected to $\mathcal{Q}$ and it looks like the minimal surfaces depicted in Fig.\,\ref{fig:3dBdyTipSym}.
The minimal surfaces in Fig.\,\ref{fig:3dBdyTipSym} are prototypical examples of the surfaces employed to find the numerical data 
corresponding to the empty circles in Fig.\,\ref{fig:bdyDropTrans}.

\begin{figure}[t!] 
\vspace{-1cm}
%\hspace{-1.1cm}
\begin{center}
\includegraphics[width=.9\textwidth]{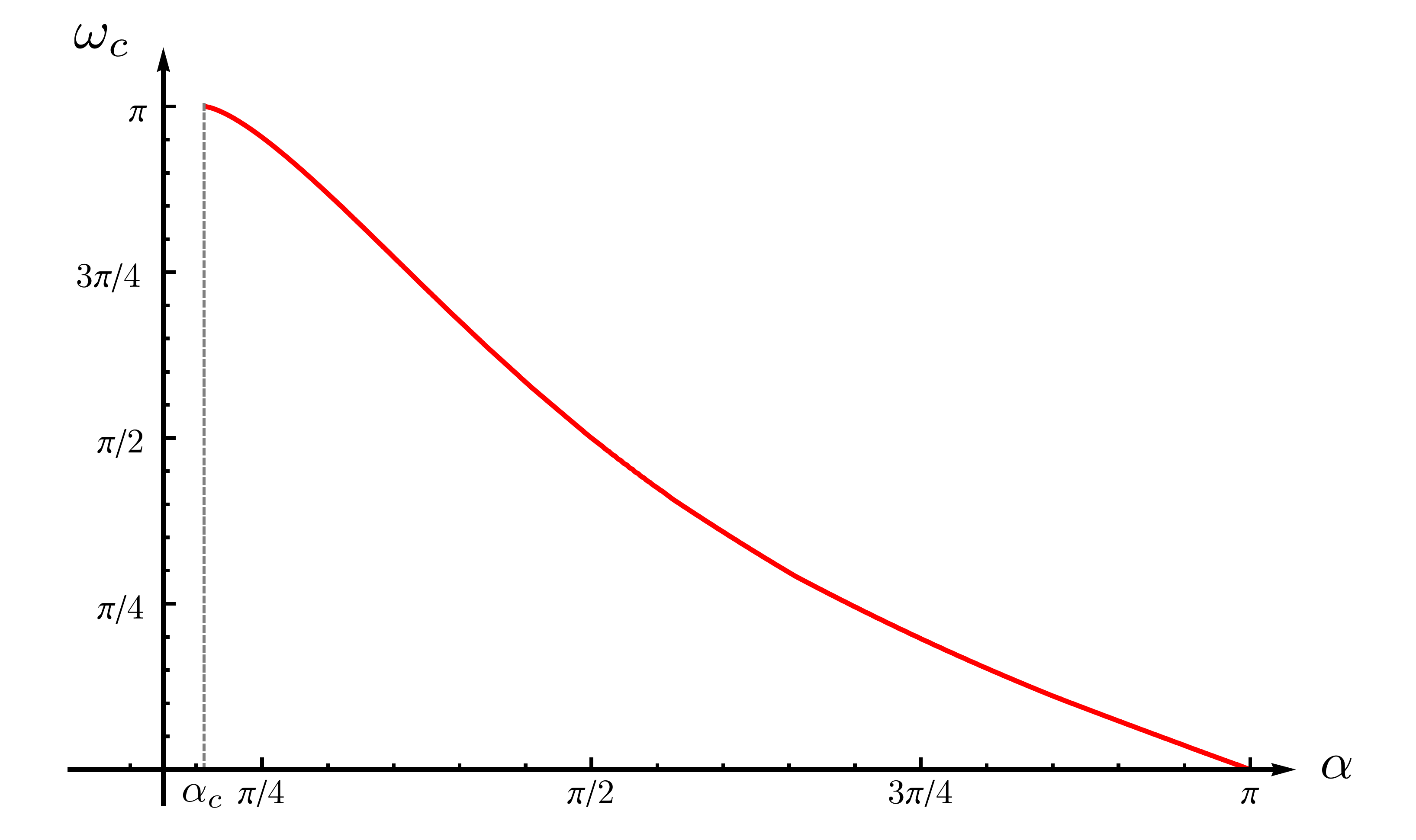}
\end{center}
\vspace{-.4cm}
\caption{\label{fig:omega_crit}
\small
Infinite wedge with only the tip on the boundary and $\tilde{\gamma} = \gamma$:
The critical opening angle $\omega_c $ as function of $\alpha \geqslant \alpha_c$.
The curve has been found by solving (\ref{omega_c def symm}) numerically. 
}
\end{figure}

By applying the remark made above about (\ref{bdy Fcal generic}) to this simpler situation, 
it could be instructive to compare  (\ref{bdy Fcal symm}) with (\ref{holog corner function W vertex symm}), 
which has been found for the analogous situation in AdS$_4$/CFT$_3$, 
as it can be observed by using the image method. 
Nonetheless, we remark again that in (\ref{bdy Fcal symm}) the parameter $\alpha$ enters in a crucial way.
By performing the same analysis done for Fig.\,\ref{fig:bdyDropTrans} setting $\alpha = \pi/2$, 
we have checked numerically the data shown in Fig.\,\ref{fig:drukker_double} are consistent with the relation (\ref{corner func at pi/2}).

In the remaining part of this section we describe the critical configurations 
corresponding to (\ref{bdy Fcal generic}) and to (\ref{bdy Fcal symm}).

Let us consider first the class of symmetric wedges where $\tilde{\gamma} = \gamma$.
From (\ref{bdy Fcal symm}), we have that the critical configuration is characterised by the opening angle $\omega_c=\omega_c(\alpha)$
which solves the following equation
\be
\label{omega_c def symm}
\widetilde{F}(\omega_c) \,=\, 2\, F_\alpha\big((\pi-\omega_c)/2\big) 
\ee

As consistency check we can set $\alpha = \pi/2$.
In this case, by employing (\ref{corner func at pi/2}) in the r.h.s. of (\ref{omega_c def symm}), 
the equation (\ref{omega_c def symm}) becomes $\widetilde{F}(\omega_c)= \widetilde{F}(\pi- \omega_c)$, whose solution is $\omega_c =\pi/2$,
as expected from the general fact the results in AdS$_4$/CFT$_3$ (see Fig.\,\ref{fig:drukker_double} for this quantity) 
are recovered in our AdS$_4$/BCFT$_3$ setup for $\alpha=\pi/2$.

We find it worth focussing also on the special value $\alpha = \alpha_c$.
By employing the characteristic property of $\alpha_c$  given by (\ref{alpha_c def zeroF})  and the fact that $\widetilde{F}(\pi) = 0$ into (\ref{omega_c def symm}),
we find
\be
\label{alpha_critic_def_omega}
\lim_{\alpha \,\to\, \alpha_c}\omega_c(\alpha)  \,=\, \pi
\ee
Since $\omega < \pi$, the limit (\ref{alpha_critic_def_omega}) tells us that, within the class of symmetric wedges with  $\tilde{\gamma} = \gamma$, 
the minimal area surface $\hat{\gamma}_A$ is always $\hat{\gamma}_{A}^{\textrm{\tiny \,dis}}$ when $\alpha \leqslant \alpha_c$.
This observation can be inferred also from  (\ref{bdy Fcal symm}) because $F_\alpha(\gamma) \leqslant 0$ for $\alpha \leqslant \alpha_c$, while $\widetilde{F}(\omega) \geqslant 0$.
Thus, when $\alpha  \leqslant \alpha_c$,
the transition from $\hat{\gamma}_A = \hat{\gamma}_{A}^{\textrm{\tiny \,dis}}$ to $\hat{\gamma}_A = \hat{\gamma}_{A}^{\textrm{\tiny \,con}}$ 
as $\omega$ increases does not occur.
The absence of this transition is a characteristic feature of the regime $\alpha \leqslant \alpha_c$ that can be detect with finite domains. 
We have not been able to get reliable numerical data from Surface Evolver for values of alpha close enough to $\alpha_c$;
therefore we have not observed (\ref{alpha_critic_def_omega}) numerically. 
Hopefully, future analysis will address this numerical issue.

\begin{figure}[t!] 
\vspace{-.8cm}
\hspace{-1cm}
%\begin{center}
\includegraphics[width=1.1\textwidth]{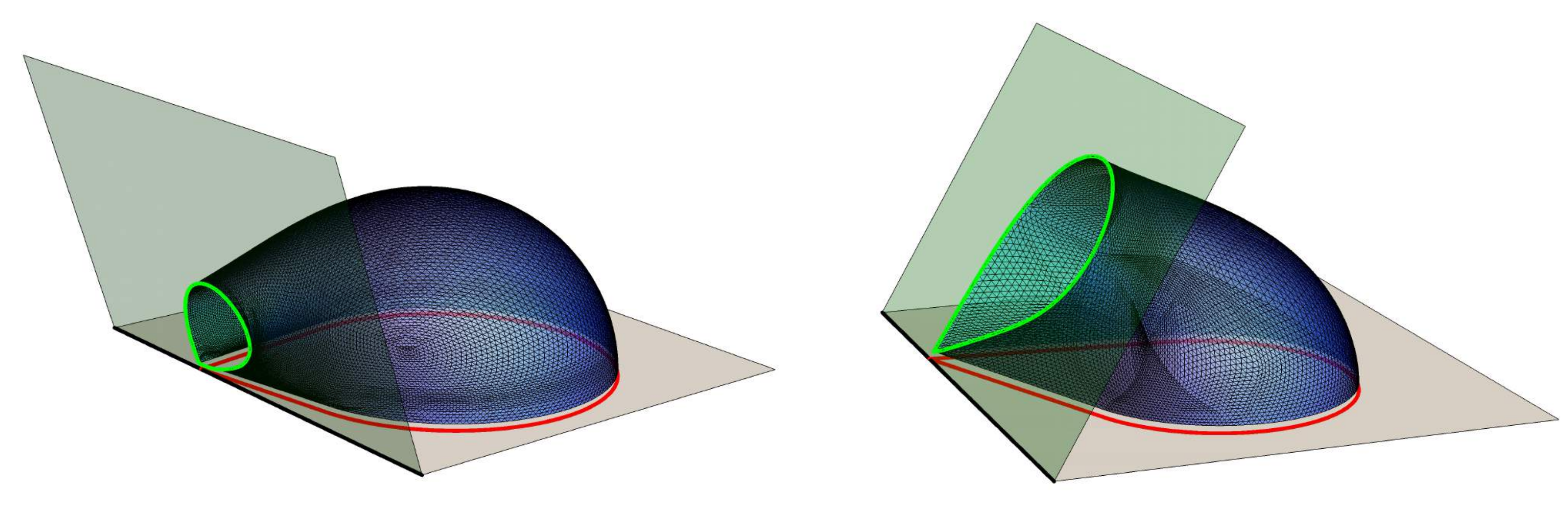}
%\end{center}
\vspace{-.1cm}
\caption{\label{fig:3dBdyTipAsym}
\small
Minimal surfaces $\hat{\gamma}_A$ obtained with Surface Evolver and corresponding to a single drop $A$ 
such that the entangling curve $\partial A$ (solid red curve in the $z=0$ grey half plane) intersects the boundary at the tip of its corner.
For these configurations of $A$, the corresponding minimal surface is the surface 
which intersects $\mathcal{Q}$ (green half plane) orthogonally along the green curve. 
In the left panel $\alpha = \pi/2.5$, $\omega = \pi/2$, $\gamma = \pi/2 - \pi/5$ and $L=0.75$,
while in the right panel $\alpha = 2\pi/3$, $\omega = \pi/3$, $\gamma = \pi/2 - \pi/5$ and $L=1$.
In both panels $\varepsilon = 0.03$.
}
\end{figure}

In Fig.\,\ref{fig:omega_crit} we show the curve $\omega_c(\alpha)$ of the critical opening angle for the symmetric wedges,
which has been obtained by solving (\ref{omega_c def symm}) numerically. 
Notice that the curve lies above the  straight line tangent to it and passing through the point $\alpha = \pi/2$.

In the general case $\tilde{\gamma} \geqslant \gamma$ and the configuration of the infinite  wedge is 
characterised by the independent angles $\gamma$ and $\omega$.
In Fig.\,\ref{fig:3dBdyTipAsym} we show the minimal area surfaces constructed with Surface Evolver which
are anchored to two different configurations of a single drop domains $A$ having the tip on the boundary and with $\tilde{\gamma} > \gamma$.
For the configurations in Fig.\,\ref{fig:3dBdyTipAsym}, the minimal area surface $\hat{\gamma}_A$ in the neighbourhood of the tip is given by $\hat{\gamma}_{A}^{\textrm{\tiny \,con}}$.

As discussed above, critical configurations exist such that the two functions involved in the maximisation procedure of (\ref{bdy Fcal generic}) have the same value.
For a given slope $\alpha$, we can equivalently characterise these configurations either by the critical value $\omega_c = \omega_c(\gamma, \alpha) $ in terms of $\gamma$ 
or by the critical value $\gamma_c = \gamma_c(\omega, \alpha) $ in terms of $\omega$.  
Choosing the former option, the critical value $\omega_c = \omega_c(\gamma, \alpha) $ is the solution of the following equation
\be
\label{eq omega_c generic}
\widetilde{F}(\omega_c) 
\,=\, 
F_\alpha(\gamma) + F_\alpha(\tilde{\gamma}) 
\qquad
\tilde{\gamma} = \pi - (\omega_c + \gamma)
\ee

\begin{figure}[t!] 
\vspace{-.9cm}
%\hspace{-1cm}
\begin{center}
\includegraphics[width=.9\textwidth]{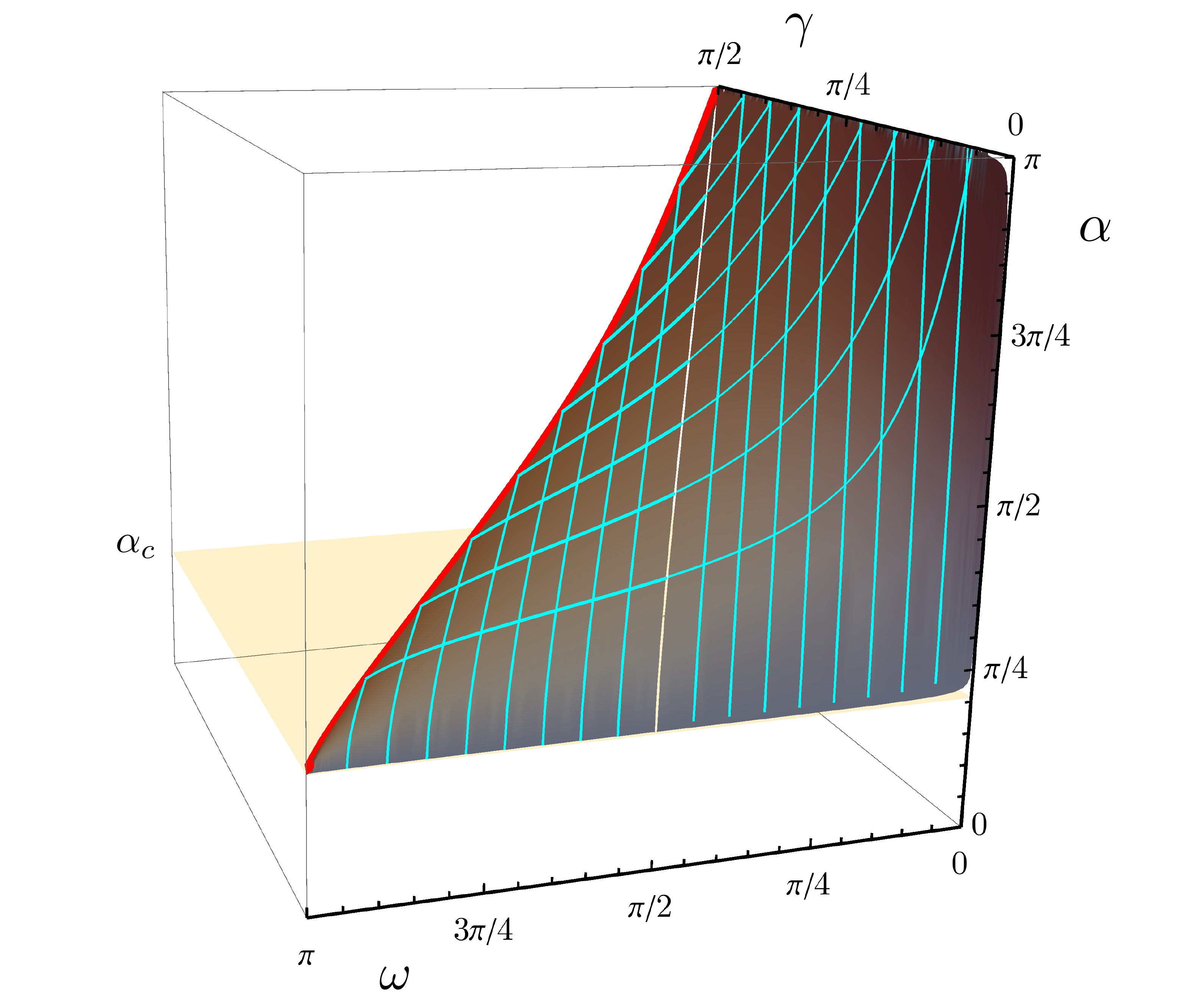}
\end{center}
\vspace{-.3cm}
\caption{\label{fig:3D_dom}
\small
Infinite wedge with only the tip on the boundary:
The surface described by the critical configurations, defined by 
(\ref{eq omega_c generic}) in the parameters space given by the angles $\omega$, $\gamma$ and $\alpha$.
The yellow plane is $\alpha = \alpha_c$.
The red curve corresponds to the symmetric configurations having $\tilde{\gamma} = \gamma$
(see Fig.\,\ref{fig:omega_crit}).
}
\end{figure}

In Fig.\,\ref{fig:3D_dom} we show the surface which characterises the critical configurations, 
obtained by solving (\ref{eq omega_c generic}) numerically. 
Notice that the surface lies in the range $\alpha > \alpha_c$, as expected from the above 
considerations. 
The red solid curve in Fig.\,\ref{fig:3D_dom} corresponds to the symmetric case $\gamma = \tilde{\gamma}$,
namely to the curve in Fig.\,\ref{fig:omega_crit}.
Furthermore, the section at $\alpha = \pi/2$ of the surface in Fig.\,\ref{fig:3D_dom}
provides the critical configurations for the symmetric domains in a CFT$_3$ 
whose coefficient of the logarithmic divergence of the corresponding holographic entanglement entropy 
is (\ref{holog corner function W vertex symm3}), which have been described in Sec.\,\ref{sec double drop ads4}.

%%%%%%%%%%%%%%%%%%%%%%%%%%%%%%%%%%%%%%%%%%%%%%%%%%%%%
\section{Conclusions}
\label{sec:conclusions}

Understanding the role of the boundary conditions in the analysis of the entanglement entropy in BCFT$_3$ 
is an interesting problem within the program of studying entanglement in quantum field theories.

Considering  a BCFT$_3$ with a flat boundary,
in this manuscript we mainly focussed on the entanglement entropy of two dimensional domains $A$ in a constant time slice 
whose boundaries $\partial A$ intersect the boundary of the BCFT$_3$.
In particular we have studied the cases where the singular points of $\partial A$ belong also to the boundary of the BCFT$_3$
(see e.g. the yellow region in the right panel of Fig.\,\ref{fig:intro}).
The expansion of the entanglement entropy of these domains as the UV cutoff $\varepsilon \to 0$ 
contains also a logarithmic divergence  whose coefficient encodes 
the characteristic features of the BCFT$_3$ through some corner functions in a non trivial way. 

In this manuscript we have studied the holographic corner functions in the 
AdS$_4$/BCFT$_3$ setup  introduced in \cite{Takayanagi:2011zk},
where the gravitational spacetime is bounded by a surface $\widetilde{\mathcal{Q}}$ anchored to the boundary of the BCFT$_3$,
which is obtained by solving certain Neumann boundary conditions.
In our simplified case where the boundary of the BCFT$_3$ is flat, 
$\widetilde{\mathcal{Q}}$ on a constant time slice is given by a half plane $\mathcal{Q}$ characterised by its slope $\alpha \in (0, \pi)$.

The holographic entanglement entropy has been computed by employing the prescription (\ref{RT formula bdy}), 
where the minimal area surface $\hat{\gamma}_A$ must be found among the surfaces $\gamma_A$ anchored to the entangling curve $\partial A \cap \partial B$.
Since the curve $\gamma_A \cap \mathcal{Q}$ can vary and restrictions are not imposed on it,
the minimisation of the area leads to the condition that $\hat{\gamma}_A$ is orthogonal to $\mathcal{Q}$ along the curve given by their intersection.

In this AdS$_4$/BCFT$_3$ setup, as preliminary simple cases we have computed the holographic entanglement entropy of infinite strips, 
both adjacent and parallel to the boundary (see also \cite{Chu:2017aab}), and of a half disk centered on the boundary. 

Our main result is the analytic expression of the corner function $F_\alpha(\gamma)$ for an infinite wedge adjacent to the boundary, 
which is given by (\ref{gamma q0 main}) and (\ref{total corner func bdy main}) in a parametric form
(see Fig.\,\ref{fig:main_plot} and Fig.\,\ref{fig:main_plot_3D}).
This result and the corner function of \cite{Drukker:1999zq} lead to the  analytic formula (\ref{bdy Fcal generic})
for the corner function $\mathcal{F}_\alpha(\omega , \gamma)$,
which corresponds to an infinite wedge having only its tip on the boundary.

Various checks have been done to test the analytic expressions of these two corner functions. 
The main one is the numerical analysis performed by employing Surface Evolver \cite{evolverpaper, evolverlink},
where minimal area surfaces corresponding to finite domains containing corners have been explicitly constructed 
in order to study the coefficient of the logarithmic divergence of their area. 
Further non trivial consistency checks have been considered by studying 
the limiting regimes $\gamma \to 0^+$ and $\gamma \to \pi/2$ of the corner function $F_\alpha(\gamma)$.
In the limit $\gamma \to 0^+$ the holographic entanglement entropy of the infinite strip adjacent to the boundary has been recovered,
while taking the limit $\gamma \to \pi/2$ we have obtained
the coefficient of the logarithmic divergence in the holographic entanglement entropy of the half disk centered on the boundary,
as expected. 

We remark that interesting transitions have been observed in the analysis of the holographic entanglement entropy for the 
various domains. 
The main one occurs in the slope $\alpha$ at the critical value given by (\ref{alpha_crit}).
This transition can be observed also through the behaviour of the corner function $F_\alpha(\gamma)$ in the regime $\gamma \to 0^+$.
We have also studied the transitions occurring in the holographic entanglement entropy 
of an infinite strip parallel to the boundary and at finite distance from it (see Fig.\,\ref{fig:2strips})
and of an infinite wedge with only the tip on the boundary (see Fig.\,\ref{fig:3D_dom}).

Some of the results mentioned above have been studied for a generic spacetime dimension. 

An interesting outcome of our analysis is the relation (\ref{rel}) found in the context of the AdS$_4$/BCFT$_3$ correspondence defined in \cite{Takayanagi:2011zk}, 
which involves the coefficient $f''_\alpha(\pi/2)$ obtained from the expansion of $F_\alpha(\gamma)$ as $\gamma \to \pi/2$
and the coefficient $A_T$ characterising the behaviour of the one point function of the stress tensor $\langle \, T_{ij} \, \rangle$ close to the boundary (see (\ref{A_T def})).
In particular, (\ref{rel}) tells us that the ratio between these coefficients is independent of $\alpha$.

Let us conclude by mentioning some open problems for future studies. 

An important conceptual issue to understand in the AdS/BCFT setup of \cite{Takayanagi:2011zk}
is the possible relation occurring between the geometrical parameter $\alpha$ and the space of the 
conformally invariant boundary conditions for the dual BCFT$_3$.
We find it important also to study different AdS/BCFT constructions \cite{Miao:2017gyt, Chu:2017aab, Astaneh:2017ghi}.
Within these setups, it is also relevant to consider a dual BCFT$_3$ with non flat boundaries,
which are not related to the flat one through a conformal transformation. 

An interesting issue that we find worth exploring is the possibility that the relation (\ref{rel}) holds for other models of BCFT$_3$.

Finally, the extension of the analysis performed in this manuscript to higher dimensions, where different kinds of singular configurations occur, 
is certainly important to improve our understanding of the holographic entanglement entropy in AdS/BCFT.

\subsection*{Acknowledgments}

It is our pleasure to thank Andrea Coser, Andrea Gambassi, Matthew Headrick, Esperanza Lopez and Mukund Rangamani
for useful discussions.
We are grateful in particular to Dmitri Fursaev, Rong-Xin Miao, Sergey Solodukhin and Tadashi Takayanagi  
for important comments or correspondence. 
We acknowledge in a special way Cristiano De Nobili for his crucial help with Surface Evolver.
ET is grateful to the IFT, Madrid, and to the University of Florence for the warm hospitality during some stages of this work.

\newpage

%%%%%%%%%%%%%%%%%%%%%%%%%%%%%%%%%%%%%%%%%%%%%%%%%%%%%
\appendix

\section{On the numerical analysis}
\label{app numerics}

Our numerical analysis is based on {\it Surface Evolver}, a multipurpose optimisation software developed by Ken Brakke \cite{evolverpaper, evolverlink}.
This tool is employed here to find minimal area surfaces embedded in the three dimensional hyperbolic space $\mathbb{H}_3$, whose metric is (\ref{AdS4 metric}).
The constraints imposed on the minimal surfaces define the ones we are interested in.

In this manuscript we deal with two qualitative different situations, depending on the occurrence of  the half plane $\mathcal{Q}$ defined by (\ref{brane-profile}).
For the corner functions in AdS$_4$/CFT$_3$ discussed in Sec.\,\ref{sec drop} we employ the standard prescription (\ref{RT formula intro}) for the holographic entanglement
entropy, which requires to construct the minimal surface $\hat{\gamma}_A$ anchored to $\partial A$ in the $z=0$ plane.
Instead, to compute the holographic entanglement entropy in AdS$_4$/BCFT$_3$ discussed in Sec.\,\ref{sec HEE bdy}, 
the minimal area surface $\hat{\gamma}_A$ belongs to the region of $\mathbb{H}_3$ defined by (\ref{AdS4 regions}) and 
it must be anchored only to the entangling curve $\partial A \cap \partial B$ in the $z=0$ half plane. 
Thus, while in the former case $\partial \hat{\gamma}_A =\partial A$, in the latter one $\partial \hat{\gamma}_A \subset \partial A$ 
and it can happen that $\partial \hat{\gamma}_A  \cap \mathcal{Q} \neq \emptyset$.
When $\partial \hat{\gamma}_A  \cap \mathcal{Q} \neq \emptyset$, the minimisation procedure implemented by Surface Evolver 
leads to surfaces which are orthogonal to $\mathcal{Q}$ along $\hat{\gamma}_A \cap \mathcal{Q}$ in the final step of the evolution.

Surface Evolver constructs surfaces as unions of triangles; therefore a smooth surface is approximated by a surface made by triangles
obtained through a particular evolution. 
The initial step of the optimisation procedure is a very simple surface, made by few triangles, which basically sets the topology.
The initial surface evolves towards a configuration which is a local minimum of the area functional 
by both increasing the number of triangles and modifying the mesh in a proper way. 
For each step of the evolution, the software provides all the elements characterising the surface, like 
the coordinates of the vertices, the way to connect them, the normal vectors, the area of each triangle, the total number of triangles and the total area of the surface.
We refer the interested reader also the appendix B of \cite{Fonda:2014cca} for another discussion on the application of Surface Evolver to find minimal area surfaces in $\mathbb{H}_3$.

 Since the area of a surface reaching the boundary at $z=0$ diverges, in our numerical analysis we have defined the entangling curve $\partial A \cap \partial B$ 
 (which coincides with $\partial A$ for the domains considered in Sec.\,\ref{sec drop} to study the corner functions in AdS$_4$/CFT$_3$) at $z=\varepsilon$
 and not at $z=0$, as required in the prescription for the holographic entanglement entropy. 
 This way to regularise the final result does not influence the coefficients of the diverging terms in the expansion of the area $\mathcal{A}[\hat{\gamma}_\varepsilon]$ as $\varepsilon \to 0^+$ \cite{Drukker:1999zq}.

 Once  the final entangling curve $\partial A\cap \partial B$ has been fixed at $z=\varepsilon$,
 let us denote by $\gamma^{\textrm{\tiny SE}}_\varepsilon$ the triangulated surface constructed by Surface Evolver at a generic step of the evolution
 and by $\tilde{\mathcal{A}}[\gamma^{\textrm{\tiny SE}}_\varepsilon]$  the corresponding numerical value for its area provided by the software. 
 We denote by $\tilde{\gamma}^{\textrm{\tiny SE}}_\varepsilon$ the  final configuration of the evolution and by 
 $\tilde{\mathcal{A}}[\tilde{\gamma}^{\textrm{\tiny SE}}_\varepsilon]$ the corresponding area given by Surface Evolver.
 The final step of the evolution depends on the required level of approximation. 
 In our analysis the typical value of the UV cutoff is $\varepsilon = 0.03$, 
 the area of the final surfaces is $O(10^{2})$ (setting $L_{\textrm{\tiny AdS}} =1$) 
 and we have stopped the evolution once the value of the area was stable up to small variations of order $O(10^{-2})$.

\begin{figure}[t] 
\vspace{-.8cm}
\hspace{-.7cm}
%\begin{center}
\includegraphics[width=1.1\textwidth]{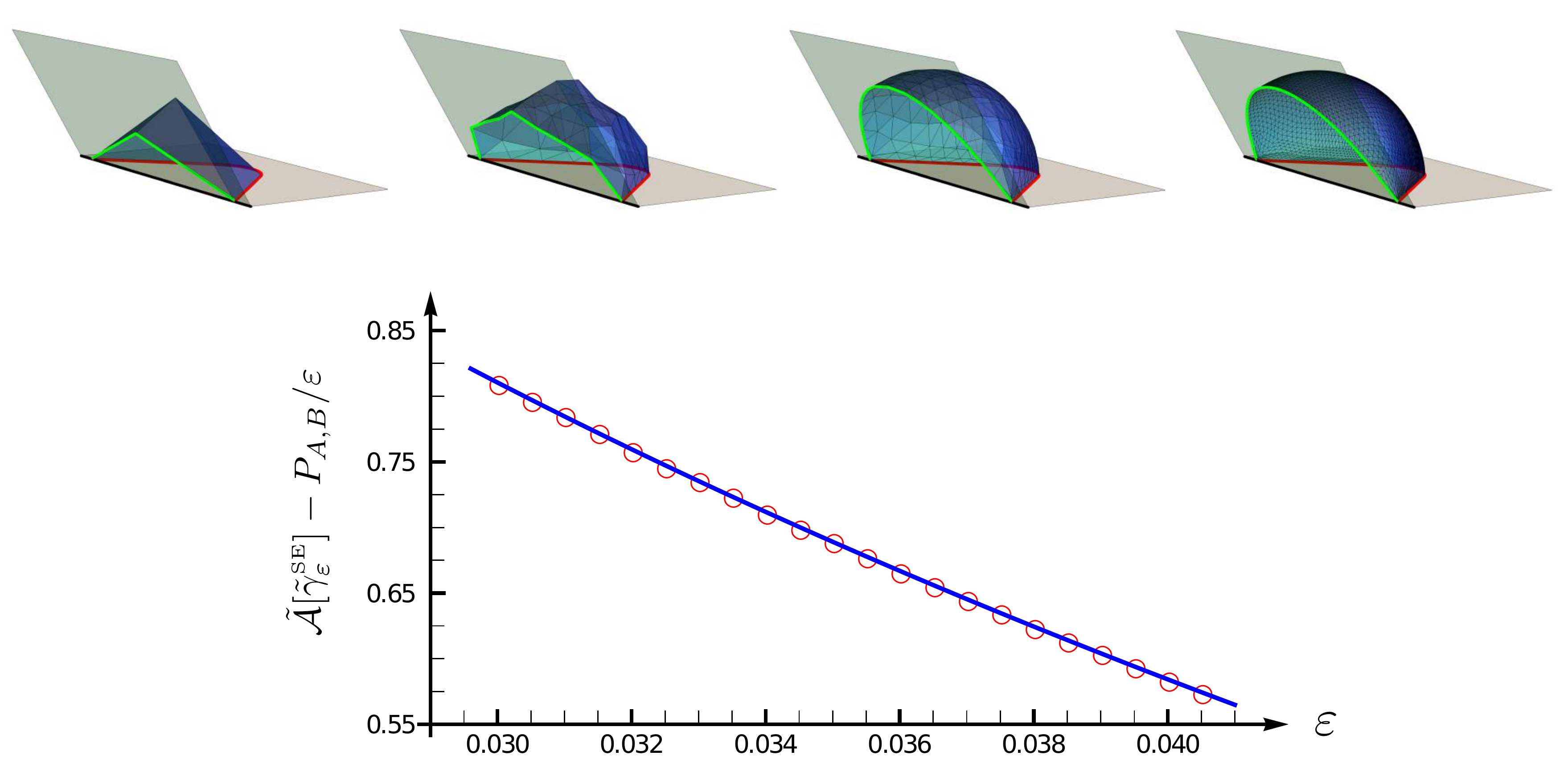}
%\end{center}
\vspace{-.4cm}
\caption{\label{fig:SEevolution}
\small 
An example of the numerical analysis of the corner functions based on Surface Evolver. 
Top: Some stages of an evolution towards the minimal area surface anchored to the entangling curve given by the red line
in the $z=0$ half plane, which has $\gamma = \pi/4$ and $L=2.5$ (see also Fig.\,\ref{fig:3dBdyDrop}). Here $\alpha = \pi/3$.
Bottom: Numerical data corresponding to the evolutions shown in the top panel for different values of $\varepsilon$. 
Fitting this data as discussed in Sec.\,\ref{app numerics}, one finds the numerical value for the corner function to compare 
with the corresponding one obtained from the analytic expression $F_\alpha(\gamma)$ given by (\ref{gamma q0 main}) and (\ref{total corner func bdy main}).
}
\end{figure}

The evolution begins from a very simple trial surface  and it develops through a number of steps which improves the triangulation of the surface towards configurations with smaller area. 
A way to improve the triangulation consists in moving the positions of the vertices without changing their total number according to a gradient descent method which decreases the total area of the surface.
Another way is to refine the mesh of the surface by splitting each edge of a facet into two new edges and then connecting them.
After a modification of this kind, a facet is partitioned into four new facets; therefore this step increases the total number of triangles.

The boundaries of the triangulated surfaces are treated differently during the evolution, depending on whether they belong to the half plane $\mathcal{Q}$ or to the section of the spacetime given by $z=\varepsilon$
(for the surfaces studied in Sec.\,\ref{sec drop} only the latter situation occurs).
 The vertices on the entangling curve $\partial A \cap \partial B$ at $z=\varepsilon$ are kept fixed although their number increases during the refinements. 
 Instead, the vertices of the curve $\partial \gamma^{\textrm{\tiny SE}}_\varepsilon \cap \mathcal{Q}$ can move freely on $\mathcal{Q}$ during the evolution.

In the top part of Fig.\,\ref{fig:SEevolution} we show some steps of an evolution made by Surface Evolver towards the minimal area surface
anchored to the entangling curve given by the red line in the $z=0$ plane (see also Fig.\,\ref{fig:3dBdyDrop}).
In this example $\partial \gamma^{\textrm{\tiny SE}}_\varepsilon \cap \mathcal{Q} \neq \emptyset$.
The initial step of the evolution is a trial surface made by 6 facets while the last step shown in the figure is a triangulated surface with 6144 facets.

The coefficient of the logarithmic divergence in the expansion of the area as $\varepsilon \to 0^+$ has been extracted as follows. 
Once the final step $\tilde{\gamma}^{\textrm{\tiny SE}}_\varepsilon$ of the evolution corresponding to a given entangling curve at $z=\varepsilon $ is reached, 
one subtracts to $\tilde{\mathcal{A}}[\tilde{\gamma}^{\textrm{\tiny SE}}_\varepsilon] $ the area law term, which is given by  either $P_A / \varepsilon$ or $P_{A,B} / \varepsilon$.
By repeating this analysis for various small values of $\varepsilon$, a list of numerical values is obtained. 
Fitting these data points through the function $a \log\varepsilon+b+c \,\varepsilon$, one finds the best fit for the parameters $a$, $b$ and $c$.
The value of $a$ is the numerical result for the coefficient of the logarithmic divergence that we have compared against the corresponding theoretical prediction.
In the bottom part of Fig.\,\ref{fig:SEevolution} we show an example of this procedure which corresponds to the domain $A$ identified by the red curve in the
top part of the same figure.

As a final technical remark, let us observe that, whenever $\tilde{\gamma}^{\textrm{\tiny SE}}_\varepsilon \cap \mathcal{Q}\ne \emptyset$  
the numerical analysis gets worse as the angle $\gamma$ decreases because of the formation of spikes at the tips of the corners.
This explains why we did not obtain reliable results for small values of $\gamma$ in Fig.\,\ref{fig:main_plot}.
The occurrence of unexpected spikes depends on $\alpha$ and it is observed for larger values of $\gamma$ as $\alpha$ decreases (see the lowest curve in Fig.\,\ref{fig:main_plot}).

%%%%%%%%%%%%%%%%%%%%%%%%%%%%%%%%%%%%%%%%%%%%%%%%%%%%%%%%%%%%%%%%%
\section{On the half disk centered on the boundary}
\label{sec app half disk}

In this appendix we report the computation of the area $\mathcal{A} [\hat{\gamma}_\varepsilon ] $, which provides the holographic entanglement entropy
of half disk of radius $R$ centered on the boundary, according to the prescription (\ref{RT formula bdy}).
The main result derived here is (\ref{area half-disk final main}), which is discussed in Sec.\,\ref{sec:half disk}.

Given the half disk $A=\{ (x,y) \in \mathbb{R}^2 \, | \,x^2+y^2\leqslant R^2 ,  \,x\geqslant 0\}$, which is centered on the boundary $x=0$,
the entangling curve $\partial A \cap \partial B$ is $\{ (x,y) \in \mathbb{R}^2 \, | \,x^2+y^2 = R^2 ,  \,x\geqslant 0\}$.
In Sec.\,\ref{sec:half disk} we have discussed for this domain $\hat{\gamma}_{A,\textrm{\tiny aux}}$ is the hemisphere $x^2+y^2+z^2 = R^2$ in $\mathbb{H}_3$ 
and that $\hat{\gamma}_A$ is just the part of $\hat{\gamma}_{A,\textrm{\tiny aux}}$ identified by the constraint (\ref{AdS4 regions}).
In Fig.\,\ref{fig:3Dhalf-disk}, the minimal surface $\hat{\gamma}_A$ is shown in a case having $\alpha < \pi/2$ and in a case where $\alpha > \pi/2$.

The holographic entanglement entropy is obtained by evaluating the area $\mathcal{A} [\hat{\gamma}_\varepsilon ] $ of the surface 
$\hat{\gamma}_A \cap \{z\geqslant \varepsilon \}$,  which is the part of $\hat{\gamma}_A$ above the yellow line in Fig.\,\ref{fig:3Dhalf-disk}.
This area  can be written as follows
\be
\label{A-split half disk}
\frac{\mathcal{A} [\hat{\gamma}_\varepsilon ] }{ L^2_{\textrm{\tiny AdS}} }
= 
\left\{\begin{array}{ll}
\mathcal{A}_{\perp} +  \mathcal{A}_{\angle}    \hspace{1cm} &  0 < \alpha \leqslant \pi/2
\\
\rule{0pt}{.5cm}
\mathcal{A}_{\perp} -  \mathcal{A}_{\angle}   &    \pi/2 \leqslant  \alpha < \pi
\end{array}
\right.
\ee
where $\mathcal{A}_{\perp} $ is the area of the half hemisphere restricted to $z\geqslant \varepsilon$  with $x\geqslant 0$ 
and $\mathcal{A}_{\angle}   \geqslant 0$ is the area of  the part of the hemisphere  restricted to $z\geqslant \varepsilon$ enclosed between the vertical half plane $x=0$ and the half plane $\mathcal{Q}$.
Notice that, in the right panel of Fig.\,\ref{fig:3Dhalf-disk}, the area $\mathcal{A}_{\angle} $ corresponds to the shaded part of $\hat{\gamma}_{A,\textrm{\tiny aux}}$.

The area $\mathcal{A}_{\perp} $ can be easily computed by adopting the usual spherical coordinates $(\theta, \phi)$, where $\theta = 0$ is the positive $z$ semi-axis and $\phi = 0$ is the positive $y$ semi-axis.
The change of coordinates between these polar coordinates and the Cartesian coordinates reads
\be
z= R\, \cos\theta
\qquad
x= R\, \sin\theta \, \sin\phi
\qquad
y= R\, \sin\theta \, \cos\phi 
\ee
In terms of the polar coordinates $(\theta, \phi)$, the induced metric on $\hat{\gamma}_A$ from $\mathbb{H}_3$ is given by
\be
ds^2\big|_{\hat{\gamma}_A}
= \frac{L^2_{\textrm{\tiny AdS}}}{(\cos\theta)^2} \big( d\theta^2 + (\sin\theta)^2 d\phi^2 \big)
\ee
By employing this metric, for $\mathcal{A}_{\perp} $ we find
\be
\label{Aperp half-disk}
\mathcal{A}_{\perp} 
= 
\int_0^{\theta_\varepsilon} d\theta \int_{0}^{\pi} d\phi\; \frac{\sin\theta}{(\cos\theta)^2}
= 
\,   \frac{\pi}{\cos\theta} \bigg|_0^{\theta_\varepsilon}
= \,
\frac{\pi R}{\varepsilon} - \pi 
\ee
where the condition defining $\theta_\varepsilon $ is $\varepsilon= R \cos \theta_\varepsilon $.

In order to compute $\mathcal{A}_{\angle} $, let us parameterise the hemisphere by employing spherical coordinates $(\theta, \phi)$, 
where $\theta = 0$ is the positive $y$ semi-axis and $\phi = 0$ is the positive $z$ semi-axis.
Now the change of coordinates is
\be
\label{polar coords spike}
z= R\, \sin\theta\, \cos \phi
\qquad
x= -\, R\, \sin\theta \, \sin\phi
\qquad
y= R\, \cos\theta
\ee
The induced metric on $\hat{\gamma}_A$ from $\mathbb{H}_3$ in terms of these polar coordinates is
\be
ds^2\big|_{\hat{\gamma}_A}
= \frac{L^2_{\textrm{\tiny AdS}}}{(\sin\theta)^2 \, (\cos\phi)^2} \big( d\theta^2 + (\sin\theta)^2 d\phi^2 \big)
\ee
From the first expression in (\ref{polar coords spike}) we obtain $ \varepsilon = R\, \sin\theta_\varepsilon \cos \phi$,
which relates the UV cutoff $\varepsilon$ to the cutoff $\theta_\varepsilon$ of the angular variable. 
This relation leads to  $\sin(\theta_\varepsilon/2) = \varepsilon [1+O(\varepsilon^2) ] /(2 R\cos\phi) $.

When $\alpha \in (0,\pi/2)$, the area $\mathcal{A}_{\angle} $ is given by the following integral
\bea
\label{integral A_angle}
\mathcal{A}_{\angle} 
&=& 
2 \int_{0}^{\pi/2-\alpha} d\phi  \int_{\theta_\varepsilon}^{\pi/2} d\theta\; \frac{1}{(\cos\phi)^2 \sin\theta}
\;=\;
-\,2   \int_{0}^{\pi/2-\alpha} d\phi  \; \frac{\log(\tan \theta_\varepsilon /2)}{(\cos\phi)^2}
\nonumber
  \\
 \rule{0pt}{.7cm}
&=& 
 2   \int_{0}^{\pi/2-\alpha} d\phi  \; \frac{\tfrac{1}{2}\log(1-[\sin(\theta_\varepsilon/2)]^2 ) - \log[\sin (\theta_\varepsilon /2)]  }{(\cos\phi)^2}
 \nonumber
  \\
 \rule{0pt}{.7cm}
 \label{step half sphere}
&=& 
2  \int_{0}^{\pi/2-\alpha} d\phi  \; \frac{1}{(\cos\phi)^2}
 \left(\,\frac{1}{2}\log\big(1- \varepsilon^2/[2R\cos\phi]^2\big) - \log(\varepsilon/R ) + \log(2\cos\phi) \right)
 +O(\varepsilon^2)
\nonumber
 \\
  \rule{0pt}{.5cm}
   \label{fin-step}
&=& 2 (\cot \alpha) \log(R/\varepsilon) + O(1)
\eea
where in (\ref{step half sphere}) the relation between $\theta_\varepsilon$ and $\varepsilon$ has been employed and 
the $O(\varepsilon^2) $ terms have been neglected. 
The $O(1)$ term in (\ref{fin-step}) can be found explicitly, but we do not report it here because we are interested only in the logarithmic divergence. 
When $\alpha \in (\pi/2,\pi)$, being $\mathcal{A}_{\angle} > 0$, the resulting integral for $\mathcal{A}_{\angle}$ is like (\ref{integral A_angle}), except for the
domain of integration for the integral in $\phi$, which is $(0,\alpha-\pi/2)$.

Summarising, the term $\mathcal{A}_{\angle} $ provides the following logarithmic divergence
\be
\label{Awedge half-disk}
 \mathcal{A}_{\angle} 
= 
\left\{\begin{array}{ll}
2 (\cot \alpha) \log(R/\varepsilon) + O(1)  
\hspace{1cm} &  0 < \alpha \leqslant \pi/2
\\
\rule{0pt}{.5cm}
- 2  (\cot \alpha) \log(R/\varepsilon) + O(1)
&    \pi/2 \leqslant  \alpha < \pi
\end{array}
\right.
\ee

Finally, by plugging (\ref{Aperp half-disk}) and (\ref{Awedge half-disk}) into  (\ref{A-split half disk}), 
we obtain the area $\mathcal{A} [\hat{\gamma}_\varepsilon ] $ given by (\ref{area half-disk final main}), which is the main 
result of this appendix. 

Let us stress that the holographic entanglement entropy for this domain provides the corner function
$F_\alpha(\pi/2)$ for the special value $\gamma = \pi/2$ and for any $\alpha \in (0,\pi)$.
This is an important benchmark for the analytic expression of the corner function $F_\alpha(\gamma)$
presented in Sec.\,\ref{sec:wedge}, whose derivation is described in the appendix\;\ref{sec app wedge}.

\section{Infinite strip adjacent to the boundary in generic dimension}
\label{sec app strip}

In this appendix we study the holographic entanglement entropy for the $d$ dimensional infinite strip 
of width $\ell$ adjacent to the boundary.
The main results of this analysis specialised to $d=2$ have been reported in Sec.\;\ref{sec:strip}.

Given a constant time slice of a BCFT$_{d+1}$, defined by $x \geqslant 0$ in proper Cartesian coordinates, 
let us consider the following spatial domain 
\be
\label{strip ddim}
 A= \{ (x,y_1, \dots, y_{d-1})\, | \; 0\leqslant x \leqslant  \ell\, , \,0\leqslant y_i \leqslant  L_\parallel \}
 \qquad
 L_\parallel \gg \ell \gg \varepsilon
\ee
The invariance under translations along the $y_i$-axis (in a strict sense, this requires $L_\parallel \to + \infty$) 
allows us to assume that the minimal surface $\hat{\gamma}_A$ is characterised by its profile obtained by sectioning 
$\hat{\gamma}_A$ through an hyperplane defined by $y_i =\textrm{const}$.
The profile of $\hat{\gamma}_A$ is given by either $x=\ell$ or by a non trivial curve $z=z(x)$.
Focussing on the latter case, let us denote by $P_\ast = (x_\ast , z_\ast)$ the intersection between the curve $z(x)$ and the section 
at $y_i =\textrm{const}$ of the half hyperplane $\mathcal{Q}$, which is a half line given by (\ref{brane-profile}).
The coordinates of $P_\ast $ are constrained by imposing that $P_\ast  \in \mathcal{Q}$ and this condition gives
\be
\label{P_ast on Q condition}
x_\ast = - \, z_\ast  \cot \alpha 
\ee
where we recall that $\alpha \in (0,\pi)$.
Since the curve  $z(x)$ characterising the extremal surface  intersects orthogonally the section at constant 
$y_i =\textrm{const}$ of the half hyperplane $\mathcal{Q}$, it is not difficult to realise that $z'(x_\ast) = \cot\alpha$.

The profile $z(x)$ can be obtained by finding the extrema of the area functional among the surfaces $\gamma_A$
anchored to the edge $x=\ell$ of the strip (\ref{strip ddim}) which are invariant under translations along the $y_i$
directions and intersect $\mathcal{Q}$ orthogonally.

Given a surface $\gamma_A$ characterised by $z(x)$, by writing the metric induced on $\gamma_A$ from the background (\ref{AdS metric ddim}), 
one obtains the following area functional
\be
\label{area RT strip 0}
\mathcal{A} [\gamma_A ] \,=\, 
L^d_{\textrm{\tiny AdS}}  \, L_\parallel^{d-1} \int_{x_\ast}^{\ell} \frac{\sqrt{1+(z')^2}}{z^d}\, dx
\ee
Since the integrand does not depend on $x$ explicitly, we can find the extremal surface $\hat{\gamma}_A$ by employing the fact that the first integral of motion is constant. 
For the functional (\ref{area RT strip 0}) this condition  tells us that $z^d \sqrt{1+(z')^2}$ is independent of $x$;
therefore we can evaluate it at any point on the extremal surface described by $z(x)$.
By choosing the point $(x_\ast , z_\ast)$, where $z'(x_\ast) = \cot\alpha$, the equation imposing the constancy of the first integral of motion reads
\be
\label{first integral strip}
z^d \sqrt{1+(z')^2} 
%= \, z_{\textrm{\tiny max}}^2 
%= \,z_\ast^2 \sqrt{1+(\cot \alpha)^2} 
= \,\frac{z_\ast^d}{\sin \alpha} 
\ee

In order to solve (\ref{first integral strip}), we find it convenient to introduce the following parameterisation 
\be
\label{z strip paramet}
z(\theta) \,=\, \frac{z_\ast}{(\sin\alpha)^{1/d}}\,(\sin \theta)^{1/d}
\hspace{.6cm}  \qquad  \hspace{.6cm} 
0\leqslant \theta \leqslant \pi -\alpha
\ee
which respects the boundary conditions $z(\pi - \alpha) = z_\ast $ and $z(0) = 0$.

Plugging (\ref{z strip paramet}) into the square of (\ref{first integral strip}), one gets $(\tfrac{dz}{dx})^2 = (\cot\theta)^2$, which gives $x'(\theta)^2 = z'(\theta)^2 (\tan\theta)^2$.
Then, by employing (\ref{z strip paramet}) into the latter differential equation, we obtain
\be
\label{x prime strip}
x'(\theta) \,=\, - \,\frac{z_\ast}{d\,(\sin \alpha)^{1/d}}\, (\sin \theta)^{1/d}
\ee
where the physical condition that $x(\theta)$ decreases for increasing values of $\theta $ has been imposed. 

The relation $(\tfrac{dz}{dx})^2 = (\cot\theta)^2$ and (\ref{z strip paramet}) leads to the geometrical meaning of the angle $\theta$:
it is the angle between the outgoing vector normal to the curve given by $P_\theta$ and the $x$ semi-axis with $x\geqslant 0 $.
Thus, from (\ref{z strip paramet}) we have that $\theta = \pi/2$ corresponds to the point of the curve $z(x)$ having the maximum value 
$ z_{\textrm{\tiny max}} = z_\ast / (\sin \alpha)^{1/d}$.

By integrating (\ref{x prime strip}) with the initial condition $x(0) = \ell$, we find
\bea
\label{x(theta)}
x(\theta) 
&=&\,
\ell - \frac{z_\ast}{d\,(\sin \alpha)^{1/d}}
\int_0^{\theta}  (\sin \tilde{\theta})^{1/d} \,d\tilde{\theta}
\\
\label{x(theta)}
\rule{0pt}{.8cm}
&=&\,
\ell  -  \frac{z_\ast}{(\sin \alpha)^{1/d}}
   \bigg[ \, \frac{\sqrt{\pi} \; \Gamma \big( \tfrac{d+1}{2d} \big)}{\Gamma \big( \tfrac{1}{2d} \big)}
   - \frac{\cos\theta}{d}\,
  _2F_1\bigg( \frac{d-1}{2d}\, , \frac{1}{2}\,; \frac{3}{2}\, ;  (\cos\theta)^2\bigg) 
   \,\bigg]
\eea

The expressions (\ref{z strip paramet}) and (\ref{x(theta)}) depend on the coordinate $z_\ast$ of the  point $P_\ast$.
We can relate $z_\ast$ to the width $\ell$ of the strip (\ref{strip ddim}) by imposing that  (\ref{x(theta)}) satisfies the  consistency condition $x(\pi -\alpha) = x_\ast$, 
where $x_\ast$ can be obtained from (\ref{P_ast on Q condition}).
This gives
\be
\ell  - \frac{z_\ast}{(\sin \alpha)^{1/d}} \,
   \bigg[ \, \frac{\sqrt{\pi} \; \Gamma \big( \tfrac{d+1}{2d} \big)}{\Gamma \big( \tfrac{1}{2d} \big)}
   + \frac{\cos\alpha}{d}\,
  _2F_1\bigg( \frac{d-1}{2d}\, , \frac{1}{2}\,; \frac{3}{2}\, ;  (\cos\alpha)^2\bigg) 
   \,\bigg]
   \,=\,
   - \,z_\ast \cot\alpha
\ee
which leads to the following relation 
\be
\label{z_ast ddim}
z_\ast \,=\,
\frac{(\sin\alpha)^{1/d}}{\mathfrak{g}_d(\alpha)}
\;\ell
\ee
where we have introduced
\be
\label{g def ddim}
\mathfrak{g}_d(\alpha) 
\,\equiv\,
 \frac{\sqrt{\pi} \; \Gamma \big( \tfrac{d+1}{2d} \big)}{\Gamma \big( \tfrac{1}{2d} \big)}
   + \frac{\cos\alpha}{d}\;
  _2F_1\bigg( \frac{d-1}{2d}\, , \frac{1}{2}\,; \frac{3}{2}\, ;  (\cos\alpha)^2\bigg) 
   - (\sin\alpha)^{1/d}  \cot\alpha
\ee
We remark that $z_\ast>0$, therefore (\ref{z_ast ddim}) is well defined only when $\mathfrak{g}_d(\alpha)  >0$, being $\alpha \in (0,\pi)$.
For $d=2$, which is the case considered through the main text, the function (\ref{g def ddim})
becomes the function $\mathfrak{g}(\alpha)  \equiv \mathfrak{g}_2(\alpha) $ in (\ref{g function def main}).

\begin{figure}[t] 
\vspace{-.8cm}
%\hspace{-.25cm}
\begin{center}
\includegraphics[width=.92\textwidth]{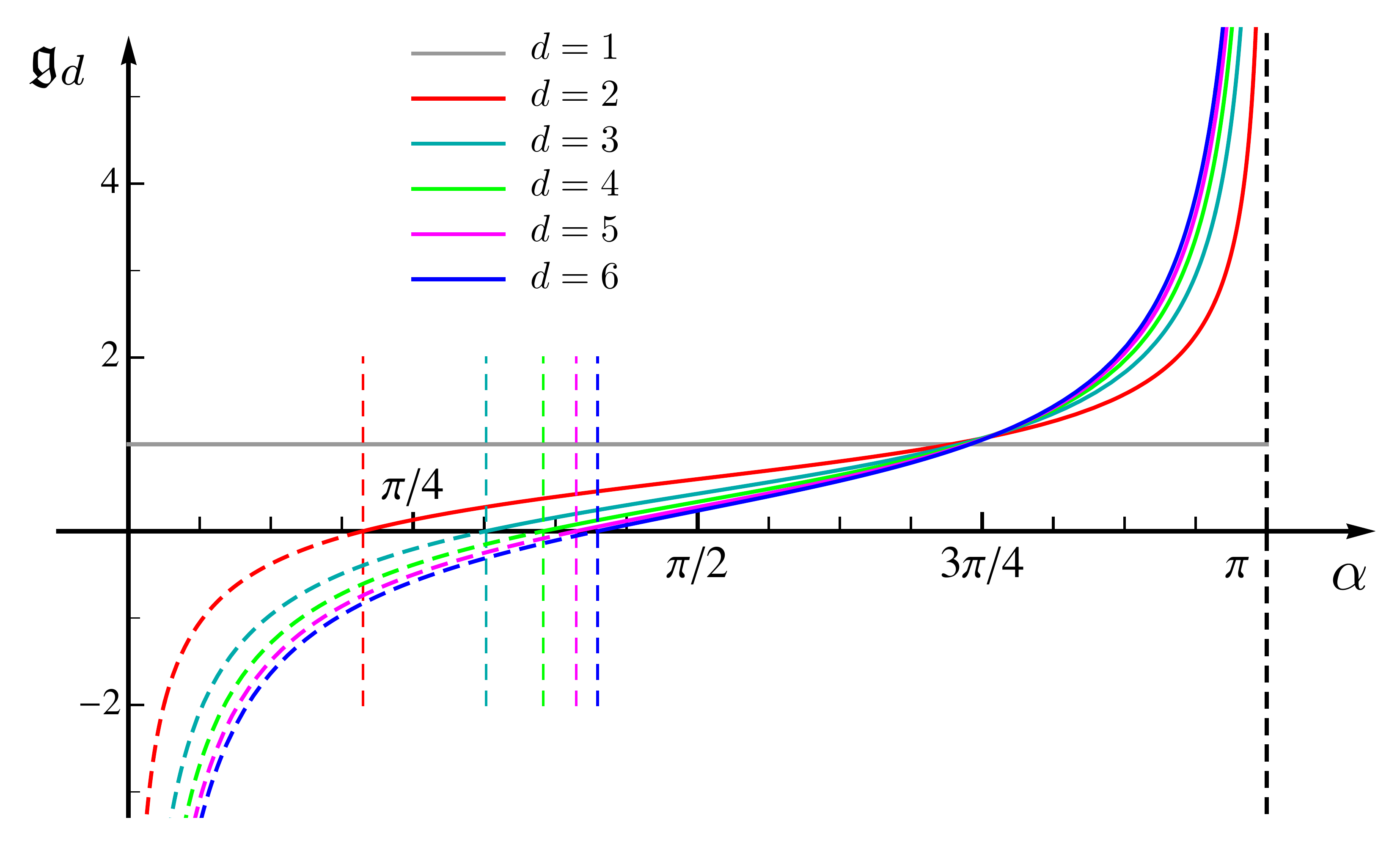}
\end{center}
\vspace{-.3cm}
\caption{\label{fig:g_ddim}
\small
The function $\mathfrak{g}_d(\alpha)$ defined in (\ref{g def ddim}) for some values of $d$.
For a given $d$, the critical value $\alpha_c(d)$ is the unique zero of $\mathfrak{g}_d(\alpha)$ (see (\ref{alpha_c def ddim}))
and it has been highlighted through a vertical dashed segment having the same colour of the corresponding curve $\mathfrak{g}_d(\alpha)$.
}
\end{figure}

\begin{figure}[t] 
\vspace{-.8cm}
%\hspace{-.25cm}
\begin{center}
\includegraphics[width=.85\textwidth]{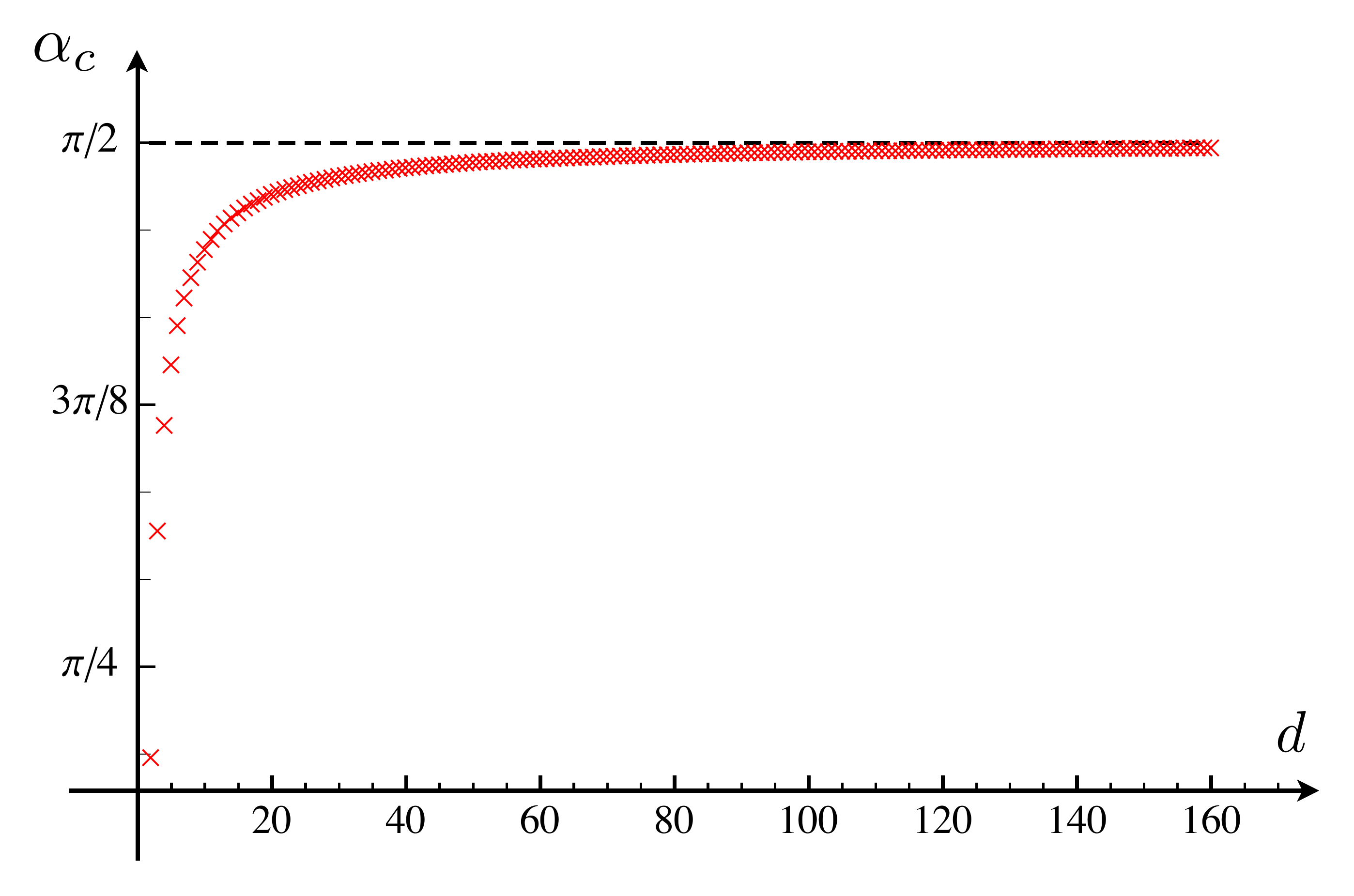}
\end{center}
\vspace{-.3cm}
\caption{\label{fig:alpha_crit}
\small
The critical slope $\alpha_c(d)$ of the half plane $\mathcal{Q}$ as function of the dimensionality parameter $d\geqslant 2$.
These points have been found by solving (\ref{alpha_c def ddim}).
The value $\alpha_c(2)$ is given by (\ref{alpha_crit}).
We find that $\alpha_c(d) \to \pi/2$ as $d\to + \infty$.
}
\end{figure}

The first derivative of $\mathfrak{g}_d(\alpha) $ with respect to $\alpha$ is very simple
\be
\label{g-prime-alpha}
\partial_\alpha \,\mathfrak{g}_d(\alpha) 
%= \frac{d-1}{d} \,(\csc\alpha)^{2-1/d}
= \bigg( 1- \frac{1}{d} \bigg) \,(\sin\alpha)^{1/d-2}
\ee
This expression tells us that $\mathfrak{g}_1(\alpha) $ is constant and, in particular, one finds $\mathfrak{g}_1(\alpha) = 1$ identically. 
When $d>1$, we have that $\mathfrak{g}'_d(\alpha) >0$ for  $\alpha \in (0,\pi)$.
Moreover, $\mathfrak{g}_d(\alpha) = -1/\alpha^{1-1/d} + O(1)$ as $\alpha \to 0^+$ and $\mathfrak{g}_d(\alpha) = 1/(\pi- \alpha)^{1-1/d} + o(1)$ as $\alpha \to \pi^-$.
These observations allow to conclude that (\ref{g def ddim}) has a unique zero $\alpha =\alpha_c$ for $d>1$, namely
\be
\label{alpha_c def ddim}
\mathfrak{g}_d(\alpha_c) = 0 
\ee

Since $z_\ast > 0$  in (\ref{z_ast ddim}), the condition (\ref{alpha_c def ddim})  defines a critical value $\alpha_c(d)$ for the slope of $\mathcal{Q}$.
Indeed, (\ref{z_ast ddim}) is well defined only for $\alpha \in (\alpha_c, \pi)$.
Moreover, from (\ref{z_ast ddim}) and (\ref{alpha_c def ddim}) we have that $z_\ast \to + \infty$ when $\alpha \to \alpha_c^+$.
These observations allow us to conclude that  for $\alpha \in (0, \alpha_c]$ the solution which intersects orthogonally the half hyperplane $\mathcal{Q}$ at a finite value of $z_\ast$ does not exist;
therefore $\hat{\gamma}_A$ is the vertical half hyperplane $x=\ell$ in this range of $\alpha$.

We remark that $\alpha_c \leqslant \pi/2$.
Indeed, for $\alpha > \pi/2$ it is straightforward to observe that the vertical half hyperplane $x=\ell$ is excluded because it
does not intersect orthogonally the half hyperplane $\mathcal{Q}$.

We find it worth considering the limit $d\to + \infty$ of (\ref{g def ddim}).
In this regime only the last term gives a non vanishing contribution and, in particular, we have $\mathfrak{g}_d(\alpha)  \to -\cot\alpha$, meaning that $\alpha_c(d) \to \pi/2$.
Thus, $\alpha_c$ tends to its  natural upper bound for large $d$.

In Fig.\,\ref{fig:g_ddim} the function $\mathfrak{g}_d(\alpha)$ is shown for $1\leqslant d\leqslant 6$.
The corresponding critical values $\alpha_c(d)$ for $d\leqslant 2$ are highlighted through vertical dashed lines. 
The value of $\alpha_c(d=3)$ has been found also in \cite{Chu:2017aab}.
In Fig.\,\ref{fig:alpha_crit} we provide the critical slope $\alpha_c(d)$ as function of the dimensionality parameter $d$.

The profile $z(x)$ of the extremal solution intersecting $\mathcal{Q}$ orthogonally at a finite value $z_\ast$ can be found
by plugging (\ref{z_ast ddim}) into (\ref{z strip paramet}) and (\ref{x(theta)}).
The result reads
\be
\label{profile strip adj ddim}
\big( x(\theta) \,, z(\theta) \big) 
=\, \frac{\ell}{\mathfrak{g}_d(\alpha) }
\left(
   \frac{\cos\theta}{d}\,
  _2F_1\bigg( \frac{d-1}{2d}\, , \frac{1}{2}\,; \frac{3}{2}\, ;  (\cos\theta)^2\bigg) 
  - \frac{\sqrt{\pi} \; \Gamma \big( \tfrac{d+1}{2d} \big)}{\Gamma \big( \tfrac{1}{2d} \big)}
+ \mathfrak{g}_d(\alpha) 
\,, \,
(\sin\theta)^{1/d}
\,\right)
\ee
It is not difficult to check that this profile satisfies the required boundary conditions. 
Indeed for $\theta=0$ and $\theta = \pi-\alpha$ we find $P_{0} = (\ell, 0)$ and $P_\ast = z_\ast ( -\cot\alpha \,, 1 )$ respectively, being $z_\ast$ given by (\ref{z_ast ddim}).
The expression of (\ref{profile strip adj ddim}) specialised to $d=2$ has been reported in (\ref{strip profile final}).

An interesting point of the curve $z(x)$ is the one where $z'(x)$ vanishes.
Denoting  its coordinates by $P_{\textrm{\tiny max}} = (x_{\textrm{\tiny max}} , z_{\textrm{\tiny max}} )$, we have that $z'(x_{\textrm{\tiny max}}) = 0$.
From the latter condition and  (\ref{first integral strip}) one finds a relation between $P_{\textrm{\tiny max}} $ and $P_\ast$ given by
\be
\label{z_max(alpha)}
z_{\textrm{\tiny max}} 
= \frac{z_\ast}{(\sin \alpha)^{1/d}} 
= \frac{\ell}{\mathfrak{g}_d(\alpha)} 
\ee
The first equality can be obtained also from (\ref{z strip paramet}) for $\theta = \pi/2$, as remarked above, 
while in the last step (\ref{z_ast ddim}) has been used. 
Notice that for $0< \alpha < \pi/2$ we have that $z_{\textrm{\tiny max}} > z_\ast$, being $\theta - \alpha \neq \pi/2$.
Instead, $P_{\textrm{\tiny max}} = P_\ast$ when $\alpha = \pi/2$, 
while  $P_{\textrm{\tiny max}}$ does not exist when $\alpha > \pi/2 $.
These features can be observed in Fig.\,\ref{fig:strip_profile} for the case $d=2$. 

We find it worth remarking that the minimal surface $\hat{\gamma}_A$ characterised by (\ref{profile strip adj ddim})
is part of an auxiliary surface $\hat{\gamma}_{A,\textrm{\tiny aux}}$  which has minimal area in the 
hyperbolic space $\mathbb{H}_{d+1}=\,$AdS$_{d+2}\big|_{t \,=\, \textrm{const}}$  and which is anchored to an infinite strip $A_{\textrm{\tiny \,aux}}$ 
of width $\ell_{\textrm{\tiny aux}}$
belonging to the boundary $z=0$ of $\mathbb{H}_{d+1}\,$.
The auxiliary infinite strip $A_{\textrm{\tiny \,aux}}$  includes $A$ and it shares with $A$ the edge at $x=\ell$.
The minimal surface $\hat{\gamma}_{A,\textrm{\tiny aux}}$ has been computed in \cite{RT}.

By employing the results of \cite{RT} and 
imposing that (\ref{z_max(alpha)}) is also the largest value assumed by the coordinate $z$ for the points of $\hat{\gamma}_{A,\textrm{\tiny aux}}$,
we find that
\be
\label{ell_aux strip ddim}
\ell_{\textrm{\tiny aux}} = 2\,
\frac{\sqrt{\pi} \, \Gamma(\tfrac{d+1}{2d})}{\Gamma(\tfrac{1}{2d})\, \mathfrak{g}_d(\alpha)}\; \ell 
\ee
In particular, $\ell_{\textrm{\tiny aux}}$ depends on $\alpha$.
As consistency check of (\ref{ell_aux strip ddim}), we observe that 
$\ell_{\textrm{\tiny aux}} = \ell - x(\pi)$, where $x(\theta)$ has been written in (\ref{profile strip adj ddim}).

In order to evaluate the area for $z\geqslant \varepsilon$ of the extremal surface characterised by the profile (\ref{profile strip adj ddim}), 
let us  compute the metric induced on this surface by the background metric of $\mathbb{H}_{d+1}\,$.
By setting $t = \textrm{const}$ into  (\ref{AdS metric ddim}) and employing the relation $x'(\theta)^2 = z'(\theta)^2 (\tan\theta)^2$ derived above
(see the text below (\ref{z strip paramet})), we find that the induced metric reads
\bea
ds^2\big|_{\hat{\gamma}_A} 
&=&\frac{L^2_{\textrm{\tiny AdS}}}{z(\theta)^2} \left[\, \frac{z'(\theta)^2}{(\cos\theta)^2}\, d\theta^2 + d\vec{y}^{\,2} \, \right]
\\
\label{induced metric strip ddim}
\rule{0pt}{.8cm}
&=&
\frac{L^2_{\textrm{\tiny AdS}} \,(\sin \alpha)^{2/d}}{z_\ast^2 \,(\sin \theta)^{2/d}} 
\left[\, \frac{z_\ast^2}{d^2\, (\sin\alpha)^{2/d} \, (\sin \theta)^{2(1-1/d)}}\, d\theta^2 + d\vec{y}^{\,2} \, \right]
\eea
where $ d\vec{y}^{\,2}=\sum_{j=1}^{d-1} dy_j^2$ and (\ref{z strip paramet}) have been used to obtain the last expression.

Let us focus on the cases with $d>1$ first. 
From (\ref{induced metric strip ddim}), for the area of $\hat{\gamma}_\varepsilon$ we find 
\bea
\frac{\mathcal{A} [\hat{\gamma}_\varepsilon ]}{L^d_{\textrm{\tiny AdS}}} 
&=&
\frac{(\sin \alpha)^{1-1/d}}{d\, z_\ast^{d-1}}\,
\int_0^{L_\parallel } dy_1 \dots dy_{d-1} \, \int_{\theta_\varepsilon}^{\pi-\alpha} \frac{d\theta}{(\sin \theta)^{2-1/d}}
\\
\label{area strip ddim full}
\rule{0pt}{.8cm}
&=&
\frac{(\sin \alpha)^{1-1/d}}{d\, z_\ast^{d-1}} \; L_\parallel^{d-1}
\left[\,
  _2F_1\bigg( \frac{3d-1}{2d}\, , \frac{1}{2}\,; \frac{3}{2}\, ;  (\cos\theta)^2\bigg) \cos\theta\,
\right]
\bigg|^{\theta_\varepsilon}_{\pi-\alpha} 
\hspace{1.1cm}
\eea
where the cutoff $\theta_\varepsilon$ is defined by imposing that $z(\theta_\varepsilon)=\varepsilon$, being $z(\theta)$ the expression in (\ref{z strip paramet}).
This gives $\theta_\varepsilon = \arcsin(\varepsilon^d \sin \alpha/z_\ast^d)$.

Taking the limit $\varepsilon \to 0^+$ in (\ref{area strip ddim full}) and neglecting  terms which vanish in this limit, we find
\bea
\label{area strip z_ast ddim}
\frac{\mathcal{A} [\hat{\gamma}_\varepsilon ]}{L^d_{\textrm{\tiny AdS}}} 
&=&
L_\parallel^{d-1} \,\Bigg\{\,
 \frac{1}{(d-1) \, \varepsilon^{d-1}}
 \\
 & &
 \hspace{1.5cm}
  - \;
\frac{(\sin \alpha)^{1-1/d}}{z_\ast^{d-1}}\, \Bigg[\,
 \frac{\sqrt{\pi} \; \Gamma \big( \tfrac{d+1}{2d} \big)}{(d-1)\, \Gamma \big( \tfrac{1}{2d} \big)}
 % \frac{\pi^{3/2}}{(d-1) \cos\big(\tfrac{\pi}{2d}\big)\, \Gamma\big(\tfrac{d-1}{2d}\big)\, \Gamma\big(\tfrac{1}{2d}\big)}
  -
  \frac{\cos\alpha}{d}\;
  _2F_1\bigg( \frac{3d-1}{2d}\, , \frac{1}{2}\,; \frac{3}{2}\, ;  (\cos\alpha)^2\bigg) 
\Bigg]
\Bigg\}
\nonumber
\eea
We remark that the divergent part of the area $\mathcal{A} [\hat{\gamma}_\varepsilon ]$ is due to the area term only. 

The above analysis extends smoothly to the whole range of $\alpha \in (0,\pi)$ 
the results of \cite{Chu:2017aab} for the infinite strip adjacent to the boundary, which hold for  $\alpha \in (0,\pi/2]$.

The finite term in  (\ref{area strip z_ast ddim})  can be written in an insightful form by considering the following identity 
\cite{bateman}
\be
\big[ (c-b) x - a \big]\,   _2F_1( a+1 , b\,; c+1 \, ;  x ) 
\,=\,
(c-a) \,   _2F_1( a , b\,; c+1 \, ;  x ) 
+
c\,(x-1) \,   _2F_1( a+1 , b\,; c \, ;  x ) \,\,
\ee
Specialising this identity to our case, we find
\be
 _2F_1\bigg( \frac{3d-1}{2d}\, , \frac{1}{2}\,; \frac{3}{2}\, ;  (\cos\alpha)^2\bigg) 
 \,=\,
 -\,\frac{1}{d-1}
 \left[\,
  _2F_1\bigg( \frac{d-1}{2d}\, , \frac{1}{2}\,; \frac{3}{2}\, ;  (\cos\alpha)^2\bigg) 
 -  d\, (\sin \alpha)^{1/d - 1}
\, \right]
\ee
By employing this result, it is straightforward to realise that the expression enclosed by the square brackets 
in (\ref{area strip z_ast ddim}) is $\mathfrak{g}_d(\alpha)/(d-1)$, being $\mathfrak{g}_d(\alpha)$ given by (\ref{g def ddim}).
This observation and (\ref{z_ast ddim}) allow us to write (\ref{area strip z_ast ddim}) in terms of the width $\ell$ of the strip $A$ as follows
\be
\label{area strip d-dim app final}
\frac{\mathcal{A} [\hat{\gamma}_\varepsilon ]}{L^d_{\textrm{\tiny AdS}}} 
\,=\,
\frac{L_\parallel^{d-1}}{d-1} \left(\,
 \frac{1}{\varepsilon^{d-1}}
-
\frac{\mathfrak{g}_d(\alpha)^d}{\ell^{d-1}}
+ O\big(\varepsilon^{d+1}\big)
\right)
\ee
The expression (\ref{area strip final main}) in the main text corresponds to (\ref{area strip d-dim app final}) specialised to $d=2$.

The other extremal surface occurring in our analysis is the half hyperplane defined by $x=\ell$.
This can be observed by considering the extrinsic curvature of a half hyperplane embedded in $\mathbb{H}_{d+1}$ whose normal vector has non vanishing components only along $z$ and $x$.
Denoting by $\theta$ the angle between this normal vector and the positive $x$ semi-axis, one finds $\textrm{Tr} K \propto \cos\theta$ for the trace of the extrinsic curvature of the half hyperplane.
This implies that the vertical hyperplane, which has $\theta = 0$, is a local minimum for the area functional. 

By introducing also an infrared cutoff $z_{\textrm{\tiny IR}}$  beside the UV cutoff $\varepsilon$,
it is straightforward to show that the portion of surface such that $\varepsilon \leqslant z \leqslant z_{\textrm{\tiny IR}} $ reads
\be
\label{area strip d-dim vertical app}
\frac{\mathcal{A} [\hat{\gamma}_\varepsilon ]}{L^d_{\textrm{\tiny AdS}}} 
\,=\,
\frac{L_\parallel^{d-1}}{d-1} \left(\,
 \frac{1}{\varepsilon^{d-1}}
-
\frac{1}{z_{\textrm{\tiny IR}}^{d-1}}
\right)
\ee
The divergent part of $\mathcal{A} [\hat{\gamma}_\varepsilon ]$ is the same one occurring in (\ref{area strip z_ast ddim}), as expected.
Let us stress that the finite term in (\ref{area strip d-dim vertical app}) vanishes as $z_{\textrm{\tiny IR}} \to \infty$.

Summarising, for $\alpha \in (0, \alpha_c]$ the minimal surface $\hat{\gamma}_A$
is the vertical half hyperplane $x=\ell$ because the surface characterised by (\ref{profile strip adj ddim}) is not well defined.
In the range $\alpha \in (\alpha_c,\pi/2]$ both the surface given by (\ref{profile strip adj ddim}) and the vertical  half hyperplane $x=\ell$ are well defined extremal solutions of the area functional and,
by comparing (\ref{area strip d-dim app final}) with (\ref{area strip d-dim vertical app}), we conclude that $\hat{\gamma}_A$ is the one characterised by (\ref{profile strip adj ddim}).
Instead, when $\alpha \in (\pi/2,\pi)$ the vertical half hyperplane is not a solution anymore of our problem because it does not intersect $\mathcal{Q}$ orthogonally;
therefore the minimal surface $\hat{\gamma}_A$ is again  the surface corresponding to (\ref{profile strip adj ddim}).

Putting these observations together, we find the following area for the restriction to $z\geqslant \varepsilon$  of the minimal surface corresponding to the strip adjacent to the boundary
\be
\label{area strip d-dim app final a}
\frac{\mathcal{A} [\hat{\gamma}_\varepsilon ]}{L^d_{\textrm{\tiny AdS}}} 
=
L_\parallel^{d-1} \bigg[\,
 \frac{1}{(d-1) \, \varepsilon^{d-1}}
+
\frac{a_{0,d}(\alpha)}{(d-1)\,\ell^{d-1}}
+ o(1)\,
\bigg]
\ee
where
\be
\label{a0 ddim def}
a_{0,d}(\alpha) \equiv
\left\{\begin{array}{ll}
-\,\mathfrak{g}_d(\alpha)^d \hspace{.4cm} &   \alpha \geqslant \alpha_c(d)
\\
\rule{0pt}{.5cm}
\;\;\;\; 0 &   \alpha  \leqslant \alpha_c(d)
\end{array}\right.
\ee
Notice that $a_{0,d}(\alpha)$ and its first derivative are continuous functions of $\alpha$.
Also the higher order derivatives of $a_{0,d}(\alpha)$ are continuous until the $d$-th derivative of $a_{0,d}(\alpha)$,
which is discontinuous at $\alpha = \alpha_c(d)$.
In (\ref{area strip final main all}) we have specialised (\ref{area strip d-dim app final a}) and (\ref{a0 ddim def}) to $d=2$.

We find it interesting to discuss separately the $d=1$ case. 
As already remarked below (\ref{g def ddim}), in this case we have that $\mathfrak{g}_1(\alpha) =1$ identically; 
therefore a critical value for $\alpha$ does not occur.
Moreover, the profile (\ref{profile strip adj ddim}) simplifies to $(x(\theta) , z(\theta) ) = \ell\, (\cos \theta\, , \sin \theta)$.
This curve is an arc of circumference of radius $\ell$; therefore it intersects orthogonally the half line $\mathcal{Q}$ given by (\ref{brane-profile})
which passes through the origin. 
We also have that $z_\ast = \ell \,\sin\alpha$, which corresponds to (\ref{z_ast ddim}) for $d=1$.

As for the length of this arc of circumference with opening angle $\pi-\alpha$ and for $z\geqslant \varepsilon$, it is straightforward to find that
\be
\frac{\mathcal{A} [\hat{\gamma}_\varepsilon ]}{L^d_{\textrm{\tiny AdS}}} 
\,=\,
 \int_{\theta_\varepsilon}^{\pi-\alpha} \frac{d\theta}{\sin \theta}
\,=\,
\log \bigg( \frac{\sin(\theta/2)}{\cos(\theta/2)}  \bigg) \bigg|_{\theta_\varepsilon}^{\pi-\alpha}
\,=\,
\log(\ell/\varepsilon) + \log\big( 2\cot(\alpha/2)\big) +O(\varepsilon^2)
\ee
where the angular cutoff $\theta_\varepsilon$ is defined by requiring that $\varepsilon =  \ell \sin \theta_\varepsilon$.
As for the extremal curve given by the half line $x=\ell$, by introducing the IR cutoff $ z_{\textrm{\tiny IR}}$,
for the length of the part of this straight line such that $\varepsilon \leqslant z \leqslant z_{\textrm{\tiny IR}} $ we find
\be
\label{area strip d=1 vertical app}
\frac{\mathcal{A} [\hat{\gamma}_\varepsilon ]}{L^d_{\textrm{\tiny AdS}}} 
\,=\,
\log(\ell/\varepsilon) + \log(z_{\textrm{\tiny IR}}/\ell)
\ee
where the term $ \log(z_{\textrm{\tiny IR}}/\ell)$ diverges when $z_{\textrm{\tiny IR}}/\ell \to +\infty$. 
Thus the minimal curve is always given by the arc of circumference.
This is consistent with the observation that a critical slope does not occur when $d=1$.

%%%%%%%%%%%%%%%%%%%%%%%%%%%%%%%%%%%%%%%%%%%%%%%%%%%%%%%%%%

\section{Infinite strip parallel to the boundary in generic dimension}
\label{sec app strip disjoint}

In this appendix we consider a strip $A$ parallel to the boundary $x=0$ and at finite distance from it. 
Let us denote by $\ell_A$ the width of the strip and by $d_A$ its distance from the boundary
(see Fig.\,\ref{fig:2strips}). 
We will focus on spacetimes having $d>1$.
For the case $d=1$ we refer the reader to \cite{Chu:2017aab}.

The main feature of the holographic entanglement entropy corresponding to this simple domain is the fact that
in some range of $\alpha$ two qualitatively different hypersurfaces are local extrema of the area functional; 
therefore the global minimum between them must be found. 
One of these candidates is the minimal area surface  in AdS$_{d+2}$ corresponding to the infinite strip found in \cite{RT} 
(see the blue solid curve in Fig.\,\ref{fig:2strips}).
Let us denote this hypersurface by $\hat{\gamma}_{A}^{\textrm{\tiny \,dis}}$, being disconnected from $\mathcal{Q}$.
The second candidate $\hat{\gamma}_{A}^{\textrm{\tiny \,con}}$ is made by the union of two disjoint hypersurfaces 
like the ones discussed in the appendix\;\ref{sec app strip}.

When $\alpha \leqslant \alpha_c$, we have that $\hat{\gamma}_{A}^{\textrm{\tiny \,con}}$ is the union of the vertical half hyperplanes defined by $x= d_A$ and $x= d_A+ \ell_A$.
Instead, for $\alpha > \alpha_c$ the hypersurface $\hat{\gamma}_{A}^{\textrm{\tiny \,con}}$ is made by two disjoint hypersurfaces characterised by the profile (\ref{profile strip adj ddim}) which depart from the edges of $A$ and intersect $\mathcal{Q}$ orthogonally
(see the green solid curves in Fig.\,\ref{fig:2strips} for a case with $\alpha > \alpha_c$).
Furthermore, when $\alpha \leqslant \pi/2$ the solution $\hat{\gamma}_{A}^{\textrm{\tiny \,dis}}$ always exists, 
while one can wonder whether this is the case also for  for $\alpha>\pi/2$, where $\mathcal{Q}$ can intersect $\hat{\gamma}_{A}^{\textrm{\tiny \,dis}}$.
This issue is discussed below.

Let us focus first on finding the regime where $\hat{\gamma}_{A}^{\textrm{\tiny \,con}}$ is the global minimum. 
The auxiliary domain corresponding to $\hat{\gamma}_{A}^{\textrm{\tiny \,con}}$ is made by two parallel and disjoint infinite strips 
$A_{\textrm{\tiny \,aux}} = A\cup A'$ in $\mathbb{R}^d$  and the corresponding minimal surface 
in $\hat{\gamma}_{A,\textrm{\tiny aux}} \subset \mathbb{H}_{d+1}$ has been studied e.g. in \cite{Tonni:2010pv}.
Denoting by $\ell'$ the width of $A'$ and by $d_{\textrm{\tiny aux}} $ the separation between $A$ and $A'$, from  Fig.\,\ref{fig:2strips} and (\ref{ell_aux strip ddim}) it is not difficult to realise that
\be
\label{ell_aux strip ddim disjoint}
\ell' = 2\,
\frac{\sqrt{\pi} \, \Gamma(\tfrac{d+1}{2d})}{\Gamma(\tfrac{1}{2d})\, \mathfrak{g}_d(\alpha)}\; \ell_A
\hspace{.3cm} \qquad \hspace{.3cm} 
d_{\textrm{\tiny aux}} 
= 2\,
\frac{\sqrt{\pi} \, \Gamma(\tfrac{d+1}{2d})}{\Gamma(\tfrac{1}{2d})\, \mathfrak{g}_d(\alpha)}\; d_A
\ee

Taking the part $z\geqslant \varepsilon$ of $\hat{\gamma}_{A}^{\textrm{\tiny \,dis}}$ and $\hat{\gamma}_{A}^{\textrm{\tiny \,con}}$, and evaluating the corresponding area as $\varepsilon \to 0^+$, 
one finds that the area law term is the same; therefore we have to compare the $O(1)$ terms to find $\hat{\gamma}_A$.
By employing (\ref{area strip d-dim app final a}) and 
the well known result for the holographic entanglement entropy of the infinite strip in AdS$_{d+2}$ \cite{RT}, 
one finds that the expansion of the area of $\hat{\gamma}_\varepsilon $ as $\varepsilon \to 0^+$ reads
\be
\label{area strip disjoint d-dim}
\frac{\mathcal{A} [\hat{\gamma}_\varepsilon ]}{L^d_{\textrm{\tiny AdS}}} 
\,=\,
\frac{L_\parallel^{d-1}}{d-1} \left(\,
 \frac{2}{\varepsilon^{d-1}}
 + \frac{1}{\ell_A^{d-1}} \;
 \textrm{min} \bigg[\,
h_d
\, , \, 
a_{0,d}(\alpha) \bigg( \frac{1}{\delta_A^{d-1}} + \frac{1}{(\delta_A + 1)^{d-1}} \bigg)
\bigg]
+ o(1)
\right)
\ee
The function $a_{0,d}(\alpha) $ has been introduced in (\ref{a0 ddim def}), while the constant $h_d$ is defined as \cite{RT}
\be
\label{h_d RT}
h_d \,\equiv\,
-\, 2^d  \pi^{d/2} \Bigg( \frac{\Gamma\big( \tfrac{d+1}{2d}\big)}{\Gamma\big( \tfrac{1}{2d}\big)}  \Bigg)^{d}
\ee
The first term in the argument of the minimisation function occurring in the  r.h.s. of (\ref{area strip disjoint d-dim}) 
corresponds to $\hat{\gamma}_{A}^{\textrm{\tiny \,dis}}$, while the second one comes from $\hat{\gamma}_{A}^{\textrm{\tiny \,con}}$.
Thus, $\hat{\gamma}_{A}=\hat{\gamma}_{A}^{\textrm{\tiny \,dis}}$ when $\delta_A \equiv d_A / \ell_A$ is large enough, 
while $\hat{\gamma}_{A}=\hat{\gamma}_{A}^{\textrm{\tiny \,con}}$ if the strip is close enough to the boundary.
We remark that (\ref{area strip disjoint d-dim}) holds for $\alpha \in (0, \pi)$.
Notice that, when $\alpha \leqslant \alpha_c$, being $h_d < 0$ and $a_{0,d}(\alpha)  =0 $, we have that $\hat{\gamma}_{A}=\hat{\gamma}_{A}^{\textrm{\tiny \,dis}}$.

The critical configurations correspond to the cases where the two terms occurring in the minimisation function of the $O(1)$ term of (\ref{area strip disjoint d-dim}) are equal.
The value $\delta_{A,c}$  of the ratio $\delta_{A}$ for these configurations can be found as solution of the following equation
\be
\label{delta_c equation ddim}
\delta_{A,c}^{d-1} \, (\delta_{A,c} + 1)^{d-1}
=\,
\tilde{a}_{0,d}(\alpha) 
\Big[
(\delta_{A,c} + 1)^{d-1} + \delta_{A,c}^{d-1} 
\Big]
\qquad
\tilde{a}_{0,d}(\alpha)  \equiv \frac{a_{0,d}(\alpha) }{ h_d } 
\ee

\begin{figure}[t] 
\vspace{-.6cm}
%\hspace{-.25cm}
\begin{center}
\includegraphics[width=.9\textwidth]{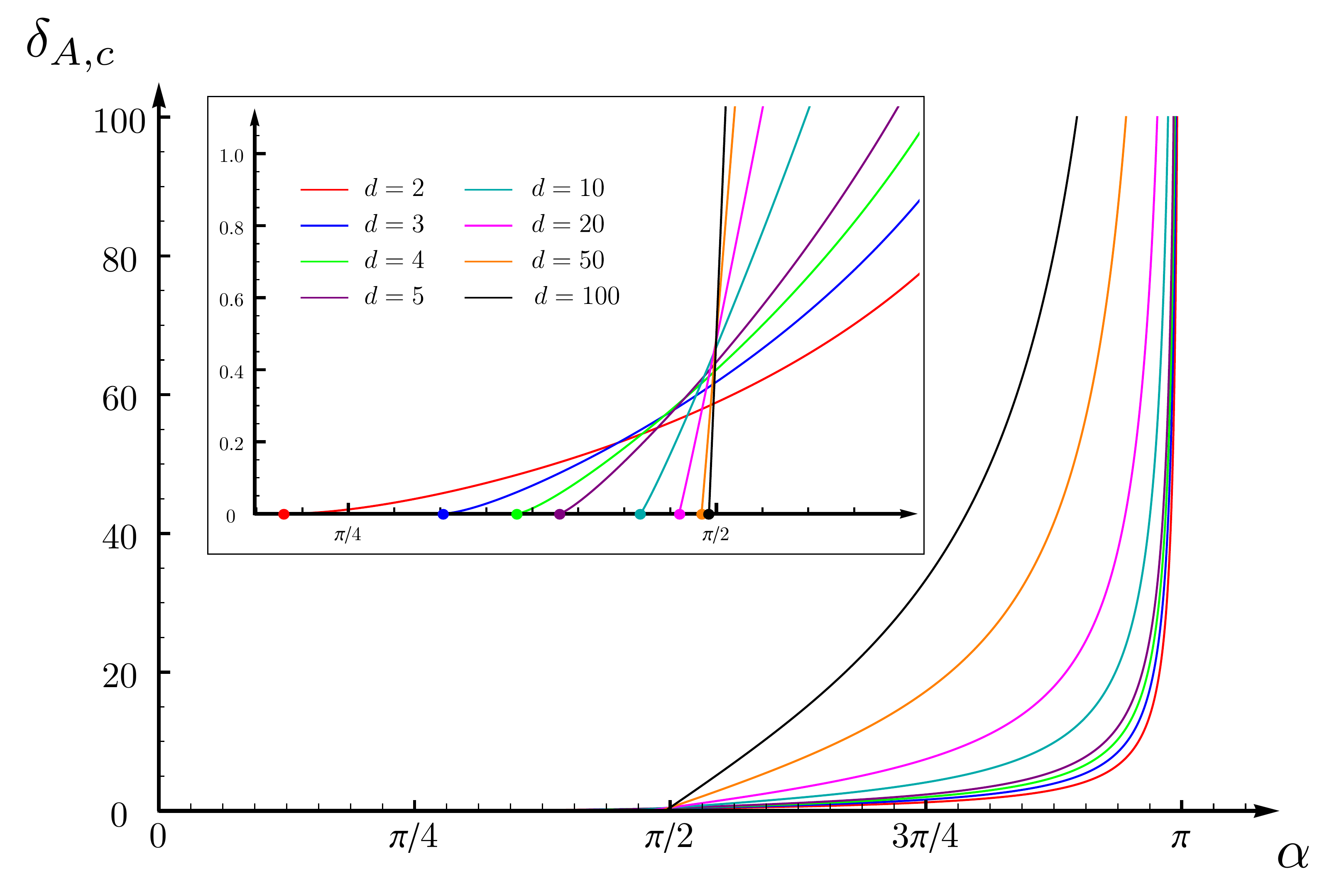}
\end{center}
\vspace{-.5cm}
\caption{\label{fig:delta_critical}
\small
Infinite strip of width $\ell_A$ parallel to the boundary at finite distance $d_A$ from it:
The ratio $\delta_{A}= d_A /\ell_A$ corresponding to the critical configurations
in terms of $\alpha \in [\alpha_c, \pi)$ for some values of $d$.
The curves are obtained by finding the unique positive root of (\ref{delta_c equation ddim}).
For $d=2$ and $d=3$ the expression of $\delta_{A,c}$ has been written analytically in (\ref{delta_c d=2}) and (\ref{delta_crit d=3}) respectively,
while for $d\geqslant 4$ the curves have been found by solving (\ref{delta_c equation ddim}) numerically.
}
\end{figure}

We remark that $\tilde{a}_{0,d}$ is a positive and non vanishing function of the slope $\alpha$ when $\alpha \in (\alpha_c, \pi)$, while $\tilde{a}_{0,d}(\alpha)   = 0$ when $\alpha \in (0, \alpha_c]$.  
This implies that a strictly positive solution of (\ref{delta_c equation ddim}) does not exist when $\alpha \leqslant \alpha_c$, 
as expected from the fact that $\hat{\gamma}_{A}=\hat{\gamma}_{A}^{\textrm{\tiny \,dis}}$.
Instead, for $\alpha > \alpha_c$ we can show that $\delta_{A,c}$  always exists and it is also unique.

The equation \eqref{delta_c equation ddim} can be written as $p(\delta_{A,c}) = 0$,
where the real polynomial $p(\delta_{A,c})$ in powers of $\delta_{A,c}$ schematically reads 
\be
\label{schema}
p(\delta_{A,c}) =
\delta_{A,c}^{2(d-1)}
+(d-1) \delta_{A,c}^{2 d-3}
+ \dots
+\big[1-2 \tilde{a}_{0,d}(\alpha)\big]\delta_{A,c}^{d-1}
- \tilde{a}_{0,d}(\alpha)  (d-1) \delta_{A,c}^{d-2}
-\dots
-\tilde{a}_{0,d}(\alpha)
\ee
The maximum number  of positive of roots of \eqref{schema} can be determined  by employing the Descartes' rule of signs.
This rule states that the maximum number of positive roots of a real polynomial is bounded by the
number of sign differences between consecutive nonzero coefficients of its powers, 
once they are set in decreasing order (the powers which do not occur must be just omitted).
Since $\tilde{a}_{0,d}(\alpha) > 0$, the expression (\ref{schema}) shows that this number is equal to one in our case; therefore we have at most one positive real root. 
Its existence  is guaranteed by the fact that $p(0)=-\,\tilde{a}_{0,d}(\alpha)<0$ and $p(\delta) \to +\infty$ as $\delta \to + \infty$.

Since $\hat{\gamma}_{A}=\hat{\gamma}_{A}^{\textrm{\tiny \,dis}}$ when $\alpha \leqslant \alpha_c$, as remarked above, 
the critical configurations exist only for $\alpha \in (\alpha_c, \pi)$.
Focussing on this range, an analytic expression for $\delta_{A,c}(\alpha)$  in terms of $\tilde{a}_{0,d} $ for a generic dimension $d$ cannot be found. 
However, we find it instructive to determine it explicitly for $d=2$ and $d=3$ because (\ref{delta_c equation ddim}) can be solved in closed form for these cases. 

When $d=2$ it is straightforward to obtain the result (\ref{delta_c d=2}) reported in the main text. 
For $d=3$ the algebraic equation (\ref{delta_c equation ddim}) has degree four. 
A shift of the variable allows to write it as follows
\be
u^4 - \frac{4 \,\tilde{a}_{0,3}(\alpha)  + 1}{2}\, u^2 + \frac{1-8 \,\tilde{a}_{0,3}(\alpha)  }{16} \,=\,0
\qquad
\delta_{A,c} = u -\frac{1}{2}
\ee
which is a biquadratic equation.
Its unique positive root reads
\be
\label{delta_crit d=3}
\delta _{A,c} = \frac{1}{2}
   \left(\sqrt{4 \tilde a_{0,3}(\alpha )
   +4\,
   \sqrt{\tilde a_{0,3}(\alpha )
   \big[\tilde a_{0,3}(\alpha)+1\big]}+1}-1\right)
\ee
For $d\geqslant 4$ the root of (\ref{delta_c equation ddim}) can be found numerically and the results for some values of $d$ are shown in Fig.\,\ref{fig:delta_critical}, 
where the curves are defined for $\alpha > \alpha_c$ (see the inset, which contains a zoom of the main plot for small values of $\delta_{A,c}$).

As briefly anticipated at the beginning of this appendix, when $\alpha > \pi/2$ one can wonder whether the brane $\mathcal{Q}$ can intersect $\hat{\gamma}_{A}^{\textrm{\tiny \,dis}}$,
leaving $\hat{\gamma}_{A}^{\textrm{\tiny \,con}}$ as unique solution.
Let us denote by $\delta_{A,0}$ the value of the ratio $d_A/\ell_{A}$ characterising the configurations of $A$ such that $\hat{\gamma}_{A}^{\textrm{\tiny \,dis}}$ is tangent to $\mathcal{Q}$.
It is not difficult to realise that $\delta_{A,0} < \delta_{A,c}$.
Indeed, when $\delta_A =  \delta_{A,0} $, the surface $\hat{\gamma}_{A}^{\textrm{\tiny \,dis}}$ can be seen as the union of two surfaces which join smoothly along the intersection
with $\mathcal{Q}$. 
Since these two surfaces connect the boundary at $z=0$ to $\mathcal{Q}$ but they are not orthogonal to $\mathcal{Q}$,
the area of $\hat{\gamma}_{A}^{\textrm{\tiny \,con}}$ is less than the area of $\hat{\gamma}_{A}^{\textrm{\tiny \,dis}}$ for $\delta_A =  \delta_{A,0} $.
This argument can be easily applied also for $\delta_A <  \delta_{A,0} $; therefore we can conclude that $\hat{\gamma}_{A}^{\textrm{\tiny \,con}}$ is always the global minimum for
$\delta_A \leqslant  \delta_{A,0} $.

We find it worth finding also the analytic expression of $\delta_{A,0}$.
The minimal surface $\hat{\gamma}_{A}^{\textrm{\tiny \,dis}}$ can be obtained by modifying the r.h.s. of \eqref{profile strip adj ddim} as follows: 
first one sets $\alpha=\pi/2$ and substitutes $\ell$ with  $\ell_A/2$, then the resulting $x(\theta)$ is replaced by $x(\theta)+d_A+\ell_A/2$.
The final result reads
\begin{equation}
\label{profile strip disconn}
\big( x(\theta) \,, z(\theta) \big) 
=\, \ell_A
\left(
\frac{\cos\theta}{(- \,h_d)^{1/d} \,d}\;
_2F_1\bigg( 
\frac{d-1}{2d}\, , \frac{1}{2}\,; \frac{3}{2}\, ;  (\cos\theta)^2\bigg) + \delta_A +\frac{1}{2}
\,, \,\frac{(\sin\theta)^{1/d}}{(- \,h_d)^{1/d}}\,\right)
\end{equation}
where $\theta \in[0,\pi]$ and $h_d$ is defined in (\ref{h_d RT}).
In order to impose that $\hat{\gamma}_{A}^{\textrm{\tiny \,dis}}$ is tangent to $\mathcal{Q}$, 
we need to find the unit vectors $V^\mu$ and $W^\mu$ which are tangent to $\hat{\gamma}_{A}^{\textrm{\tiny \,dis}}$ and $\mathcal{Q}$ respectively,
being $\mu \in \{ z, x\}$ because of translation invariance along the remaining directions. 
The vector $V^\mu$ can be easily written by setting $\alpha=\pi/2$ in \eqref{z strip paramet} and \eqref{x prime strip}, obtaining
$V^\mu = \tfrac{1}{z}(-\cos \theta , \sin\theta)$.
As for the vector $W^\mu$, it reads $W^\mu = \tfrac{1}{z}(\sin\alpha, -\cos\alpha)$.
Notice that the intersection is possible only where both $\alpha > \pi/2$ and $\theta > \pi/2$.
By requiring that $V^\mu=W^\mu$, we find that $\theta_0=3\pi/2-\alpha$ is the value of the parameter $\theta$ corresponding to the intersection
between $\hat{\gamma}_{A}^{\textrm{\tiny \,dis}}$ and $\mathcal{Q}$.
By employing this value, $\delta_{A,0}$ can be written by imposing that $\hat{\gamma}_{A}^{\textrm{\tiny \,dis}}$ intersects $\mathcal{Q}$.
This condition gives  $z(\theta_0)=- \,x(\theta_0) \tan\alpha$,
where $z=z(\theta)$ and $x=x(\theta)$ are given in \eqref{profile strip disconn}.
By solving this equation for $\delta_A$, we find 
\begin{equation}
\label{deltaA0}
\delta_{A,0}=\frac{1}{(- h_d)^{1/d}}\left[ \frac{\sin\alpha}{d}\;
_2F_1\bigg( \frac{d-1}{2d}\, , \frac{1}{2}\,; \frac{3}{2}\, ;  (\sin\alpha)^2\bigg)-\cot\alpha\left( -\cos \alpha \right)^{1/d}  \right]-\frac{1}{2}
\end{equation}
A numerical comparison between this analytic expression for $\delta_{A,0}$ and the curves for $\delta_{A,c}$ 
 leads us to conclude that $\delta_{A,0}<\delta_{A,c}$ for any value of $d$, as expected from the argument discussed above.

%%%%%%%%%%%%%%%%%%%%%%%%%%%%%%%%%%%%%%%%%%%%%%%%%%%%%%%%%%%%

\section{On a modification for the holographic entanglement entropy}
\label{sec app ssa}

In Sec.\,\ref{sec HEE bdy} the prescription (\ref{RT formula bdy}) for the holographic entanglement entropy in AdS/BCFT 
employed throughout this manuscript has been discussed.
Considering the minimal area surface anchored to the entangling surface $\partial A \cap \partial B$ in the $z=0$ half hyperplane,
since the spacetime has also the boundary $\mathcal{Q}$, it can happen that part of 
$\partial \hat{\gamma}_A$ belongs to $\mathcal{Q}$.
Let us denote this hypersurface by $\partial_{\mathcal{Q}} \hat{\gamma}_A \equiv \hat{\gamma}_A \cap \mathcal{Q}$ and by 
$\partial_{\mathcal{Q}} \hat{\gamma}_\varepsilon$ its restriction to $z\geqslant \varepsilon$.
By minimising the area functional in presence of $\mathcal{Q}$ without imposing particular boundary conditions
on $\partial_{\mathcal{Q}} \hat{\gamma}_A$,
one finds that $\hat{\gamma}_A$ is the minimal area surface which intersects $\mathcal{Q}$ orthogonally. 
The prescription (\ref{RT formula bdy}) adopted throughout this manuscript  
requires to compute the area of such minimal surface $\hat{\gamma}_A$ restricted to $z\geqslant \varepsilon$.

In this appendix we consider a possible modification of the prescription for the holographic entanglement entropy
suggested in \cite{FarajiAstaneh:2017hqv}.
This proposal includes also the area of $\partial_{\mathcal{Q}} \hat{\gamma}_\varepsilon$ as follows
\be
\label{RT formula bdy solo}
S_A = \frac{1}{4 G_{\textrm{\tiny N}}}
\Big(  \mathcal{A}[\hat{\gamma}_\varepsilon ]  + a_{\mathcal{Q}} \,L_{\textrm{\tiny AdS}} \,\mathcal{A}[\partial_{\mathcal{Q}} \hat{\gamma}_\varepsilon ]  \Big)
\ee
where $a_{\mathcal{Q}} $ is a dimensionless parameter which might depend on $\mathcal{Q}$ (i.e. on $\alpha$ in our case) but is independent on the region $A$.
We remark that in \cite{FarajiAstaneh:2017hqv} the slope $\alpha$ is fixed to $\alpha =\pi/2$, but in the following discussion we keep $\alpha$ generic. 

Our first observation is that for some domains $A$ the expression (\ref{RT formula bdy solo}) leads to a discontinuous holographic entanglement entropy 
as the size of $A$ changes. 
In particular, this discontinuity occurs whenever two local minima of the area functional compete to determine the global minimum $\hat{\gamma}_A$
 and only one of them intersects $\mathcal{Q}$.
At the critical configuration both of them provide the same value for the area $\mathcal{A}[\hat{\gamma}_\varepsilon ] $.
Thus, deforming $A$ in a smooth way passing through the critical configuration, 
we have that on one side of the transition the term $\mathcal{A}[\partial_{\mathcal{Q}} \hat{\gamma}_\varepsilon ] \neq 0$ 
because $\hat{\gamma}_A$ intersects $\mathcal{Q}$, while on the other side of the transition $ \mathcal{A}[\partial_{\mathcal{Q}} \hat{\gamma}_\varepsilon ] = 0$ 
because $\hat{\gamma}_A \cap \mathcal{Q} = \emptyset$.
Thus, if we require that the holographic entanglement entropy must be a continuous function in terms of the size of the region $A$, then $a_{\mathcal{Q}} $ should vanish.

In the following discussion we employ the strong subadditivity of the entanglement entropy to show that $a_{\mathcal{Q}} = 0$.
Our argument is based on a choice of domains made by infinite strips which are adjacent  to the boundary or parallel to the boundary at a finite distance from it.

The minimal area surface $\hat{\gamma}_A$ for these strips has been described in the appendices\;\ref{sec app strip} and \ref{sec app strip disjoint}.
Here we have to consider also $\mathcal{A}[\partial_{\mathcal{Q}} \hat{\gamma}_\varepsilon ] $ for these domains. 
In particular, from the discussions reported in the appendices\;\ref{sec app strip} and \ref{sec app strip disjoint},
it is straightforward to realise that the only quantity we miss is $\mathcal{A}[\partial_{\mathcal{Q}} \hat{\gamma}_\varepsilon ] $ for a strip adjacent to the boundary when $\alpha > \alpha_c$.
For these values of $\alpha$ we have that $\partial_{\mathcal{Q}} \hat{\gamma}_A \equiv \hat{\gamma}_A \cap \mathcal{Q}$ 
is characterised by the point $P_\ast = (x_\ast , z_\ast)$  in the two dimensional space described by the
coordinates $(x,z)$. 
In particular, notice that $\partial_{\mathcal{Q}} \hat{\gamma}_A \equiv \hat{\gamma}_A \cap \mathcal{Q}$ does not reach $z=0$ because we are dealing with infinite strips.

The metric induced on $\partial_{\mathcal{Q}} \hat{\gamma}_\varepsilon $ from (\ref{AdS metric ddim}) is given by 
\be
\label{induced metric z_ast}
ds^2 = \frac{L^2_{\textrm{\tiny AdS}}}{z_\ast^2} \,\big(dy_1^2 + \dots + dy_{d-1}^2 \big)
\ee
This leads to
\be
\label{intersection strip/Q}
\mathcal{A}[\partial_{\mathcal{Q}} \hat{\gamma}_\epsilon] 
\,=\,
\frac{1}{z_\ast^{d-1}}\, L_{\textrm{\tiny AdS}}^{d-1} L_\parallel^{d-1}
=
L_{\textrm{\tiny AdS}}^{d-1} L_\parallel^{d-1}\, \frac{a_{d,\ast}}{(d-1)\,\ell^{d-1}}
\qquad
a_{d,\ast}(\alpha) 
\equiv
\frac{(d-1)\,\mathfrak{g}_d(\alpha)^{d-1}}{(\sin\alpha)^{1-1/d}}
\ee

Let us assume that $\alpha > \alpha_c$.
Furthermore, we need (\ref{RT formula bdy solo}) for a strip adjacent and parallel to the boundary.

Given an infinite strip $A$ of width $\ell$ which is adjacent to the boundary,
from (\ref{area strip d-dim app final a}) and (\ref{intersection strip/Q}) we find that (\ref{RT formula bdy solo}) becomes
\be
\label{area strip adjacent solo}
 \frac{\mathcal{A}[\hat{\gamma}_\varepsilon ]  + a_{\mathcal{Q}} \,L_{\textrm{\tiny AdS}} \,\mathcal{A}[\partial_{\mathcal{Q}} \hat{\gamma}_\varepsilon ]  }{ L^d_{\textrm{\tiny AdS}} }
=
\frac{ L_\parallel^{d-1}}{d-1} 
\left(\,
 \frac{1}{\varepsilon^{d-1}}
+
\frac{ a_{\mathcal{Q}} \,a_{d,\ast}(\alpha)  - \mathfrak{g}_d(\alpha)^d  }{\ell^{d-1}}
+ o(1)
\right)
\ee

As for an infinite  strip $A$ of width $\ell_A$ parallel to the boundary at a distance $d_A$ from it (i.e. whose points have $d_A \leqslant x \leqslant d_A + \ell_A$, as shown in Fig.\,\ref{fig:2strips}), 
from (\ref{area strip disjoint d-dim}) and (\ref{area strip adjacent solo}) we obtain
\bea
\label{area strip disjoint solo}
 \frac{\mathcal{A}[\hat{\gamma}_\varepsilon ]  + a_{\mathcal{Q}} \,L_{\textrm{\tiny AdS}} \,\mathcal{A}[\partial_{\mathcal{Q}} \hat{\gamma}_\varepsilon ]  }{ L^d_{\textrm{\tiny AdS}} }
\, = 
& &
\\
\rule{0pt}{.8cm}
& & \hspace{-3.9cm}
=\,
\frac{ L_\parallel^{d-1}}{d-1} 
\left(\,
 \frac{2}{\varepsilon^{d-1}}
 + 
 \textrm{min} \bigg[\,
\frac{ h_d}{ \ell_A^{d-1} }
\, , \, 
\big( a_{\mathcal{Q}} \,a_{d,\ast}(\alpha)  - \mathfrak{g}_d(\alpha)^d \, \big)
\bigg( \, \frac{1}{d_A^{d-1}} + \frac{1}{(d_A + \ell_A)^{d-1}} \bigg)
\bigg]
+ o(1)
\right)
\nonumber
\eea

\begin{figure}[t] 
\vspace{-.9cm}
\hspace{-.7cm}
%\begin{center}
\includegraphics[width=1.1\textwidth]{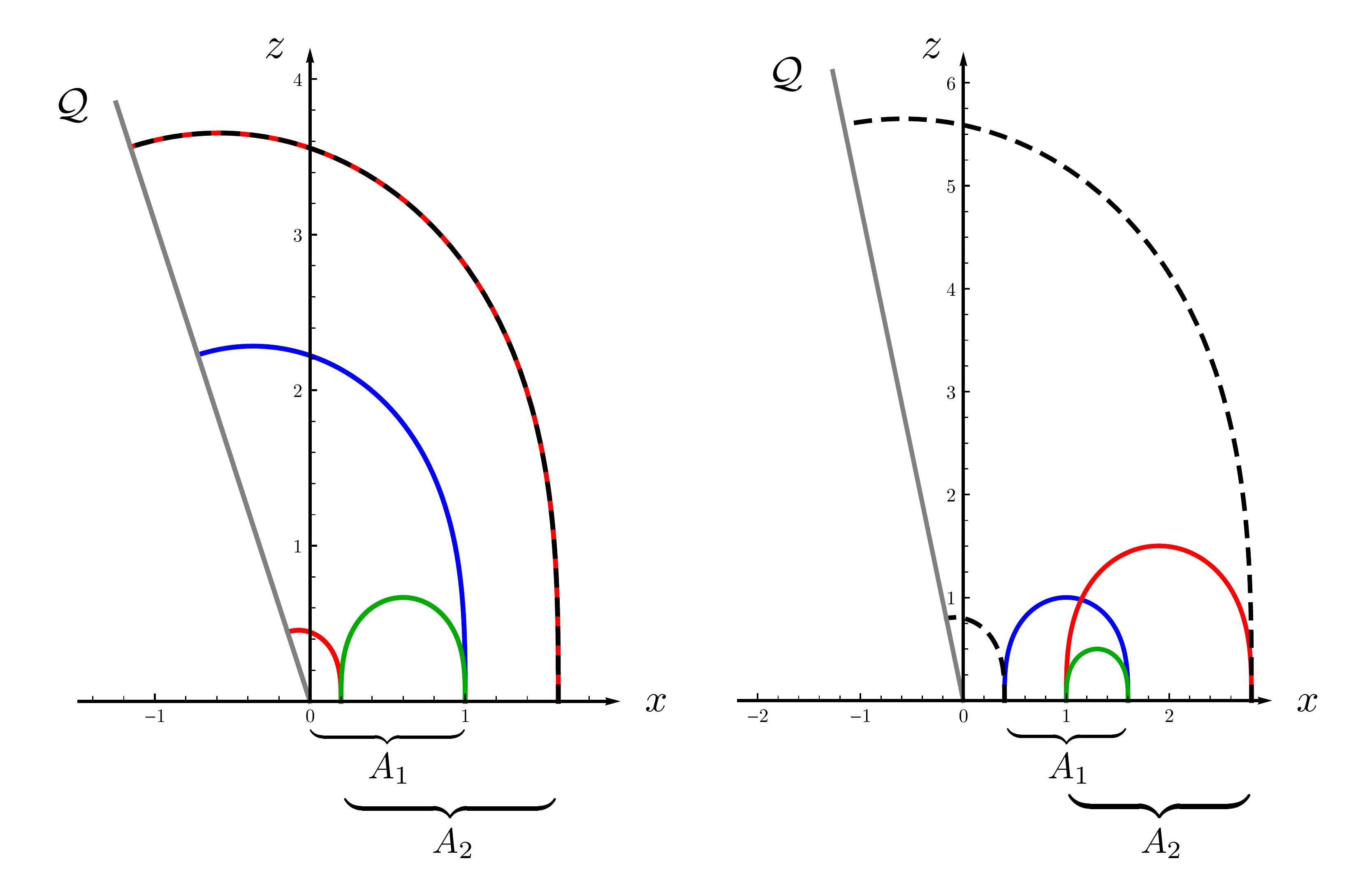}
%\end{center}
\vspace{-.7cm}
\caption{\label{fig:length_term}
\small
Configurations of two infinite strips $A_1$ and $A_2$ such that $A_1 \cap A_2 \neq \emptyset$, with the corresponding minimal surfaces,
which have been employed in Sec.\,\ref{sec app ssa}.
The blue solid curve corresponds to $\hat{\gamma}_{A_1}$,
the red solid curve (which is made by two components in the left panel) to $\hat{\gamma}_{A_2}$,
the green solid curve to $\hat{\gamma}_{A_1 \cap A_2}$
and the black dashed curve (which is made by one component in the left panel and by two components in the right panel) to $\hat{\gamma}_{A_1 \cup A_2}$.
Left: $A_1$ is adjacent to the boundary while $A_2$ is parallel to the boundary at a finite distance from it. 
Right: Both $A_1$ and $A_2$ are parallel to the boundary at finite distances from it. 
}
\end{figure}

Our argument is based on the strong subadditivity of the entanglement entropy, 
which states that, given spatial domains $A_1$ and $A_2$ such that $A_1 \cap A_2 \neq \emptyset$, 
the following inequality must hold
\cite{Lieb-Ruskai}
 \be
 \label{SSAstrip-app}
 S_{A_1}+S_{A_2}\geqslant S_{A_1 \cup A_2}+S_{A_1\cap A_2}
 \ee
By employing this inequality for special configurations where $A_1$ and $A_2$ are two  infinite strips 
and by assuming that the holographic entanglement entropy is given by (\ref{RT formula bdy solo}), 
in the following we show that $a_{\mathcal{Q}} = 0$.
Let us denote by $\ell_j$  the width of $A_j$ and by $d_j$ its distance from the boundary 
 (namely the points of $A_j$ have $d_j \leqslant x \leqslant d_j + \ell_j$), being $j \in \{1,2\}$.
 We can assume $d_1 < d_2$ without loss of generality. 
 Notice that the area law terms of the four terms involved in the inequality (\ref{SSAstrip-app}) always simplify.

In the first configuration of infinite strips that we consider, $d_1 = 0$ and the remaining parameters are such that the 
corresponding configuration of minimal surfaces in the bulk is like the one depicted in the left panel of Fig.\,\ref{fig:length_term}.
This means that $d_2 / \ell_2 \leqslant \delta_{A,c}$ and $d_2 / (\ell_1 - d_2) \geqslant \delta_{A,c}$, 
being $ \delta_{A,c}$ the unique positive root
of (\ref{delta_c equation ddim}) (see Fig.\,\ref{fig:delta_critical}).
Moreover, we also have the geometrical constraints  given by $d_2 <\ell_1  < d_2 + \ell_2$.
By assuming that the prescription (\ref{RT formula bdy solo}) holds, let us consider the inequality (\ref{SSAstrip-app}) for the configuration 
depicted in the left panel of Fig.\,\ref{fig:length_term}.
By employing (\ref{area strip adjacent solo}) and (\ref{area strip disjoint solo}), we find 
\be
\label{ssa-config-1}
 \frac{ a_{\mathcal{Q}} \,a_{d,\ast}(\alpha)  - \mathfrak{g}_d(\alpha)^d }{\ell_1^{d-1}}
+ 
 \frac{ a_{\mathcal{Q}} \,a_{d,\ast}(\alpha)  - \mathfrak{g}_d(\alpha)^d }{d_2^{d-1}}
\,\geqslant\,
\frac{ h_d}{ (\ell_1 - d_2 )^{d-1} } 
\ee
where we remark that the term coming from $S_{A_1 \cup A_2}$ simplifies with the term originated from the component of 
$\hat{\gamma}_{A_2}$ anchored to $x=d_2 + \ell_2$.
This cancellation occurs between the terms corresponding to the two curves which overlap in Fig.\,\ref{fig:length_term}.
Isolating $a_{\mathcal{Q}}$ on one side of the inequality (\ref{ssa-config-1}), one finds
\be
\label{ssa-config-1-b}
 \frac{ h_d}{ (\ell_1 - d_2 )^{d-1} } 
 +
  \mathfrak{g}_d(\alpha)^d \,
\bigg(
\frac{1}{\ell_1^{d-1}} + \frac{1}{d_2^{d-1}}
\bigg)
\,\leqslant\,
a_{\mathcal{Q}} \,a_{d,\ast}(\alpha) \, 
\bigg(
\frac{1}{\ell_1^{d-1}} + \frac{1}{d_2^{d-1}}
\bigg)
\ee
By using (\ref{area strip d-dim app final}), we recognise in the l.h.s. of this inequality the combination of terms which provides the critical configuration
for the strip $A_1 \cap A_2$ parallel to the boundary. 
Furthermore, since $d_2 / (\ell_1 - d_2) \geqslant \delta_{A,c}$ for the configuration we are considering, the l.h.s. of (\ref{ssa-config-1-b}) is negative. 
In particular, by choosing the parameters such that $d_2 / (\ell_1 - d_2)\to  \delta_{A,c}^+$,
we find that the l.h.s. of (\ref{ssa-config-1-b}) vanishes in the limit. 
Being $a_{d,\ast}(\alpha)  > 0$, we have that $a_{\mathcal{Q}} \geqslant 0$.

A similar analysis can be performed by considering the configuration depicted in the right panel of Fig.\,\ref{fig:length_term}, where $A_1$ and $A_2$ are two infinite strips parallel to the boundary
and $d_1 > 0$.
The geometrical constraint for this configuration is $d_2 < d_1 + \ell_1 < d_2 + \ell_2$.
Instead, in order to find the configuration of minimal surfaces shown in the right panel of Fig.\,\ref{fig:length_term},
we must require that $d_1 / \ell_1 \geqslant \delta_{A,c}\,$, 
$d_2 / \ell_2 \geqslant \delta_{A,c}\,$,
$d_1 / (d_2 + \ell_2 - d_1) \leqslant \delta_{A,c}\,$ and also
$d_2 / (d_1 + \ell_1 - d_2) \geqslant \delta_{A,c}\,$.
By employing (\ref{area strip adjacent solo}) and (\ref{area strip disjoint solo}),
the strong subadditivity inequality (\ref{SSAstrip-app}) for the configuration of infinite strips in the right panel of Fig.\,\ref{fig:length_term}
provides the following inequality
\be
\label{ssa-config-2}
 \frac{ h_d}{ \ell_1^{d-1} } 
 +
  \frac{ h_d}{ \ell_2^{d-1} } 
\,\geqslant\,
  \frac{ h_d}{ (d_1 + \ell_1 - d_2)^{d-1} } 
  +
   \big(\, a_{\mathcal{Q}} \,a_{d,\ast}(\alpha)  - \mathfrak{g}_d(\alpha)^d \,\big)
   \bigg(\,
\frac{1}{d_1^{d-1}} + \frac{1}{(d_2 + \ell_2)^{d-1}}
\bigg)
\ee
which can be conveniently rearranged as 
\bea
\label{ssa-config-2-b}
h_d \,\bigg(  \,
\frac{1}{ \ell_1^{d-1} } + \frac{ 1}{ \ell_2^{d-1} } - \frac{1}{ (d_1 + \ell_1 - d_2)^{d-1} } 
\bigg)
+
  \mathfrak{g}_d(\alpha)^d \,
   \bigg(\,
\frac{1}{d_1^{d-1}} + \frac{1}{(d_2 + \ell_2)^{d-1}}
\bigg)
& &
\\
\rule{0pt}{.7cm}
& &
\hspace{-4cm}
\,\geqslant\,
   a_{\mathcal{Q}} \,a_{d,\ast}(\alpha) 
   \bigg(\,
\frac{1}{d_1^{d-1}} + \frac{1}{(d_2 + \ell_2)^{d-1}}
\bigg)
\nonumber
\eea
Considering the special case of $\ell_1 = \ell_2$ first and then taking the limit $d_2 \to d_1^{\, +}$, we have that $A_1$ and $A_2$ tends to overlap. 
In this limit, the r.h.s. of (\ref{ssa-config-2-b}) becomes the combination occurring in the holographic entanglement entropy of an infinite strip parallel to the boundary
(see (\ref{area strip disjoint d-dim}) and (\ref{a0 ddim def})).
In particular, given the configuration in the right panel of Fig.\,\ref{fig:length_term}, the l.h.s. of (\ref{ssa-config-2-b}) in this limit is a positive quantity
which is proportional to the combination that appears in \eqref{delta_c equation ddim}. 
Since in this limit the constraints introduced above \eqref{ssa-config-2} saturate, the l.h.s. of (\ref{ssa-config-2-b}) becomes arbitrarily closed to $0^+$.
Combining this observation with $a_{d,\ast}(\alpha)  > 0$, we obtain $a_{\mathcal{Q}} \leqslant 0$.

Thus, by employing the strong subadditivity inequality (\ref{SSAstrip-app}) for the configurations depicted in Fig.\,\ref{fig:length_term}, 
we can conclude that $a_{\mathcal{Q}} = 0$ in (\ref{RT formula bdy solo}).

\newpage

%%%%%%%%%%%%%%%%%%%%%%%%%%%%%%%%%%%%%%%%%%%%%%%%%%%%%%%
\section{On the infinite wedge adjacent to the boundary}
\label{sec app wedge}

In this appendix we provide the technical details underlying the computation of the holographic entanglement entropy 
of  the infinite wedge $A$ adjacent to the boundary. 
The main results have been collected and discussed in Sec.\;\ref{sec:wedge}.

In the half plane  $\{(x,y) \in \mathbb{R}^2 \,, \, x\geqslant 0\}$, let us introduce the polar coordinates $(\rho, \phi)$ such that $\phi=0$ is the half line given by $x=0$ and $y>0$, namely
\be
\label{polar coord def}
x= \rho \, \sin\phi
\qquad
y= \rho \, \cos\phi 
\ee
In terms of these coordinates, the  infinite wedge $A$ having one of its two edges on the boundary $x=0$ can be described without loss of generality as follows
\be
\label{domain wedge bdy}
A = \big\{ (\rho, \phi) \,|\, 0\leqslant \phi \leqslant \gamma \,, \,\rho \leqslant L \big\}
\qquad
L \gg \varepsilon
\ee

In order to study the holographic entanglement entropy (\ref{RT formula bdy}) of the infinite wedge $A$ within the AdS$_4$/BCFT$_3$ setup described in Sec.\,\ref{sec HEE bdy},
let us consider the surfaces anchored to the edge $\{(\rho,\phi)\,|\, \phi=\gamma \}$ of $A$ and embedded in the region of $\mathbb{H}_3$ defined by (\ref{AdS4 regions}).
The symmetry under dilatations tells us that $\hat{\gamma}_A$ belongs to the class of surfaces $\gamma_A$ described by (\ref{cusp profile main}) with $q(\phi)>0$.
The metric induced on $\gamma_A$ from $\mathbb{H}_3$, whose metric is (\ref{AdS4 metric}), reads
\be
\label{induced metric cusp}
ds^2\big|_{\gamma_A}
=\, L^2_{\textrm{\tiny AdS}}
\left(
\frac{1 + q^2}{\rho^2}\, d\rho^2 - \frac{2\, q'}{\rho \, q}\, d\rho \, d\phi +\frac{(q')^2 + q^4}{q^2}\, d\phi^2
\right)
\ee

\begin{figure}[t] 
\vspace{-.5cm}
\hspace{-.9cm}
%\begin{center}
\includegraphics[width=1.1\textwidth]{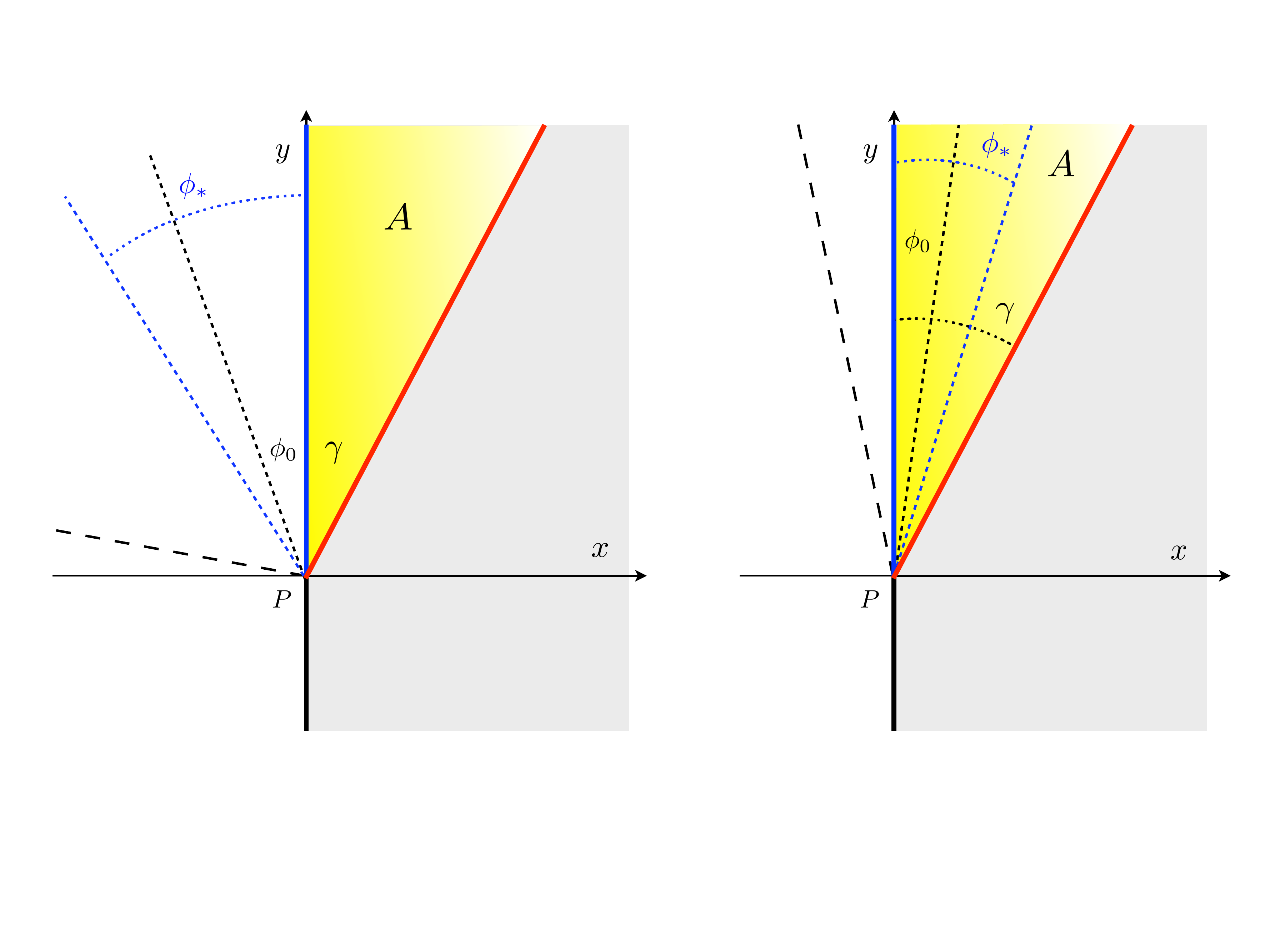}
%\end{center}
\vspace{-.5cm}
\caption{\label{fig_app:wedge}
\small
The opening angles occurring in the construction of the minimal surface $\hat{\gamma}_A$ 
anchored to the infinite wedge $A$ adjacent to the boundary with opening angle $\gamma$,
which has been discussed in Sec.\,\ref{sec:wedge} and in the appendix\;\ref{sec app wedge}. 
In the left panel  $\alpha \in (0,\pi/2]$ and in the right panel $\alpha \in [\pi/2, \pi)$.
The wedge $A$ is the yellow region, whose edges are the red and the blue solid half lines, 
given by $\phi=\gamma$ and $\phi=0$ respectively. 
The auxiliary wedge $A_{\textrm{\tiny aux}}$ is the infinite wedge in $\mathbb{R}^2$ containing $A$ whose tip is $P$ and whose edges are the
red half line  and the black dashed line with the largest dashing. 
The black dashed line with the smallest dashing at $\phi=\phi_0$ corresponds to the bisector of $A_{\textrm{\tiny aux}}$.
The blue dashed half line at $\phi = \phi_\ast$ corresponds to the projection of $\hat{\gamma}_A \cap \mathcal{Q}$ in the $z=0$ plane.
}
\end{figure}

Our analysis heavily relies on \cite{Drukker:1999zq}, where the authors have found the minimal area surface in $\mathbb{H}_3$ anchored to both the edges of an infinite wedge.
Indeed, we study $\hat{\gamma}_A$ by introducing an auxiliary wedge $A_{\textrm{\tiny aux}}$ in the $z=0$ boundary of $\mathbb{H}_3$ such that  $A \subsetneq A_{\textrm{\tiny aux}}$
and $\{(\rho,\phi)\,|\, \phi=\gamma\}$ is a common edge of both $A$ and $A_{\textrm{\tiny aux}}$.
Considering  the minimal area surface $\hat{\gamma}_{A,\textrm{\tiny aux}}$  in $\mathbb{H}_3$ anchored to the edges of $A_{\textrm{\tiny aux}}$,
the minimal area surface $\hat{\gamma}_A$ anchored to the edge $\{(\rho,\phi)\,|\, \phi=\gamma\}$ of $A$ and intersecting $\mathcal{Q}$ orthogonally 
is the part of $\hat{\gamma}_{A,\textrm{\tiny aux}}$  identified by the constraint (\ref{AdS4 regions}).
Thus, finding $\hat{\gamma}_A$ corresponds to find the proper $\hat{\gamma}_{A,\textrm{\tiny aux}}$.

In Fig.\,\ref{fig_app:wedge} we show the relevant angles occurring in our construction, by distinguishing the two cases of $\alpha \in (0,\pi/2]$ (left panel) and $\alpha \in [\pi/2, \pi)$ (right panel).
The infinite wedge $A$ adjacent to the boundary $x=0$ is the yellow region, which is embedded into the grey half plane $x\geqslant 0$.
The edges of the auxiliary wedge $A_{\textrm{\tiny aux}}$ are the red half line $\{(\rho,\phi)\,|\, \phi=\gamma\}$ and the half line denoted by the black large dashing.
The bisector of $A_{\textrm{\tiny aux}}$ is the black dashed half line at $\phi=\phi_0$; therefore the opening angle of $A_{\textrm{\tiny aux}}$  is $2(\gamma - \phi_0)$.
The half line corresponding to the black small dashing is the bisector of the auxiliary wedge, 
while the blue dashed half line is the projection on the $z=0$ plane of the half line given by $\hat{\gamma}_A \cap \mathcal{Q}$.

\subsection{Minimal surface condition}
\label{sec app wedge min surf}

The metric (\ref{induced metric cusp}) induced on the surfaces $\gamma_A$  leads to the following area functional 
\be
\label{area functional wedge app}
\frac{\mathcal{A}[\gamma_A]}{L^2_{\textrm{\tiny AdS}} }
=   
\int_{\gamma_A} \frac{1}{\rho} \,\sqrt{q'^2+q^2+q^4} \,d\phi \,d \rho  
\,=
\int_{\gamma_A} \frac{1}{\rho}\, \mathcal{L} \; d\phi \,d \rho  
\qquad
 \mathcal{L} \equiv \sqrt{q'^2+q^2+q^4}
\ee
The functions $q(\phi)$ characterising the extrema of this functional can be found by observing that its integrand is independent of $\phi$.
The first integral associated to this invariance provides a quantity which is independent of $\phi$. 
It reads
\be
\label{first-integral-0}
\frac{\partial  \mathcal{L}}{\partial q'}\, q' -  \mathcal{L} 
\;\propto\;
\frac{q^4+q^2}{\sqrt{(q')^2 + q^4 + q^2}}
\ee
Let us introduce the angle $\phi_0$ such that
\be
\label{phi0 def}
q'(\phi_0) = 0
\qquad
q(\phi_0)  \equiv q_0 
\qquad
q_0 > 0
\ee
The angle $\phi_0$ provides the bisector of the auxiliary wedge $A_{\textrm{\tiny aux}}$.

%Notice that we do not fix $\phi_0$ here. 
By employing (\ref{phi0 def}) into the condition that (\ref{first-integral-0}) is independent of $\phi$,
one obtains the following first order differential equation
\be
\label{first-integral}
\frac{q^4+q^2}{\sqrt{(q')^2 + q^4 + q^2}} = \sqrt{q_0^4 + q_0^2}
\ee
Taking the square of this expression, one gets
\be
\label{first-integral-squared}
\frac{(q')^2}{q^2} = (q^2+1) \left(  \frac{q^4 + q^2}{q_0^4 + q_0^2}  - 1\right)
\qquad
q \geqslant q_0
\ee
Separating the variables in (\ref{first-integral-squared}), one finds $d\phi = \mathcal{P}_\phi(q,q_0)\, dq $.
Then, by integrating the latter expression, we get 
\be
\label{Pintegration def}
 \big| \phi  - \phi_0 \big| =\int_{q_0}^q 
\mathcal{P}_\phi(\hat{q},q_0)\, d\hat{q} = P(q,q_0) 
\qquad
\mathcal{P}_\phi(q,q_0)
\equiv
\frac{\sqrt{q_0^4 + q_0^2}}{q\,\sqrt{(q^2 + 1)(q^2 - q_0^2)(q^2 + q_0^2+1)}}
\ee
where $q\geqslant q_0$ and $P(q,q_0) $ has been written in (\ref{Pfunction def main}).
From (\ref{Pintegration def}), it is straightforward to realise that $P(q_0, q_0) =0 $ and that the function $P(q, q_0) \geqslant 0$ is an increasing function of $q\geqslant q_0$.
The minimal area surface $\hat{\gamma}_A$ is described by (\ref{cusp profile main}) with the proper $q(\phi)$ obtained by inverting (\ref{Pintegration def}).

The opening angle of the auxiliary wedge $A_{\textrm{\tiny aux}}$ is $2(\gamma - \phi_0 )$, 
as already observed above from Fig.\,\ref{fig_app:wedge}.
This angle can be found from (\ref{Pintegration def}) as follows
\be
\label{gamma + phi0}
\gamma - \phi_0 
=
 \int_{\phi_0}^\gamma d\phi\;
= \int_{q_0}^{\infty} \mathcal{P}_\phi(\tilde{q},q_0)\, d\tilde{q} \;
= \lim_{q\to + \infty} P(q,q_0)
\,\equiv \, 
P_0(q_0) 
\ee
Equivalent expressions of $P_0(q_0)$ have been reported in (\ref{theta_V expression}) and  (\ref{P0 expression v2}).

The next step of our analysis consists in studying the intersection $\hat{\gamma}_A \cap \mathcal{Q}$ and the opening angle of $A_{\textrm{\tiny aux}}$.

\subsection{Intersection between the minimal surface and the brane}
\label{sec app wedge intersection}

In order to find the extremal surface $\hat{\gamma}_A$ anchored to  the edge $\{(\rho,\phi)\,|\, \phi=\gamma\}$ of $A$ and ending on the half plane $\mathcal{Q}$, 
beside the differential equation (\ref{first-integral}) we also have to impose that $\hat{\gamma}_A$ and $\mathcal{Q}$ intersect orthogonally.

By writing the equation (\ref{brane-profile}) for $\mathcal{Q}$ in terms of the polar coordinates (\ref{polar coord def}) 
and intersecting the resulting expression with the ansatz (\ref{cusp profile main}) for $\gamma_A$, we find
\be
\label{intersection-cond}
 q_\ast \, \sin \phi_\ast= - \cot\alpha 
 \qquad
 q_\ast \equiv q(\phi_\ast)
\ee 
This relation defines the angle $\phi=\phi_\ast$ at which $\gamma_A$ and $\mathcal{Q}$ intersect. 
Thus, $\gamma_A \cap \mathcal{Q}$ is the half line whose points have coordinates $(z,\rho,\phi) = (\rho/q_\ast , \rho , \phi_\ast)$, with $\rho > 0$.
Since $ q_\ast > 0$, from (\ref{intersection-cond}) we have that $\phi_\ast \leqslant 0$ when $\alpha \in (0,\pi/2]$, 
and $\phi_\ast \geqslant 0$ when $\alpha \in [\pi/2, \pi)$.
This is shown in Fig.\,\ref{fig_app:wedge}, where the blue dashed half line corresponds to the projection of $\hat{\gamma}_A \cap \mathcal{Q}$ on the $z=0$ plane.
The relation (\ref{intersection-cond}) tells us that $\phi_\ast=0$ when $\alpha =\pi/2$, as expected.

In order to impose that $\gamma_A$ and $\mathcal{Q}$ intersect orthogonally along the half line at $\phi=\phi_\ast$, 
we have to find the unit vector normal to $\gamma_A$ and the unit vector normal to the $\mathcal{Q}$.
The surfaces $\gamma_A$ described by the ansatz (\ref{cusp profile main}) can be equivalently written as $\mathcal{C}  =0$, with $\mathcal{C} \equiv z -\rho/q(\phi)$.
Thus, the unit vector normal to $\gamma_A$ is
\be
\label{n_mu vector}
n_\mu
=
\frac{\partial_\mu \mathcal{C}}{\sqrt{g^{\alpha\beta} \,\partial_\alpha \mathcal{C} \, \partial_\beta \mathcal{C}}}  
\,=\, 
\frac{L_{\textrm{\tiny AdS}}}{z\, \sqrt{(q')^2 + q^4 + q^2}}\,
 \big( q^2 \,, -q\, , \,  \rho \,q' \big)
\ee
where the components of the vector have been ordered according to $\mu \in \{z,\rho,\phi\}$.
As for the half plane $\mathcal{Q}$, its definition in (\ref{brane-profile}) can be written as $\mathcal{C}_{\mathcal{Q}} = 0$, with 
$\mathcal{C}_{\mathcal{Q}} \equiv  z + \rho  \sin\phi \tan \alpha$, where the first relation in (\ref{polar coord def}) has been used. 
This tells us that the unit vector normal to the half plane  $\mathcal{Q}$ is 
\be
\label{b_mu vector}
b_\mu 
=
\frac{\partial_\mu \mathcal{C}_{\mathcal{Q}}}{\sqrt{g^{\alpha\beta} \,\partial_\alpha \mathcal{C}_{\mathcal{Q}} \, \partial_\beta \mathcal{C}_{\mathcal{Q}}}}  
=
\frac{L_{\textrm{\tiny AdS}} \cos\alpha}{z}\,
 \big( 1\,,\,  \sin\phi \tan \alpha \, , \,  \rho  \cos\phi \tan \alpha \big)
\ee

Given the unit vectors (\ref{n_mu vector}) and (\ref{b_mu vector}),  we have to impose that they are orthogonal (namely $g^{\mu\nu} n_\mu  b_\nu = 0$)
along the half line $\gamma_A \cap \mathcal{Q}$ at $\phi = \phi_\ast$. 
This requirement leads to the following relation
\be
q_\ast^2 +\big[ \, q'_\ast \cos\phi_\ast - q_\ast \sin\phi_\ast  \big] \tan\alpha = 0
\qquad
q'_\ast  \equiv q'(\phi_\ast)
\ee
which can be written also as
\be
\label{q_prime step}
\frac{q'_\ast}{q_\ast} = \tan\phi_\ast - \frac{q_\ast}{\cos\phi_\ast} \cot\alpha
\ee
Taking the square of (\ref{q_prime step}) first and then employing (\ref{first-integral-squared}) to write $(q'_\ast / q_\ast)^2$ in terms of $q_\ast$ and $q_0$, we have
\be
\label{orto-rel eq1}
\bigg( \tan\phi_\ast - \frac{q_\ast}{\cos\phi_\ast} \cot\alpha \bigg)^2
=\,
(q_\ast^2+1) \left(  \frac{q_\ast^4 + q_\ast^2}{q_0^4 + q_0^2}  - 1\right)
\ee
This expression can be simplified by using (\ref{intersection-cond}) to rewrite $q_\ast$ in terms of $\phi_\ast$, finding
\be
(\tan\phi_\ast )^2
\left[\frac{(\cot\alpha)^2}{(\sin\phi_\ast)^2} + 1\right] 
=
 \frac{1}{q_0^4 + q_0^2}\,  \frac{(\cot\alpha)^2}{(\sin\phi_\ast)^2} \left[\frac{(\cot\alpha)^2}{(\sin\phi_\ast)^2} + 1\right]  - 1 
\ee
This relation leads to the following biquadratic equation
\be
\label{eq1}
q_0^4 + q_0^2 
= \left[\,1+\frac{(\cot\alpha)^2}{(\sin\phi_\ast)^2}\,\right] (\cos\alpha)^2 \,(\cot\phi_\ast )^2
\ee
which has only one positive root in terms of $q_0^2$.
This solution allows us to write $q_0$ in terms of $\phi_\ast$ as follows
\be
\label{q0 phi_ast}
q_0 = \frac{1}{\sqrt{2}}\left(\sqrt{1+4 \big[ 1+(\cot\alpha)^2 (\csc\phi_\ast)^2 \big] (\cos\alpha)^2 \,(\cot\phi_\ast )^2 } - 1\right)^{1/2}
\ee

Instead of $\phi_\ast$, we prefer to adopt $q_0$ as fundamental parameter;
therefore let us consider the biquadratic equation in terms of $\sin\phi_\ast$ obtained from (\ref{eq1}), namely
\be
\left[1+\frac{q_0^4 + q_0^2 }{(\cos\alpha)^2}\right] (\sin\phi_\ast)^4 
-\big[ 1-(\cot\alpha)^2 \big](\sin\phi_\ast)^2
-(\cot\alpha)^2
=0
\ee
whose positive solution for $(\sin\phi_\ast)^2 \equiv s_\ast(\alpha,q_0)^2$ reads
\bea
\label{s_ast def}
s_\ast(\alpha,q_0)^2 \,  = & &
\\
& &  
\hspace{-1.5cm}
=\,
\frac{1}{2}
 \left( 1+\frac{q_0^4 + q_0^2 }{(\cos\alpha)^2} \right)^{-1}
\left[ \,1-(\cot\alpha)^2 +\sqrt{\Big[ 1-(\cot\alpha)^2 \Big]^2+4 \left( 1+\frac{q_0^4 + q_0^2 }{(\cos\alpha)^2} \right) (\cot\alpha)^2}\;\, \right]
\nonumber
\eea
Notice that $s_\ast(\pi- \alpha,q_0)^2= s_\ast(\alpha,q_0)^2$.
We denote by $s_\ast(\alpha,q_0) > 0$ the positive root of (\ref{s_ast def}), which has been written explicitly in (\ref{s_ast def main}).
Plugging $s_\ast(\alpha,q_0) $ into (\ref{intersection-cond}), one obtains (\ref{qast(q0,alpha) main}).

Since $\phi_\ast  \leqslant \phi_0  \leqslant 0$  when $\alpha \in (0, \pi/2]$, while  $0  \leqslant \phi_0  \leqslant \phi_\ast $  when $\alpha \in [\pi/2, \pi)$ (see Fig.\,\ref{fig_app:wedge}),
we find it convenient to introduce $\eta_\alpha \equiv -  \, \textrm{sign}(\cot\alpha)$, as done in (\ref{s_ast def main}).
Then, the expression for $\phi_\ast = \phi_\ast(q_0, \alpha)$ in (\ref{phi_ast q0 main}) can be written straightforwardly.
Furthermore, (\ref{Pintegration def}) leads to
\be
 \big| \phi_\ast  - \phi_0 \big|
= \int_{q_0}^{q_\ast} \mathcal{P}_\phi(q,q_0)\, dq
 = P(q_\ast,q_0) 
 =\left\{\begin{array}{ll}
\phi_0  - \phi_\ast   \hspace{.7cm} &  0 < \alpha \leqslant \pi/2
\\
\rule{0pt}{.5cm}
\displaystyle
\phi_\ast  - \phi_0    &    \pi/2 \leqslant  \alpha < \pi
\end{array}
\right. 
\ee
This provides the angle $\phi_0 = \phi_0(q_0, \alpha)$  as follows
\be
\label{phi0 explicit 0}
\phi_0  
\,=\,
\phi_\ast(q_0, \alpha) - \eta_\alpha \, P\big(q_\ast(\alpha,q_0) ,q_0\big)   
\,=\,
\eta_\alpha \big(
\arcsin [  s_\ast(\alpha, q_0) ] - P(q_\ast,q_0) 
\big)
\ee
where the last step has been obtained by using $\phi_\ast(q_0, \alpha)$ in (\ref{phi_ast q0 main}).
Notice that $\phi_0$ characterises the opening angle of the auxiliary wedge $A_{\textrm{\tiny aux}}$.

Finally, an expression for the opening angle $\gamma$ in terms of $\alpha$ and $q_0$ can be written.
Indeed, from (\ref{gamma + phi0}) one first finds that $\gamma = P_0(q_0) + \phi_0$; then (\ref{phi0 explicit 0}) can be used to get (\ref{gamma q0 main}).

Summarising, we have determined the angles $\phi_\ast$, $\phi_0$ and $\gamma$ as functions of $\alpha$ and $q_0$. They are given in
(\ref{phi_ast q0 main}), (\ref{phi0 explicit 0}) and (\ref{gamma q0 main}) respectively.

\subsection{Area of the minimal surface}
\label{sec app wedge area}

The minimal surface $\hat{\gamma}_A$ anchored to the edge $\{(\rho,\phi)\,|\, \phi=\gamma\}$ of the infinite wedge adjacent to the boundary given by (\ref{domain wedge bdy})
is non compact; therefore we have to compute the area of its restriction $\hat{\gamma}_\varepsilon $ to $z\geqslant \varepsilon$.
We stress that $\hat{\gamma}_A\subsetneq \hat{\gamma}_{A,\textrm{\tiny aux}}$ is the part of the auxiliary minimal surface $\hat{\gamma}_{A,\textrm{\tiny aux}}$ 
identified by the constraint (\ref{AdS4 regions}), as discussed in Sec.\,\ref{sec:wedge} and in the beginning of the appendix\;\ref{sec app wedge} (see also Fig.\,\ref{fig_app:wedge}).
The auxiliary infinite wedge $A_{\textrm{\tiny aux}}$ and the corresponding minimal surface $\hat{\gamma}_{A,\textrm{\tiny aux}}$ 
have been obtained through the analysis of the appendices\;\ref{sec app wedge min surf} and \ref{sec app wedge intersection}.
The area of  $\hat{\gamma}_{\varepsilon,\textrm{\tiny aux}} \equiv \hat{\gamma}_{A,\textrm{\tiny aux}} \cap \{z\geqslant \varepsilon\}$ has been computed in \cite{Drukker:1999zq}.

We compute $\mathcal{A}[\hat{\gamma}_\varepsilon] $ by considering two parts of $\hat{\gamma}_{A,\textrm{\tiny aux}} $,
that we denote by $\hat{\gamma}^{\infty}_{A,\textrm{\tiny aux}} $ and $\hat{\gamma}^\ast_{A,\textrm{\tiny aux}}$.

The surface $\hat{\gamma}^{\infty}_{A,\textrm{\tiny aux}} $ corresponds to the part of $\hat{\gamma}_{A,\textrm{\tiny aux}} $ such that
with $\phi_0 \leqslant \phi \leqslant \gamma$.
We remark that $\hat{\gamma}^{\infty}_{A,\textrm{\tiny aux}} $ reaches the half plane at $z=0$ along the edge at $\phi = \gamma$
and it corresponds to half of  $\hat{\gamma}_{A,\textrm{\tiny aux}} $.
The surface $\hat{\gamma}^\ast_{A,\textrm{\tiny aux}}$ is  the part of  $\hat{\gamma}_{A,\textrm{\tiny aux}} $ having $\phi_\ast \leqslant \phi \leqslant  \phi_0$ when $\alpha \in ( 0, \pi/2 ]$ 
and $\phi_0 \leqslant \phi \leqslant  \phi_\ast$ when $\alpha \in [ \pi/2 , \pi )$ (see respectively the left and right panel of Fig.\;\ref{fig_app:wedge}). 
Notice that $\hat{\gamma}^\ast_{A,\textrm{\tiny aux}} =\emptyset$ when $\alpha=\pi/2$.

The restrictions of $\hat{\gamma}^{\infty}_{A,\textrm{\tiny aux}}$ and $\hat{\gamma}^\ast_{A,\textrm{\tiny aux}}$ to $z\geqslant \varepsilon$ 
provide $\hat{\gamma}^\infty_{\varepsilon,\textrm{\tiny aux}}$ and $\hat{\gamma}^\ast_{\varepsilon,\textrm{\tiny aux}}$ respectively,
and we denote their areas by $ L^2_{\textrm{\tiny AdS}} \,\mathcal{A}_{\infty} $ and $ L^2_{\textrm{\tiny AdS}} \, \mathcal{A}_{\ast}$ respectively.
From (\ref{area functional wedge app}), one finds
\be
\label{area splitted integrals}
\mathcal{A}_{\infty} 
\equiv 
\int_{\hat{\gamma}_\varepsilon^{\textrm{\tiny $\infty$}} } 
\frac{1}{\rho} \,\sqrt{q'^2+q^2+q^4} \, d\phi \,d \rho  
\qquad
 \mathcal{A}_{\ast} 
 \equiv 
 \int_{\hat{\gamma}_\varepsilon^{\ast}} 
\frac{1}{\rho} \,\sqrt{q'^2+q^2+q^4} \, d\phi \,d \rho  
\ee
which give the area of $\hat{\gamma}_\varepsilon$ as follows
\be
\label{area wedge splitting}
\frac{\mathcal{A}[\hat{\gamma}_\varepsilon] }{L^2_{\textrm{\tiny AdS}}}
=
\left\{\begin{array}{ll}
\mathcal{A}_{\infty} +  \mathcal{A}_{\ast}    \hspace{1cm} &  0 < \alpha \leqslant \pi/2
\\
\rule{0pt}{.5cm}
\mathcal{A}_{\infty} -  \mathcal{A}_{\ast}    &    \pi/2 \leqslant  \alpha < \pi
\end{array}
\right.
\ee

By using (\ref{first-integral}) and (\ref{Pintegration def}), the angular part of the integrands in (\ref{area splitted integrals}) 
can be written  as
\be
\sqrt{q'^2+q^2+q^4} \, \frac{dq}{|q'|}
\,=\,
\frac{q^4+q^2}{\sqrt{q_0^4+q_0^2}} \;\, \mathcal{P}_\phi(q,q_0) \, dq
\,=\,
\frac{\sqrt{q^4 + q^2 }}{\sqrt{q^4 + q^2 - q_0^4 - q_0^2}}\; dq
\ee
which leads us to introduce the following function
\be
\label{Gdef integral}
\int_{q_0}^q \frac{\sqrt{\hat{q}^4 + \hat{q}^2 }}{\sqrt{\hat{q}^4 + \hat{q}^2 - q_0^4 - q_0^2}}\; d\hat{q}
\,\equiv\,
\mathcal{G}(q,q_0) 
\qquad
q\geqslant q_0
\ee 
Performing explicitly this integral, we obtain
\be
\label{poe def}
\mathcal{G}(q,q_0) \,\equiv \,
- \,\textrm{i} \, \sqrt{q_0^2+1}\;\, \mathbb{E} \bigg(  \textrm{i} \, \textrm{arcsinh} \sqrt{\frac{q^2-q_0^2}{1+2q_0^2}}\;\bigg|\, \frac{2q_0^2+1}{q_0^2+1} \bigg)
\ee
which satisfies the condition $\mathcal{G}(q_0,q_0) =0$, as expected from (\ref{Gdef integral}).
By employing the following identity \cite{abramowitz}
\bea
\mathbb{E}( \textrm{i} \psi\vert m )
&= & 
\textrm{i} \, \mathbb{F}\big(\arctan(\sinh\psi) \,\big|\, 1-m \,\big) 
- \textrm{i} \,\mathbb{E}\big(\arctan(\sinh\psi) \,\big|\, 1-m\,\big)
\nonumber
\\
\rule{0pt}{.7cm}
& & +\, \textrm{i} \, \sqrt{1-(1-m)\tanh^2\psi}\; \sinh\psi
\eea
we can write  (\ref{poe def}) in a form which does not contain the imaginary unit, finding the real expression reported in (\ref{mathcalG def}).

Since $\hat{\gamma}^{\infty}_{A,\textrm{\tiny aux}} $ is half of $\hat{\gamma}_{A,\textrm{\tiny aux}}$,
the area $\mathcal{A}_{\infty} $ has been already computed in \cite{Drukker:1999zq}.
First we have to expand (\ref{poe def}) for large $q$, finding
\be
\label{G large q expansion}
\mathcal{G}(q,q_0) = q - F(q_0) +O(1/q^3)
\qquad
q\gg1
\qquad
q\gg q_0
\ee
where $F(q_0)$ has been explicitly written in (\ref{F(q0) def main}).
In order to get the area $\mathcal{A}_{\infty}$, a large cutoff $\rho_{\textrm{\tiny max}} \gg 1$ in the radial direction must be introduced. 
Then, we have
\be
\label{rho minimax}
\varepsilon = \frac{\rho_{\textrm{\tiny min}}}{q_0} \,,
\qquad
\varepsilon = \frac{\rho_{\textrm{\tiny max}}}{q(\gamma - \delta_\varepsilon)} \,,
\qquad
L = \rho_{\textrm{\tiny max}} \cos \delta_\varepsilon \,,
\ee
where $\delta_\varepsilon \sim 0^+$ is the angle between the edge of $A$ at  $\phi=\gamma $ and the straight line 
in the $z=0$ half plane connecting the tip of the wedge to the intersection point between the circumference given by $\rho= \rho_{\textrm{\tiny max}} $ 
and the projection of $\partial\hat{\gamma}_\varepsilon \cap \{ z = \varepsilon  \}$ on the $z=0$ half plane.
By employing the expansion (\ref{G large q expansion}) and (\ref{rho minimax}), the area $\mathcal{A}_{\infty} $ is obtained as follows \cite{Drukker:1999zq}
\bea
\label{int area plus}
\mathcal{A}_{\infty} 
&=&
\int_{\rho_{\textrm{\tiny min}}}^{\rho_{\textrm{\tiny max}}}
\frac{d \rho }{\rho} \,
\int_{q_0}^{\rho/\varepsilon}
\frac{\sqrt{q^4 + q^2 }}{\sqrt{q^4 + q^2 - q_0^4 - q_0^2}}\; dq
\; = 
\int_{\rho_{\textrm{\tiny min}}}^{\rho_{\textrm{\tiny max}}}
\frac{\mathcal{G}(\rho/\varepsilon ,q_0) }{\rho} \,d \rho  
\nonumber
\\
\rule{0pt}{.7cm}
&=&
\int_{\rho_{\textrm{\tiny min}}}^{\rho_{\textrm{\tiny max}}}
\frac{1}{\rho} \left[  \,\frac{\rho}{\varepsilon} - F(q_0) +O\big((\varepsilon/\rho)^3\big) \right]
d \rho  
\; = \;
\frac{\rho_{\textrm{\tiny max}}- \rho_{\textrm{\tiny min}}}{\varepsilon} 
- F(q_0) \,\log(\rho_{\textrm{\tiny max}}/\rho_{\textrm{\tiny min}})
+ \dots
\nonumber
\\
\label{int area inf fin}
\rule{0pt}{.7cm}
&=&
\frac{L}{\varepsilon}  - F(q_0) \,\log(L/\varepsilon)+ \dots
\eea
where the dots correspond to finite terms for $\varepsilon \to 0^+$.
We remark that $\mathcal{A}_{\infty} $ provides the expected linear divergence (area law term) whose coefficient is the length of the entangling curve $\partial A \cap \partial B$.
Furthermore, the coefficient of the subleading logarithmic divergence is half of the corresponding coefficient (\ref{cusp exact result}) found for the wedge in AdS$_4$/CFT$_3$,
as expected, being $\hat{\gamma}^{\infty}_{A,\textrm{\tiny aux}} $ half of $\hat{\gamma}_{A,\textrm{\tiny aux}}$.

The computation of the surface integral $\mathcal{A}_{\ast}$ in (\ref{area splitted integrals}) is similar to the one of $\mathcal{A}_{\infty}$, with 
a crucial difference in the angular integral. 
In particular, we find
\bea
\mathcal{A}_{\ast}  
&=&
\int_{\rho_{\textrm{\tiny min}}}^{\rho_{\textrm{\tiny max}}}
\frac{d \rho }{\rho} \,
\int^{q_\ast}_{q_0}
\frac{\sqrt{q^4 + q^2 }}{\sqrt{q^4 + q^2 - q_0^4 - q_0^2}}\; dq
\; = 
\int_{\rho_{\textrm{\tiny min}}}^{\rho_{\textrm{\tiny max}}}
\frac{\mathcal{G}( q_\ast ,q_0) }{\rho} \,d \rho  
\nonumber
\\
\label{int area ast fin}
\rule{0pt}{.6cm}
&=&\;
\mathcal{G}( q_\ast ,q_0)  \, \log(\rho_{\textrm{\tiny max}} / \rho_{\textrm{\tiny min}}) 
\; = \;
\mathcal{G}( q_\ast ,q_0)  \, \log(L / \varepsilon)  + \dots
\eea
Notice that the double integral in $\mathcal{A}_\ast$ factorises into the product of two integrals that can be computed separately. 
This simplification does not occur in the computation of  $\mathcal{A}_{\infty} $.

Finally, plugging (\ref{int area inf fin}) and (\ref{int area ast fin}) into (\ref{area wedge splitting}), we find the total corner function $F_\alpha$ 
in terms of $\alpha$ and $q_0$, whose explicit expression has been reported in (\ref{total corner func bdy main}).
Combining this formula with (\ref{gamma q0 main}), we obtain $F_\alpha(\gamma)$ parametrically through the real parameter $q_0 > 0$.
This function is the main result of this manuscript. 
It is shown in Fig.\;\ref{fig:main_plot} and Fig.\;\ref{fig:main_plot_3D}.

A considerable simplification occurs in the expressions obtained above  when  $\alpha=\pi/2$.
Indeed, being  $q_\ast > 0$, the relation (\ref{intersection-cond})  tells us that $\phi_\ast = 0$.
Then, since $0 \leqslant |\phi_0| \leqslant |\phi_\ast |$, we have that $\phi_\ast = \phi_0 = 0$, and this implies $q_\ast = q_0$.
By substituting $\phi_0 = 0$ into  (\ref{gamma + phi0}), we can conclude that $\gamma = P_0(q_0)$ in this special case.
As for the corner function, the condition $q_\ast = q_0$ tells us that $\mathcal{G}(q_\ast, q_0) = \mathcal{G}(q_0, q_0)=0$.
Plugging this result in (\ref{total corner func bdy main}), we find that  $F_{\pi/2} = F(q_0)$.
Thus, when $\alpha=\pi/2$ we have that the minimal surface $\hat{\gamma}_A $ is half of the minimal surface found in \cite{Drukker:1999zq}, 
namely $\hat{\gamma}_A = \hat{\gamma}^{\infty}_{A,\textrm{\tiny aux}}$ as expected.
This is also stated in (\ref{limit alpha = pi/2}).

\subsection{On the limiting regimes of the corner function}
\label{app wedge limits}

We find it worth considering some interesting regimes of the corner function $F_\alpha(\gamma)$, whose analytic expression is given by (\ref{gamma q0 main}) and  (\ref{total corner func bdy main}). 
In particular, we focus on the limits $\gamma \to 0$ and $\gamma \to \pi/2$, which correspond to $q_0 \to +\infty$ and $q_0 \to 0$ respectively. 
The main results derived in the following are discussed also in Sec.\,\ref{sec:limiting regimes}.

In order to expand $\gamma(q_0)$ in (\ref{gamma q0 main}) for small and large values of $q_0$, 
we find it convenient to write it as follows
\be
\label{gamma integral}
\gamma \,=\, 
P_0(q_0) + \int_{\pi/2}^\alpha \partial_{\tilde{\alpha}} \gamma \, d\tilde{\alpha}
\ee
where (\ref{limit alpha = pi/2}) has been used and $P_0(q_0)$ is given by (\ref{theta_V expression}) or (\ref{P0 expression v2}).
From (\ref{gamma q0 main}) we have that the integrand in (\ref{gamma integral}) reads
\be
\label{partial_alpha gamma}
\partial_{\alpha} \gamma
=
\eta_\alpha \Big(
\partial_{\alpha} \arcsin [  s_\ast(\alpha, q_0) ] 
-
\partial_\alpha P\big(q_\ast(\alpha,q_0) , q_0\big)
\Big)
\ee
where $s_\ast(\alpha, q_0) $ in the first term is given by (\ref{s_ast def main}).
Then,  (\ref{Pintegration def}) tells us that 
$P(q_\ast(\alpha,q_0) ,q_0)$ depends on $\alpha$ only through its first argument $q_\ast(\alpha,q_0)$, which is also 
the upper extremum in the integral defining $P(q,q_0)$.

Thus, for the second term in (\ref{partial_alpha gamma}) with $\alpha \in (0,\pi)$ we find 
\bea
\label{partial_alpha P}
& & \hspace{-1cm}
\partial_\alpha P\big(q_\ast(\alpha,q_0) ,q_0\big)
=\,
 \partial_\alpha \big(  q_\ast(\alpha,q_0)  \big)\,
 \mathcal{P}_\phi\big( q ,q_0\big) \big|_{q= q_\ast(\alpha,q_0)}
\\
\rule{0pt}{.75cm}
& & \hspace{-.4cm}
\hspace{1.2cm}
= \;
\frac{\sqrt{q_0^4 + q_0^2}}{\sqrt{(q^2_\ast(\alpha,q_0) + 1)(q^2_\ast(\alpha,q_0) - q_0^2)(q^2_\ast(\alpha,q_0) + q_0^2+1)}}
\; \frac{\partial_\alpha \big( q_\ast(\alpha,q_0)  \big) }{ q_\ast(\alpha,q_0) }
\nonumber
\\
\rule{0pt}{.85cm}
& & \hspace{-.4cm}
\hspace{1.2cm}
= \,-\,\eta_\alpha\;
\frac{\sqrt{q^2_\ast(\alpha,q_0) -(\cot\alpha)^2}}{(q^2_\ast(\alpha,q_0)+1) \,q_\ast(\alpha,q_0) }
\; \partial_\alpha \big( q_\ast(\alpha,q_0)  \big) \tan\alpha
\nonumber
\eea
We remark that the combination $(\tan\alpha) \, \partial_\alpha q_\ast(\alpha,q_0)  $ in the last expression is regular when $\alpha \to \pi/2$.
Similarly, for the first term in (\ref{partial_alpha gamma}) we find
\be
\label{Ped}
\partial_{\alpha} \arcsin [  s_\ast(\alpha, q_0) ] 
\,=\,
\partial_{\alpha} \arcsin\left [ \, \frac{|\cot\alpha|}{ q_\ast(\alpha,q_0)}\, \right]
=\,\eta_\alpha\, 
\frac{
\cot \alpha \; \partial_\alpha q_\ast(\alpha,q_0) +(\csc\alpha)^2 \, q_\ast (\alpha,q_0)
}{
q_\ast (\alpha ,q_0) \, \sqrt{q^2_\ast (\alpha ,q_0)-(\cot\alpha)^2 }}
\ee
It is important to observe that the factor $\eta_\alpha$ in (\ref{partial_alpha P}) and (\ref{Ped}) simplify with the analogous one in (\ref{partial_alpha gamma}).
Thus, it becomes evident that (\ref{partial_alpha gamma}) is smooth for $\alpha \in (0,\pi)$.

To study $\gamma$ for small and large values of $q_0$, we first employ (\ref{partial_alpha gamma}) and (\ref{partial_alpha P}) for the integrand in  (\ref{gamma integral});
then we expand the resulting expression in the regime we are interested in and only at the end we integrate the coefficients of the expansion.

The corner function $F_\alpha(q_0)$ in (\ref{total corner func bdy main}) can be treated in the same way. 
First, by employing (\ref{limit alpha = pi/2}), we write $F_\alpha$ as 
\be
\label{F_alpha integral}
F_\alpha 
\,=\,
F(q_0) - 
\int^{\pi/2}_{\alpha} \partial_{\tilde{\alpha}} F_{\tilde{\alpha}} \, d \tilde{\alpha}
\ee
Then, from the derivative of (\ref{total corner func bdy main}) with respect to $\alpha$, the integral representation of $\mathcal{G}( q ,q_0) $ in (\ref{Gdef integral}) and the expression of $q_\ast(\alpha,q_0)$ in (\ref{qast(q0,alpha) main}),
we find that
\bea
\label{partial_alpha F}
\partial_\alpha F_\alpha 
&=&
 \,\eta_\alpha \big( \partial_\alpha q_\ast(\alpha,q_0) \big)\,
\partial_q \mathcal{G}( q ,q_0) \big|_{q= q_\ast(\alpha,q_0)}
\\
\rule{0pt}{.75cm}
&=&
\eta_\alpha\,
\frac{\sqrt{q^4_\ast(\alpha,q_0) + q^2_\ast(\alpha,q_0) }}{\sqrt{q^4_\ast(\alpha,q_0) + q^2_\ast(\alpha,q_0) - q_0^4 - q_0^2}}
\; \partial_\alpha \big( q_\ast(\alpha,q_0)\big)
\nonumber \\
\rule{0pt}{.8cm}
&=&
\eta_\alpha\,
\frac{| \sec\alpha\,|\, q_\ast(\alpha,q_0) }{\sqrt{1+q^2_\ast(\alpha,q_0)}} 
\,\partial_\alpha \big( q_\ast(\alpha,q_0)\big)
=
\; -\, \frac{q_\ast(\alpha,q_0)}{\cos\alpha\,\sqrt{1+q^2_\ast(\alpha,q_0)}} \,
\partial_\alpha \big( q_\ast(\alpha,q_0)\big)
\nonumber
\eea
where, like in (\ref{partial_alpha P}), we observe again the occurrence of $(\partial_\alpha q_\ast(\alpha,q_0)) / \cos\alpha$, which is finite and regular when $\alpha = \pi/2$.
Thus, also in this case $\eta_\alpha$ simplifies; 
therefore it becomes evident that $\partial_\alpha F_\alpha $ is a smooth function in $\alpha \in (0,\pi)$.

By plugging (\ref{partial_alpha F}) into (\ref{F_alpha integral}), we obtain an expression which can be easily expanded for $q_0 \to 0$ and $q_0 \to +\infty$.
Only at the end one integrates the coefficients of the expansion as prescribed in the r.h.s. of (\ref{F_alpha integral}).
We remark that the analysis presented here holds for any $\alpha \in (0,\pi)$.

\begin{figure}[t] 
\vspace{-.5cm}
\hspace{-.5cm}
%\begin{center}
\includegraphics[width=1.1\textwidth]{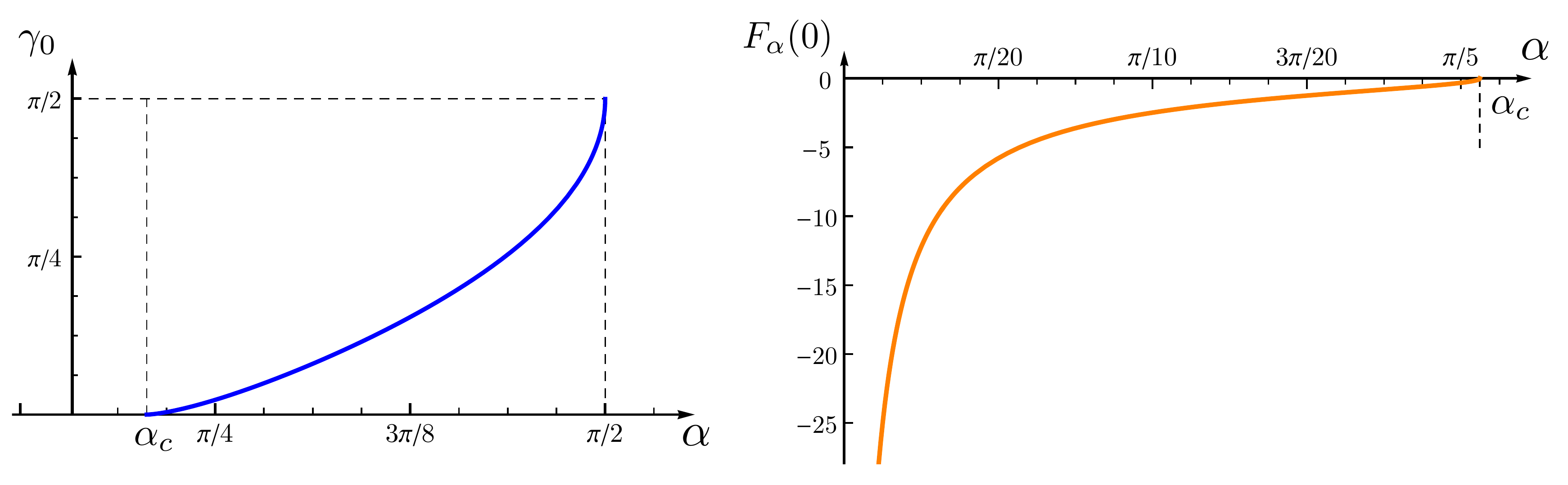}
%\end{center}
\vspace{-.5cm}
\caption{\label{fig_app:zeros}
\small
Left: The function $\gamma_0(\alpha)$ for $\alpha \in [\alpha_c , \pi/2]$, being  $\gamma_0$ defined by $F_\alpha(\gamma_0) = 0$.
Right: The function $F_\alpha(0)$ in terms of $\alpha \leqslant \alpha_c$.
}
\end{figure}

Before considering the regimes $\gamma \to 0$ and $\gamma \to \pi/2$ of the corner function, we find it worth
remarking that when $\alpha \in [\alpha_c , \pi/2]$ the corner function $F_\alpha(\gamma)$ has a unique zero
(see Fig.\,\ref{fig:main_plot}), as already discussed in Sec.\,\ref{sec:wedge}.
Denoting by $\gamma_0$ the value of $\gamma$ such that $F_\alpha(\gamma_0) = 0$, the function
$\gamma_0(\alpha)$ in terms of $\alpha \in [\alpha_c , \pi/2]$ can be obtained numerically and the result is shown in the
left panel of Fig.\,\ref{fig_app:zeros}.

\subsubsection{Large $q_0$ regime}
\label{app large q0}

Let us consider the limit $q_0 \to +\infty$ of the opening angle $\gamma(q_0)$ written in the form (\ref{gamma integral}).
For the first term, which is given by (\ref{theta_V expression}) or (\ref{P0 expression v2}), we find
%\footnote{The identity $\mathbb{E}(-1) - \sqrt{\pi}\, \Gamma(\tfrac{5}{4})/\Gamma(\tfrac{3}{4}) = \Gamma(\tfrac{3}{4})^2 / \sqrt{2\pi} $ could be useful.}
\be
\label{P0(q0) expansion q0=inf}
P_0(q_0) 
\,=\,  
\frac{1}{\sqrt{2\pi}\; \Gamma(\tfrac{3}{4})^2}
\left(\,
\frac{\Gamma(\tfrac{3}{4})^4}{q_0} 
+ \frac{\pi^2 - 6\, \Gamma(\tfrac{3}{4})^4}{24\; q_0^3}
+ \frac{16\, \Gamma(\tfrac{3}{4})^4 - 5  \pi^2}{160\; q_0^5}
+ O(1/q_0^7)
\right)
\ee
As for the second term in (\ref{gamma integral}), the integrand can be expanded by employing (\ref{partial_alpha gamma}), obtaining
\be
\label{partial_alpha gamma expansion q0=inf}
\partial_{\alpha} \gamma
=
\frac{(\csc\alpha)^{3/2}}{2\, q_0}
- 
\frac{(1+\csc\alpha)\, (\csc\alpha)^{3/2}}{8\, q_0^3}
+O(1/q_0^5)
\ee
Finally,  by plugging (\ref{P0(q0) expansion q0=inf}) and (\ref{partial_alpha gamma expansion q0=inf}) into (\ref{gamma integral}), and integrating separately the coefficients of the 
resulting expansion, one finds the first expression in  (\ref{gamma expansion q0=inf main}).

The limit $q_0 \to +\infty$ of the corner function $F_\alpha(q_0)$ can be studied in a similar way, starting from (\ref{F_alpha integral}).
As for the first term, whose explicit expression has been reported in (\ref{F(q0) def main}), its expansion reads
\be
\label{Fq0 expansion q0=infty}
F(q_0) 
=
\frac{1}{\sqrt{2\pi}\; \Gamma(\tfrac{3}{4})^2}
\left(
\Gamma(\tfrac{3}{4})^4 \, q_0 
- \frac{\pi^2 - 2\, \Gamma(\tfrac{3}{4})^4}{8\, q_0}
+ \frac{\pi^2}{32\, q_0^3}
+ O(1/q_0^5)
\right)
%\qquad
%q_0 \to +\infty
\ee
The second term in (\ref{F_alpha integral}) can be addressed by using (\ref{partial_alpha F}), whose expansion is
\be
\label{partial_alpha F expansion}
\partial_\alpha F_\alpha 
=
\frac{(\csc\alpha)^{3/2}}{2}\, q_0 
+
\frac{(3 \csc\alpha + 1) \, (\csc\alpha)^{3/2} }{8\, q_0}
+
\frac{\big(3 \cos(2\alpha) -12 \sin \alpha +7\big)  (\csc\alpha)^{7/2} }{128\, q_0^3}
+ O(1/q_0^5)
\ee
The coefficient of the leading term in this expansion coincides with the coefficient of the leading term in the expansion (\ref{partial_alpha gamma expansion q0=inf}),
while  the subleading terms  are different. 
By inserting the expansions (\ref{Fq0 expansion q0=infty}) and (\ref{partial_alpha F expansion}) into (\ref{F_alpha integral}) first and then integrating the coefficient of the leading term
of the resulting expression, one obtains the second expression in (\ref{gamma expansion q0=inf main}).

As discussed in detail in Sec.\,\ref{sec:limiting regimes}, a peculiar feature of the corner function $F_\alpha(\gamma)$ as $\gamma \to 0^+$ is that $F_\alpha(\gamma) \to + \infty$ when $\alpha > \alpha_c$,
while it tends to a finite value $F_\alpha(\gamma) \to F_\alpha(0)$ when $\alpha \leqslant \alpha_c$.
The function $F_\alpha(0)$ in terms of $\alpha \leqslant \alpha_c$ can be obtained numerically and the result is shown in the right panel of Fig.\,\ref{fig_app:zeros}.
In particular, (\ref{alpha_c def zeroF}) holds for the critical slope $\alpha_c$, and this feature has been employed to get (\ref{alpha_critic_def_omega}) for 
an infinite wedge which has only its tip on the boundary.

We find it worth discussing also the behaviour of the angle $\phi_\ast$ characterising the half line $\hat{\gamma}_A \cap \mathcal{Q}$ as $\gamma \to 0^+$. 
When $\alpha>\alpha_c$, 
from the expansion of (\ref{phi_ast q0 main}) as $q_0 \rightarrow +\infty$ and (\ref{gamma expansion q0=inf main}) we find that
\be
\label{phi_ast exp gamma=0}
\phi_\ast = -\frac{\cos\alpha \,\sqrt{\csc\alpha}}{\mathfrak{g}(\alpha)} \; \gamma + \dots
\qquad
\alpha>\alpha_c
\ee
which implies that $\phi_\ast  \to 0$ when $\gamma \to 0^+$.
Instead, when $\alpha\leqslant \alpha_c$, we have to consider the value $\hat{q}_0$ introduced in Sec.\,\ref{sec:limiting regimes} and plug it into (\ref{phi_ast q0 main}).
The result is a negative and increasing function of $\alpha$ which takes the value $-\pi/2$ for $\alpha \to 0^+$ and vanishes for $\alpha = \alpha_c$.

\subsubsection{Small $q_0$ regime}
\label{app small q0}

The method described in the appendix\;\ref{app large q0} can be adapted to study the limit $q_0 \to 0^+$ 
of the functions $\gamma(q_0)$ and $F_\alpha(q_0)$, once they are written in the form given by  (\ref{gamma integral}) and (\ref{F_alpha integral}) respectively.

Considering the opening angle $\gamma$, for the first term of (\ref{gamma integral}) we find
\be
\label{P0 q0=0 expansion}
P_0(q_0) = \frac{\pi}{2} -\frac{\pi}{2}\, q_0  + \frac{3\pi}{8}\, q_0^3  - \frac{61\pi}{128}\, q_0^5 + O(q_0^7)  
\ee
As for the expansion of the integrand in (\ref{gamma integral}), we find
\be
\label{partial_alpha gamma expansion}
\partial_{\alpha} \gamma
=
q_0 
+ \frac{5 \cos(2\alpha) - 3}{4}\, q_0^3
+ \frac{63 \cos(4\alpha) - 132 \cos(2\alpha) + 61}{64}\, q_0^5
+ O(q_0^7)
\ee
Plugging (\ref{P0 q0=0 expansion}) and (\ref{partial_alpha gamma expansion}) into (\ref{gamma integral}) first
and then integrating the coefficients of the resulting expansion, we find that
\be
\label{gamma expansion q0 final}
\gamma = 
\frac{\pi}{2} 
- (\pi-\alpha) \,q_0 
+ \frac{3(\pi-\alpha)+ 5 \sin \alpha \cos\alpha}{4} \,q_0^3
+ O(q_0^5)
\ee
which can be inverted obtaining 
\be
\label{q0 at gamma=pi/2}
q_0 = 
\frac{\pi/2 - \gamma}{\pi -\alpha} 
- \frac{6(\pi-\alpha) + 5 \sin (2\alpha)}{8(\pi-\alpha)^4}\, (\pi/2 - \gamma)^3
+ O\big((\pi/2 - \gamma)^5\big)
\qquad
\gamma \to \frac{\pi}{2}
\ee

The limit $q_0 \to 0^+$ of the corner function $F_\alpha(q_0)$ in the form (\ref{F_alpha integral}) can be studied in the same way.
The first term in the r.h.s. of (\ref{F_alpha integral}) is (\ref{F(q0) def main}) and its expansion reads
\be
\label{Fq0 expansion}
F(q_0) = \frac{\pi}{4}\, q_0^2 - \frac{7\pi}{32}\, q_0^4 + O(q_0^6)
\ee
As for the integrand occurring in (\ref{F_alpha integral}), from (\ref{partial_alpha F}) we obtain
\be
\label{partial_alpha F expansion q0=0}
\partial_\alpha F_\alpha 
=
\frac{1}{(\sin\alpha)^2} -\frac{1}{2}\, q_0^2 +\frac{7-15 \cos(2\alpha)}{16}\, q_0^4 + O(q_0^6)
\ee
By inserting (\ref{Fq0 expansion}) and (\ref{partial_alpha F expansion q0=0}) into (\ref{F_alpha integral}) first 
and then integrating separately the coefficients of the resulting expansion, 
we find
\be
\label{F expansion q0=0 final}
F_\alpha = 
-\, \cot\alpha 
+\frac{\pi - \alpha}{2}\, q_0^2 
-\frac{7(\pi-\alpha) +15 \cos \alpha \sin \alpha}{16}\, q_0^4 + O(q_0^6)
\ee
Finally, by employing (\ref{q0 at gamma=pi/2}) into (\ref{F expansion q0=0 final}), we obtain
\be
\label{limit gamma=pi/2 app}
F_\alpha(\gamma) 
\,=\, 
- \cot\alpha 
+ \frac{(\pi/2-\gamma)^2}{2(\pi-\alpha)} 
+ \frac{5(\pi-\alpha + \cos \alpha \, \sin \alpha) }{16(\pi-\alpha)^4} \, (\pi/2-\gamma)^4
+ O\big((\pi/2-\gamma)^6\big) 
\ee
which is one of our main results.
In (\ref{limit gamma=pi/2 main}) the first two terms of (\ref{limit gamma=pi/2 app}) have been reported.

\subsection{A relation between the infinite wedge and the infinite strip}
\label{app map strip-cusp}

In the expansion (\ref{limit gamma=0 main}) of the holographic corner function $F_\alpha(\gamma)$ for $\gamma \to 0$ 
and in the $O(1)$ term of the holographic entanglement entropy of the infinite strip adjacent to the boundary in (\ref{area strip final main all}),
the same function $\mathfrak{g}(\alpha)$ given by (\ref{g function def main}) occurs. 
In the following we explain this connection by exploiting a conformal map which relates the infinite wedge adjacent to the boundary in the half plane 
and the infinite strip adjacent to a border of a half cylinder. 
This analysis has been done by adapting to our case in a straightforward way
the analogue relation in absence of the boundary, which involves the infinite wedge in $\mathbb{R}^2$ and the infinite strip on the surface of an
infinite cylinder \cite{stripcusp, Myers:2012vs, Bueno:2015xda}.

\begin{figure}[t] 
\vspace{-.7cm}
%\hspace{-.5cm}
\begin{center}
\includegraphics[width=.5\textwidth]{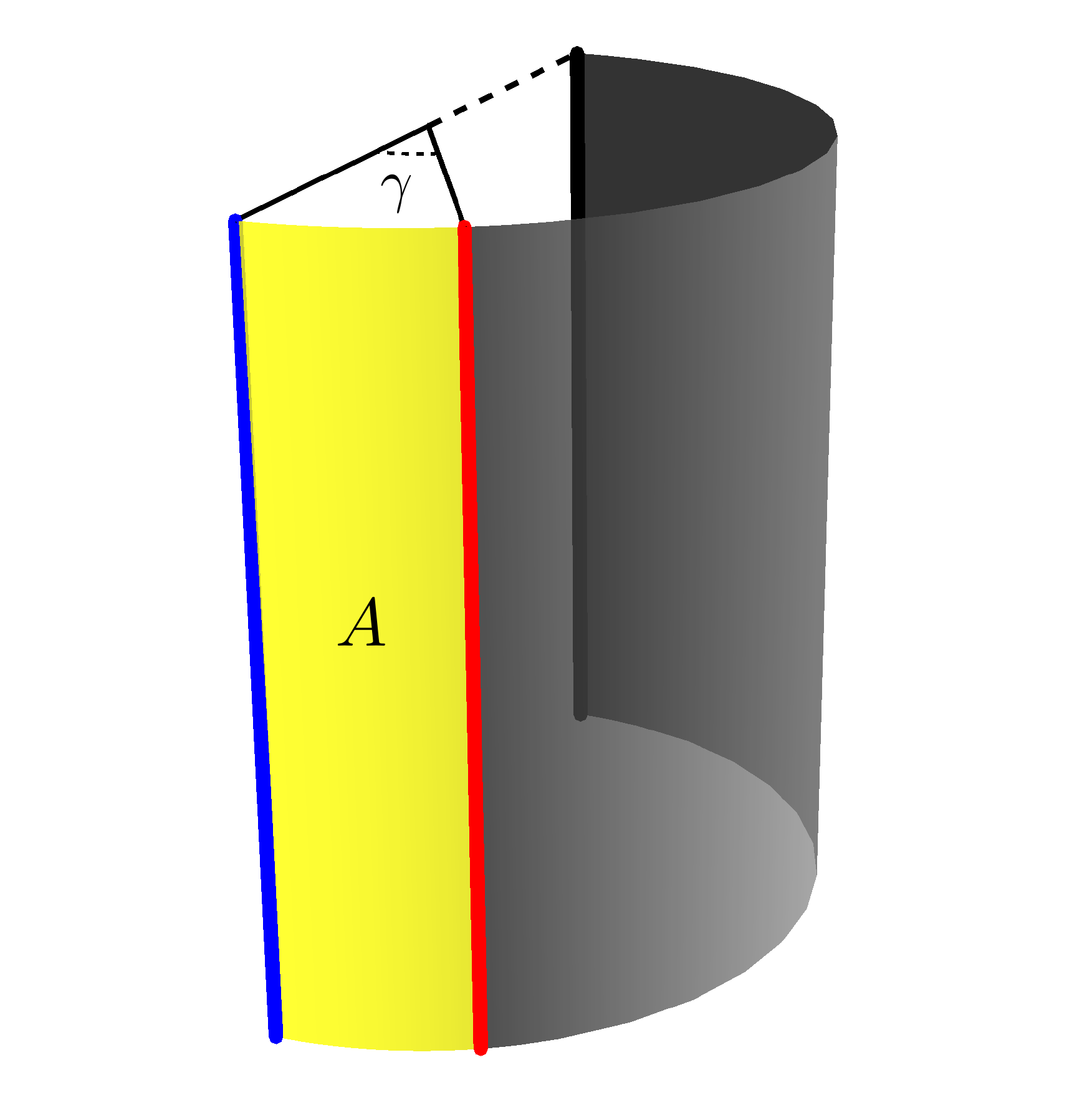}
\end{center}
\vspace{-.3cm}
\caption{\label{fig_cylinder}
\small
Part of the surface of a half cylinder, introduced in the appendix\;\ref{app map strip-cusp}
(see $d\tilde{s}^2$ in (\ref{backg1}) for $\chi=\pi/2$).
This surface corresponds to the conformal boundary of the gravitational spacetimes depicted in Fig.\,\ref{fig_cylinder_AdS}.
The yellow region is an infinite strip $A$ adjacent to the boundary. 
}
\end{figure}

Consider a BCFT$_3$ defined on $\mathds{R}^3_{+}\equiv\{ (t_{\textrm{\tiny E}}, x,y)\in \mathds{R}^3\,|\,x\geqslant 0\}$ endowed 
with the usual Euclidean metric $ ds^2=dt_{\textrm{\tiny E}}^2+dx^2+dy^2 $.
By adopting  the polar coordinates introduced in (\ref{polar coord def}), where we recall that $0\leqslant \phi \leqslant \pi$,
this metric becomes $ds^2=dt_{\textrm{\tiny E}}^2+d\rho^2+\rho^2 d\phi^2$.
We define $t_{\textrm{\tiny E}} = 0$ the slice containing  the infinite wedge $A$ adjacent to the boundary introduced in  Sec.\,\ref{sec:wedge},
whose edges are given by $\phi=0$ and $\phi=\gamma$.
By introducing the coordinates $(\tilde{r},\chi)$ through the relations $t_{\textrm{\tiny E}}= \tilde{r} \cos\chi$ and $\rho=\tilde{r} \sin \chi$, 
where $\tilde{r} \geqslant 0$ and $0\leqslant \chi \leqslant \pi$, 
the flat metric becomes
\be
\label{metric a-xi}
ds^2=d\tilde{r}^2+\tilde{r}^2\big( d \chi^2+(\sin \chi)^2 d\phi^2\big)
\ee
Let us define the coordinate $\tau \in \mathbb{R}$ as $\tilde{r}=L_0 \, e^{\tau/L_0}$.
The tip of the wedge $A$ corresponds to $\tau \to -\infty$, being $\rho=0 = \tilde{r}$ in the previous coordinates. 
In terms of the coordinates $(\tau, \chi, \phi)$, the metric (\ref{metric a-xi})  reads
\be
\label{backg1}
ds^2 = e^{2\tau/L_0} d\tilde{s}^2 
\qquad
d\tilde{s}^2 \equiv d\tau^2+L_0^2 \big( d \chi^2+(\sin \chi)^2 d\phi^2\big)
\ee
i.e.  the flat metric on $\mathds{R}^3_{+}$ is conformally equivalent to $d\tilde{s}^2$, which is the metric $\mathbb{R} \times S^2_+$, being $S^2_+$ 
a two dimensional hemisphere whose radius is $L_0$.
The condition $t_{\textrm{\tiny E}} = 0$ corresponds to $\chi=\pi/2$ and the metric induced on this slice from $d\tilde{s}^2 $ is given by
$d\tilde{s}^2 |_{\chi=\pi/2} = d\tau^2 + L_0^2 \,d\phi^2$, which characterises the external surface of a half cylinder of radius $L_0$, 
whose boundaries are defined by $\phi=0$ and $\phi=\pi$ (see Fig.\,\ref{fig_cylinder}).
Thus, on this surface, the infinite wedge $A$ corresponds to
the infinite strip adjacent to the boundary and enclosed by the generatrices given by $\phi=0$ and $\phi=\gamma$
(the yellow region in Fig.\,\ref{fig_cylinder}).
The width of this infinite strip measured along the surface of the cylinder is $\ell = L_0 \gamma$.

In terms of the coordinates $(\rho, \phi)$ in $\mathds{R}^3_{+}|_{t_{\textrm{\tiny E}} =0}$, 
the entanglement entropy of the infinite wedge $A$ adjacent to the boundary can be written as
\be
\label{ee cusp bdy app}
 S_A
 \,=\, 
 b\, \frac{\rho_{\textrm{\tiny max}}-\rho_{\textrm{\tiny min}}}{\varepsilon}
 - f_{\alpha}(\gamma) \, \log ( \rho_{\textrm{\tiny max}} / \rho_{\textrm{\tiny min}} ) + O(1)
\ee
where $\rho_{\textrm{\tiny max}} = L$ and  $\rho_{\textrm{\tiny min}} = \varepsilon$, being $L \gg \varepsilon$ the infrared regulator introduced in the beginning of Sec.\,\ref{sec:wedge}.
We remark that (\ref{ee cusp bdy app}) is a special case of the general expression (\ref{ee bcft3 corner intro}) (see (\ref{area wedge v0 main}) for the holographic case). 
Since at $\chi = \pi/2$ we have that $\rho =  \tilde{r} = L_0\, e^{\tau/L_0}$,
in terms of this coordinate $\tau$ one finds that (\ref{ee cusp bdy app}) becomes
\be
\label{ee cusp bdy app_r}
 S_A
 \,=\, 
 b\,L_0\, \frac{e^{\tau_+/L_0} - e^{\tau_-/L_0}}{\varepsilon}
 - f_{\alpha}(\gamma) \, \frac{\tau_+ - \tau_-}{L_0}  + O(1)
\ee
where we used that $\rho_{\textrm{\tiny max}} = L =L_0\, e^{\tau_+/L_0}$ and $\rho_{\textrm{\tiny min}} = \varepsilon =L_0\, e^{\tau_-/L_0}$.
These relations and the condition $L/\varepsilon \gg 1$ imply that $(\tau_+ - \tau_-)/L_0 \gg 1$.

In order to relate (\ref{ee cusp bdy app_r}) to the expansion of the entanglement entropy of the infinite strip adjacent to the boundary, 
we take $L_0 \to +\infty$  and $\gamma \to 0^+$ such that $\ell = L_0 \gamma$ is kept constant. 
Notice that the width $L_0(\pi - \gamma)$ of the complementary region $B$ in the half cylinder of Fig.\,\ref{fig_cylinder} diverges in this limit. 
Moreover, since $L_0 \to +\infty$ we have that $L_0(e^{\tau_+/L_0} - e^{\tau_-/L_0}) \to \tau_+ - \tau_-$ in the r.h.s. of (\ref{ee cusp bdy app_r}).
Thus, in this regime (\ref{ee cusp bdy app_r}) becomes 
\be
\label{ee cusp bdy app_r L0=inf}
 S_A
 \,=\, 
 b\, \frac{L_\parallel}{\varepsilon}
 + A_0 \, L_\parallel + O(1)
 \qquad
 \tau_+ - \tau_- = L_\parallel \gg L_0
\ee
where $O\big((\tau^2_+ - \tau^2_-)/L_0^2\big)$ term has been neglected and $A_0$ is defined as follows
\be
\label{A0 def limit}
- \frac{f_{\alpha}(\gamma)}{L_0} \,\to \, A_0
\qquad
 \hspace{.6cm} \textrm{as} \hspace{.5cm}
\left\{
\begin{array}{l}
L_0 \to +\infty \\
\gamma \to 0^+ \\
L_0 \gamma = \ell
\end{array}
\right.
\ee
The expression (\ref{ee cusp bdy app_r L0=inf}) in a BCFT$_3$ corresponds to the entanglement entropy of an infinite strip ($L_\parallel \gg \varepsilon$) of width $\ell$ adjacent to the boundary.

The above discussion holds for any BCFT$_3$ with a flat boundary. 
In the following we focus on the case of AdS$_4$/BCFT$_3$, where this relation between the infinite wedge and the infinite strip adjacent to the boundary
can be explicitly checked.

\begin{figure}[t] 
\vspace{-.7cm}
\hspace{-.6cm}
%\begin{center}
\includegraphics[width=1.1\textwidth]{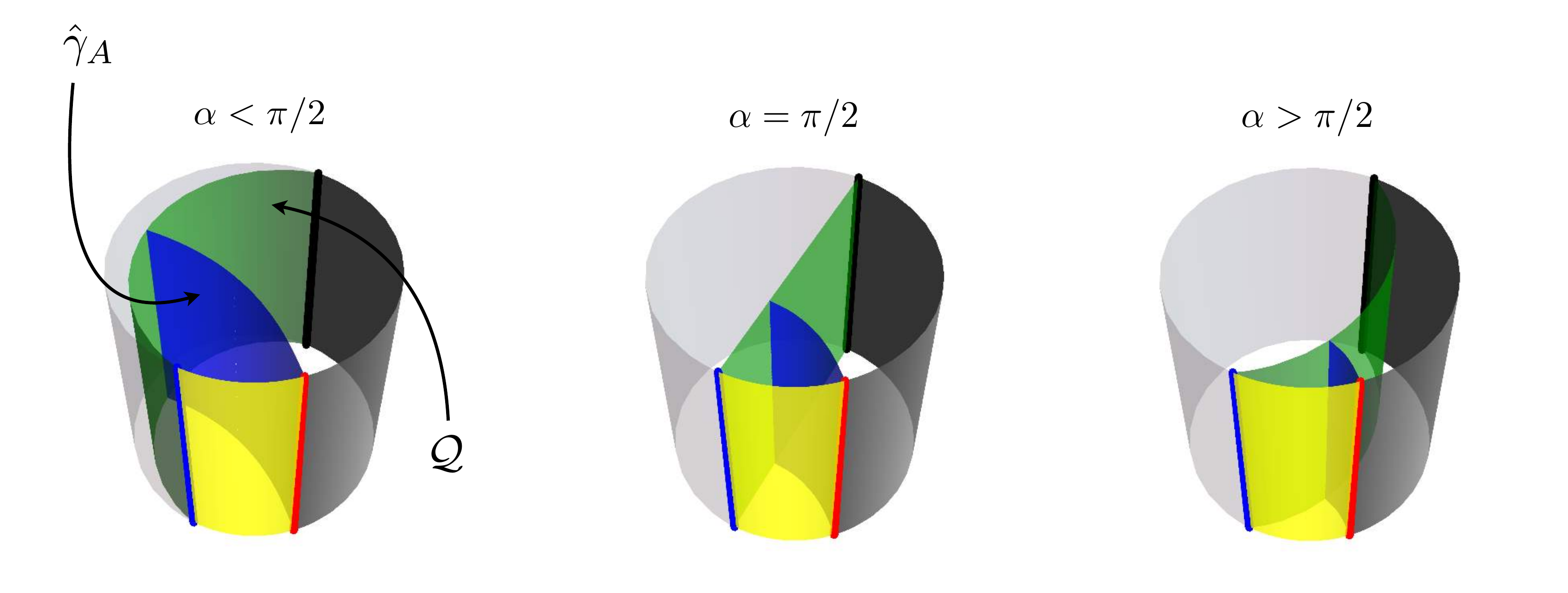}
%\end{center}
\vspace{-.7cm}
\caption{\label{fig_cylinder_AdS}
\small
The spacetime (\ref{bulkcyl}), whose boundary is the union of the surface in Fig.\,\ref{fig_cylinder} 
and of the green surface $\mathcal{Q}$ in (\ref{brane_eq_cylinder}), which depends on the parameter $\alpha \in (0,\pi)$.
The blue surface is the minimal area surface $\hat{\gamma}_A$ corresponding to the infinite strip adjacent to the boundary (yellow region).
The parameter $\alpha$ changes in the various panels:
$\alpha =\pi/10$ (left), $\alpha =\pi/2$ (middle) and  $\alpha =3\pi/4$ (right).
}
\end{figure}

In order to address the holographic case, let us consider a part of the Euclidean AdS$_4$ spacetime in global coordinates, 
whose spacetime interval reads
\be
\label{global AdS4}
ds^2=\frac{dr^2}{1+r^2/L^2_{\textrm{\tiny AdS}}}+ \big(1+ r^2/L^2_{\textrm{\tiny AdS}}\big)d\tau^2 + r^2 \big( d\chi^2 + (\sin \chi)^2 d\phi^2 \big)
\ee
where $\tau \in \mathbb{R}$, $\chi \in [0,\pi]$, $r\geqslant 0$ and $\phi \in [0,2\pi) $,
but the ranges of the last two coordinates are influenced by the occurrence of a constraint coming from (\ref{AdS4 regions}).
Indeed, we have that the conformal boundary corresponds to $r\to +\infty$ and $0\leqslant \phi \leqslant \pi$.
On the $\chi =\pi/2$ slice, the induced metric is given by
\be
\label{bulkcyl}
ds^2=\frac{dr^2}{1+r^2/L^2_{\textrm{\tiny AdS}}}+ \big(1+ r^2/L^2_{\textrm{\tiny AdS}}\big)d\tau^2 + r^2 d\phi^2 
\ee
By introducing the coordinates $(z, \rho)$ as follows
\be
\label{coord_transf}
	r\,=\, L_{\textrm{\tiny AdS}} \,\frac{\rho}{z}
	\qquad
	\tanh \big(\tau/L_{\textrm{\tiny AdS}}\big) 
	= \frac{z^2 +\rho^2 -L_{\textrm{\tiny AdS}}^2}{z^2+\rho^2 +L_{\textrm{\tiny AdS}}^2}	
\ee
one finds that (\ref{bulkcyl}) becomes
\be
\label{poincare}
ds^2=\frac{L_{\textrm{\tiny AdS}}^2}{z^2} \left(dz^2+d\rho^2+\rho^2 d\phi^2  \right) 
\ee
which is the metric of $\mathbb{H}_3$ (see (\ref{AdS4 metric})) in terms of the polar coordinates (\ref{polar coord def}), 
whose conformal boundary corresponds to $z\rightarrow 0^+$.

From the definition (\ref{brane-profile}) of  half plane $\mathcal{Q}$ written in terms of the polar coordinates (\ref{polar coord def})
and the first expression of (\ref{coord_transf}), 
we find that the position of $\mathcal{Q}$ in the spacetime (\ref{bulkcyl}) is given by 
\be
\label{brane_eq_cylinder}
\mathcal{Q}\;:
\hspace{.7cm}
r \,=\, - \, L_{\textrm{\tiny AdS}}\,\frac{\cot\alpha}{\sin\phi}
\hspace{.6cm} \qquad \hspace{.6cm}
\Bigg\{
 \begin{array}{ll} 
 \pi\leqslant \phi \leqslant 2\pi \hspace{.6cm} & \alpha \in (0,\pi/2)
 \\
 \rule{0pt}{.55cm}
 0\leqslant \phi \leqslant \pi &  \alpha \in (\pi/2,\pi)
\end{array}
\ee
In Fig.\,\ref{fig_cylinder_AdS} the spacetime defined by (\ref{bulkcyl}) and constrained by (\ref{brane_eq_cylinder}) 
is the internal part of the cylinder enclosed by the green surface, which corresponds to $\mathcal{Q}$ and the darker half of the cylindrical surface,
which is the conformal boundary of the spacetime (\ref{bulkcyl}) (see also Fig.\,\ref{fig_cylinder}).
Since the conformal boundary is defined by $r\rightarrow +\infty$, 
in Fig.\,\ref{fig_cylinder_AdS} the radial variable $\hat r= (2 L_{\textrm{\tiny AdS}} / \pi)\arctan r$ has been employed. 
Notice that for $\alpha=\pi/2$ half of the global AdS$_4$ must be considered, as expected.

Close to this boundary the second expression of (\ref{coord_transf}) becomes
\be
\tanh \big(\tau/L_{\textrm{\tiny AdS}}\big)  =\frac{\rho^2-L_{\textrm{\tiny AdS}}^2}{\rho^2+L_{\textrm{\tiny AdS}}^2}
\hspace{.8cm} \Longleftrightarrow \hspace{.8cm} 
\rho=L_{\textrm{\tiny AdS}} \,e^{\tau/L_{\textrm{\tiny AdS}}}
\ee 
i.e. we recover the exponential change of coordinate reported in the text above (\ref{ee cusp bdy app_r}), once the identification $L_{\textrm{\tiny AdS}} = L_0$
is assumed.

In AdS$_4$/BCFT$_3$, by computing the holographic entanglement entropy of 
the infinite wedge adjacent to the boundary (Sec.\,\ref{sec:wedge}), 
we have found that (\ref{ee cusp bdy app}) holds with $b = L^2_{\textrm{\tiny AdS}} / (4 G_{\textrm{\tiny N}}) $ 
and $ f_{\alpha}(\gamma) = \tfrac{L^2_{\textrm{\tiny AdS}}}{4 G_{\textrm{\tiny N}}}  \,F_{\alpha}(\gamma)$,
being  $F_{\alpha}(\gamma)$ given by (\ref{gamma q0 main}) and (\ref{total corner func bdy main}). 
By employing the results discussed in Sec.\,\ref{sec:limiting regimes} for the regime $\gamma \to 0^+$ of $F_{\alpha}(\gamma)$
(namely (\ref{limit gamma=0 main}) and the fact that $F_{\alpha}(\gamma) \to F_{\alpha}(0)$ when $\alpha \leqslant \alpha_c$)
and the identification $L_{\textrm{\tiny AdS}} = L_0$, 
we find that (\ref{A0 def limit}) in this case gives
\be
\label{A0 case hee}
A_0 = 
\frac{L^2_{\textrm{\tiny AdS}}}{4 G_{\textrm{\tiny N}}}
\;\frac{a_0(\alpha)}{\ell}
\ee
where  $\ell = L_{\textrm{\tiny AdS}} \,\gamma$ and $a_0(\alpha)$ has been defined in (\ref{area strip final main all}).
Plugging these results into (\ref{ee cusp bdy app_r L0=inf}), we recover the holographic entanglement entropy (\ref{area strip final main all})
of the infinite strip adjacent to the boundary, as expected.

As further consistency check of the relation between the infinite wedge and the infinite strip 
adjacent to the boundary, we find it worth considering the quantity $L_{\textrm{\tiny AdS}}\, \phi_\ast$ 
in the limit defined in (\ref{A0 def limit}) with $L_0 = L_{\textrm{\tiny AdS}}$.
By employing (\ref{phi_ast exp gamma=0}) and the corresponding result for $\alpha  \leqslant \alpha_c$, we find
\be
\label{xstar}
\left\{\,
\begin{array}{ll}
\,\displaystyle{
L_{\textrm{\tiny AdS}} \,\phi_\ast 
= - \,\frac{\cos\alpha\,\sqrt{ \csc\alpha}}{\mathfrak{g}(\alpha)} \; L_{\textrm{\tiny AdS}}\,\gamma  + \dots
=- \,\frac{\cos\alpha\,\sqrt{ \csc\alpha}}{\mathfrak{g}(\alpha)} \;\ell + \dots
}
\hspace{.4cm} 
&   \alpha > \alpha_c
\\
\rule{0pt}{.6cm}
L_{\textrm{\tiny AdS}}\,\phi_\ast 
\rightarrow -\infty 
&   \alpha  \leqslant \alpha_c
\end{array}
\right.
\ee
In Sec.\,\ref{sec:strip} we have found that $x_\ast = - \, z_\ast \cot\alpha $ when $\alpha > \alpha_c$, with $z_\ast$ given by (\ref{z_ast strip main}), 
while $x_\ast \to - \infty$ when $\alpha \leqslant \alpha_c$. 
Comparing these results with (\ref{xstar}), we have that $x_\ast = L_{\textrm{\tiny AdS}}\,\phi_\ast $ in the limit that we are considering.
This identification allows to interpret the transition between $\hat{\gamma}_{A}^{\textrm{\tiny \,con}}$ and $\hat{\gamma}_{A}^{\textrm{\tiny \,dis}}$ 
at $\alpha = \alpha_c$ for the infinite strip adjacent to the boundary (see Sec.\,\ref{sec:strip})
in terms of the behavior of $\phi_\ast$ for $\gamma \to 0$. 
Indeed, when $\alpha>\alpha_c$, from (\ref{phi_ast exp gamma=0}) we have $\phi_\ast \rightarrow 0$ as $\gamma\rightarrow 0$,
therefore $x_\ast$ remains finite and the minimal surface for the infinite strip is $\hat{\gamma}_{A}^{\textrm{\tiny \,con}}$.
Instead, when $\alpha \leqslant \alpha_c$ the angle $\phi_\ast$ remains finite and negative, as discussed below (\ref{phi_ast exp gamma=0}).
This means that $x_\ast \rightarrow -\infty$ for large $L_{\textrm{\tiny AdS}}$, which tells us that the minimal area surface for the infinite strip 
adjacent to the boundary is the vertical half plane $\hat{\gamma}_{A}^{\textrm{\tiny \,dis}}$.

\newpage

%%%%%%%%%%%%%%%%%%%%%%%%%%%%%%%%%%%%%%%%%%%%%%%%%%%%%%%

\section{The coefficient $A_T$ from holography}
\label{app A_T}

In this appendix we describe the details of the holographic computation of the coefficient $A_T$ defined in (\ref{A_T def})
in the AdS$_{d+2}$/BCFT$_{d+1}$ setup of \cite{Takayanagi:2011zk}.
The main result of our analysis is the analytic expression of $A_T$ for arbitrary $d>1$.
The special case of $d=2$ has been reported in Sec.\,\ref{sec stress}.

The AdS$_{d+2}$/BCFT$_{d+1}$ construction of \cite{Takayanagi:2011zk} has been described by employing the following metric
\be
\label{tak-metric}
ds^2 = d\xi^2 +\big[ \cosh(\xi / L_{\textrm{\tiny AdS}})\, \big]^2 
\left(
L^2_{\textrm{\tiny AdS}}\, \frac{-\,dt^2 + d\zeta^2 + d\vec{y}^{\, 2}}{\zeta^2}
\, \right)
\qquad
\zeta > 0
\ee
where $d\vec{y}^{\, 2}$ is the Euclidean flat metric of $\mathbb{R}^{d-1}$. 
If $\xi \in \mathbb{R}$, then the metric (\ref{tak-metric}) describes AdS$_{d+2}$.
Indeed, the change of coordinates
\be
\label{z-x map}
z=\frac{\zeta}{\cosh(\xi / L_{\textrm{\tiny AdS}})}
\qquad
x = -\; \zeta \, \tanh(\xi / L_{\textrm{\tiny AdS}})
\ee
brings the metric (\ref{tak-metric}) into the usual form (\ref{AdS metric ddim}) in terms of the Poincar\'e coordinates. 
Notice that on a generic $\xi = \textrm{const}$ slice of (\ref{tak-metric}) the induced metric is the Poincar\'e metric of AdS$_{d+1}$.
In terms of the coordinates occurring in (\ref{tak-metric}), the half hyperplane $\mathcal{Q}$ corresponds to a particular $\xi = \textrm{const}$ slice.
From (\ref{z-x map}), we have that the conformal boundary where the BCFT$_{d+1}$ is defined is defined by taking $\xi \to -\infty$ and $\zeta\to 0^+$,
keeping the product $\zeta \,\xi$ fixed.

In order to make contact with the coordinates mainly employed throughout this manuscript, 
we find it convenient to introduce the angular coordinate $\psi \in (0, \pi)$ as follows
\be
\label{theta-xi}
\cot \psi \,=\, - \,\sinh(\xi / L_{\textrm{\tiny AdS}})
\ee
From (\ref{z-x map}), it is straightforward to observe that 
\be
\label{z/x ratio}
\frac{z}{x}  = - \frac{1}{\sinh(\xi / L_{\textrm{\tiny AdS}})} = \tan \psi
\ee
In terms of the angular coordinate $\psi \in (0,\pi)$ defined in (\ref{theta-xi}), the metric (\ref{tak-metric}) becomes
\be
\label{theta-metric}
ds^2 = 
\frac{L^2_{\textrm{\tiny AdS}}}{(\sin\psi)^2}
\left( 
d\psi^2
+
\frac{-\,dt^2 + d\zeta^2 + d\vec{y}^{\, 2}}{\zeta^2}
\, \right)
\qquad
\zeta > 0
\ee

By employing the metric (\ref{theta-metric}) in the AdS$_{d+2}$/BCFT$_{d+1}$ setup described in Sec.\,\ref{sec HEE bdy},
where the boundary of the BCFT$_{d+1}$ is a flat hyperplane,
we have that the half hyperplane $\mathcal{Q}$ in (\ref{brane-profile}) is given by $\psi = \pi - \alpha$, with $\alpha \in (0,\pi)$,
and the spacetime of the BCFT$_{d+1}$ corresponds to the limit $\psi \to 0^+$.
Indeed, (\ref{z/x ratio}) tells us that the limit $z \to 0^+$ for fixed $x>0$ corresponds to $\psi \to 0^+$.

In order to find $A_T$ for the AdS$_{d+2}$/BCFT$_{d+1}$ construction proposed in \cite{Takayanagi:2011zk},
one introduces a non vanishing extrinsic curvature $k_{ij}$ for the boundary of the BCFT$_{d+1}$
and solves the Einstein equations with the Neumann boundary condition $K_{ij} = (K-T) h_{ij}$ 
proposed by \cite{Takayanagi:2011zk} perturbatively in $k_{ij}$, considering only the first order in the perturbation.

Since we consider the first non trivial order in the curvature of the boundary,
the metric of the BCFT$_{d+1}$ close to the boundary can be written as the follows
\be
\label{metric-bdy-gauss}
ds^2 = dx^2 + \big(\, \eta_{ij} - 2 \,x \, k_{ij} + \dots\big) \,dY^i dY^j
\ee
where $Y^i = (t, \vec{y}\,)$ and $\eta_{ij}$ is the $d$ dimensional Minkowski metric.
The dots denote higher order terms in the extrinsic curvature and in the distance $x$.
In the literature this gauge choice is sometimes called geodesic slicing.

In order to find the bulk metric corresponding to (\ref{metric-bdy-gauss}), 
in the following we employ the ansatz recently suggested in \cite{Miao:2017aba} written in the coordinates adopted in (\ref{theta-metric}).
Also in \cite{Nozaki:2012qd} a similar analysis has been performed.
In particular, let us consider the perturbation of (\ref{theta-metric}) given by 
\be
\label{theta-metric-pert}
ds^2 = 
\frac{L^2_{\textrm{\tiny AdS}}}{(\sin\psi)^2}
\,\bigg( 
d\psi^2
+
\frac{d\zeta^2 + \big(\eta_{ij}-2\,\zeta \cos\psi \; p_d(\psi) \, \kappa_{ij} \big)  dY^i dY^j} {\zeta^2}
\, \bigg)
+ O(k^2)
\ee
where $\kappa_{ij} = k_{ij} - (k / d) \eta_{ij}$ is the traceless part of the extrinsic curvature
and the boundary condition $p_d(0) = 1$ is imposed to recover (\ref{metric-bdy-gauss}) for the BCFT$_{d+1}$.

The metric (\ref{theta-metric-pert}) is a solution of the Einstein equations with negative cosmological constant up to $O(k^2)$ terms when
%$k_{ij} = \kappa_{ij}$ and  
$p_d(\theta)$ solves the following ordinary differential equation
\be
\label{eq p(theta)}
\sin(2\psi ) \, p_d''(\psi) -2 \big[ \, (d-2) (\cos\psi)^2 + 2 \,\big] \, p_d'(\psi) \,=\,0
\ee

We remark that in (\ref{theta-metric-pert}) $\kappa_{ij}$ occurs in the perturbation (and not $k_{ij}$) without loss of generality.
Indeed, if we start with a metric like  (\ref{theta-metric-pert}) where $\kappa_{ij}$ is replaced by $k_{ij}$
and $\eta_{ij}$ by $\eta_{ij} [1+\zeta \cos\psi \,q_d(\psi)\, k \,] $, being $k$ the trace of $k_{ij}$,
we find that the Einstein equations at the first perturbative order in the extrinsic curvature provide again the equation (\ref{eq p(theta)}) for $p_d(\psi)$
besides another equation for the function $q_d(\psi)$.
Otherwise, if we start with an ansatz like  (\ref{theta-metric-pert}) with $\kappa_{ij}$ just replaced by $k_{ij}$,
the Einstein equations to this order lead to (\ref{eq p(theta)}), as expected, and also the condition that $k=0$.

The general solution of (\ref{eq p(theta)}) reads
\be
\label{p(theta) sol-gen}
p_d(\psi) = B_d + \frac{C_d}{\cos\psi} \; _2F_1 \big( -1/2\,, (1-d)/2\,; 1/2\,; (\cos\psi)^2 \,\big)
\ee
where $B_d$ and $C_d$ are integration constants. 
The requirement that (\ref{p(theta) sol-gen}) satisfies the boundary condition $p_d( 0) = 1 $ leads to
\be
\label{q_sol_0}
B_d = 1 - \frac{\sqrt{\pi}\;\Gamma(\tfrac{d+1}{2})}{\Gamma(\tfrac{d}{2})}\, C_d
\ee
Thus, the solution of (\ref{eq p(theta)}) fulfilling the constraint $p_d( 0) = 1 $ can be written as
\be
\label{q_final_C1}
p_d(\psi) = 1+ C_d \, \mathcal{P}_d(\psi)
\ee
where 
\be
\label{calP_d}
\mathcal{P}_d(\psi)
\equiv
\frac{1}{\cos\psi} \;  _2F_1 \big( -1/2\,, (1-d)/2\,; 1/2\,; (\cos\psi)^2 \,\big)
-
\frac{\sqrt{\pi}\;\Gamma(\tfrac{d+1}{2})}{\Gamma(\tfrac{d}{2})}
\ee
We find important to remark that the combination $p_d(\psi) \cos\psi$ occurring in the metric is smooth for $\psi \in (0, \pi)$.

In the following we show that the constant $C_d$ in (\ref{q_final_C1}) can be fixed in order to have that
 the half hyperplane $\mathcal{Q}$ given by $\psi = \pi - \alpha $ is a solution of 
the Neumann boundary conditions $K_{ab} = (K-T) h_{ab}$ of \cite{Takayanagi:2011zk} up to $O(k^2)$ terms.

Considering the metric $ds^2 = g_{\mu\nu} dx^\mu dx^\nu$ defined in (\ref{theta-metric-pert}),
the outward unit normal vector of the half hyperplane $\mathcal{Q}$ is $n^\mu= L^{-1}_{\textrm{\tiny AdS}} (\,\sin\alpha, \vec{0}\,)$.
As for the extrinsic curvature of $\mathcal{Q}$, we find that its non vanishing components are given by
\bea
K_{\zeta\zeta}
&=&
\left.
\frac{L_{\textrm{\tiny AdS}}\, \sin\psi}{2\,\zeta^2} \;
\partial_\psi \left(\frac{1}{\sin^2\psi}\right)\right|_{\,\psi\,=\,\pi-\alpha}
\\
\rule{0pt}{.7cm}
K_{Y^i Y^j}
&=&
\left. \frac{L_{\textrm{\tiny AdS}} \, \sin\psi}{2}  \; 
\partial_{\psi} \left(\,
\frac{1}{\zeta^2  \sin^2\psi}\, \delta_{ij}
-
\frac{2\, p_d(\psi)\cos\psi}{\zeta \sin^2\psi}\; k_{ij} 
\right)\right|_{\,\psi\,=\,\pi-\alpha}
\eea

Taking the trace of the Neumann boundary conditions, it is straightforward to observe that they can be written as $K_{ab} = (T/d) h_{ab}$.
Since $T= (d / L_{\textrm{\tiny AdS}}) \cos\alpha$ for the half hyperplane $\mathcal{Q}$, 
the condition to impose in order to get the solution given by $\mathcal{Q}$ 
becomes $K_{ab} = (\cos\alpha / L_{\textrm{\tiny AdS}}) h_{ab}$ at $\psi = \pi - \alpha$.
At $O(k)$, the component having $(a,b)=(\zeta,\zeta)$ is identically satisfied, while the components with $(a,b) = (Y^i , Y^j)$ lead to the following equation
\be
\label{eq for C_d}
(\cot \alpha ) \, p_d'(\pi -\alpha )+p_d(\pi -\alpha   ) \,= \,0
\ee
Plugging (\ref{q_final_C1}) into (\ref{eq for C_d}), we obtain an equation for the integration constant $C_d$ which can be easily solved.
For $\alpha \in (0,\pi)$, we find
\bea
\label{C_d sol ddim}
\frac{1}{C_d}
&=&  - \,\mathcal{P}_d(\pi -\alpha) - \cot\alpha \; \partial_\psi \mathcal{P}_d(\psi) \big|_{\,\psi \,=\, \pi - \alpha}
\\
\rule{0pt}{.5cm}
&=& 
 \frac{1}{\cos\alpha} \;  _2F_1 \big( -1/2\,, (1-d)/2\,; 1/2\,; (\cos\alpha)^2 \,\big)
 -
 \frac{(\sin \alpha)^{d-1}}{\cos\alpha} 
+
\frac{\sqrt{\pi}\;\Gamma(\tfrac{d+1}{2})}{\Gamma(\tfrac{d}{2})}
\nonumber
\eea
Let us observe that $C_d = 1 /(\pi-\alpha)^{d-1} + \dots$ when $\alpha \to \pi$ and also that
\be
\label{Cd derivative}
\partial_\alpha \big( 1/C_d\big) \,=\, - \,(d-1) (\sin\alpha)^{d-2}
\ee

Comparing (\ref{Cd derivative}) with (\ref{g-prime-alpha}) it is straightforward to observe that $\partial_\alpha \big( 1/C_d\big)   = \partial_\alpha \,\mathfrak{g}_{1/d} $.
This observation suggests to perform a  direct comparison between (\ref{C_d sol ddim}) and (\ref{g def ddim}), which provides the following intriguing relation
\be
\label{C-g rel}
\frac{1}{C_d(\alpha)} = \mathfrak{g}_{1/d}(\alpha) 
\ee
It would be interesting to explore whether this observation leads to some physical insights.

We find it worth considering the special cases of $d=2$ and $d=3$ explicitly. 
\\
In AdS$_4$/BCFT$_3$, the expressions (\ref{calP_d}) and (\ref{C_d sol ddim}) give respectively 
\be
\mathcal{P}_2(\psi) = \tan \psi - \psi 
\ee
and
\be
C_2 = \frac{1}{\pi -\alpha}
\ee
In the case of AdS$_5$/BCFT$_4$ we have that (\ref{calP_d}) simplifies to
\be
\mathcal{P}_3(\psi) = \cos\psi + \sec\psi -2
\ee
and (\ref{C_d sol ddim}) leads to
\be
C_3 = \frac{1}{2\,( 1+ \cos\alpha)} 
%= \frac{1}{4 [\, \cos(\alpha/2)\,]^2}
\ee

In the remaining part of this appendix, we show that the constant $C_d$ is proportional to the constant $A_T$ defined by (\ref{A_T def}).

According to the holographic prescription of \cite{deHaro:2000vlm},
the expansion close to the boundary
of the one point function $\langle \, T_{ij} \, \rangle$ of the stress tensor in the BCFT$_3$ is given by
\be
\label{holog stress}
\langle \, T_{ij} \, \rangle
\,=\,
\frac{(d+1)L^{d}_{\textrm{\tiny AdS}}}{16\pi G_{\textrm{\tiny N}}}\,
\lim_{z\,\to\, 0}\, 
\frac{g^{\textrm{\tiny $(1)$}}_{ij}}{z^{d-1}}
\qquad
x \to 0^+
\ee
being $g^{\textrm{\tiny $(1)$}}_{ij}$ the $O(k)$ perturbation, which can be read from (\ref{theta-metric-pert}), finding
\be
g^{\textrm{\tiny $(1)$}}_{ij} 
= 
 -\, \frac{2 \cos\psi \, p_d(\psi)}{(\sin \psi)^2 \,\zeta}\, \kappa_{ij}
\ee
where $p_d(\psi)$ is (\ref{q_final_C1}) with the constant $C_d$ given by (\ref{C_d sol ddim})

In order to recover the expression (\ref{A_T def}) from (\ref{holog stress}), 
we have to exploit the relations among the various coordinates. 
In particular, from (\ref{theta-xi}) we have that $\xi \to - \infty$ as $\psi \to 0^+$.
Furthermore, taking this limit in the second expression in (\ref{z-x map}), one finds $\zeta \to  x $.
By considering the $O(\psi^{d+1})$ term in the expansion of $p_d(\psi)$ for $\psi \to 0$,
we obtain 
\be
\label{delta_g limit}
\lim_{z\,\to\, 0}
\, \frac{ g^{\textrm{\tiny $(1)$}}_{ij} }{z^{d-1}}
%\,=\,
%\lim_{z\,\to\, 0}
%\; \frac{1}{\theta^2\, \zeta \,z^{d-1}} \left( \frac{-\,2\, C_d\, \theta^{d+1}}{d+1} \, \kappa_{ij}\right)
\,=\,
-\,  \frac{2\, C_d}{(d+1)\, x^d} \,\kappa_{ij}
\ee
where we used that $z/x = \psi$ and $\zeta = x$ when $\psi \to 0^+$.

Finally, by plugging (\ref{delta_g limit}) into (\ref{holog stress}), we find that
\be
\label{A_T - C rel}
\langle \, T_{ij} \, \rangle
\,=\,
 \frac{A_T}{x^d} \, \kappa_{ij} + \dots 
 \qquad
 x \to 0^+
 \hspace{.5cm} \qquad  \hspace{.5cm} 
 A_T  = - \frac{L^{d}_{\textrm{\tiny AdS}}}{8\pi G_{\textrm{\tiny N}}}\, C_d
\ee
which corresponds to the expected BCFT$_{d+1}$ behaviour (\ref{A_T def}).
The proportionality relation between $A_T$ and the integration constant $C_d$ comes from
the dual gravitational description of the BCFT$_{d+1}$ at strong coupling.

We can write $A_T$ explicitly by employing the expression of $C_d$ that can be read from (\ref{C_d sol ddim}).
The result is
\be
\label{A_T ddim}
A_T 
%\,=\,
%-\, \frac{L^{d}_{\textrm{\tiny AdS}}}{8\pi G_{\textrm{\tiny N}}}\, C_d
\,=\,
-\, \frac{L^{d}_{\textrm{\tiny AdS}}}{8\pi G_{\textrm{\tiny N}}}
\left[\,
 \frac{1}{\cos\alpha} \;  _2F_1 \big( -1/2\,, (1-d)/2\,; 1/2\,; (\cos\alpha)^2 \,\big)
 -
 \frac{(\sin \alpha)^{d-1}}{\cos\alpha} 
+
\frac{\sqrt{\pi}\;\Gamma(\tfrac{d+1}{2})}{\Gamma(\tfrac{d}{2})}
\,\right]^{-1}
\ee
We find worth remarking that $A_T$ can be written also in terms of the function $\mathfrak{g}_{d}(\alpha) $ 
defined in (\ref{g def ddim}).
From (\ref{A_T - C rel}) and the relation (\ref{C-g rel}), we obtain
\be
\label{A_T ddim-g}
A_T 
\,=\,
-\, \frac{L^{d}_{\textrm{\tiny AdS}}}{8\pi G_{\textrm{\tiny N}}} \;\frac{1}{\mathfrak{g}_{1/d}(\alpha) }
\ee

 The function $A_T(\alpha) $ is negative and decreasing function in the range $\alpha \in (0,\pi)$,
Indeed, for $\alpha = 0$ we find
\be
A_T \big|_{\,\alpha \,=\,0}
\, = \, 
\frac{L^{d}_{\textrm{\tiny AdS}}}{8\pi G_{\textrm{\tiny N}}} \,
\bigg(
\frac{2\,\sqrt{\pi}\;\Gamma(\tfrac{d+1}{2})}{\Gamma(\tfrac{d}{2})}
-\delta_{d,1}
\bigg)^{-1}
\ee
which is negative for every value of $d$.
Moreover, from (\ref{Cd derivative}) it is straightforward to observe that
\be
\partial_\alpha A_T(\alpha) 
\,=\, 
\frac{L^{d}_{\textrm{\tiny AdS}}}{8\pi G_{\textrm{\tiny N}}}\; C_d^2\, \partial_\alpha(1/C_d) 
\,=\,
-\,\frac{L^{d}_{\textrm{\tiny AdS}}}{8\pi G_{\textrm{\tiny N}}} \,(d-1) (\sin\alpha)^{d-2} \, C_d^2
\ee
which implies $\partial_\alpha A_T(\alpha) \leqslant 0$ for $\alpha \in (0,\pi)$.
Furthermore, let us notice that the behaviour of $C_d$ as $\alpha \to \pi$ leads to conclude that $A_T(\alpha) = - \tfrac{L^{d}_{\textrm{\tiny AdS}}}{8\pi G_{\textrm{\tiny N}}} (\pi - \alpha)^{-(d-1)} \,$ in this limit. 

In the special case of $d=2$, the expression (\ref{A_T ddim}) of $A_T$ simplifies to (\ref{A_T holog})
and this result is crucial to observe the relation (\ref{rel}), which holds for $\alpha \in (0,\pi)$.

The computation described above has been recently done for $d=2$ and $d=3$ also in \cite{Miao:2017aba}
and non smooth expressions for $A_T$ have been found in the regime $\alpha \in (0,\pi)$.

%%%%%%%%%%%%%%%%%%%%%%%%%%%%%%%%%%%%%%%%%%%%%%%%%%%%%%%

\section{Check of the constraints for the corner functions}
\label{app constraints}

In this appendix we check that the holographic corner functions derived in Sec.\,\ref{sec:wedge} and Sec.\,\ref{sec wedge tip bdy} 
for AdS$_4$/BCFT$_3$ satisfy the constraints found in Sec.\,\ref{sec constraints}.

The corner function (\ref{bdy Fcal generic}) fulfils the inequality (\ref{ssa config 1}) in a trivial way. 
Indeed, whenever the maximisation procedure selects $\widetilde{F}(\omega)$ 
(namely for either $\alpha \leqslant \alpha_c$ or $\omega \leqslant \omega_c$ when $\alpha >\alpha_c$), 
this constraints simply tells us that the corner function $\widetilde{F}(\omega)$  is convex. 
The property $\tilde{f}''(\theta) \geqslant 0 $ for the generic corner function $\tilde{f}(\theta)$  has been derived from the strong subadditivity in \cite{Hirata:2006jx}
and, in the special case of the holographic corner function $\widetilde{F}(\omega)$ found in \cite{Drukker:1999zq}, the convexity is evident from its plot
(see the solid curve in Fig.\,\ref{fig:drukker_data}).
When the second function in the r.h.s. of (\ref{bdy Fcal generic}) is selected, the inequality (\ref{ssa config 1}) is saturated, as one can straightforwardly 
observe by using that $\tilde{\gamma} = \pi - (\omega +\gamma)$.

As for the constraint obtained from the configuration shown in the middle panel of Fig.\,\ref{fig:ssa}, we find it worth specialising
the inequality (\ref{ssa_2_step0}) to the holographic corner functions. By employing (\ref{bdy Fcal generic}), we find
\bea
\label{ssa2_holog}
& &
F_\alpha(\omega_1+\omega_2 + \gamma) - F_\alpha( \omega_1+ \gamma)
\,\geqslant
 \\
\rule{0pt}{.6cm}
& & \hspace{.5cm}
 \textrm{max}\,\Big\{  \widetilde{F}(\omega_1+\omega_2)\,, \, F_\alpha(\gamma) + F_\alpha(\omega_1+\omega_2 + \gamma)  \Big\}
 - \textrm{max}\,\Big\{  \widetilde{F}(\omega_1)\,, \, F_\alpha(\gamma) + F_\alpha(\omega_1 + \gamma)  \Big\}
 \nonumber
\eea
For the configurations such that in both the maximisations occurring in the r.h.s. of (\ref{ssa2_holog})
the second function is selected, $F_\alpha(\gamma) $ simplifies in the r.h.s. and this inequality becomes a trivial identity.
As for other configurations, the inequality (\ref{ssa2_holog}) is a non trivial inequality. 
We checked numerically for some cases that it is verified but, unfortunately, we do not have a general proof. 

The last  constraint to check is (\ref{ssa config 3}).
Specifying this  inequality for the holographic corner function (\ref{bdy Fcal generic}), we obtain
\be
F_{\alpha}(\gamma+\omega) + F_{\alpha}(\gamma) 
\,\leqslant \,
 \textrm{max}\, \Big\{  \widetilde{F}(\omega)\,, \, F_\alpha(\gamma) + F_\alpha(\omega + \gamma)  \Big\}
\ee
It is straightforward to observe that this inequality is trivially true.

\bibliographystyle{nb}
%\bibliography{Ref_latitude}

%%%%%%%%%%%%%%%%%%%%%%%%%%%%%%%%%%%%%%%%%%%%%%

\end{document}